a.dem
\documentclass{aa}  

\usepackage{graphicx}
\usepackage{txfonts}
\usepackage{color}

\newcommand{\red}{\textcolor{red}}

\begin{document}

\title{LOFAR observations of galaxy clusters in HETDEX}

   \subtitle{Extraction and self-calibration of individual LOFAR targets}

   \author{R.~J.~van~Weeren \inst{1}
           \and T.~W.~Shimwell \inst{2,1}
           \and A.~Botteon \inst{1}
           \and G.~Brunetti \inst{3}
           \and M.~Br\"uggen \inst{4}
           \and J.~M.~Boxelaar \inst{1}
           \and R.~Cassano \inst{3}
           \and G.~Di~Gennaro \inst{1}
           \and F.~Andrade-Santos \inst{5}
           \and E.~Bonnassieux  \inst{6}
           \and A.~Bonafede \inst{6,3}
           \and V.~Cuciti \inst{4}
           \and D.~Dallacasa  \inst{3,6}
           \and F.~de~Gasperin \inst{4}
           \and F.~Gastaldello \inst{7}
           \and M.~J.~Hardcastle \inst{8}
           \and M.~Hoeft \inst{9}
           \and R~.P.~Kraft  \inst{5}
           \and S.~Mandal \inst{1}
           \and M.~Rossetti \inst{7}
           \and H.~J.~A.~R\"ottgering \inst{1}
           \and C.~Tasse \inst{10,11,12}
           \and A.~G.~Wilber \inst{13,4}
          }

   \institute{Leiden Observatory, Leiden University, PO Box 9513, 2300 RA Leiden, The Netherlands\\
              \email{rvweeren@strw.leidenuniv.nl}
   \and
   ASTRON, The Netherlands Institute for Radio Astronomy, Postbus 2, 7990 AA Dwingeloo, The Netherlands
   \and
   INAF-Istituto   di   Radioastronomia,   Via   Gobetti   101, I-40129 Bologna, Italy
   \and   
   University  of  Hamburg,  Hamburger  Sternwarte,  Gojenbergsweg 112, 21029 Hamburg, Germany
   \and
   Harvard-Smithsonian Center for Astrophysics, 60 Garden Street, Cambridge, MA 02138
   \and
   Dipartimento di Fisica e Astronomia, Universit\`a di Bologna, via P. Gobetti 93/2, I-40129 Bologna, Italy   
   \and
   INAF-IASF Milano, Via A. Corti 12, I-20133, Milano, Italy
   \and
   Centre for Astrophysics Research, University of Hertfordshire, College Lane, Hatfield, AL10 9AB, UK
   \and   
   Th\"uringer Landessternwarte, Sternwarte 5, 07778 Tautenburg, Germany
   \and 
   GEPI, Observatoire de Paris, CNRS, Universit\'e Paris Diderot, 5 place Jules Janssen, 92190 Meudon, France
   \and
   Centre for Radio Astronomy Techniques and Technologies, Department of Physics and Electronics, Rhodes University, Grahamstown 6140, South Africa
   \and
   USN, Observatoire de Paris, CNRS, PSL, UO, Nan\c{c}ay, France
   \and
    Curtin Institute of Radio Astronomy, 1 Turner Avenue, Technology Park, Bentley WA 6102, Australia\\
             }

   \date{}

 
\abstract{Diffuse cluster radio sources, in the form of radio halos and relics, reveal the presence of cosmic rays and magnetic fields in the intracluster medium (ICM).  These cosmic rays are thought to be (re-)accelerated through ICM turbulence and shock waves generated by cluster merger events. Here we characterize the presence of diffuse radio emission in known galaxy clusters in the HETDEX Spring Field, covering 424~deg$^2$. For this, we developed a method to extract individual targets  from LOFAR observations processed with the LoTSS DDF-pipeline. This procedure enables improved calibration and joint imaging and deconvolution of multiple pointings of selected targets. The calibration strategy can also be used for LOFAR Low-Band Antenna (LBA) and international-baseline observations.

The fraction of Planck PSZ2 clusters with any diffuse radio emission apparently associated with the ICM is $73\pm17\%$.
We detect a total of 10 radio halos and 12 candidate halos in the  HETDEX Spring Field. Five clusters host radio relics. The fraction of radio halos in Planck PSZ2 clusters is $31\pm11\%$, and $62\pm15\%$ when including the candidate radio halos. Based on these numbers, we expect that there will be at least $183 \pm 65$ radio halos found in the LoTSS survey in PSZ2 clusters, in agreement with predictions. The integrated flux densities for the radio halos were computed by fitting exponential models to the radio images. From these flux densities, we determine the cluster mass (M$_{500}$) and Compton Y parameter (Y$_{500}$) 150~MHz radio power (P$_{\rm{150~MHz}}$) scaling relations for Planck PSZ2-detected radio halos. We find that the slopes of these relations are steeper than those determined from the 1.4~GHz radio powers. However, considering the uncertainties this is not a statistically significant result.

}


   \maketitle

\section{Introduction}

Radio observations have revealed the presence of Mpc-scale radio sources associated with intracluster medium (ICM)  in a growing number of galaxy clusters. This indicates that the ICM is filled with cosmic ray (CR) electrons and magnetic fields. Diffuse cluster radio sources are commonly divided into radio relics (radio shocks), giant halos, and mini-halos \citep[for reviews see][]{2012A&ARv..20...54F,2014IJMPD..2330007B,2019SSRv..215...16V}. Importantly, the short lifetime ($\sim10^{7-8}$~yrs) of the CR electrons implies that some form of in-situ particle (re-)acceleration is required to explain the Mpc extent of these sources.

\emph{Giant radio halos} are Mpc-size sources that approximately follow the X-ray emission from the thermal ICM. They are predominantly found in merging clusters \citep{2010ApJ...721L..82C}. The radio power of giant halos correlates with cluster mass \citep[e.g.,][]{2013ApJ...777..141C} and often used mass proxies are the cluster's X-ray luminosity ($L_{\rm{X}}$) or integrated Compton Y parameter. The upper limits derived for clusters that are dynamically relaxed are underluminous with respect to these correlations \citep[e.g.,][]{2007ApJ...670L...5B,2013ApJ...777..141C}.  The  fraction of clusters with radio halos is about 30\% \citep{2008A&A...484..327V,2013A&A...557A..99K,2015A&A...579A..92K} for $L_{\rm{X, 0.1-2.4~keV}} > 5\times 10^{44}$~erg~s$^{-1}$ clusters in the range $0.2<z<0.4$. For the most massive ($\sim10^{15}$~M$_{\odot}$) clusters the occurrence fraction is as high as $\sim 80\%$, and there is evidence that this fraction decreases for lower mass clusters \citep{2015A&A...580A..97C}.

Two main classes of models have been proposed for the origin of radio halos. In the turbulent  re-acceleration model, particles are re-accelerated by merger induced magneto-hydrodynamical turbulence \citep[][]{2001MNRAS.320..365B, 2001ApJ...557..560P,2007MNRAS.378..245B,2015ApJ...800...60M}. In the hadronic model, the radio emission is produced by secondary electrons that arise from hadronic collisions \citep[e.g.,][]{1980ApJ...239L..93D, 1999APh....12..169B, 2000A&A...362..151D}. Most observational evidence nowadays is in favor of the turbulent re-acceleration model. This evidence includes the discovery of radio halos with ultra-steep radio spectra \citep[USSRH; e.g.,][]{2008Natur.455..944B} and the non-detection of gamma ray emission from the Coma cluster at the level that would be necessary to generate the observed radio emission  \citep{2012MNRAS.426..956B, 2013A&A...558A..52B,2016ApJ...819..149A,2017MNRAS.472.1506B}. Models for radio halos invoking a combination of turbulent re-acceleration and generation of secondary particles that are consistent with gamma ray limits have also been proposed \citep[e.g.,][]{2011MNRAS.410..127B,2017MNRAS.465.4800P,2017MNRAS.472.1506B}.

\emph{Radio mini-halos} are smaller sized halos ($\sim$200--500~kpc) that are exclusively found in relaxed cool-core clusters. Recently, \cite{2017ApJ...841...71G} found that mini-halos are rather common in massive cool-core clusters; about 80\% of such clusters host them.
The radio emission from mini-halos surrounds the central AGN associated with the brightest cluster galaxy (BCG). Similar to giant radio halos some form of in-situ acceleration is required to power them \citep[e.g.,][]{2004A&A...417....1G,2014ApJ...781....9G}. Mini-halos have been explained by hadronic scenarios \citep[e.g.,][]{2004A&A...413...17P,2007ApJ...663L..61F,2010ApJ...722..737K,2013MNRAS.428..599F}, or by turbulent re-acceleration induced by gas sloshing motions \citep{2008ApJ...675L...9M,2013ApJ...762...78Z}. 

\emph{Radio relics} are polarized, elongated sources found in galaxy cluster outskirts \citep[e.g.,][]{1998A&A...332..395E}. They have sizes that can extend up to about 2~Mpc. High-resolution observations show that radio relics often have filamentary morphologies \citep[e.g.][]{2006Sci...314..791B,2018ApJ...865...24D,2020A&A...636A..30R,2018ApJ...852...65R}. These sources trace ICM shock waves with low to moderate Mach numbers \citep[e.g.,][]{2010ApJ...715.1143F, 2013PASJ...65...16A,2013MNRAS.433.1701O,2015MNRAS.449.1486S}. The physical mechanisms by which particles are (re-)accelerated at shocks are still being debated. One possibility to accelerate particles at shocks is via the diffusive shock acceleration mechanism \citep[DSA; e.g.,][]{1987PhR...154....1B,1991SSRv...58..259J, 2001RPPh...64..429M}. However, this mechanism is thought to be rather inefficient for weak ICM shocks if particles are accelerated from the thermal pool and it fails to explain the observed radio power and spectral indices in a number of cases \citep[e.g.,][]{2013MNRAS.435.1061P,2014MNRAS.437.2291V,2016MNRAS.459...70V,2016ApJ...818..204V,2020A&A...634A..64B}. It has therefore been proposed that some form of re-acceleration of pre-existing fossil CR electrons takes place \citep[e.g.,][]{2005ApJ...627..733M,2008A&A...486..347G,2011ApJ...734...18K,2012ApJ...756...97K,2016ApJ...818..204V,2016MNRAS.460L..84B}. Observations do provide support for this model, where the fossil CRs originate from the tails and lobes of radio galaxies \citep{2014ApJ...785....1B,2015MNRAS.449.1486S,2017NatAs...1E...5V}. It should be noted, however, that for some relics DSA of thermal pool electrons seems to be sufficient to explain their luminosity \citep[e.g.,][]{2016MNRAS.463.1534B, 2020MNRAS.496L..48L}.

Low-frequency observations provide important information about particle acceleration processes. The turbulent re-acceleration model for radio halos predicts that the occurrence rate of halos should be higher at low frequencies \citep{2010AA...517A..10C,2012AA...548A.100C}. 
Radiative losses of CR electrons limit the acceleration by turbulence via second order Fermi mechanisms, causing a break in the energy spectrum of these electrons \citep{2004JKAS...37..589C,2005MNRAS.357.1313C,2006MNRAS.369.1577C,2008Natur.455..944B}. This in turn results in a synchrotron spectrum that becomes steeper in the situations where the amount of turbulent energy is smaller (less powerful mergers) or the energy losses are stronger (higher redshift).   

Another important role of low-frequency observations is to probe the connection between re-acceleration processes and fossil radio plasma. Because of synchrotron and inverse Compton (IC) losses, as relativistic electrons age they emit almost exclusively at low frequencies. These seed fossil CR particles are a critical ingredient in both shock and turbulent re-acceleration models. Possible examples of  revived fossil plasma, by compression \citep{2001A&A...366...26E,2002MNRAS.331.1011E} or other mechanisms \citep[e.g.,][]{2005ApJ...627..733M}, have been been detected in some clusters \citep[e.g.,][]{2001AJ....122.1172S,2017NatAs...1E...5V,2017SciA....3E1634D,2020A&A...634A...4M}. However, this is probably just the tip of the iceberg.

Finally , recent LOFAR observations have entered uncharted territories, discovering the existence of radio bridges that connect pairs of massive and pre-merging clusters \citep{2019Sci...364..981G,2020MNRAS.499L..11B}. One possibility is that these bridges originate from second order Fermi acceleration mechanisms powered by turbulence filling these vast regions \citep{2020PhRvL.124e1101B}.

The LOFAR Two-metre Sky Survey \citep[LoTSS][]{2017A&A...598A.104S} is a deep 120--168~MHz survey that will cover the entire northern sky when completed. This survey is carried out with the High Band Antenna (HBA) stations of LOFAR \citep{2013A&A...556A...2V}. The nominal sensitivity of this survey is 0.1~mJy~beam$^{-1}$ at a resolution of 6\arcsec. The first data release (DR1) covers  424 deg$^2$ in the region of the HETDEX Spring field \citep{2019A&A...622A...1S}.  Given its unprecedented survey depth at low frequencies, resolution, and sky coverage, LoTSS will play an important role in determining the statistics of diffuse cluster sources. In this paper we present the first results on a full sample of galaxy clusters in the LoTSS DR1-area using improved calibration techniques. This is required to properly study extended low-surface brightness cluster sources. The outline of this paper is as follows. In Sect.~\ref{sec:observationsandselfcal} we describe the procedure we developed to extract and re-calibrate targets of interest from the LoTSS data products. The sample selection is described in Sect.~\ref{sec:sample}. The results are presented in Sect.~\ref{sec:results} and we end with a discussion and conclusions in Sects.~\ref{sec:discussions} and~\ref{sec:conclusions}.

Throughout this paper we assume a $\Lambda$CDM cosmology with $H_{0} = 70$~km~s$^{-1}$~Mpc$^{-1}$, $\Omega_{m} = 0.3$, and $\Omega_{\Lambda} = 0.7$. All images are in the J2000 coordinate system.

\section{Extraction and re-calibration}  
\label{sec:observationsandselfcal}   
Here we utilize LOFAR observations that were taken as part of the LoTSS survey. These are typically 8~hr observations. The pointing centers of these observations are placed with the aim of obtaining close to uniform sensitivity coverage of the northern sky. The LoTSS survey design and the observations are discussed in detail in \cite{2017A&A...598A.104S,2019A&A...622A...1S}.

\subsection{Extraction}
\label{sec:extraction}
The LoTSS processing pipeline ({\tt DDF-PIPELINE})  delivers images of the full field of view (FoV) of the Dutch LOFAR HBA stations \citep[][]{2014arXiv1410.8706T,2014A&A...566A.127T,2019A&A...622A...1S}. These images have a resolution of 6\arcsec and an r.m.s. noise level of the order of 100~$\mu$Jy~beam$^{-1}$. This is achieved by correcting for the direction dependent effects (DDE) in the LOFAR data (due to the ionosphere and imperfect station beam models). For LoTSS, DDE corrections are applied towards 45~directions (facets). The {\tt DDF-PIPELINE} is optimized to create images of LOFAR's full FoV and carry out survey science. The latest version 2 of the {\tt DDF-PIPELINE}, which we employ in this work, is described in \cite{tasse}. Version 2 of the pipeline is a major improvement compared to version 1 which was used for public Data Release 1. The images from DR1 are generally not suitable to study extended low-surface brightness cluster sources \citep[see Section 3.7 in][]{2019A&A...622A...1S}.

The individual images produced by the {\tt DDF-PIPELINE} \footnote{\url{https://github.com/mhardcastle/ddf-pipeline}} are very large, $20,000 \times 20,000$ pixels with a pixel size of 1.5\arcsec. This makes re-imaging with different settings, e.g., uv-ranges, weighting schemes, and deconvolution algorithms, expensive. The tessellation of the sky into 45 calibration facets is done in a fully automated way. The DDE calibration takes all sources in a facet into account, assuming there are no DDE inside a facet. This means that for certain specific targets of interest the facet layout is not optimal. Experience with various faceting schemes has shown that it is often possible to further improve the quality of the DDE calibration for a specific target (sometimes at the expense of other sources). 

To allow flexible re-imaging and optimize calibration towards targets of interest, we extended the {\tt DDF-PIPELINE} with an optional step. In this step all sources are subtracted, apart from those in a specific user defined region, from the visibility data and after applying their DDE calibration solutions. The model visibility prediction is done  with the {\tt DDFacet} imager \citep{2018A&A...611A..87T}.
An important requirement for extraction is that region of interest needs to be quite small, roughly similar to the size of the original facet, but ideally smaller. This is directly related to the fact that we want to improve the DDE calibration, which was limited by the original facet size and the ``incorrect'' assumption that the DDE correction is constant across that entire facet. This requirement competes with another requirement that there is sufficient flux density available for calibration. A smaller extraction region will have less flux density available for the calibration.

\subsubsection{Phase shifting and averaging}
After subtracting all sources (clean components) from the uv-data, apart from those in the region of interest, we phase shift the uv-data. The new phase center is placed at the center of the region of interest, i.e., the region where the sources were kept. In practice we use {\tt DS9} \citep{2003ASPC..295..489J} region files which can be easily generated. 
After phase shifting, the data are averaged in time and frequency. The extracted region has a small angular extent and thus bandwidth and time smearing are not an issue. By default, the data are averaged to 16~s and 0.39~MHz. With these  averaging parameters, the size of the dataset is reduced by a factor of~8. The visibility data are compressed using {\tt Dysco} to further reduce data size by about a factor of four \citep{2016A&A...595A..99O}. Additional averaging is often allowed from the point of bandwidth and time smearing, but we found that more averaging can hinder ionospheric calibration. Optionally, additional radio frequency interference (RFI) flagging is carried out with {\tt AOFlagger} on the output data  \citep{2010ascl.soft10017O, 2010MNRAS.405..155O,2012A&A...539A..95O}. 

When the data is shifted to the new phase center we also correct for the LOFAR station beam response in this direction. Additionally, the visibility weights set by the {\tt DDF-pipeline} \citep[for details see][]{2018A&A...615A..66B}  are updated. They are multiplied by a factor inversely proportional to the station beam response. In this way, we can optimally combine observations from multiple pointing centers with joint imaging and deconvolution. Combining visibility data in this way is allowed because the beam correction is close to constant across the small region that is extracted\footnote{If that approximation does not hold we use the Image Domain Gridding algorithm which  corrects each phase rotated dataset with the correct beam response \citep{2018A&A...616A..27V}.}.

\subsection{Self-calibration}
\label{sec:selfcalibration}
\begin{figure}
\centering
\includegraphics[width=1.0\columnwidth]{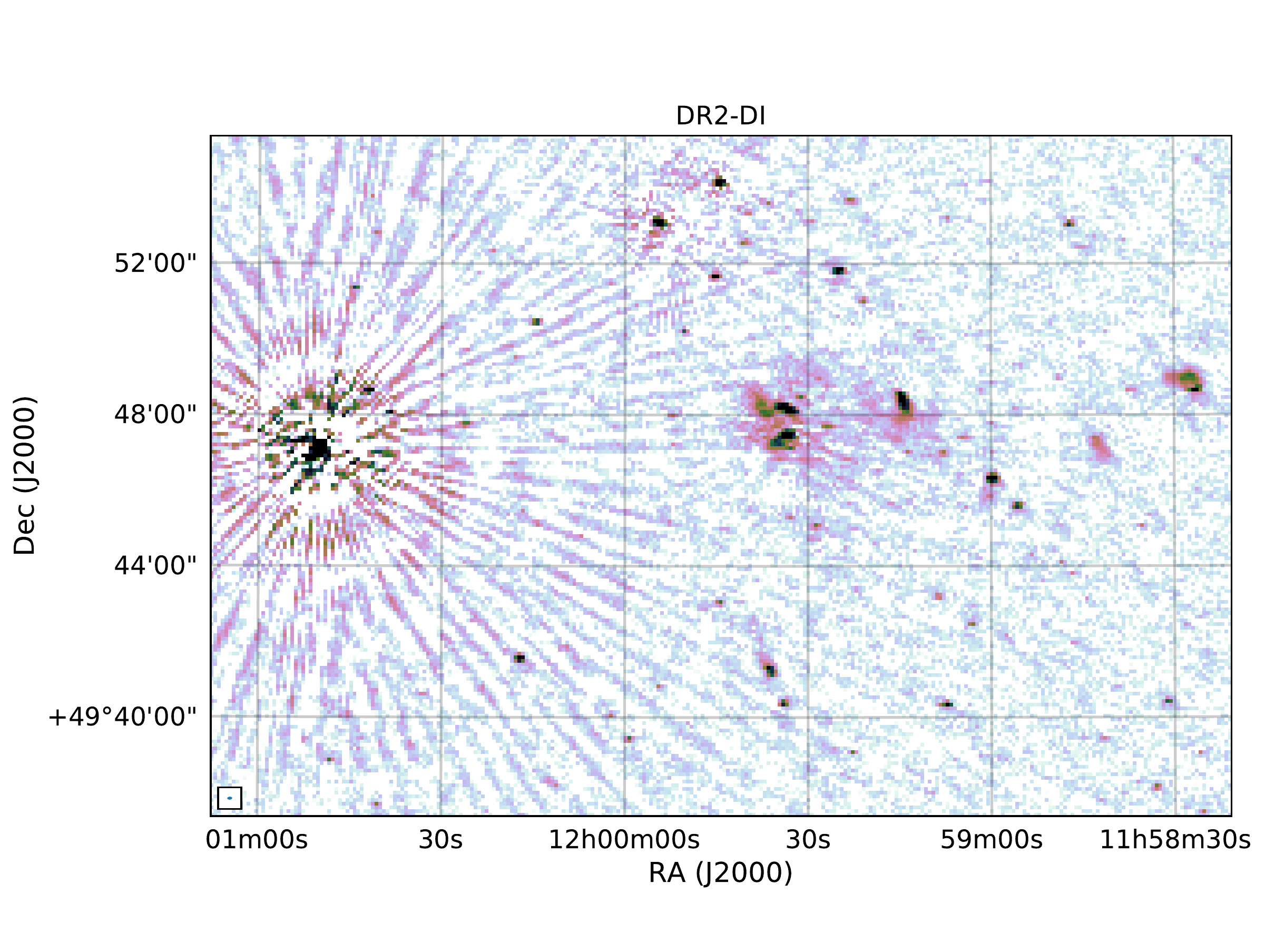}
\includegraphics[width=1.0\columnwidth]{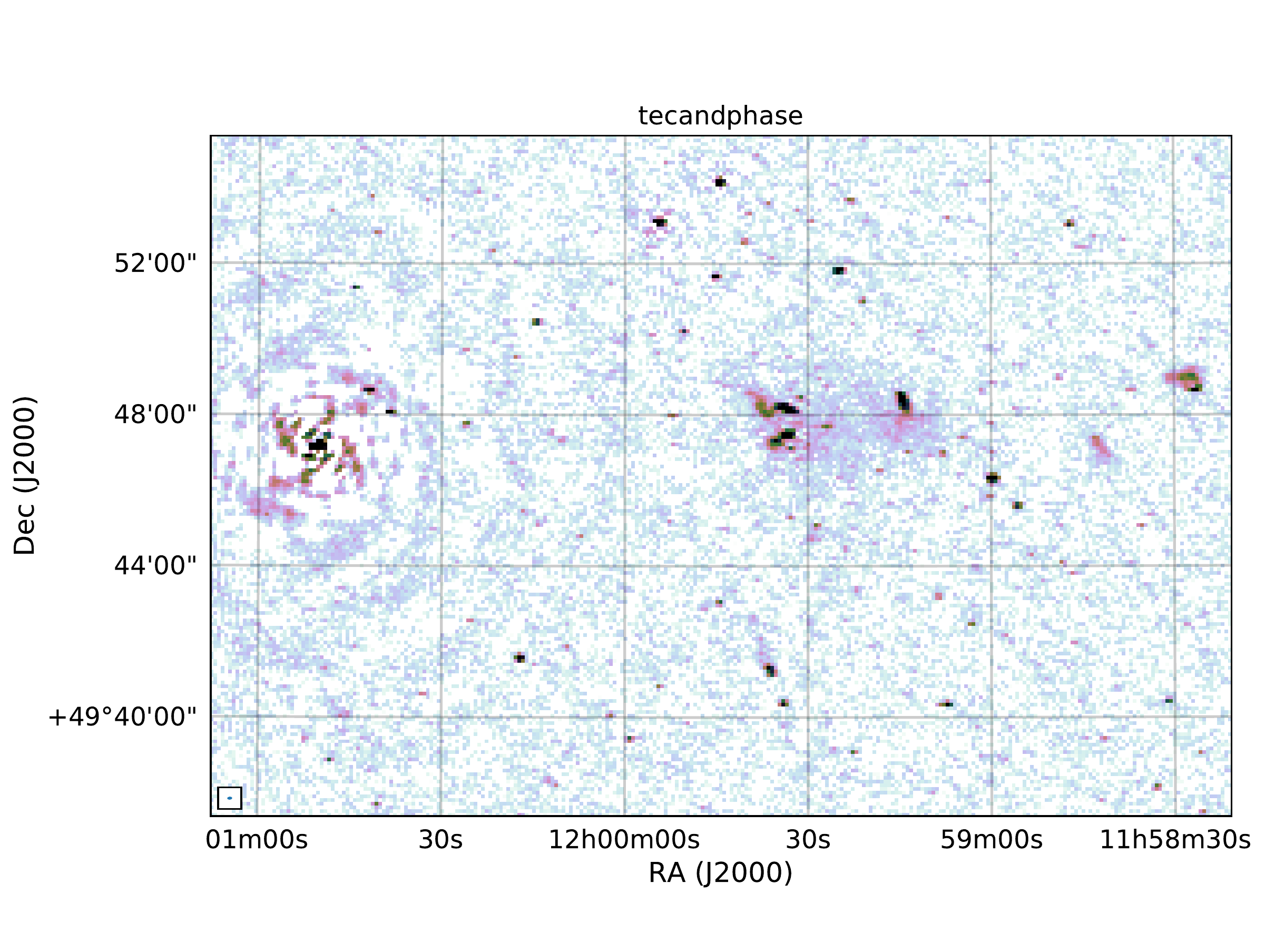}
\includegraphics[width=1.0\columnwidth]{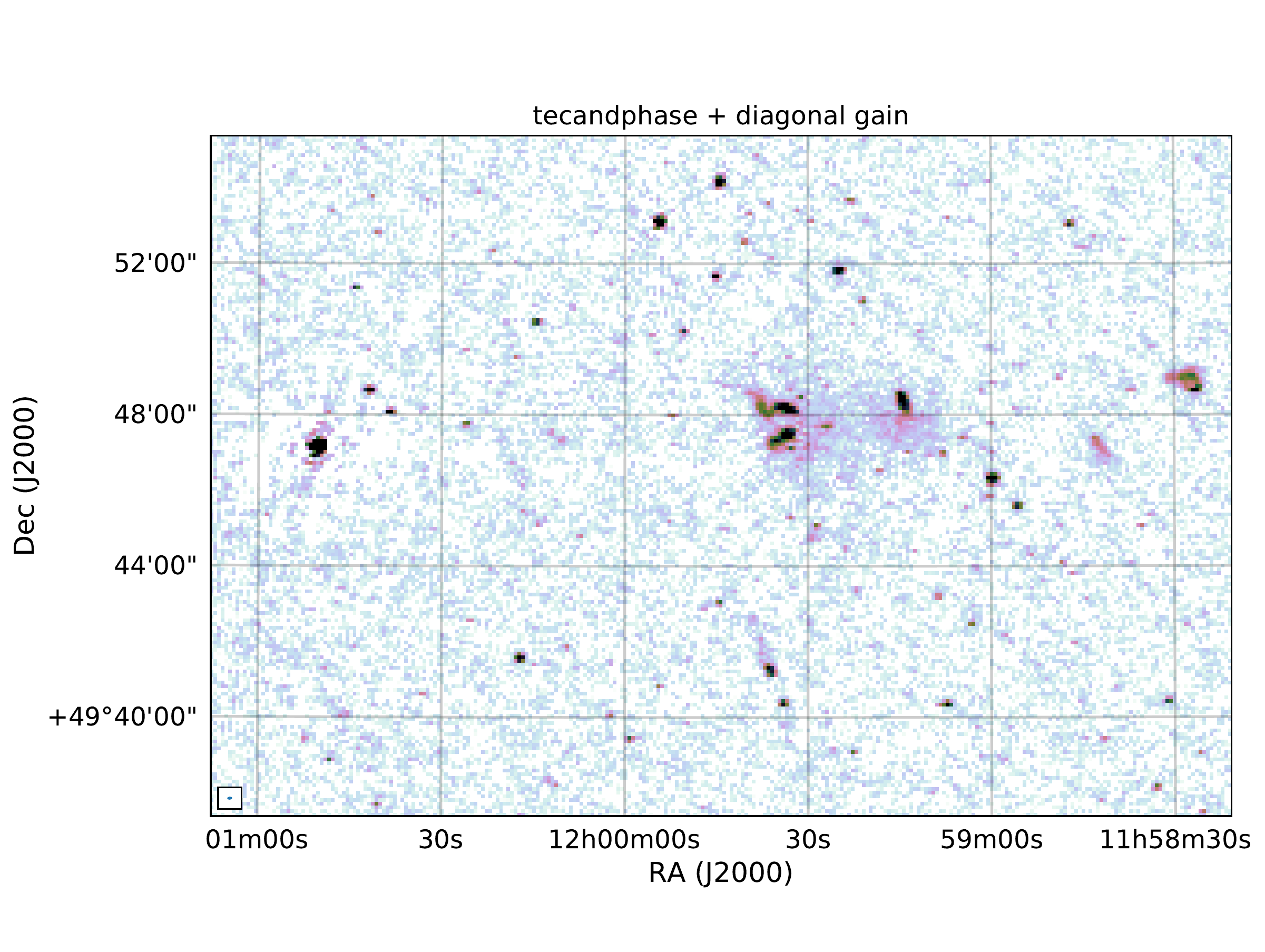}
     \caption{Images showing the subsequent improvement during self-calibration on the extracted data of the cluster Abell\,1430. The top panel shows the starting point of this process: the {\tt DDF-pipeline} (v2) direction independent (DI) calibration. The middle panel displays the results after {\tt DPPP} `tecandphase' calibration. The bottom panel shows the final image after {\tt DPPP} diagonal gain calibration.}
     \label{fig:selfcalsteps}
\end{figure}

The next step in the ``extraction'' process for a target of interest is to self-calibrate the data. The starting point of this step are one or more phase shifted and averaged datasets for the target of interest, each one corresponding to a different observation with a potentially different pointing center. The direction independent (DI) full Jones calibrations are carried over from the {\tt DDF-pipeline}. We do not carry over the DDE solutions to avoid the issue with the ``negative haloes'' described in \cite{tasse}.

The self-calibration steps consist of three rounds of `tecandphase' calibration with {\tt DPPP} \citep{2018ascl.soft04003V}. This is followed by several rounds of diagonal (i.e., XX and YY) gain calibration using a longer solution interval. The shorter timescale `tecandphase' solutions are pre-applied when solving for the diagonal gains. This scheme somewhat mimics the facet-calibration scheme \citep{2016ApJS..223....2V} and the {\tt DDF-pipeline} which also pre-applies fast phase/TEC solutions before solving for slow gain solutions. The diagonal gain solutions are filtered for outliers with {\tt LoSoTo} \citep{ 2019A&A...622A...5D}.
The solution intervals are automatically determined based on the amount of apparent compact source flux in the extracted region. Solutions intervals for the `tecandphase' calibration are between 16~s and 48~s. Solution intervals for the diagonal gains are between 16~min and 48~min. All solution intervals are set per observation since the apparent flux in the target of interest region can differ.
Solution intervals along the frequency axis for the diagonal gains are between 2~MHz and 6~MHz. For HBA data, a minimum of 0.3~Jy compact (apparent) source flux is needed for the self-calibration steps to converge well. If needed, all solution intervals can be manually controlled by the user. An overview of the parameters used is given in Table~\ref{tab:selfcalparameters}.

The imaging is done with {\tt WSClean} \citep{2014MNRAS.444..606O} with wideband joint deconvolution mode (`channelsout' 6 default), optionally with multi-scale clean \citep{2017MNRAS.471..301O}. The {\tt DDFacet} imager \citep{2018A&A...611A..87T} can also be used instead of {\tt WSClean}. As default Briggs weighting $-0.5$ is used. The imaging includes automatic clean masking from the {\tt DDF-pipeline}, with a default $5\sigma$ threshold. Baseline based averaging is also employed when imaging for performance.  An inner uv-cut of $80\lambda$ is used in the imaging. For the calibration, a default inner uv-cut of $350\lambda$ (corresponding to 0.65\degr) is employed. This value is a compromise between the number of baselines in the calibration and the increased difficulty of modelling very extended structures.

In Figure~\ref{fig:selfcalsteps} we show an example of the self-calibration process. The top panel shows the first image before any self-calibration. Note that here  two different observations (L254483 and L719874), with different pointing centers,  are jointly imaged and deconvolved. The only calibration that is applied here is the DI calibration from the {\tt DDF-pipeline}. In the middle panel we show the result after three rounds of `tecandphase' self-calibration. The radial patterns, caused by the ionosphere disappear after this calibration. In the bottom panel we show the results after additional diagonal gain calibration. The diagonal gain calibration corrects for the imperfect knowledge of the station beam response or other slowly varying gain errors. 

\begin{figure*}
\centering
\includegraphics[width=1.0\columnwidth]{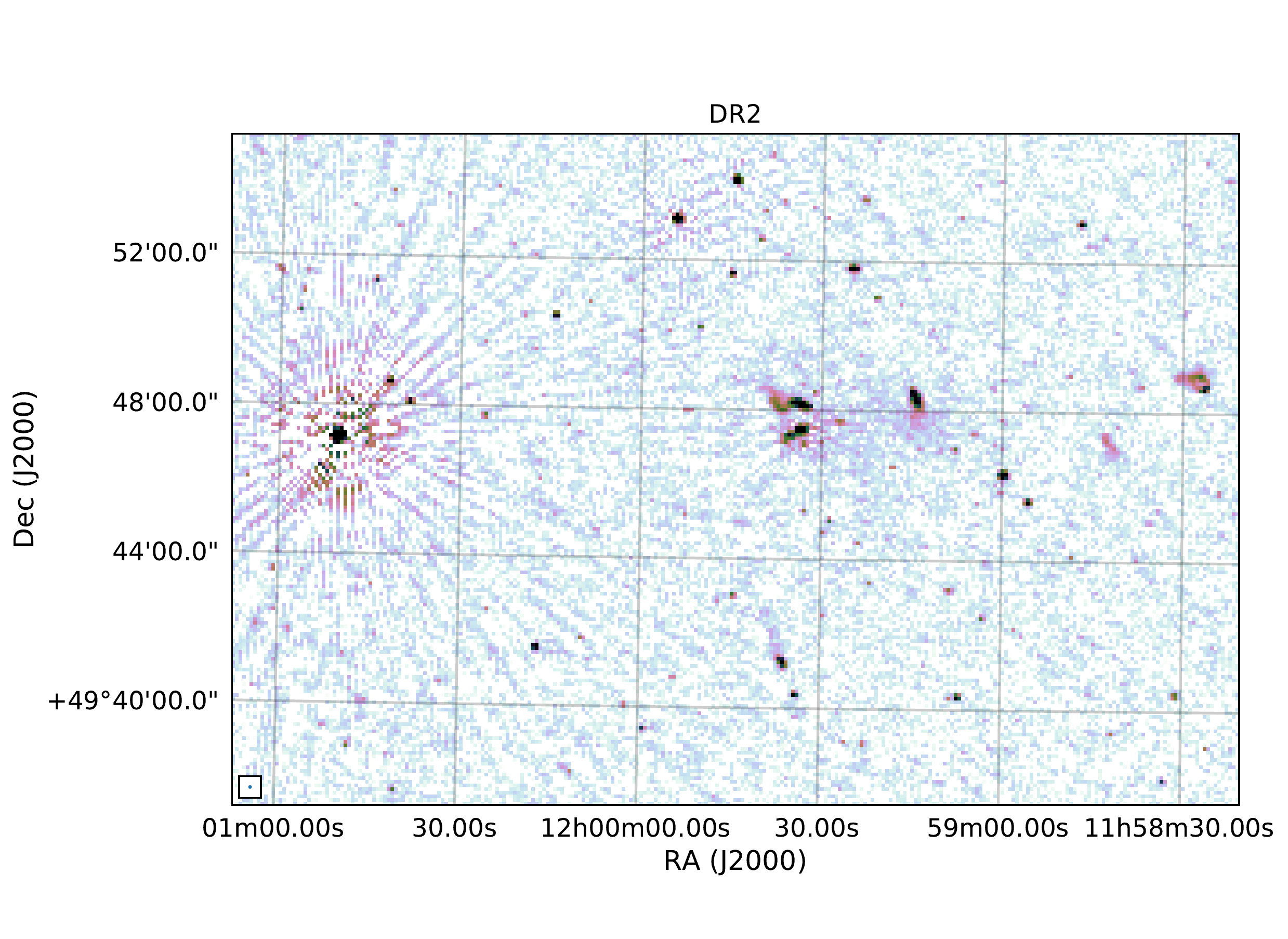}
\includegraphics[width=1.0\columnwidth]{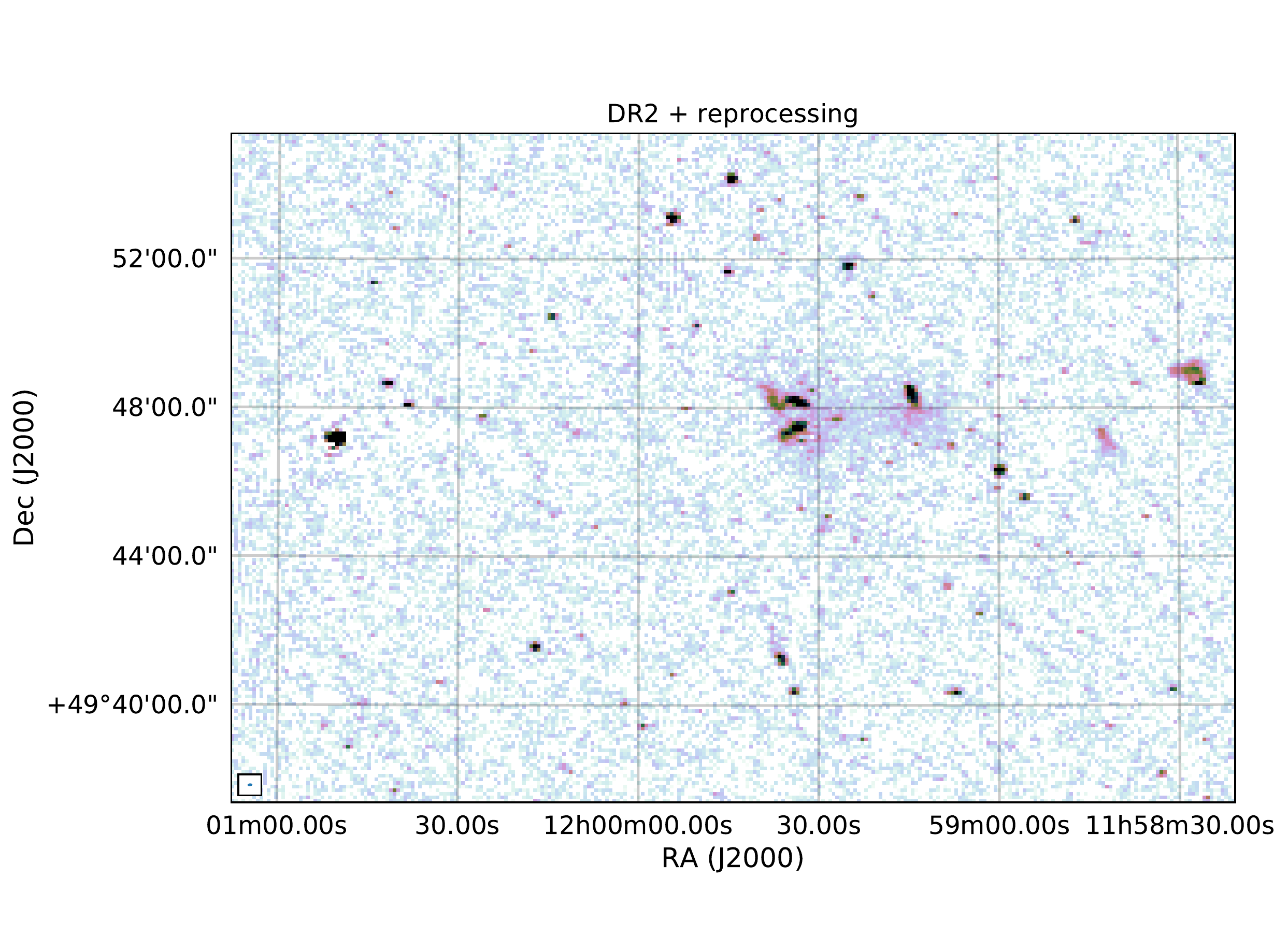}
\includegraphics[width=1.0\columnwidth]{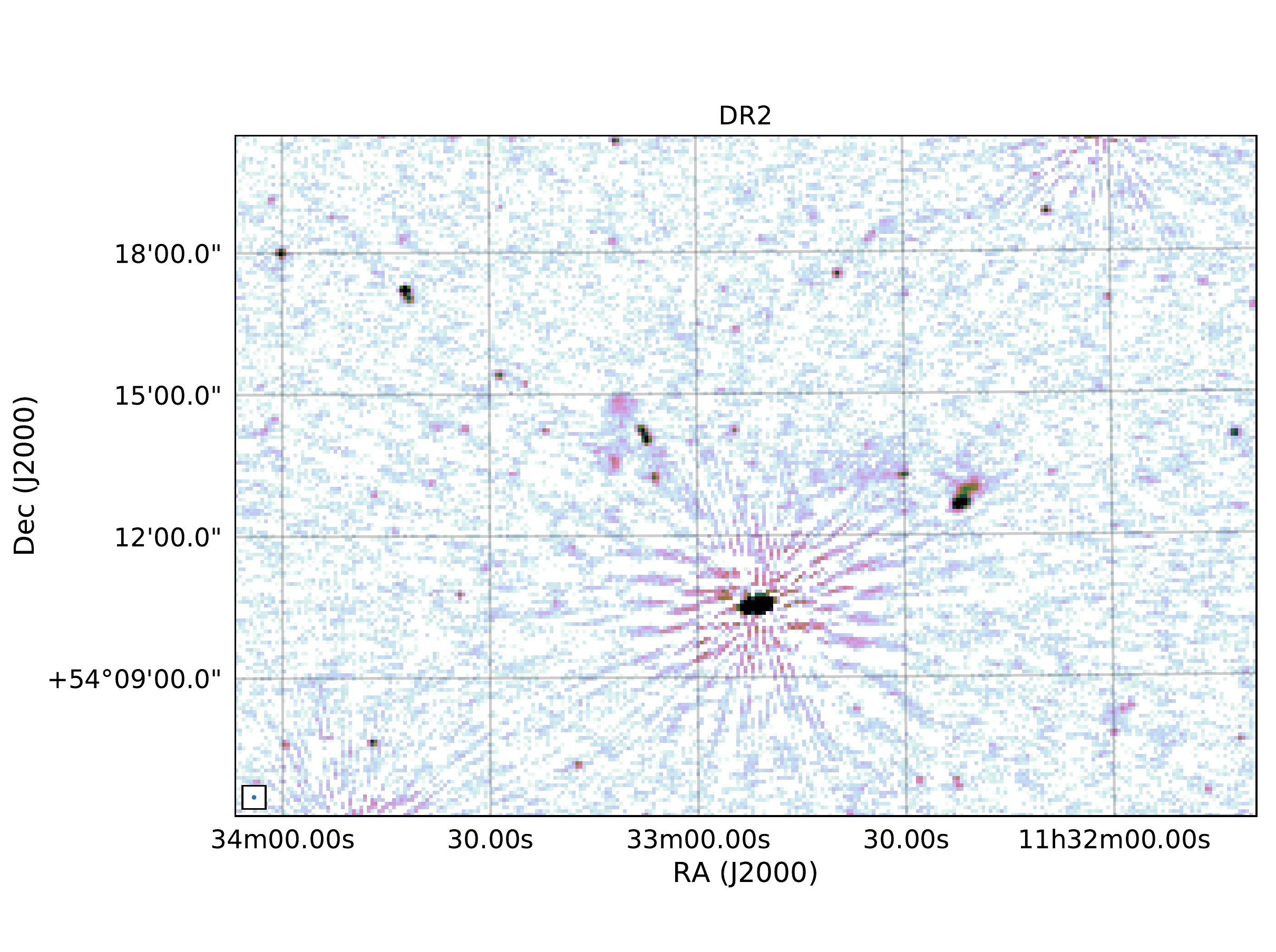}
\includegraphics[width=1.0\columnwidth]{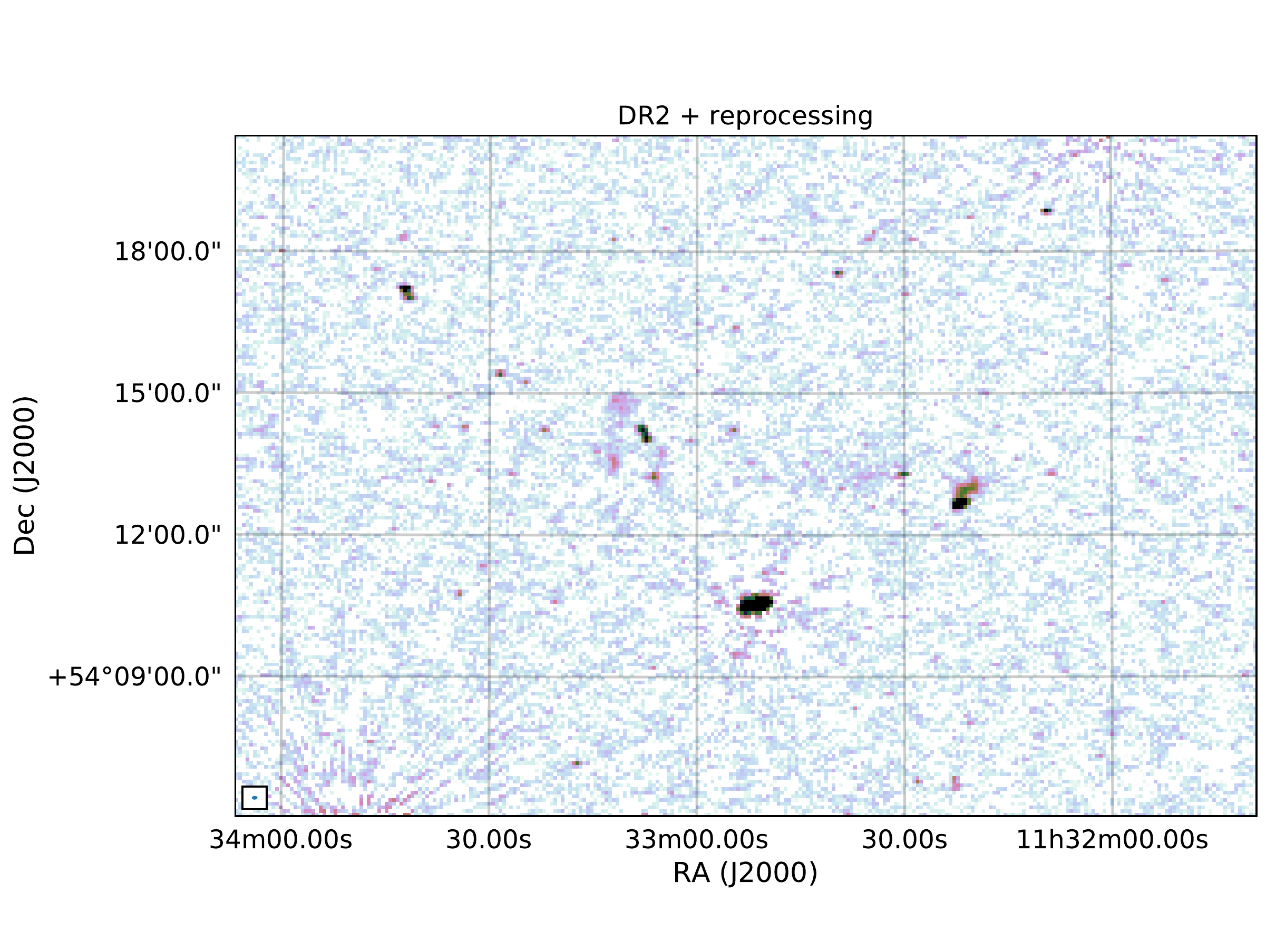}
     \caption{Comparison between the LoTSS DR2 mosaics and the extracted and re-calibrated images for the clusters Abell\,1430 (top panels) and Abell\,1294 (bottom panels). For both clusters optimizing the calibration towards a nearby bright compact source improved the image quality. These errors in the LoTSS DR2 mosaics were caused by the spatially varying ionosphere. Note that the DR2 pipeline could have achieved more or less the same results with a different facet layout (but then at the cost of reducing image quality in other directions).}
     \label{fig:A1430comp}
\end{figure*}

In Figure~\ref{fig:A1430comp} we display two comparisons between the default LoTSS mosaics and  newly extracted and re-calibrated targets (in this case the clusters Abell\,1430 and~1294). As can be seen, the calibration for the targets of interest has improved. The improvement is the result of shrinking the size of the calibration region so that a nearby bright compact source is better calibrated. 
The extraction and self-calibration scheme has already been successfully applied in various recent works \citep[e.g.,][]{2020MNRAS.499L..11B,2020ApJ...897...93B,2020A&A...634A...4M,2019MNRAS.488.3416H,2019ApJ...881L..18C,2019NatAs...3..838G}. 

\subsubsection{Other usage}
\begin{figure*}
\centering
\includegraphics[width=0.285\paperwidth]{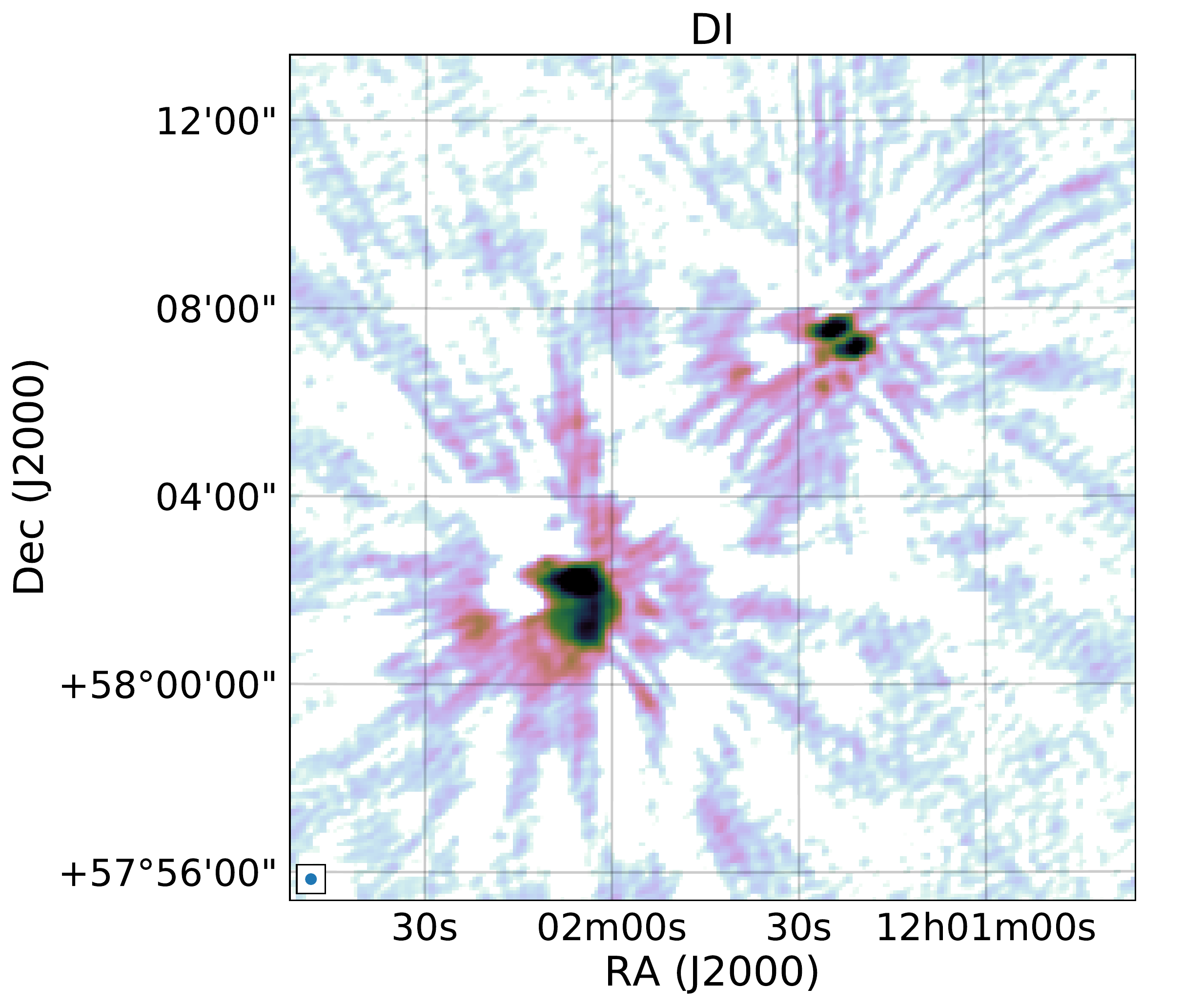}
\includegraphics[width=0.285\paperwidth]{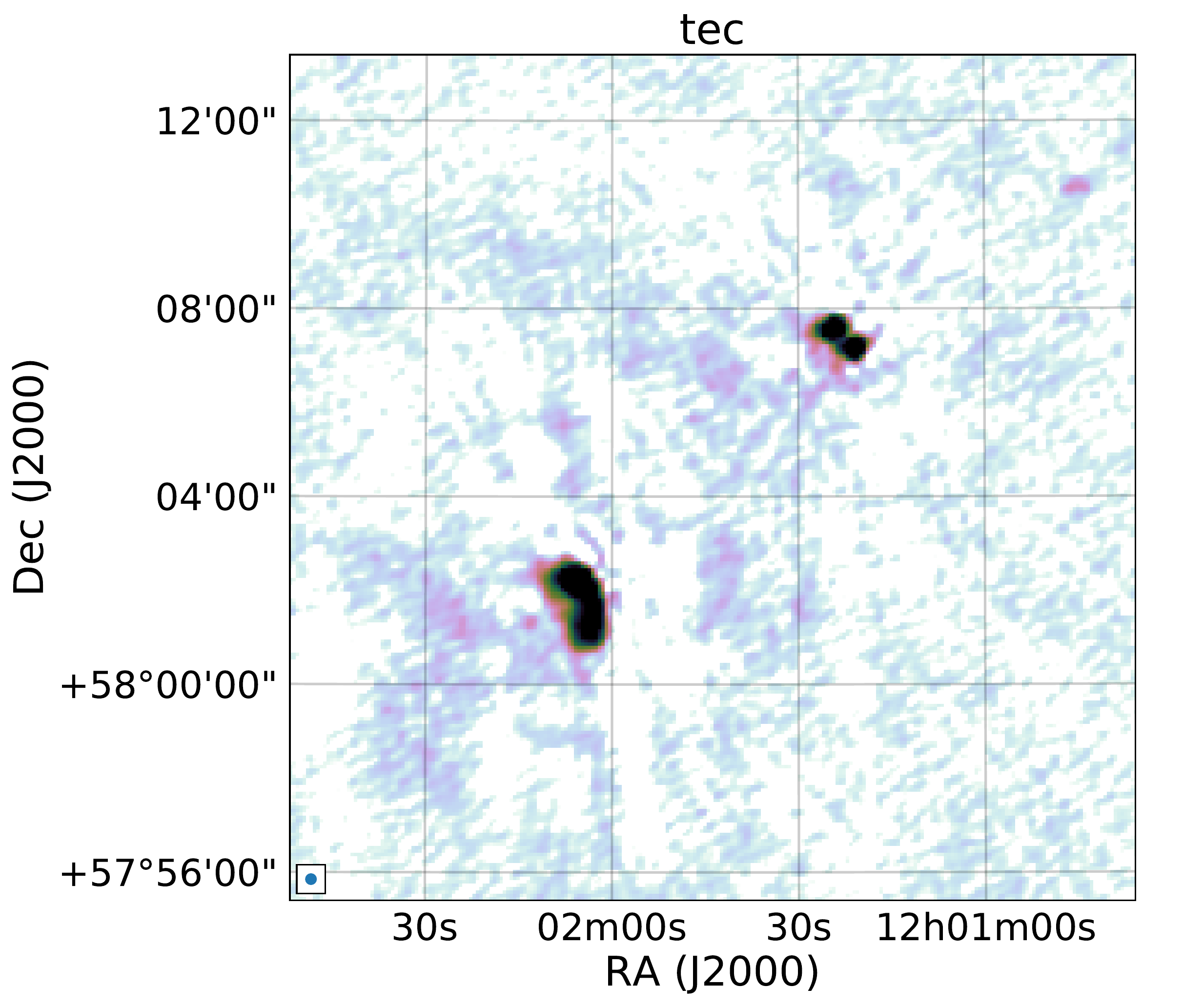}
\includegraphics[width=0.285\paperwidth]{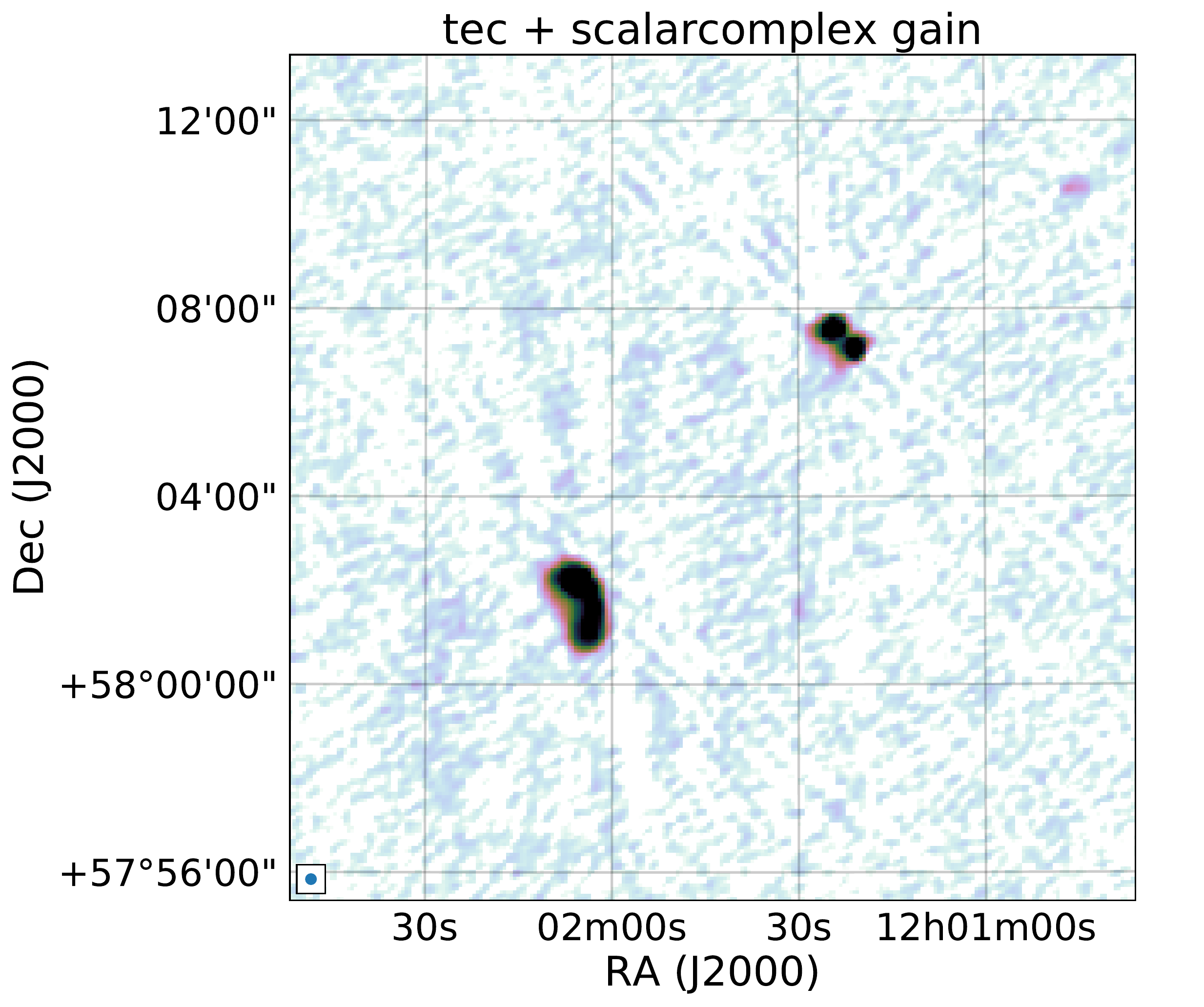}
\includegraphics[width=0.285\paperwidth]{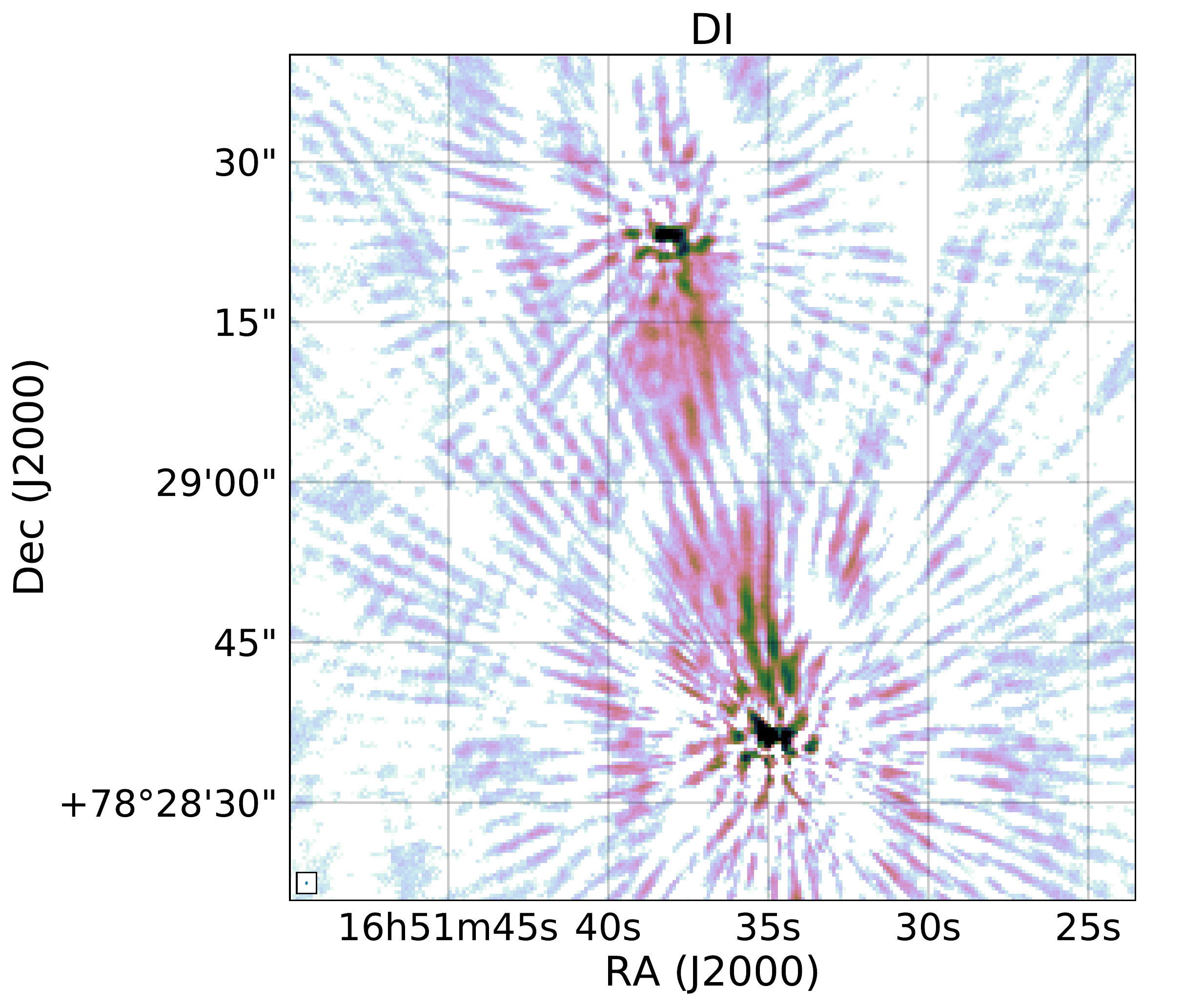}
\includegraphics[width=0.285\paperwidth]{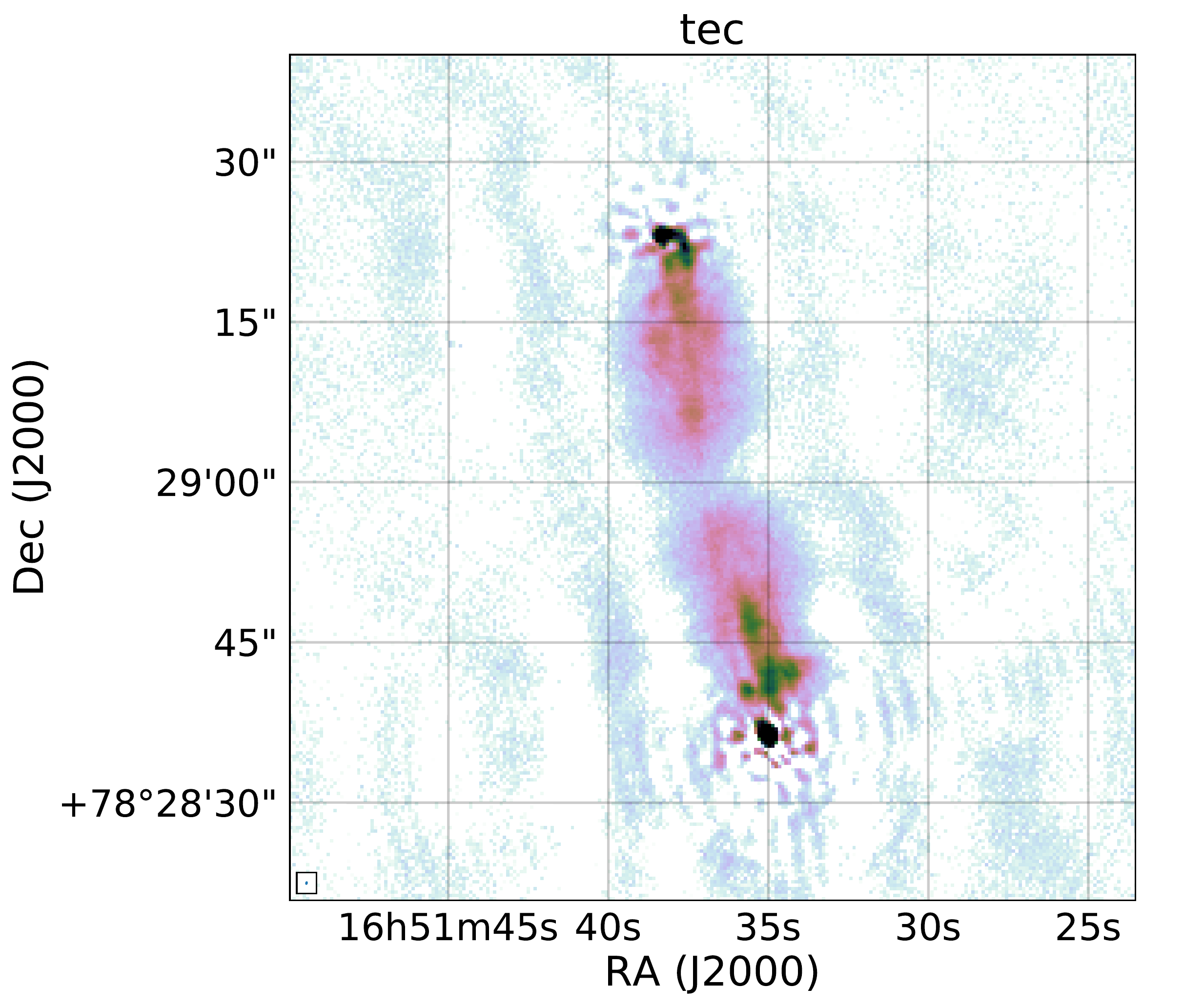}
\includegraphics[width=0.285\paperwidth]{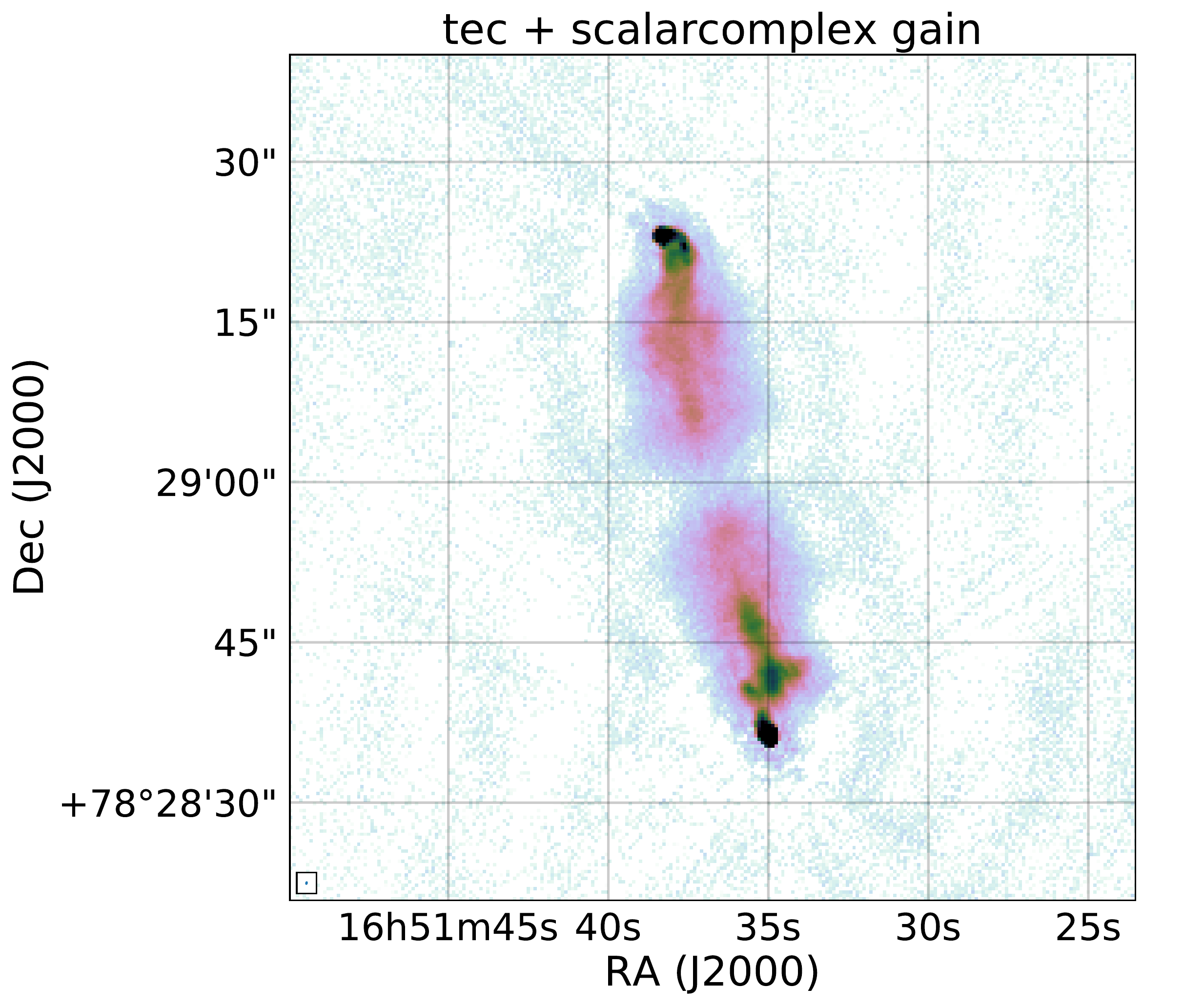}
     \caption{\textit{Top row:} Images showing the subsequent improvement during self-calibration on extracted 43--67~MHz LBA data. The left panel shows the starting point of this process: a direction independent calibrated image. The middle panel displays the results after {\tt DPPP} `tec' calibration, with all LOFAR core stations forced (``antenna constraint'' solve) to having the same solution. The right panel shows the final image after {\tt DPPP} scalarcomplex gain calibration. \textit{Bottom row:} The same as the top row images but for LOFAR international baseline 120--168~MHz HBA data and without using an ``antenna constraint'' solve.}
     \label{fig:LBA}
\end{figure*}

The self calibration step has been designed in such a way that it can also be carried out on other (non-LoTSS) LOFAR extracted datasets. It offers full flexibility for the various effects that can be solved for in {\tt DPPP}, including constraint solves. For example, it can enforce the gains to be smooth along the frequency axis or enforce the same solutions for certain stations. The self calibration step has been successfully used on LOFAR HBA international-baseline data and on the Low Band Antennas (LBA) observations. Figure~\ref{fig:LBA} shows an example of it working on LBA and international baseline data.

\begin{table}
\caption{Default imaging and calibration parameters for HBA extraction}         
\label{tab:selfcalparameters}
\centering                                      
\begin{tabular}{l l l}          
\hline\hline                        
parameter & value &  \\    
\hline                                   
inner uv-range (calibration) & 350\tablefootmark{a} &  \\      
inner uv-range (imaging, $\lambda$) & 80 &  \\
clean mask threshold (sigma) & 5 &  \\
pixelsize (arcsec) & 1.5 &  \\
Briggs robust weighting & $-0.5$ & \\
channelsout\tablefootmark{b} & 6 \\
\hline      
\end{tabular}
\tablefoot{
   \tablefoottext{a}{Increased to $750\lambda$ for targets with extended low-surface brightness emission.}
   \tablefoottext{b}{{\tt WSClean} wideband deconvolution setting}
   }
\end{table}

\section{HETDEX-DR1 area galaxy cluster sample}
\label{sec:sample}
\begin{figure}
\centering
\includegraphics[width=1.0\columnwidth]{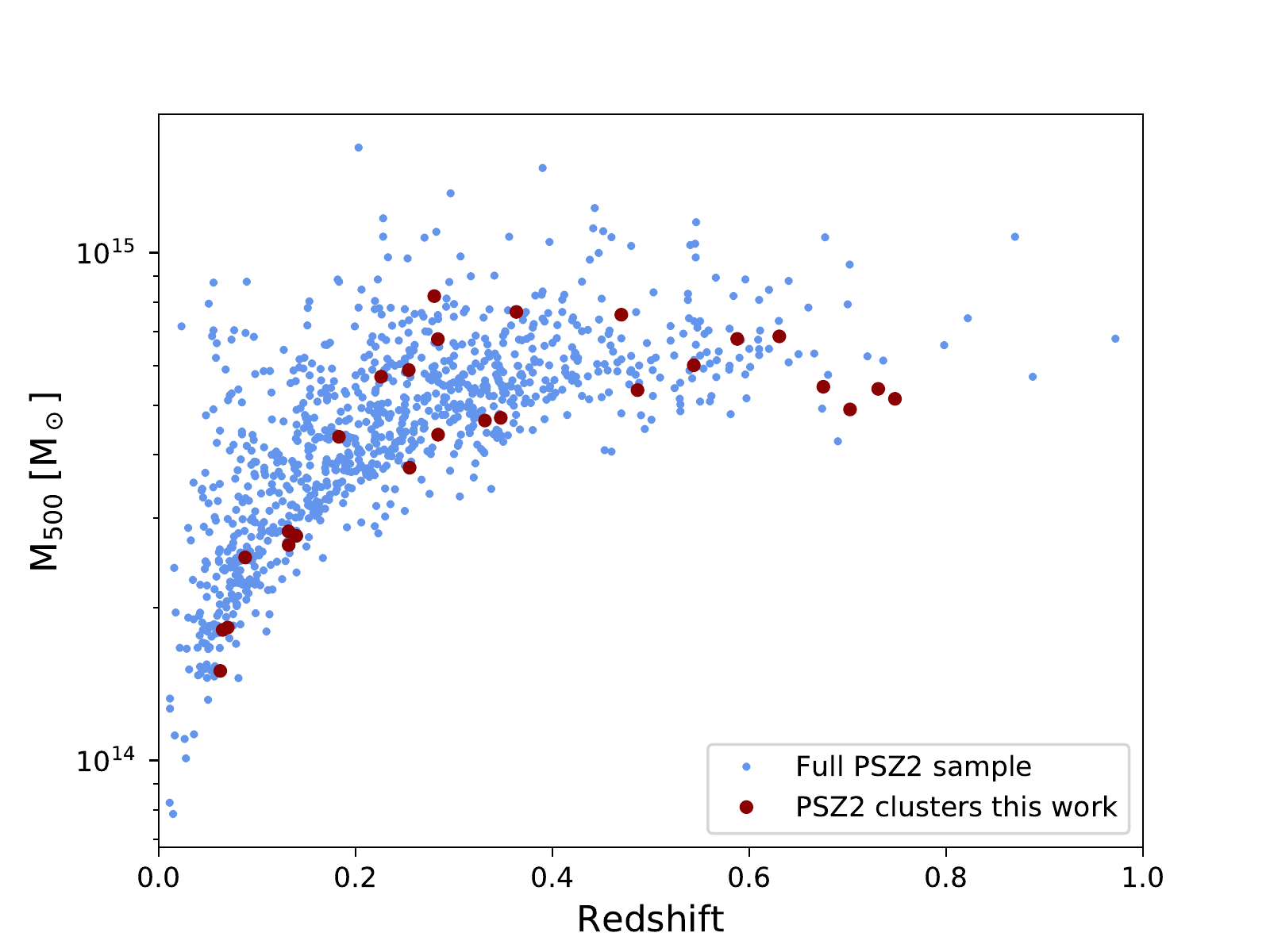}
     \caption{The redshift-mass distribution of PSZ2 clusters. Clusters that are located in the HETDEX DR1 area are indicated with dark red points.}
     \label{fig:zm}
\end{figure}

We applied the extraction scheme described in Section~\ref{sec:extraction} to a sample of galaxy clusters that fall inside the LOFAR Data Release 1 area \citep{2017A&A...598A.104S,2019A&A...622A...1S}. 
We selected  clusters from the all-sky PSZ2 Planck catalog \citep{2016A&A...594A..27P} of Sunyaev-Zel'dovich (SZ) sources, see Figure~\ref{fig:zm}.  This sample of 26~clusters serves as our primary sample for our statistical investigations and is listed in Table~\ref{tab:clustersample1}. 

In addition to the SZ-selected sample, we compiled a secondary sample by visually inspecting the LoTSS images at the location of known clusters for the presence of diffuse emission and possible revived fossil plasma sources. Clusters in this sample with (candidate) extended radio emission are from the Abell and Zwicky \citep{1958ApJS....3..211A,1961cgcg.book.....Z,1974AJ.....79.1356C}, MCXC \citep{2011A&A...534A.109P}, GMBCG \citep{2010ApJS..191..254H}, MaxBCG  \citep{2007ApJ...660..239K}, 
and WHL \citep{2012ApJS..199...34W} catalogs.

To search for diffuse radio emission we produced images with the emission from compact sources subtracted. This was done by first imaging the calibrated datasets with a uv-cut corresponding to a physical scale of 0.4~Mpc. This model was subsequently subtracted from the visibility data.

\subsection{Flux density measurements}
The flux density measurements for all sources, except radio halos, were done manually by placing a polygon around the sources and integrating the flux density. The uncertainty on the flux density is given by
\begin{equation}
\label{eq:sigmas}
 \sigma_S^2 = N_{\rm{beams}}\sigma_{\rm{rms}}^2 + \sigma_{\rm{sub}}^2 +  \left(f\times S\right)^{2} \mbox{ ,}
\end{equation}
where $f=0.2$ is the absolute flux-scale uncertainty \citep{2019A&A...622A...1S}, $N_{\rm{beams}}$ the number of beams covering the source, $\sigma_{\rm{rms}}$ the map noise, and $\sigma_{\rm{sub}}$ the uncertainty due to compact source subtraction. The first two terms of Eq.~\ref{eq:sigmas} represent the statistical uncertainty on the flux density measurement.
The uncertainty on the compact source subtraction is given by
\begin{equation}
     \sigma_{\rm{sub}}^2 = \sum_{i}
     N_{\rm{beams,i}}\sigma_{\rm{rms}}^2  \mbox{ ,}
\end{equation}
where the sum is taken over all $i$ sources that were subtracted in the polygon.

The radio halo integrated flux densities were determined by fitting exponential profiles \citep{2009A&A...499..679M} to the radio images of the form
\begin{equation}
	I(r)=I_0e^{-r/r_e} \mbox{ ,}
	\label{eq:murgia}
\end{equation}
where $r_e$ is a characteristic $e$-folding radius and $I_0$ the central surface brightness. We used the image where compact sources were subtracted.  Also, all images were smoothed to a beam size corresponding to a physical scale of 50~kpc at the cluster's redshift. For the fitting we used {\tt Halo-FDCA}\footnote{\url{https://github.com/JortBox/Halo-FDCA}} described in \cite{boxelaar}. 
Extended sources, such as large tailed radio galaxies, were masked during the fitting (in case they were not fully subtracted). The  flux density was integrated to a radius of $3r_e$, as proposed by \cite{2009A&A...499..679M}. Integrating up to a radius of $3r_e$ results in 80\% of the flux density when compared to using an infinite radius. This also avoids to some extent the uncertainties related to the extrapolation of the profile to large radii where the radio surface brightness is below the detection limit. The fitting uses a Markov Chain Monte–Carlo method to determine the radio halo parameters and associated uncertainties.
For some clusters, when specifically mentioned in Section~\ref{sec:results}, we fitted an elliptical exponential model \citep[see][]{boxelaar}. In these cases, a major and minor characteristic $e$-folding radius and position angle are determined. This is done for radio halos that clearly display an elongated, rather than circular, shape. In this work we did not use the skewed (asymmetric) models that are available in {\tt Halo-FDCA}. A full exploration of radio halo fitting models and methods is beyond the scope of this work.

For the total uncertainty on the radio halo integrated flux density we add the uncertainty from the MCMC (statistical errors) and flux calibration uncertainty in quadrature.

\subsection{X-ray observations}
We searched the \textit{Chandra} and \textit{XMM-Newton} archives for the available X-ray observations of the clusters in the sample. When available, data were retrieved and processed with CIAO 4.11 using CalDB v4.8.2 and SAS v16.1.0 following standard data reduction recipes. We used the time periods of the observations cleaned by anomalously high background to produce cluster exposure-corrected images in the $0.5-2.0$ keV band. These images are used in the paper to investigate the connection between the thermal and non-thermal components in the ICM.

\onecolumn
\begin{longtable}{llllrrl}
\caption{\label{tab:clustersample1} HETDEX DR1-area  cluster sample}\\
\hline\hline
Cluster&  alternative name(s) &  RA & DEC & redshift & M$_{500, SZ}$ \\
\hline
\endfirsthead
\caption{continued.}\\
\hline\hline
Cluster& alternative name(s)  & RA & DEC & redshift & M$_{500, SZ}$   \\
\hline
\endhead
\hline
\endfoot
PSZ2\,G080.16+57.65 & Abell\,2018 &15 01 08 & +47 16 37  & 0.0878&2.51$^{+0.20}_{-0.21}$  \\
PSZ2\,G084.10+58.72 & &14 49 01 & +48 33 24 & 0.7310 & 5.40$^{+0.62}_{-0.62}$  \\
PSZ2\,G086.93+53.18 &WHL\,J228.466+52.8333 & 15 14 00& +52 48 14 & 0.6752 & 5.45$^{+0.50}_{-0.52}$ \\
PSZ2\,G087.39+50.92 & [WH2015]\,0986 &15 26 33 & +54 09 08& 0.7480& 5.16$^{+0.53}_{-0.60}$&\\
PSZ2\,G088.98+55.07 & & 14 59 01 &+52 49 01 & 0.7023 & 4.92$^{+0.60}_{-0.64}$ \\
PSZ2\,G089.52+62.34 & Abell\,1904 & 14 22 13 & +48 29 54 & 0.0701 & 1.83$^{+0.19}_{-0.20}$  \\
& RX\,J1422.1+4831 \\

PSZ2\,G095.22+67.41 & RXC\,J1351.7+4622 &13 51 45 & +46 22 00 &  0.0625 & 1.50$^{0.21+}_{-0.22}$ \\
&  MCXC J1351.7+4622  \\

PSZ2\,G096.14+56.24 & Abell\,1940 &14 35 21 & +55 08 29 & 0.1398 & 2.77$^{+0.24}_{-0.26}$ \\
&RX\,J1435.4+5508 \\
PSZ2\,G098.44+56.59 & Abell\,1920 & 14 27 25& +55 45 02 &0.1318 & 2.83$^{+0.28}_{-0.26}$ \\
& RX\,J1427.4+5545 \\

PSZ2\,G099.86+58.45  & WHL\,J141447.2+544704  & 14 14 43 & +54 47 01 & 0.6160  & 6.85$^{+0.48}_{-0.49}$ \\ 
& WHL\,J213.697+54.7844 \\

PSZ2\,G106.61+66.71 & & 13 30 29 & +49 08 48 & 0.3314 & 4.67$^{+0.55}_{-0.57}$ \\

PSZ2\,G107.10+65.32 & Abell\,1758 & 13 32 35& +50 29 09 &0.2799 &8.22$^{+0.27}_{-0.28}$ \\

PSZ2\,G111.75+70.37 &Abell\,1697 & 13 13 03 & +46 16 52 & 0.1830 & 4.34$^{+0.32}_{-0.33}$  \\
& RXC\,J1313.1+4616 \\

PSZ2\,G114.31+64.89	  & Abell\,1703 & 13 15 05 & +51 49 02&0.2836 &6.76$^{+0.36}_{-0.38}$ \\

PSZ2\,G114.99+70.36 & Abell\,1682 & 13 06 50 & +46 33 27 &0.2259 &5.70$^{+0.35}_{-0.35}$ \\

PSZ2\,G118.34+68.79 & ZwCl\,1259.0+4830 & 13 01 24 & +48 14 31 &0.2549 & 3.77$^{+0.45}_{-0.52}$ \\
& [WH2015]\,0746 \\

PSZ2\,G123.66+67.25 & Abell\,1622 & 12 49 41 & +49 52 18 & 0.2838 & 4.38$^{+0.50}_{-0.52}$ \\
& ZwCl\,1247.2+5008 \\

PSZ2\,G133.60+69.04 & Abell\,1550 & 12 29 02 & +47 37 21 & 0.2540 & 5.88$^{+0.38}_{-0.42}$ \\

PSZ2\,G135.17+65.43 & WHL\,J121912.2+505435 & 12 19 12 & +50 54 35 & 0.5436 & 6.00$^{+0.57}_{-0.62}$ \\

PSZ2\,G136.92+59.46 & Abell\,1436 & &&0.0650 & 1.80$^{+0.17}_{-0.16}$ \\ 
&  RXC\,J1200.3+5613 \\

PSZ2\,G143.26+65.24	& Abell\,1430 &11 59 17 & +49 47 37  &0.3634 & 7.65$^{+0.42}_{-0.44}$ \\
& RXC\,J1159.2+4947 \\
& ZwCl\,1156.4+5009 \\

PSZ2\,G144.33+62.85 & Abell\,1387& 11 49 05&+51 35 08 & 0.1320 & 2.66$^{+0.33}_{-0.38}$\\
& RXC\,J1149.0+5135 \\

PSZ2\,G145.65+59.30 & Abell\,1294  & 11 32 42 & +54 13 12 & 0.3475 & 4.73$^{+0.59}_{-0.64}$ \\
& ZwCl\,1129.6+5430 \\

PSZ2\,G150.56+58.32 & MACS\,J1115.2+5320 & 11 15 11 & +53 19 39 & 0.4660 & 7.55$^{+0.50}_{-0.52}$  \\
& RXC\,J1115.2+5320  \\

PSZ2\,G151.62+54.78  & RX\,J105453.3+552102 & 10 54 52 & +55 21 13 & 0.4864 & 5.37$^{+0.68}_{-0.75}$   \\
& [WH2015]\,0472 \\
& 1RXS\,J105453.3+552102 \\

PSZ2\,G156.26+59.64 & [WH2015] 0485 & 11 08 30 & +50 16 02 & 0.6175 & 6.77$^{+0.59}_{-0.60}$  \\

\hline
\hline

Abell\,1156  & & 11 04 56& +47 25 15& 0.2091 & \ldots \\

Abell\,1314  & & 11 34 49& +49 04 40& 0.0335 &   \ldots\\

Abell\,1615  & & 12 47 43 & +48 51 57 & 0.2106 &  \ldots \\

GMBCG\,J211.77332+55.09968 & &14 06 55 &+55 04 02 &0.2506 &   \ldots \\

MaxBCG\,J173.04772+47.81041 & & 11 32 11 & +47 48 38 & 0.2261 &  \ldots \\

NSC\,J143825+463744 && 14 38 46& +46 39 56 &0.0357 &  \ldots  \\

RXC\,J1053.7+5452 & MCXC J1053.7+5452 & 10 53 44 &+54 52 20 & 0.0704 & \ldots  \\

WHL\,J125836.8+440111 & &  12 58 37& +44 01 11 & 0.5339 &  \ldots  \\
{WHL\,J122418.6+490549} & & 12 24 19& +49 05 50 &0.1004 &  \ldots\\
{WHL\,J124143.1+490510}&&12 41 43 &+49 05 10 & 0.3707& \ldots \\

{WHL\,J132226.8+464630} &&13 22 27& +46 46 30& 0.3718 &  \ldots\\
{WHL\,J132615.8+485229} & &13 26 16 &+48 52 29 & 0.2800 &  \ldots\\
{WHL\,J133936.0+484859} &&13 39 36 &+48 48 59 & 0.3265 &  \ldots \\
{WHL\,J134746.8+475214} & &13 47 47& +47 52 15&0.1695 & \ldots \\

\end{longtable}

\twocolumn

\section{Results}  
\label{sec:results}   
Below we first describe the LOFAR results for the individual clusters. By default images were made using Robust weighting $-0.5$ \citep{1995AAS...18711202B}. In addition, images at lower resolution were produced using a Gaussian taper to down weight the visibilities from longer baselines. Images with the emission from compact sources subtracted are also shown for some clusters. 

The classification of diffuse sources is summarised in Table~\ref{tab:classification}. Descriptions and images of clusters for which no diffuse emission was detected are given in Appendix \ref{sec:nodiffuse}. For optical overlays we use Pan-STARRS \emph{gri} images \citep{2016arXiv161205560C}.

\onecolumn
\begin{longtable}{llllrrl}
\caption{\label{tab:classification} Cluster radio properties}\\
\hline\hline
Cluster&   classification & LLS & $S_{144}$ \\ 
 & &Mpc & mJy \\ 
\hline
\endfirsthead
\caption{continued.}\\
\hline\hline
Cluster&   classification & LLS & $S_{144}$ \\
\hline
\endhead
\hline
\endfoot
PSZ2\,G080.16+57.65 & cHalo, Relic & (R) 1.1, (H) $\sim1$  & (H) $92\pm32$, (R) $55.8 \pm 11.6$ \\

PSZ2\,G084.10+58.72 & cHalo & 0.5 & $4.2\pm1.2$\\

PSZ2\,G086.93+53.18 & Halo & 0.6 & $12.4\pm3.6$\\

PSZ2\,G087.39+50.92 & AGN-no diffuse & --- & ---\\

PSZ2\,G088.98+55.07 & AGN-no diffuse & --- & ---\\

PSZ2\,G089.52+62.34 & (A) Relic  & (A) 0.23 & (A+C) $74.6 \pm 15.0$ \\
         &   (B) cRelic (A+B: cdRelic)             & (B) 0.17 &  (B) $12.2\pm2.6$  \\

PSZ2\,G095.22+67.41 & cRelic & 0.4 & $6.8\pm 1.5$ \\

PSZ2\,G096.14+56.24 &  AGN-no diffuse  & --- & ---\\

PSZ2\,G098.44+56.59  & AGN-no diffuse & --- & ---\\
PSZ2\,G099.86+58.45  &   Halo & 1.0 & $14.7\pm3.2$\\

PSZ2\,G106.61+66.71 &  cHalo & 0.5 & $20\pm4$ \\

PSZ2\,G107.10+65.32 &  dHalo & (Hn) 2.0, (Hs) 1.3 & (Hn) $123\pm25$, (Hs) $63\pm14$ \\
                            & Relic (s) & 0.5 & $20.9\pm4.3$ \\
                            & cFossil (n: S1, S2) &  (S1) 0.4, (S2) 0.23  &  (S1) $79\pm17$, (S2) $17.5\pm3.6$         \\

PSZ2\,G111.75+70.37 & Relic & 0.7  & (R) $106.7\pm21.4$\\
                   &  Halo &  $\sim0.6$ & $27.7\pm6.3$  \\

PSZ2\,G114.31+64.89	  & Halo & 0.5 & $91.7\pm18.6$\\

PSZ2\,G114.99+70.36 & cHalo, cFossil & --- & $^{\star}$  \\

PSZ2\,G118.34+68.79 &  cHalo, Fossil & (H) $\sim0.4$, (F) 0.5 & (H) $28\pm13$, (F) $67\pm13$  \\

PSZ2\,G123.66+67.25 &  cFossil & 0.26 & $7.5\pm1.6$\\

PSZ2\,G133.60+69.04 &  Halo, cdRelic/cFossil & (H) 0.9 & (H) $129\pm26$	\\

PSZ2\,G135.17+65.43 &  cHalo & 0.5 & $29.0\pm6.6$  \\
                   &  cRelic/cFossil & 0.4 & $9.14 \pm 1.9$  \\ 

PSZ2\,G136.92+59.46 &  AGN-no diffuse & --- & ---\\ 

PSZ2\,G143.26+65.24	&  Halo & 1.5 & (H) $29.8\pm6.6$\\
PSZ2\,G144.33+62.85 &  AGN-no diffuse & --- & ---\\

PSZ2\,G145.65+59.30  & cHalo/cFossil & 0.4 & $6.7\pm1.7$ \\

PSZ2\,G150.56+58.32  & Halo & 1.0 & $71.2\pm14.5$ \\

PSZ2\,G151.62+54.78  & AGN-no diffuse & --- & ---\\

PSZ2\,G156.26+59.64 &  cHalo & 0.5 & $7.9\pm 3.7$ \\

\hline
\hline
RXC\,J1053.7+5452  & Relic & 0.75 & $214\pm43$ \\

Abell\,1156   & cHalo & 0.7  & $15.7\pm6.8$ \\
Abell\,1314   & Fossil (central) & 0.44 & $136.9 \pm 27.4$\\

Abell\,1615   & AGN-no diffuse & --- & ---\\
MaxBCG\,J173.04772+47.81041  & AGN-no diffuse & --- & --- \\
NSC\,J143825+463744 & Unclassified & 0.4 & ---\\
GMBCG\,J211.77332+55.09968  & AGN-no diffuse & --- & ---\\
WHL\,J125836.8+440111   & Halo & 0.8 &$58.4\pm11.7$\\

WHL\,J122418.6+490549 &  Fossil & 0.37 & $209 \pm42$\\

WHL\,J124143.1+490510   & cHalo & 1.2 & $^{\star}$\\
WHL\,J132226.8+464630   & cHalo & 0.5 & $^{\star}$\\
WHL\,J132615.8+485229   & AGN-no diffuse & 0.32 & $^{\star}$\\
WHL\,J133936.0+484859   & cHalo & $\sim0.3$& $^{\star}$\\
WHL\,J134746.8+475214   & AGN-no diffuse & --- & ---\\
\end{longtable}
H = Halo, R=Relic, d=double, s=South, n=North, LLS=largest linear size\\
$^{\star}$ radio emission is blended with other sources and therefore no reliable flux density measurement can be obtained
\twocolumn

\subsection{PSZ2\,G080.16+57.65, Abell\,2018}
\begin{figure*}
\centering
   \includegraphics[width=0.285\paperwidth]{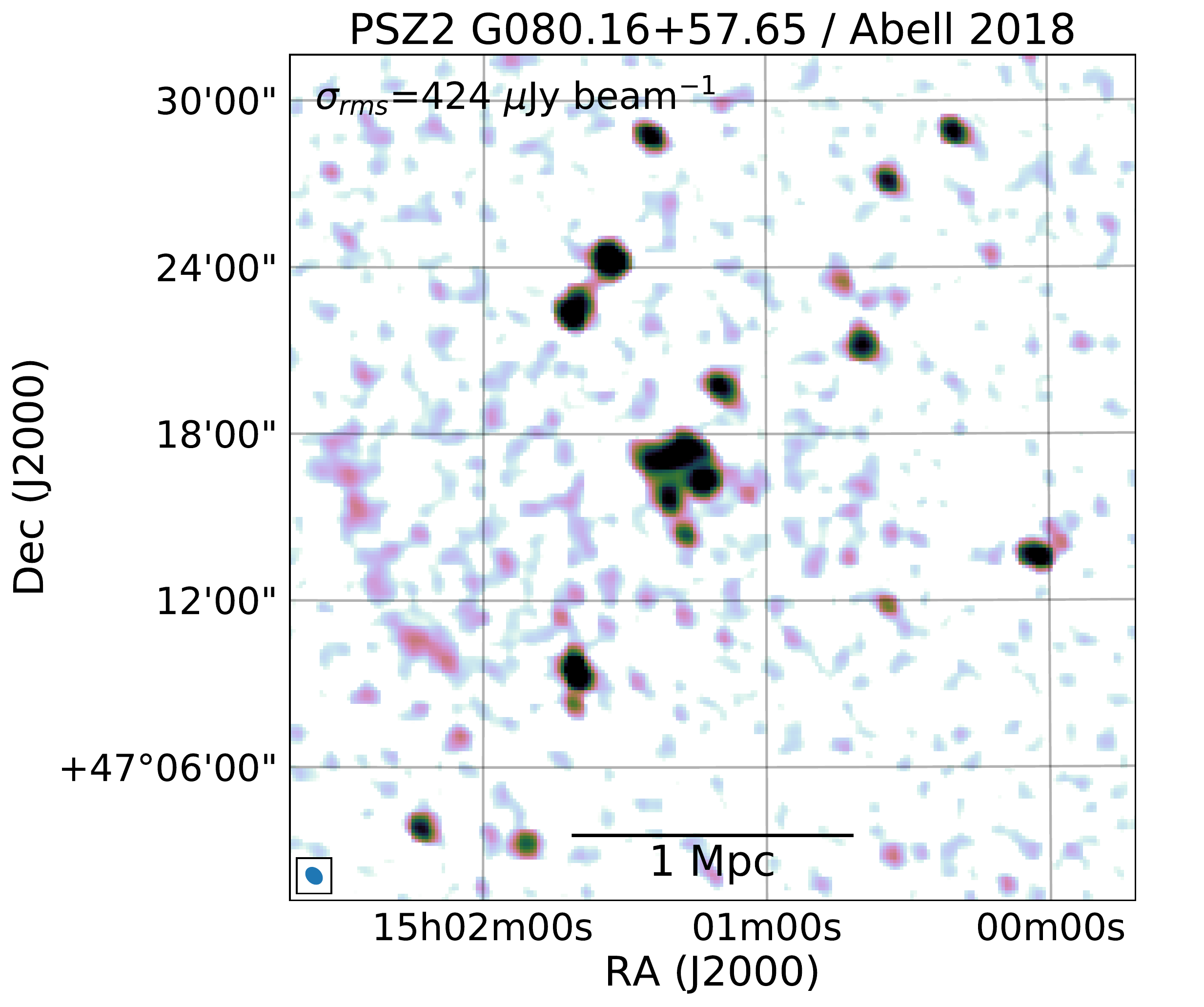}
   \includegraphics[width=0.285\paperwidth]{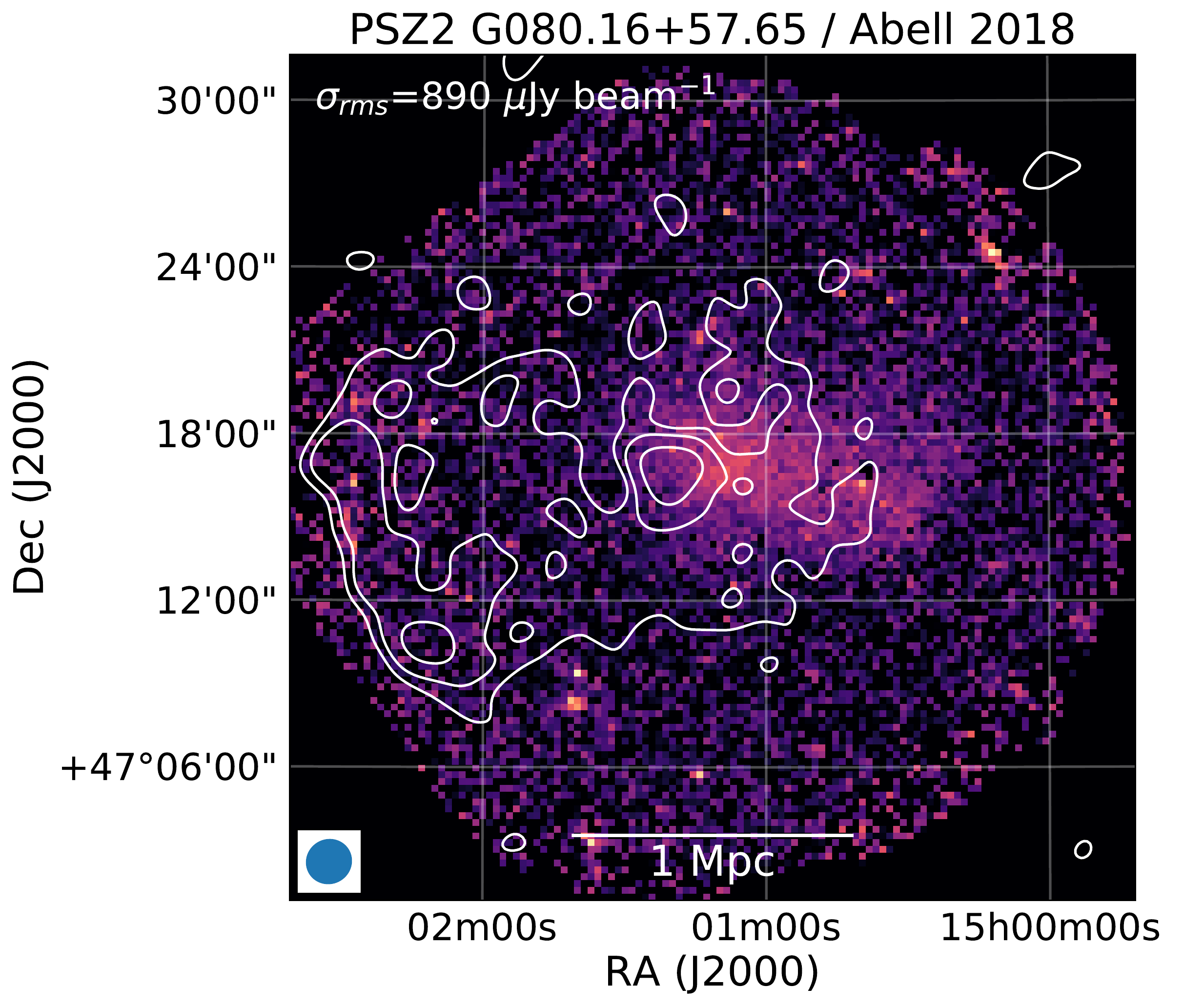}
   \includegraphics[width=0.285\paperwidth]{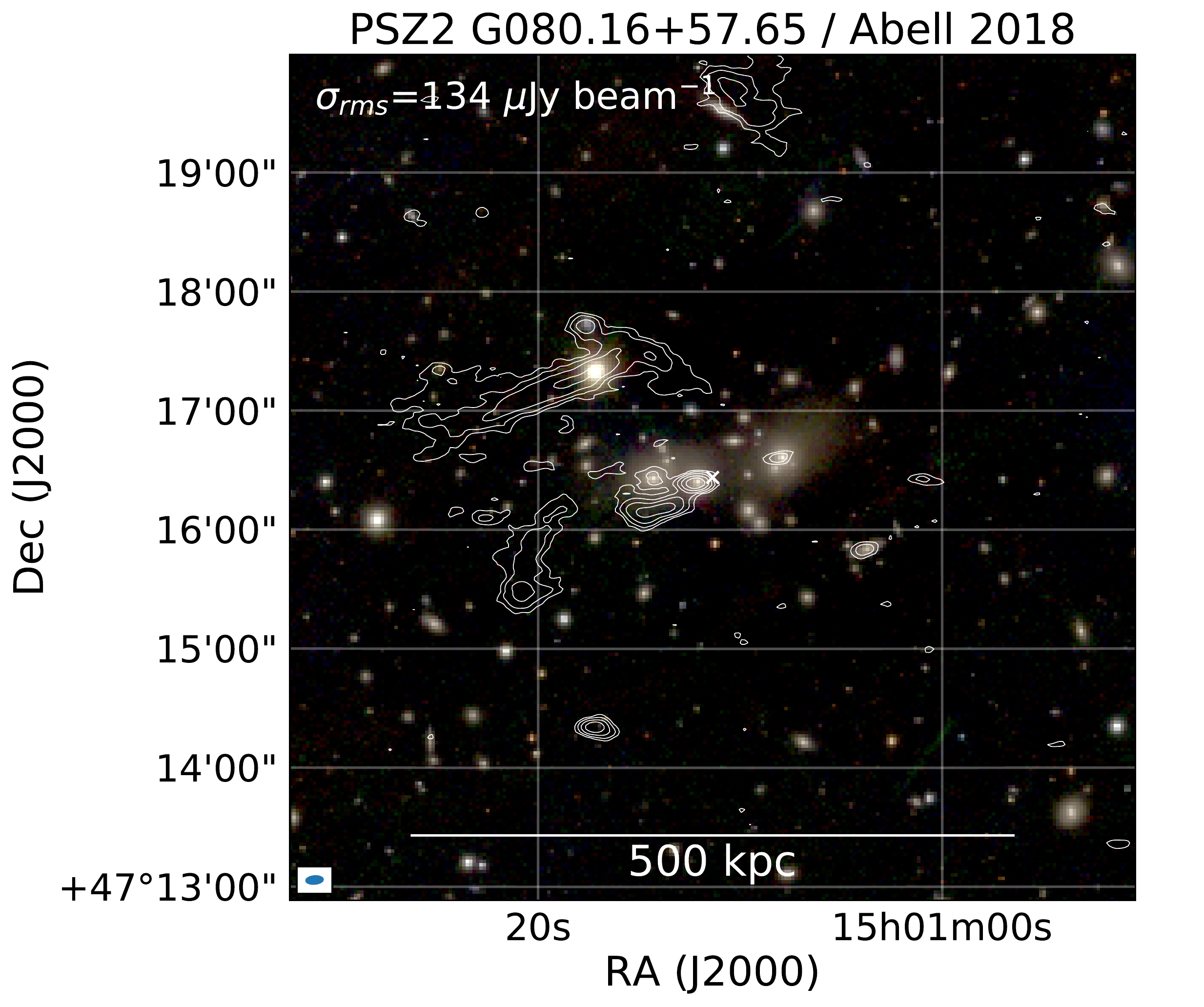}
   \caption{PSZ2\,G080.16+57.65 / Abell\,2018. Left: Low-resolution 144~MHz radio image made with a 30\arcsec~taper. Middle: XMM-Newton image with radio contours from a 90\arcsec~tapered image overlaid. Compact sources were subtracted in this image. Right: Pan-STARRS \emph{gri} color image overlaid with radio contours from the robust $-0.5$ weighted image. The cluster center position is marked with a white `\rm{X}'. Radio contours on both the X-ray and optical images are drawn at $3\sigma_{\rm{rms}}\times [1,2,4,\ldots]$, with $\sigma_{\rm{rms}}$ the r.m.s. map noise.}
   \label{fig:A2018}
\end{figure*}

 Abell\,2018 is relatively nearby cluster located at $z=0.0878$. The Pan-STARRS \emph{gri} image reveals a number of central galaxies, without a clear dominant BCG. The cluster has a low PSZ2 mass of $M_{500}= 2.51^{+0.20}_{-0.21} \times 10^{14}$~M$_{\odot}$. The high-resolution radio image, see Figure~\ref{fig:A2018}, shows a complex central region with several AGN being present as well as more extended emission.  A low-surface brightness arc-like structure is found about 1.3~Mpc to the east of the optical center of the cluster. This source has an LLS of about 1.1~Mpc. In a low-resolution image (made with a 90\arcsec~taper), additional faint diffuse emission spanning the central region of the cluster is found. This emission extends all the way to the arc-like structure to the east. This emission has a size of approximately 1.0~Mpc by 1.8~Mpc.
 The XMM image reveals an EW merging cluster with the main component located at the optical center of the cluster. We classify the arc-like structure to the east as a giant radio relic and the large-scale centrally placed emission as a candidate radio halo. The nature of the central brighter diffuse emission near the AGN remains unclear. It could be related to the radio halo or be revived fossil plasma from the AGNs. For the candidate radio halo we determine an integrated flux density of  $S_{144}=92\pm32$~mJy, where we masked the central brighter diffuse emission around the AGN and the extension towards the radio relic in the fitting. We note that this cluster falls above the correlation between cluster mass and radio power (see Section~\ref{sec:scaling}). Deeper observations are required to shed more light on this point and on the origin of the extension from the central emission towards the relic.

\subsection{PSZ2\,G084.10+58.72}
\begin{figure*}
\centering
   \includegraphics[width=0.285\paperwidth]{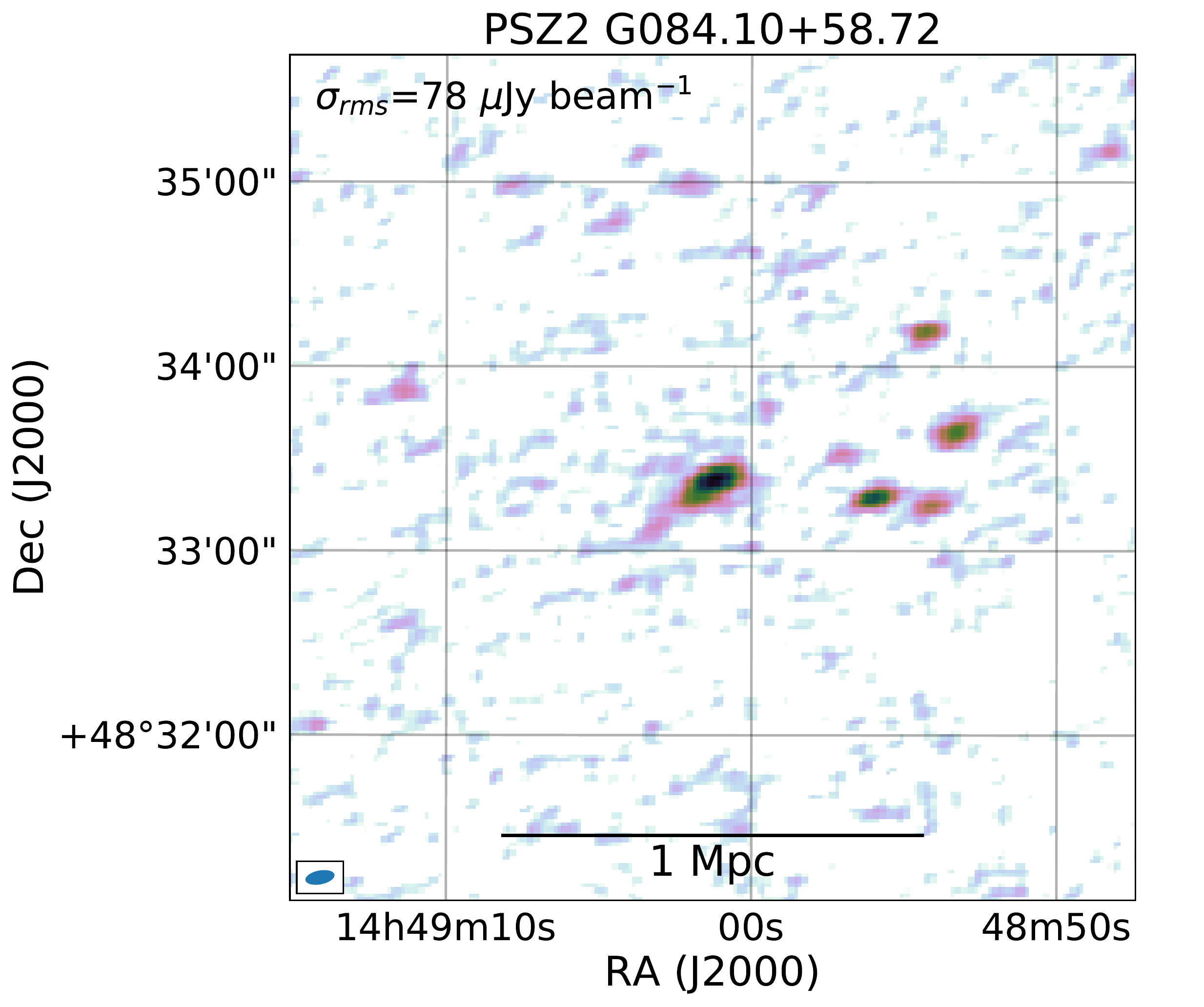}
   \includegraphics[width=0.285\paperwidth]{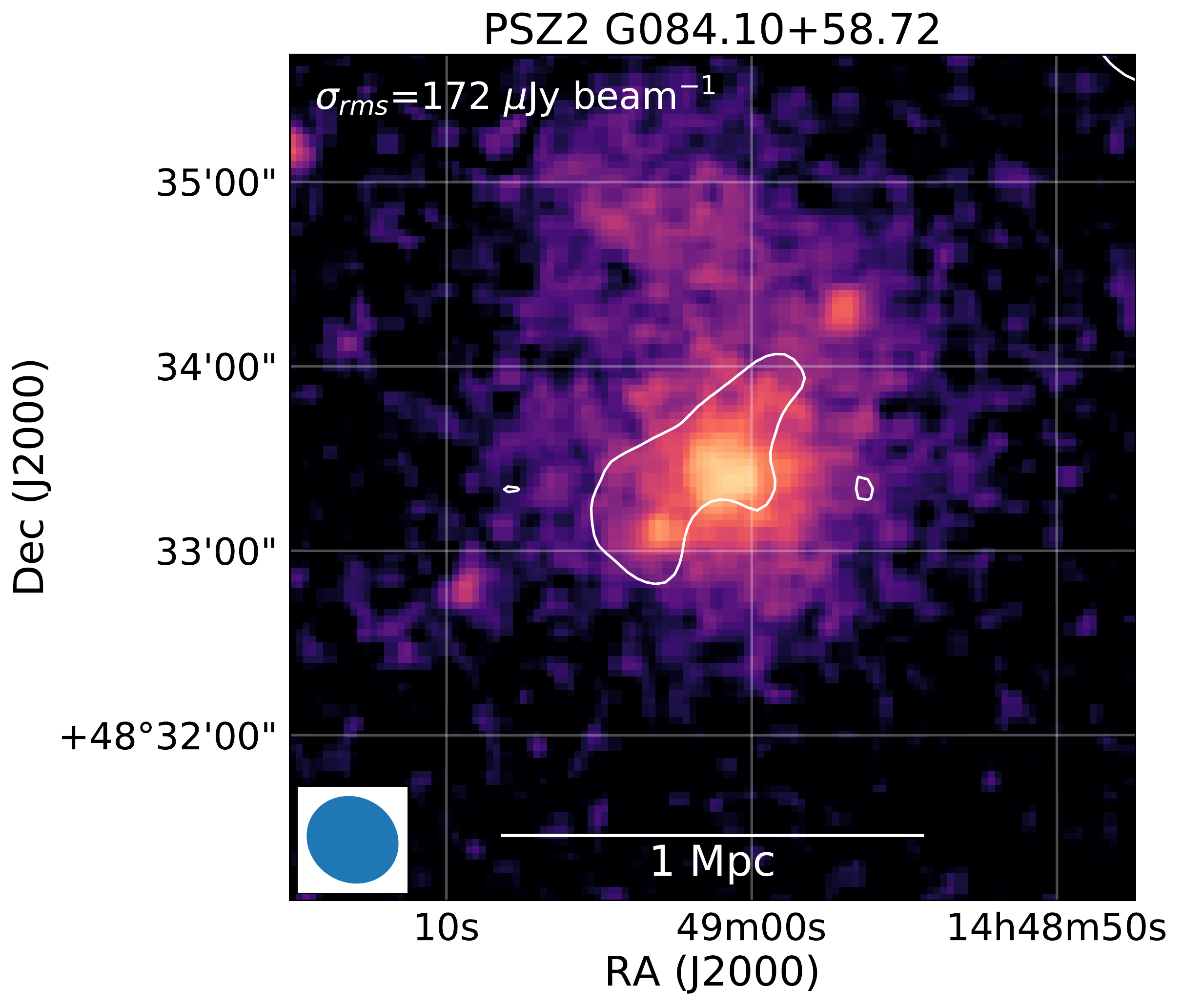}
   \includegraphics[width=0.285\paperwidth]{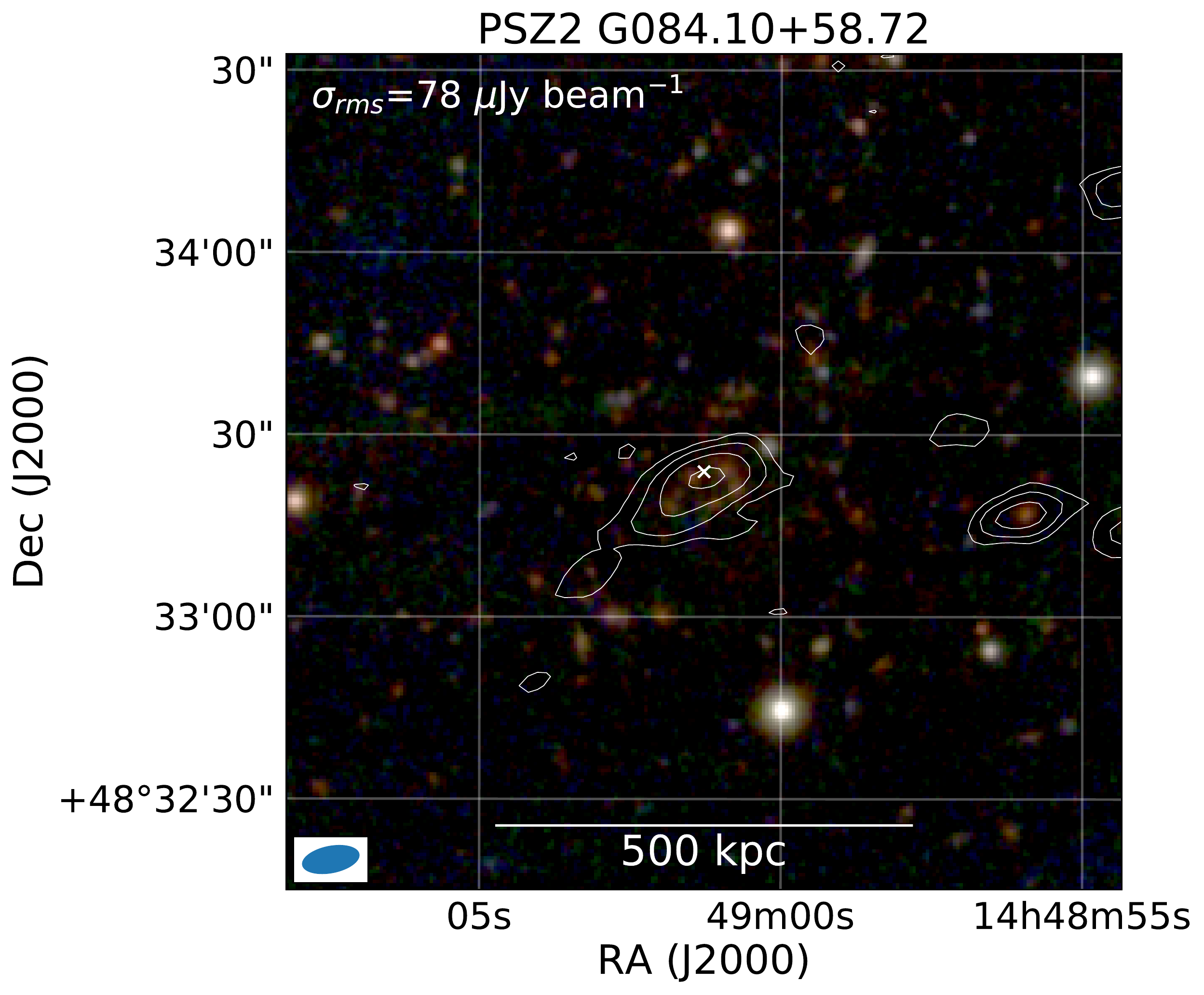}
   \caption{PSZ2\,G084.10+58.72. Left: Robust $-0.5$ radio image. Middle: XMM-Newton X-ray image with 15\arcsec~tapered radio contours (compact sources were subtracted). Right: Optical image with Robust $-0.5$ image radio contours. For more details see the caption of Figure~\ref{fig:A2018}.}
   \label{fig:PSZ2G08410}
\end{figure*}

The XMM-Newton image of this distant ($z=0.73$) cluster shows a disturbed system with a 1~Mpc bar-like structure north of the cluster center, see Figure~\ref{fig:PSZ2G08410}. A hint of centrally located diffuse emission is seen in our image with compact sources subtracted. This emission extends on scales of about 0.5~Mpc. We therefore classify this as a candidate radio halo with $S_{144}=4.2\pm1.2$~mJy. \cite{gennaro} also reported a hint of diffuse emission in this cluster.

\subsection{PSZ2\,G086.93+53.18}
\begin{figure*}
\centering
   \includegraphics[width=0.285\paperwidth]{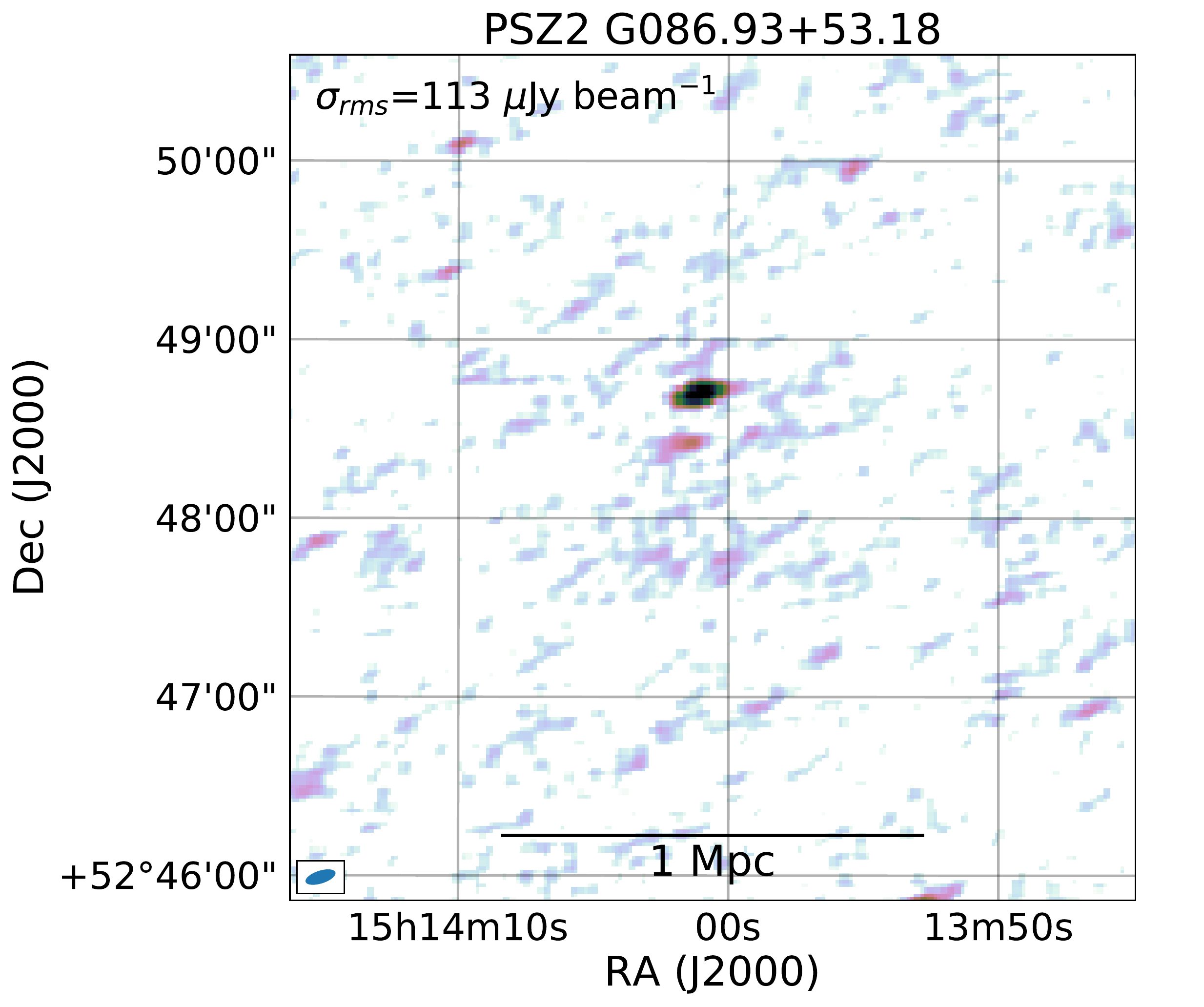}
   \includegraphics[width=0.285\paperwidth]{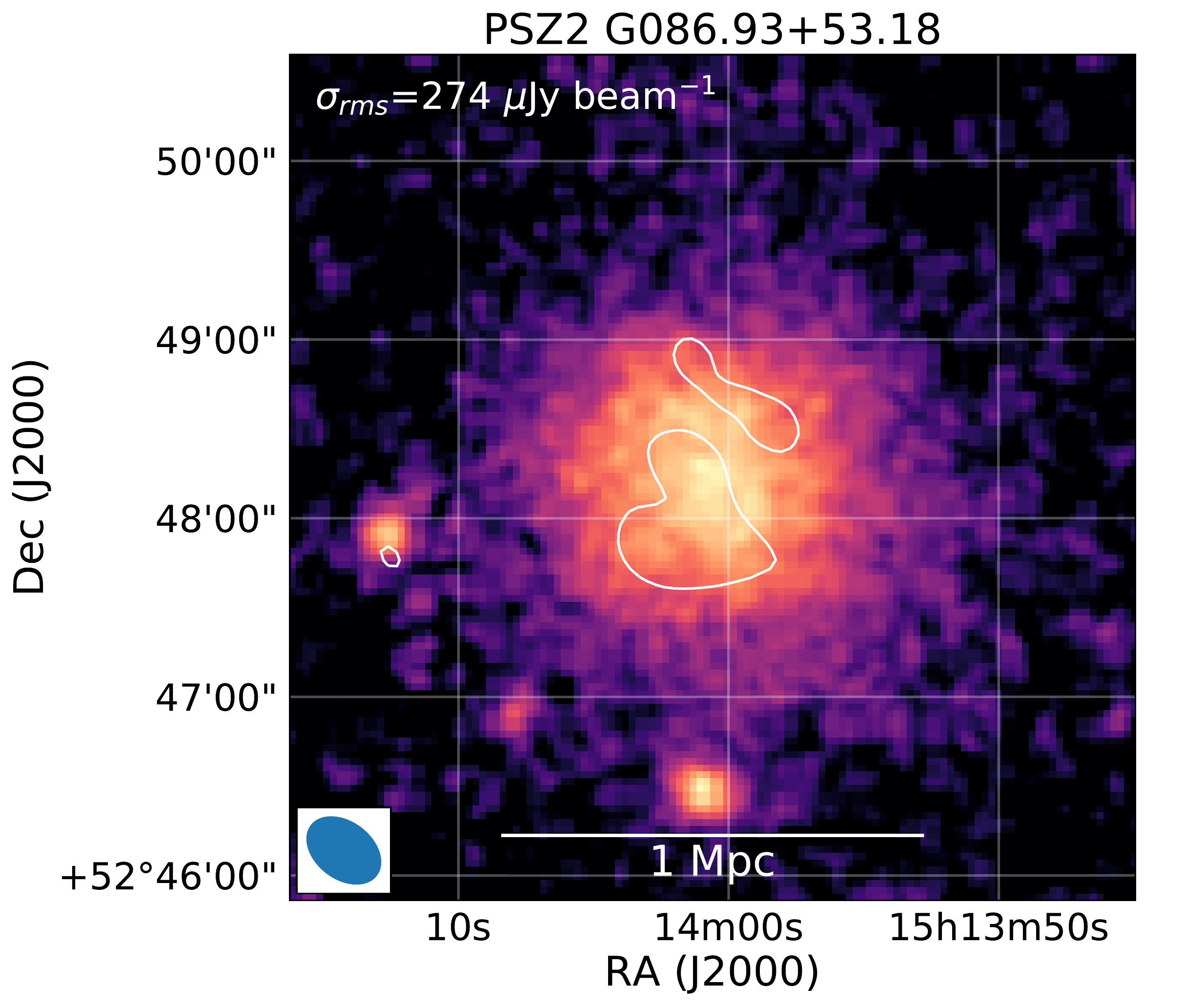}
   \includegraphics[width=0.285\paperwidth]{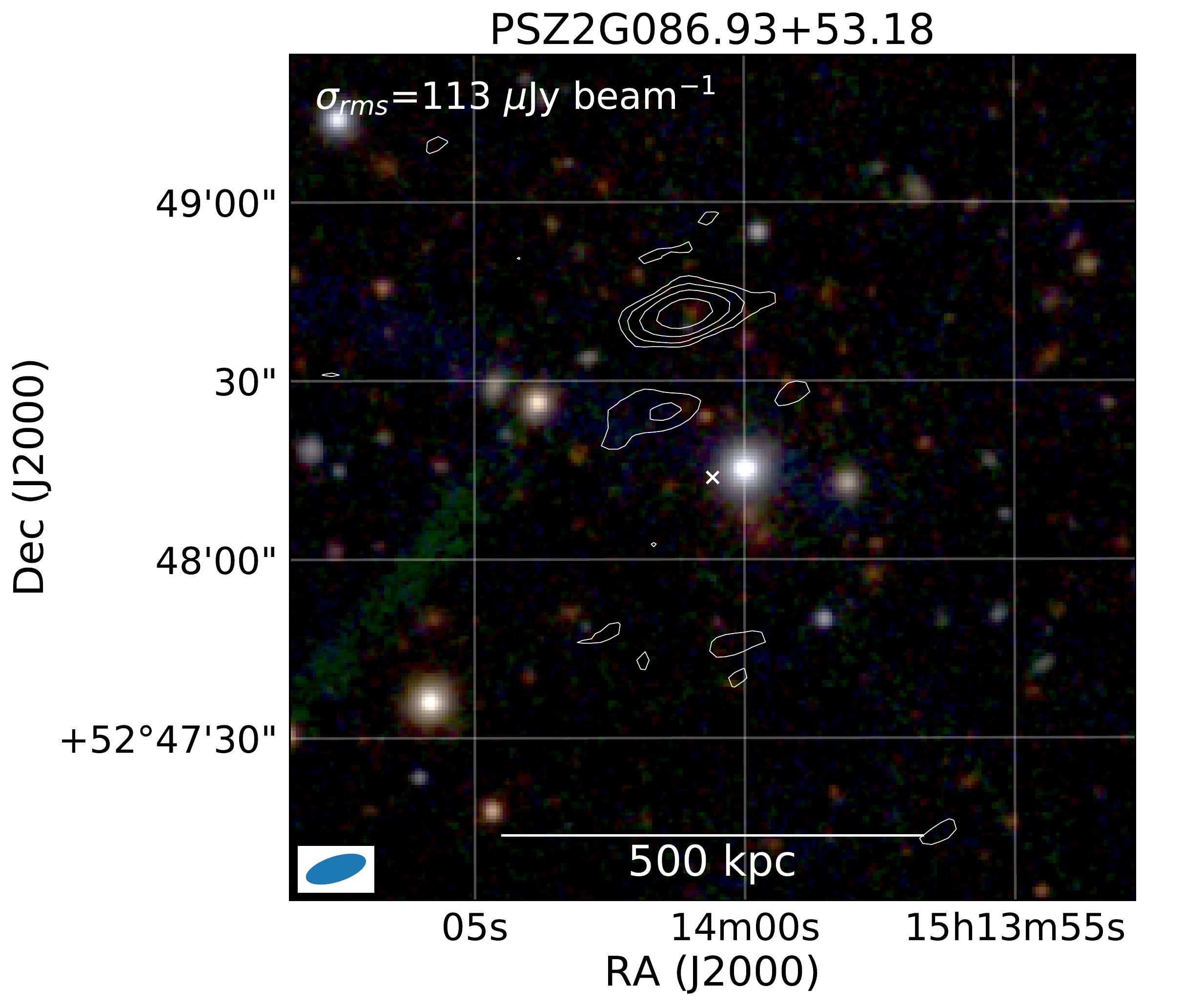}
   \caption{PSZ2\,G086.93+53.18. Left: Robust $-0.5$ radio image. Middle: XMM-Newton X-ray image with 15\arcsec~tapered radio contours (compact sources were subtracted). Right: Optical image with Robust $-0.5$ image radio contours. For more details see the caption of Figure~\ref{fig:A2018}.}
   \label{fig:PSZ2G08693}
\end{figure*}

Our low-resolution LOFAR image reveals diffuse emission with a largest extent of about 0.6~Mpc in this $z=0.6752$ cluster, see Figure~\ref{fig:PSZ2G08693}. This emission is centrally located and we therefore classify it as a radio halo with a flux density of $12.2\pm3.6$~mJy. This is consistent within the uncertainties with the value reported by \cite{gennaro} who presented the discovery of a radio halo in this cluster. \cite{2020MNRAS.497.5485Y} classify the cluster as a disturbed system based on Chandra observations.

\subsection{PSZ2\,G089.52+62.34, Abell\,1904}
\begin{figure*}
\centering
  \includegraphics[width=0.285\paperwidth]{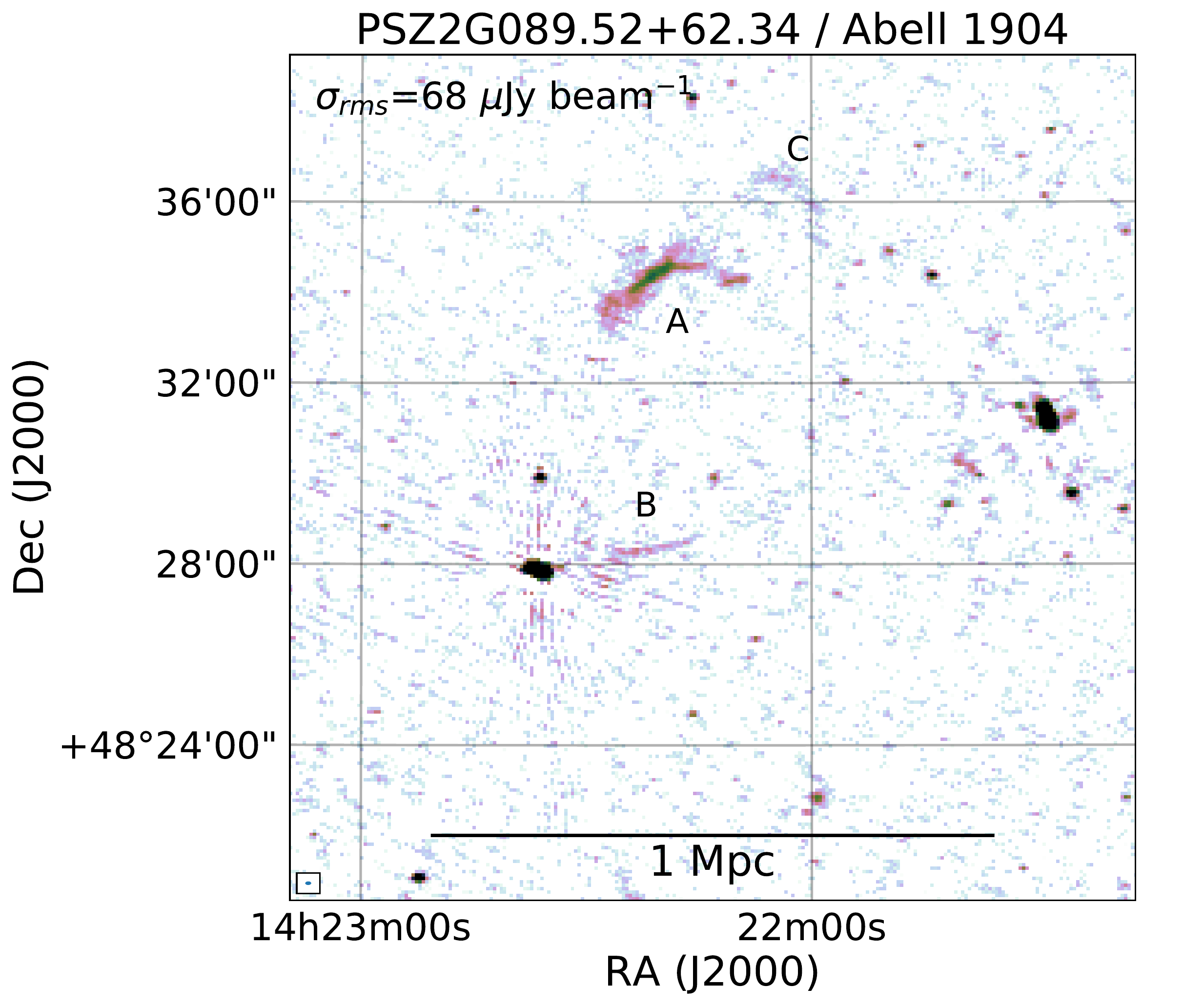}
  \includegraphics[width=0.285\paperwidth]{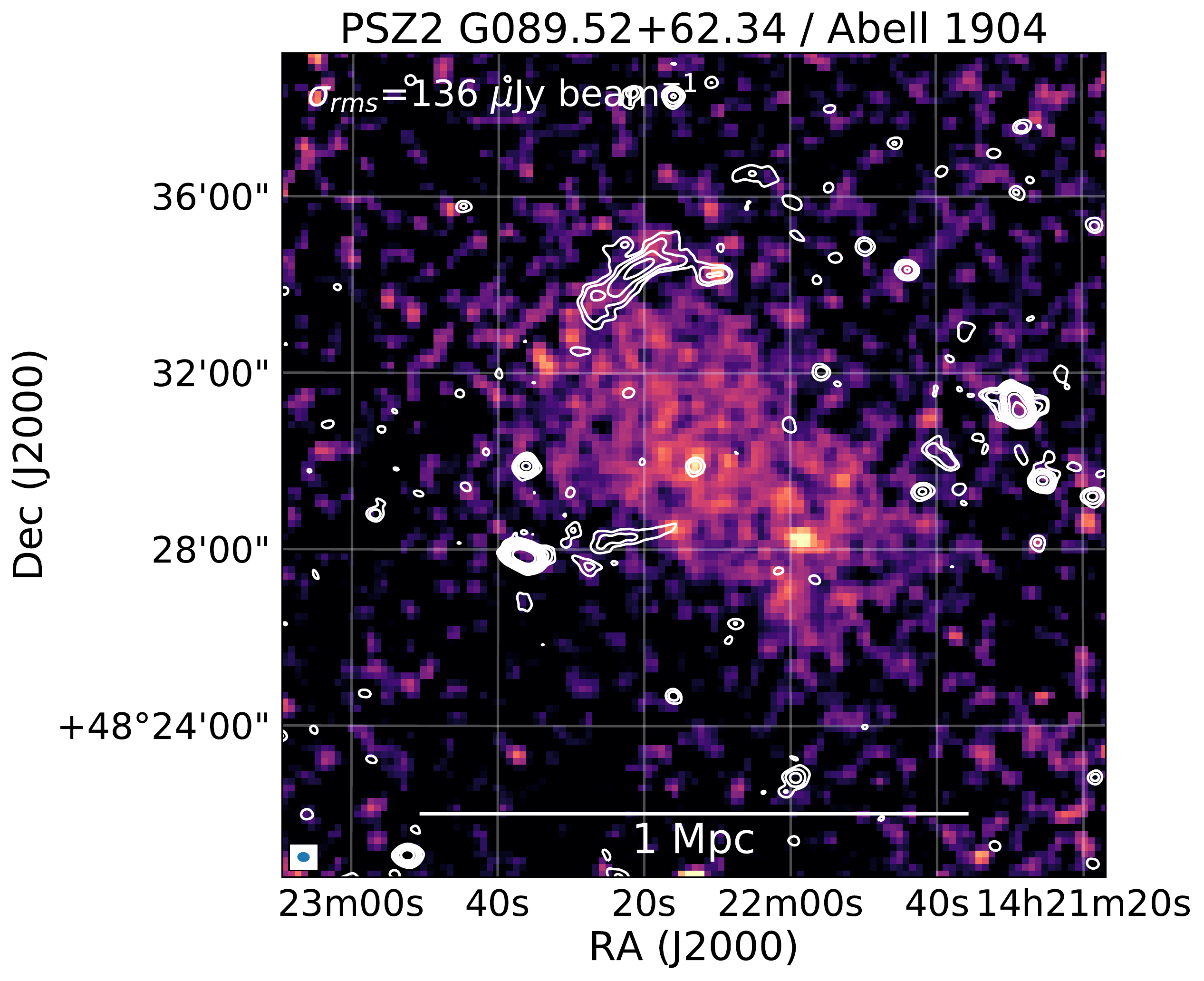}
  \includegraphics[width=0.285\paperwidth]{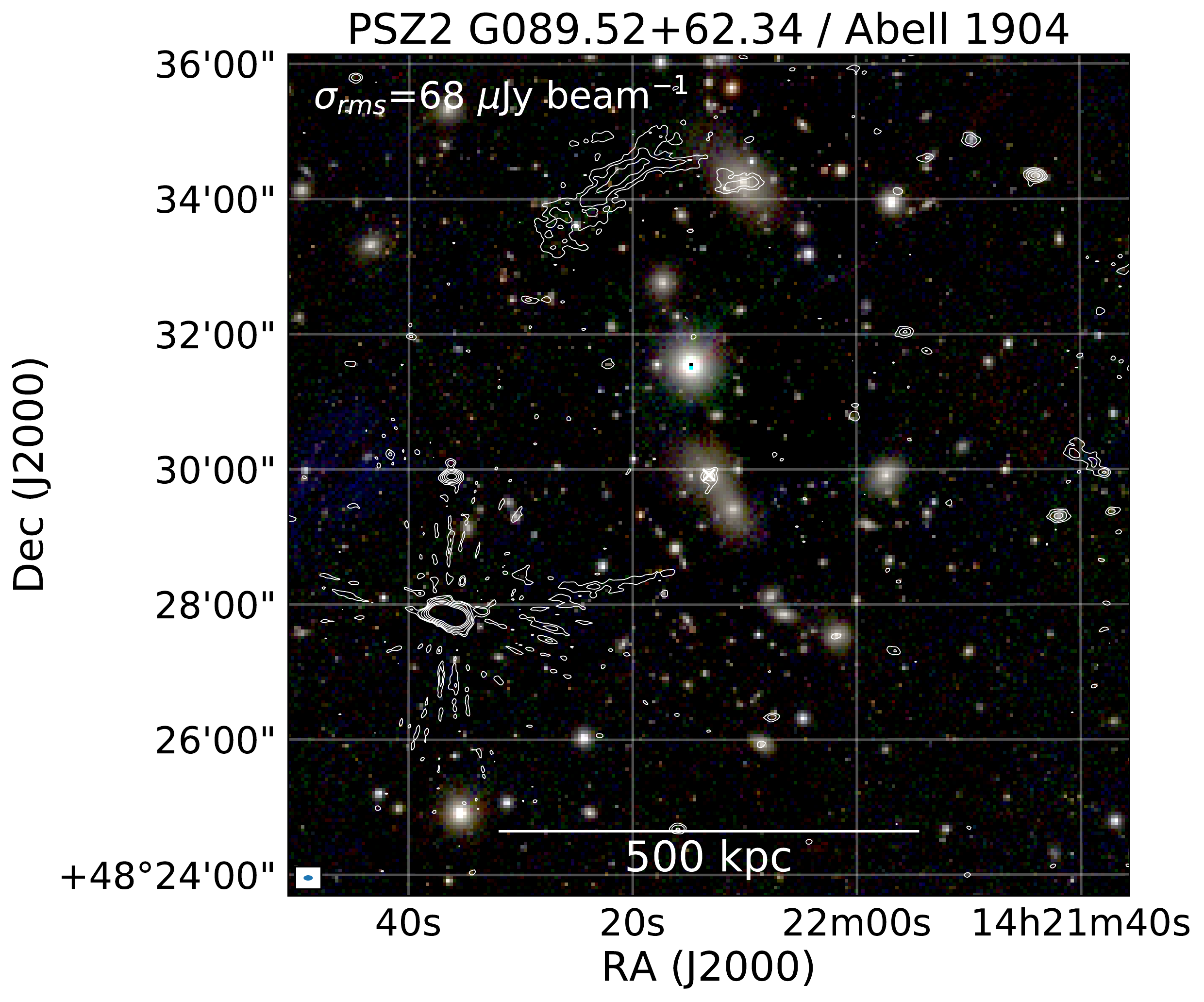}
   \caption{PSZ2\,G089.52+62.34 / Abell\,1904. Left: Robust $-0.5$ radio image. Middle: XMM-Newton X-ray image with 10\arcsec~tapered radio contours. Right: Optical image with Robust $-0.5$ image radio contours. For more details see the caption of Figure~\ref{fig:A2018}.}
   \label{fig:A1904}
\end{figure*}

The XMM-Newton image reveals a disturbed cluster, with the ICM being elongated in the NE-SW direction, see Figure~\ref{fig:A1904}. In the LOFAR image we detect two arc-like radio sources to the north (A) and SE (B) of the cluster center. At the redshift of the cluster ($z=0.0701$), these sources have an LLS of 230 and 170~kpc, respectively. Sources~A and~B are located at projected distances to cluster center of 350 and 150~kpc, respectively. Despite these relatively small distances, we consider these sources to be peripheral given the ICM distribution. 
A much fainter elongated source (C) is visible NW of source A. No optical counterparts are detected for these sources. A compact double lobed radio galaxy is located just west of source~A. Given the peripheral location of source~A with respect to the ICM distribution and elongated ICM shape, we classify A as a radio relic. The location of B is somewhat peculiar with respect to the overall ICM elongation. Given that  source~B is also affected by calibration artefacts from a nearby bright source, we classify source~B as a candidate radio relic. If the classification of source~B is confirmed this cluster hosts a double radio relic. The (candidate) relics in Abell\,1904 are relatively small compared to other well-studied relics, although we note that, for example, the western relic in  \object{ZwCl\,0008.8+5215} also has a small LLS of 290~kpc \citep{2019ApJ...873...64D}.


\subsection{PSZ2\,G095.22+67.41, RXC\,J1351.7+4622}
\begin{figure*}
\centering
   \includegraphics[width=1.0\columnwidth]{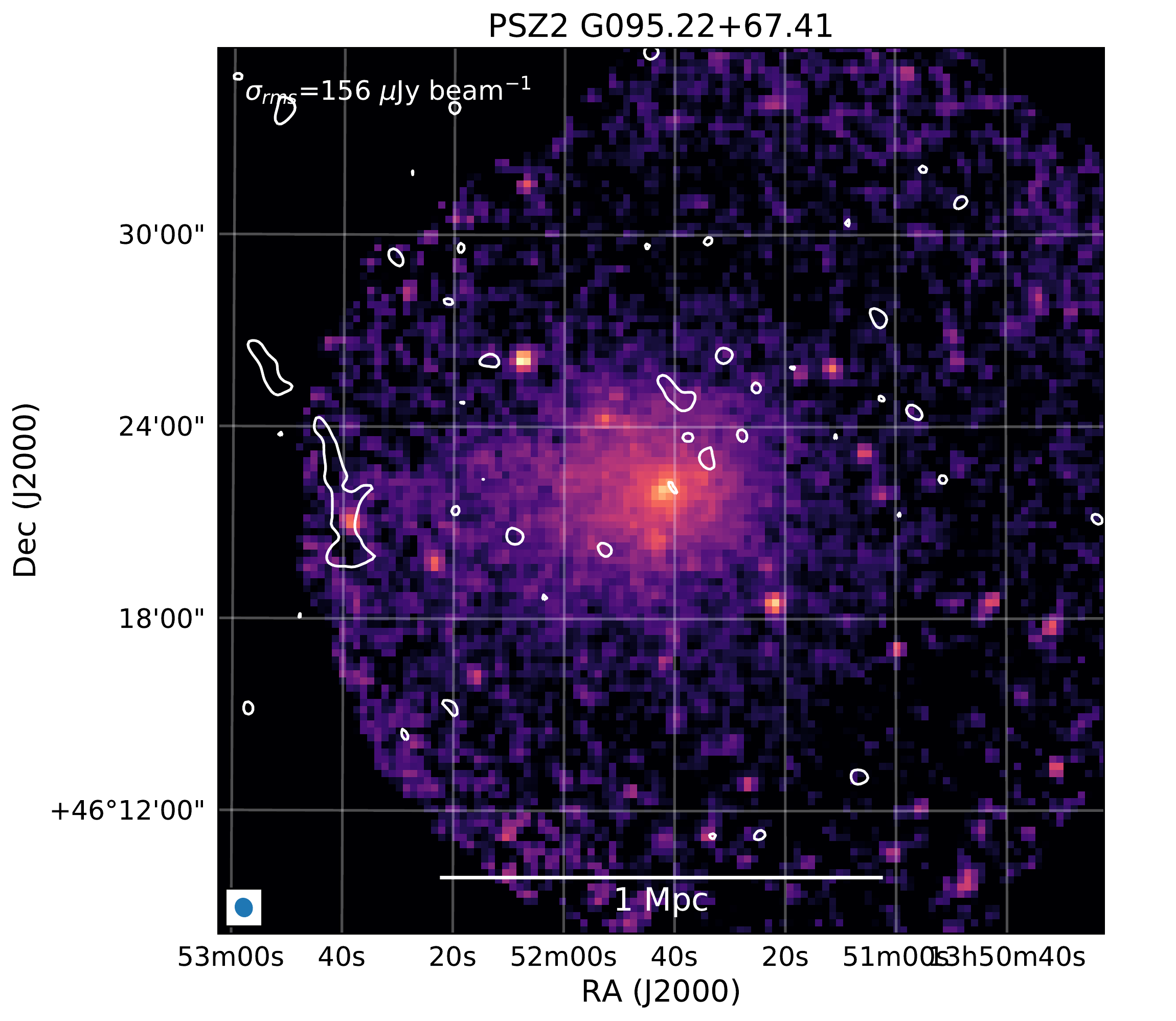}
   \includegraphics[width=1.0\columnwidth]{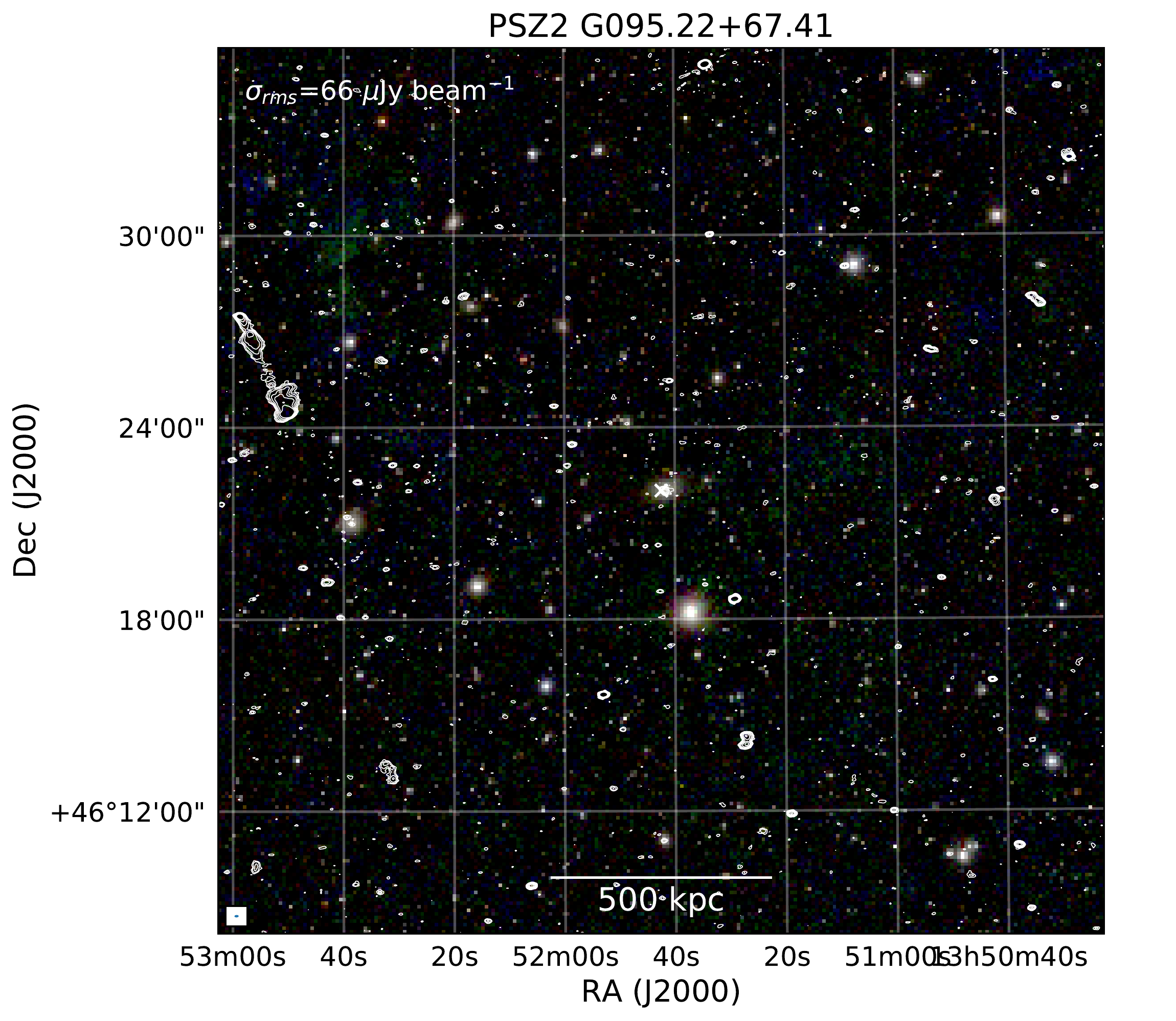}
   \caption{PSZ2\,G095.22+67.41, RXC\,J1351.7+4622. Left: XMM-Newton X-ray image with 30\arcsec~tapered radio contours (compact sources were removed). Right: Optical image with Robust $-0.5$ image radio contours. For more details see the caption of Figure~\ref{fig:A2018}.}
   \label{fig:PSZ2G09522}
\end{figure*}

PSZ2\,G095.22+67.41 is a relatively nearby cluster located at $z=0.0625$.
The XMM-Newton image reveals that the ICM peak coincides with the location of the BCG, see Figure~\ref{fig:PSZ2G09522}. A faint X-ray extension is visible  to the east. This suggests that the  cluster is not fully relaxed and is undergoing a merger event in the EW direction.

No central diffuse radio emission is found in our LOFAR image. However, we detect a NS elongated  source about 0.8~Mpc to the east of the cluster center. The source has an LLS of about 0.4~Mpc and is located south of a brighter radio galaxy.
We classify the source as a relic tracing a shock that could have originated from the EW merger event. This is consistent with the NS extent, highly elongated shape, and peripheral location of the radio source. An alternative explanation is that this source traces (old) AGN plasma from a bright elliptical galaxy (\object{MCG+08-25-051, $z=0.0623$}) that is located near this source. A faint ($\sim 0.5$~mJy) compact radio counterpart is detected in our LOFAR image for MCG+08-25-051. However, there is no evidence of jets or lobes originating from this source which then suggests that the emission resulted from a previous episode of AGN activity.

\subsection{PSZ2\,G099.86+58.45}
\begin{figure*}
\centering
   \includegraphics[width=0.285\paperwidth]{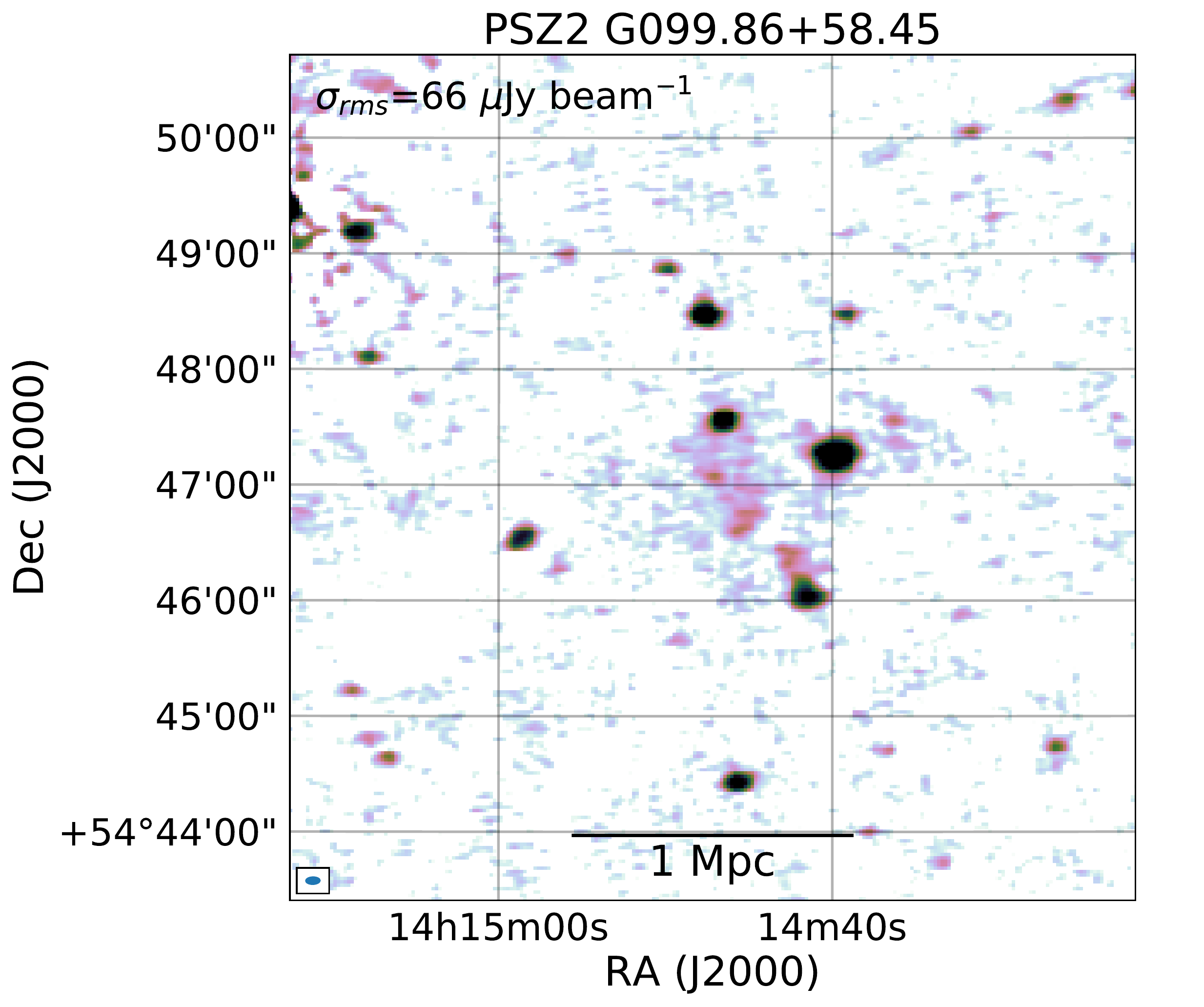}
   \includegraphics[width=0.285\paperwidth]{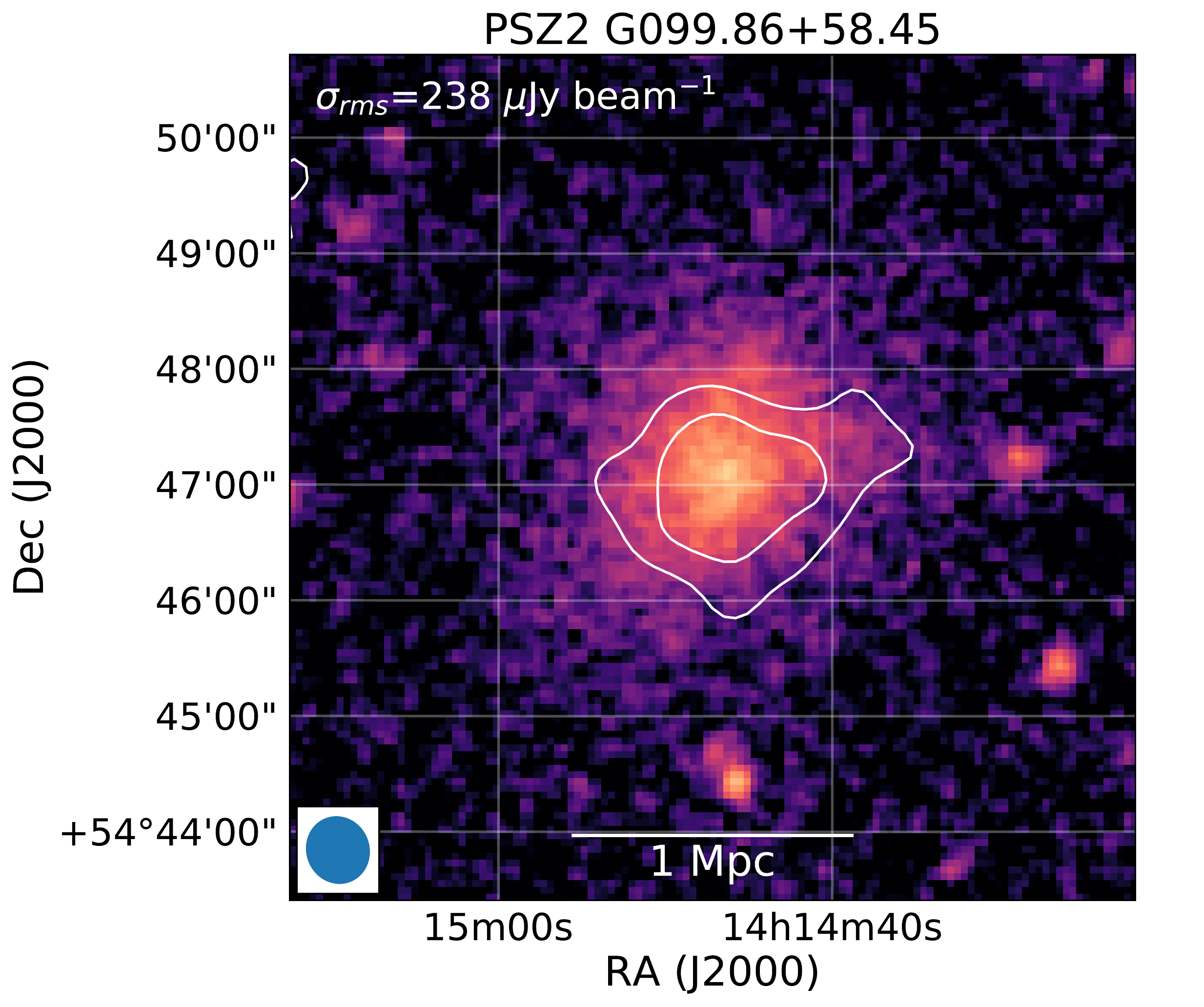}
   \includegraphics[width=0.285\paperwidth]{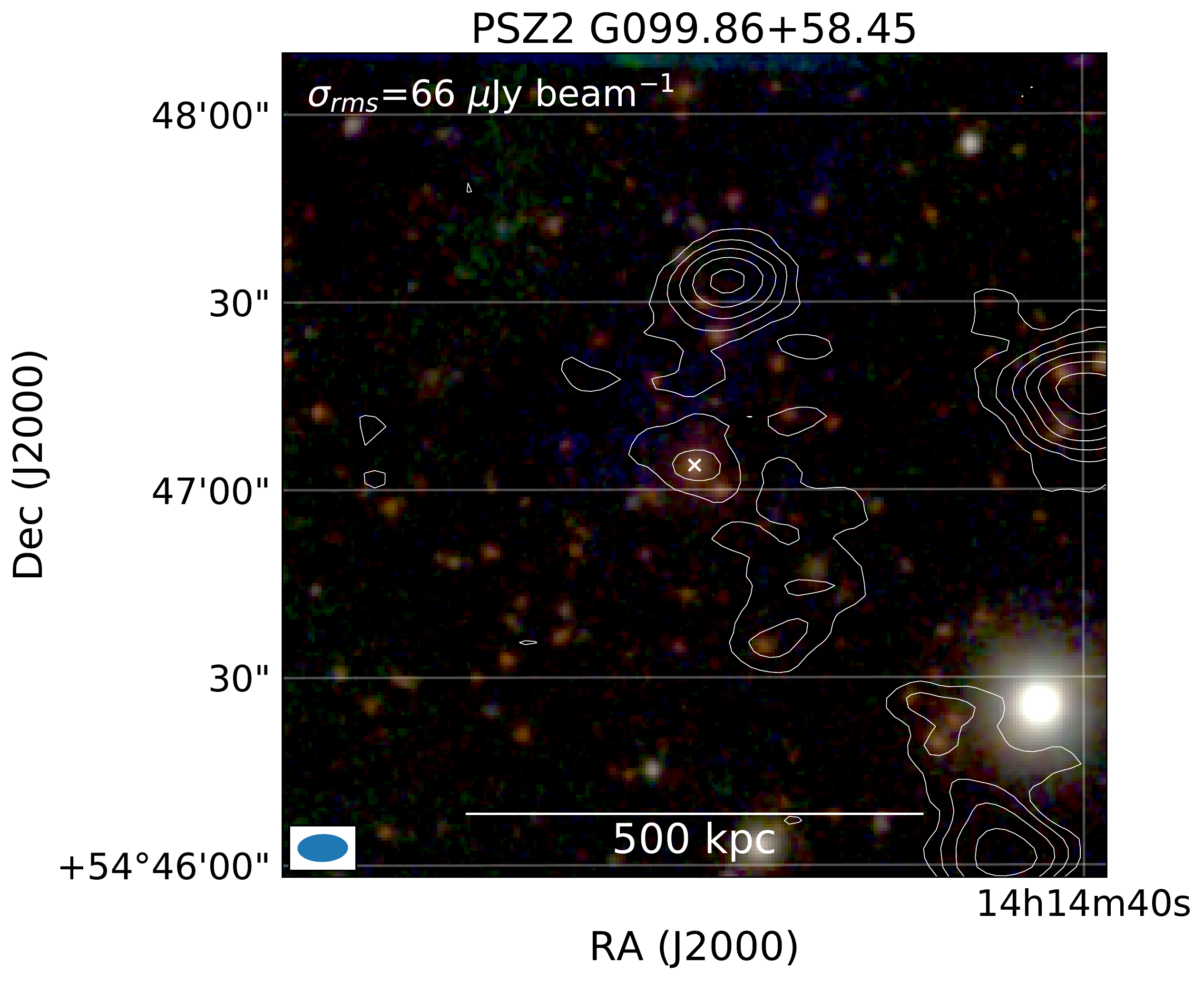}
   \caption{PSZ2\,G099.86+58.45. Left: Robust $-0.5$ radio image. Middle: XMM-Newton X-ray image with 30\arcsec~tapered radio contours (compact sources were subtracted). Right: Optical image with Robust $-0.5$ image radio contours. For more details see the caption of Figure~\ref{fig:A2018}.}
   \label{fig:PSZ2G09986}
\end{figure*}

PSZ2\,G099.86+58.45 is a massive cluster (M$_{500}=6.85^{+0.48}_{-0.49} \times 10^{14}$~M$_{\odot}$) with a global temperature of 8.9$^{+2.8}_{-1.1}$~keV \citep{2018NatAs...2..744S}. The cluster is known to be undergoing a merger event. The discovery of a 1~Mpc radio halo in this distant ($z=0.616$) cluster, see Figure~\ref{fig:PSZ2G09986}, was reported by \cite{2019ApJ...881L..18C} based on LOFAR data. From an elliptical fit we determine $S_{144}=14.7\pm3.2$~mJy. This number somewhat lower than the $25.3\pm5.7$ reported by \cite{2019ApJ...881L..18C}. This difference is caused by a small extension of the radio halo around compact source~B (discussed by \cite{2019ApJ...881L..18C}). If this region is  included, the measurements are consistent with each other given the uncertainties.

\subsection{PSZ2\,G106.61+66.71}
\begin{figure*}
\centering
   \includegraphics[width=0.285\paperwidth]{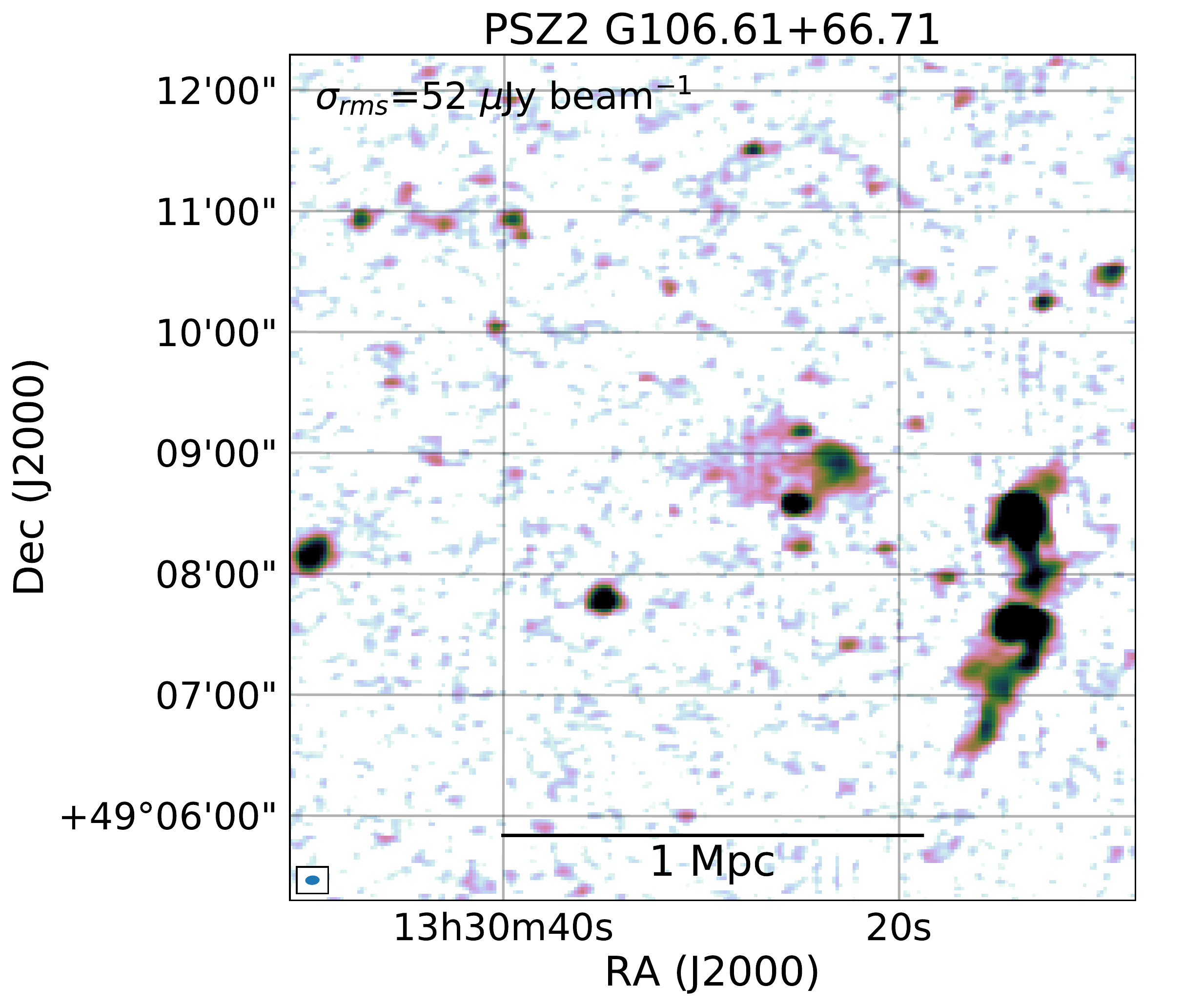}
   \includegraphics[width=0.285\paperwidth]{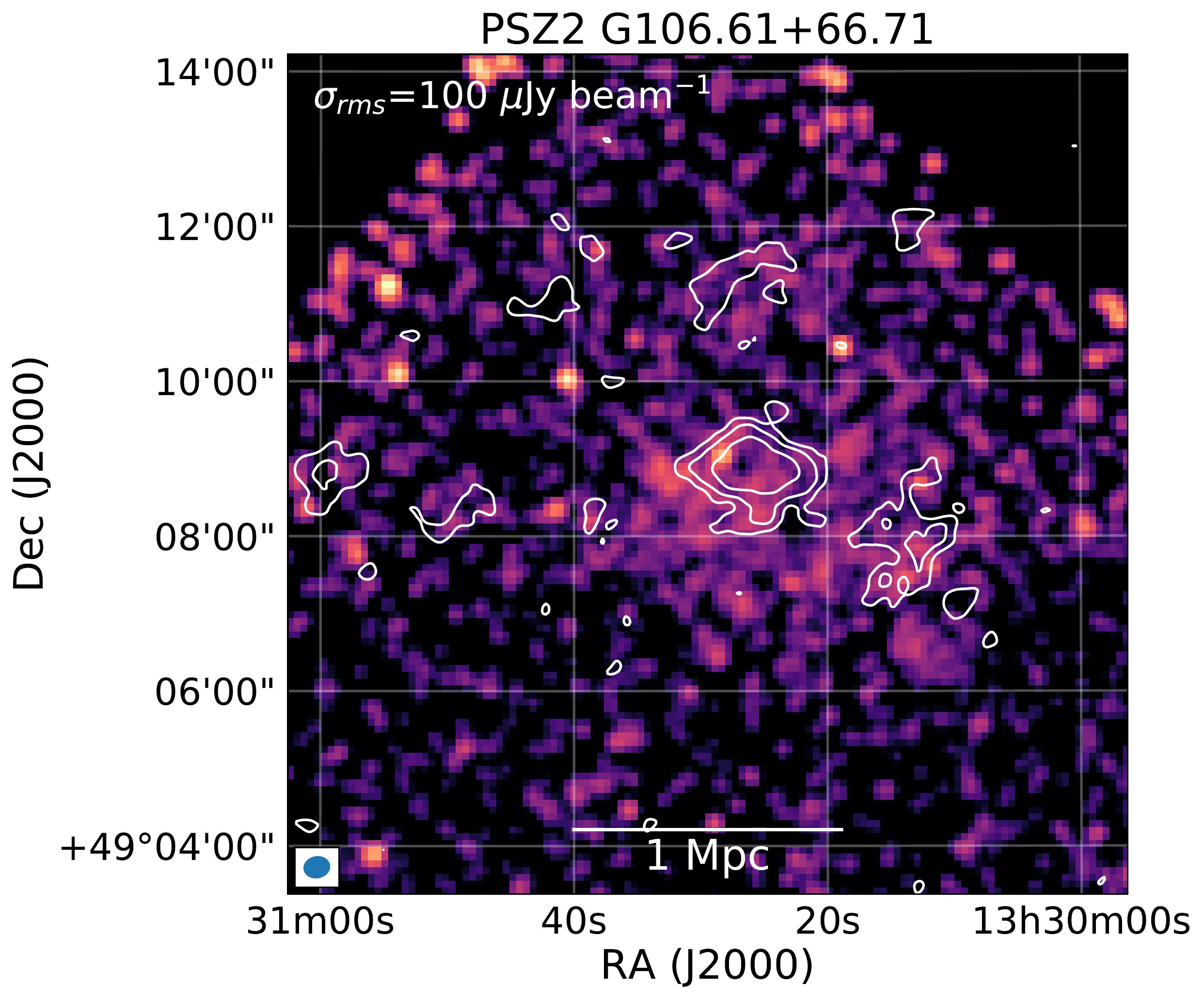}
   \includegraphics[width=0.285\paperwidth]{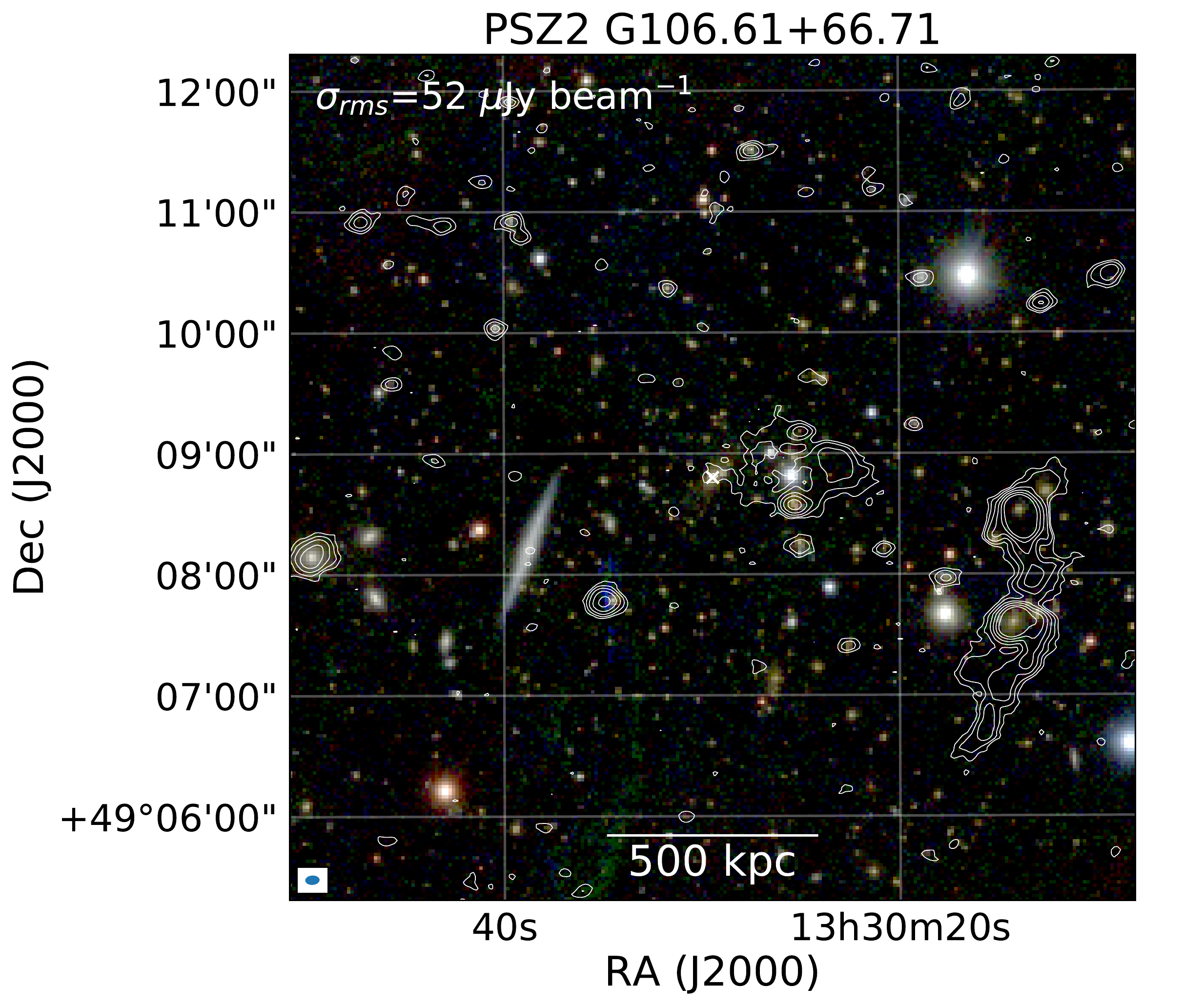}
   \caption{PSZ2\,G106.61+66.71. Left: Robust $-0.5$ radio image. Middle: Chandra X-ray image with 10\arcsec~tapered radio contours (compact sources were subtracted). Right: Optical image with Robust $-0.5$ image radio contours. For more details see the caption of Figure~\ref{fig:A2018}.}
   \label{fig:PSZ2G10661}
\end{figure*}

The LOFAR image displays two bright, partly blended tailed radio galaxies in the western part of the cluster, see Figure~\ref{fig:PSZ2G10661}. The Chandra X-ray image shows a cluster without a strongly peaked core. In addition, the cluster is elongated in the direction of the two tailed radio galaxies. In the center of the cluster we detect diffuse radio emission with an LLS of about 0.5~Mpc. Within this diffuse emission we find two compact AGN associated with the cluster, including the BCG. We consider a mini-halo origin unlikely given the lack of centrally peaked X-ray emission.  The western part of the central diffuse emission has a rather high surface brightness, suggesting a link with a cluster AGN. On the other hand, some of the emission (in particular the eastern part) has a smooth morphology somewhat similar to a radio halo. For the above reasons we list the source as candidate radio halo. 


\subsection{PSZ2\,G107.10+65.32, Abell 1758}
\begin{figure*}
\centering
   \includegraphics[width=0.285\paperwidth]{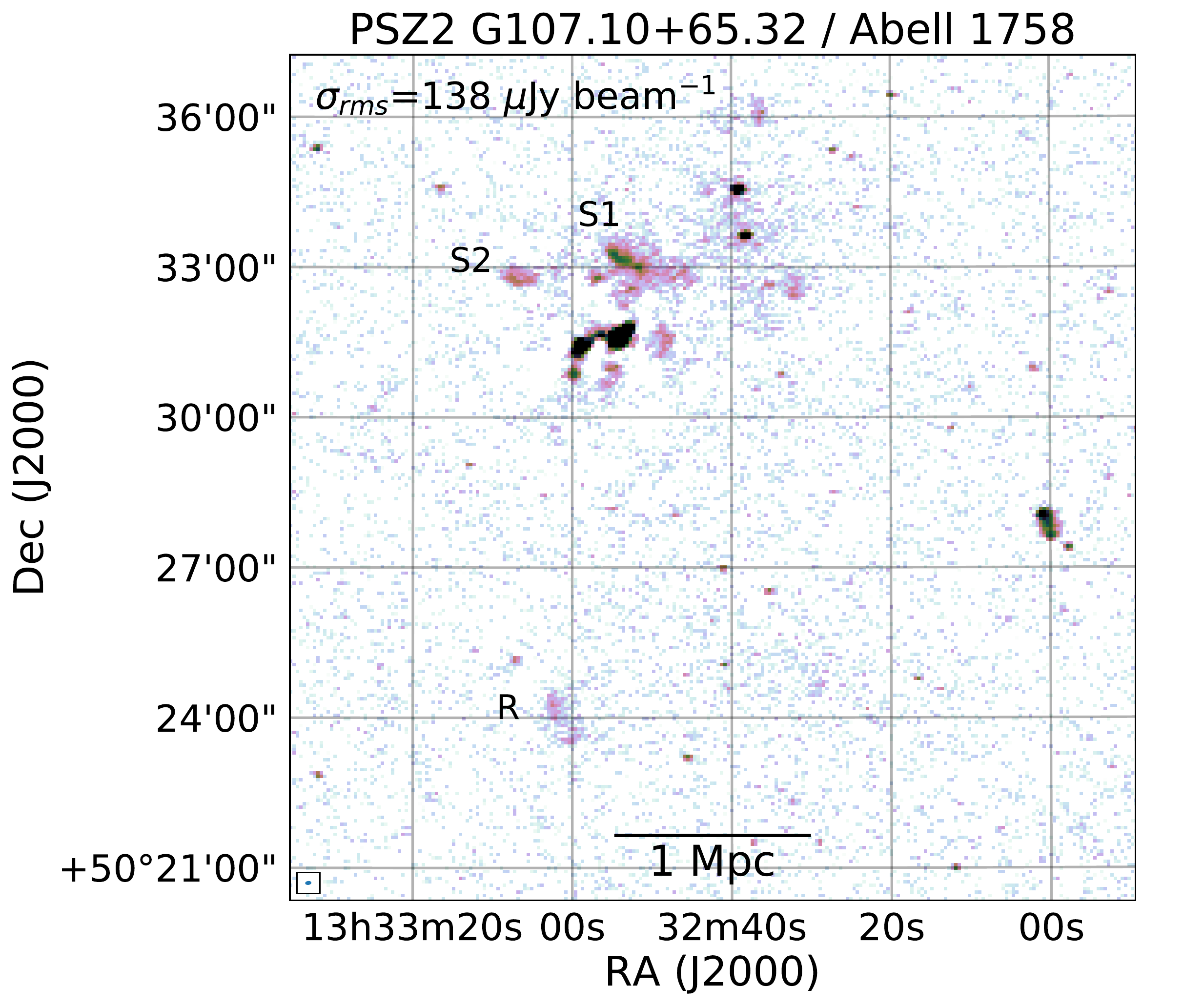}
   \includegraphics[width=0.285\paperwidth]{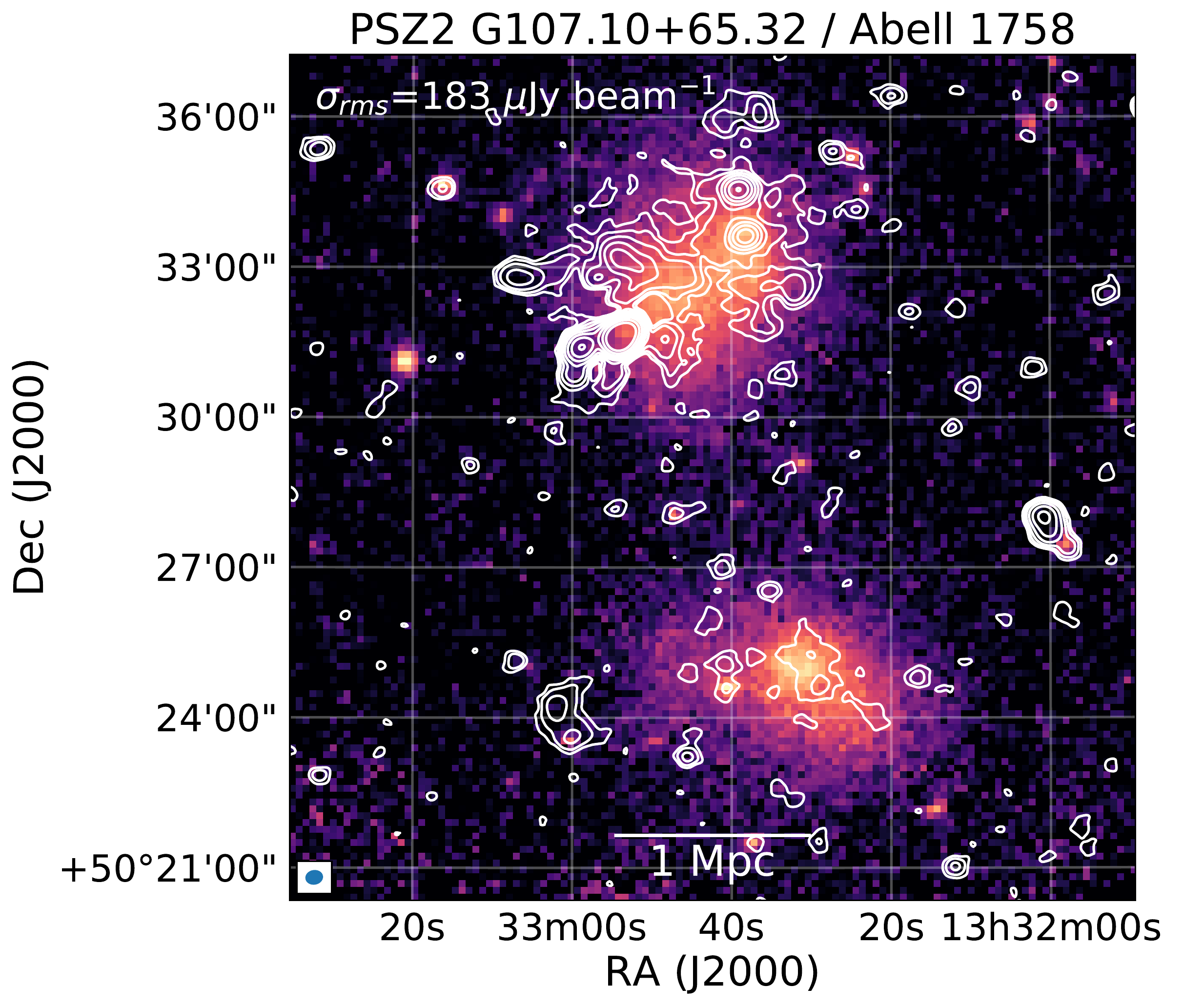}
   \includegraphics[width=0.285\paperwidth]{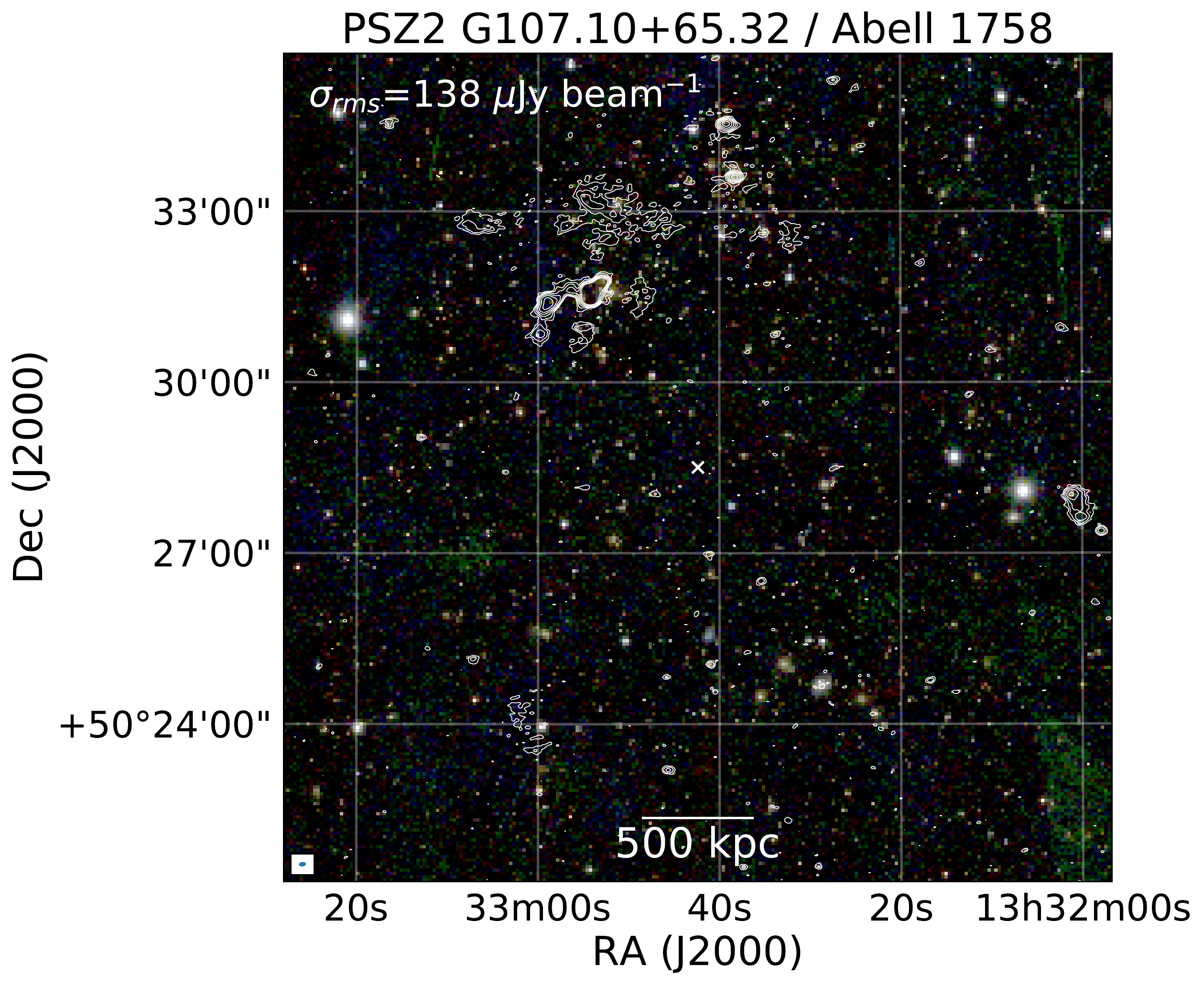}
   \caption{PSZ2\,G107.10+65.32 / Abell\,1758. Left: Robust $-0.5$ radio image. Middle: Chandra X-ray image with 10\arcsec~tapered radio contours. Right: Optical image with Robust $-0.5$ image radio contours. For more details see the caption of Figure~\ref{fig:A2018}.}
   \label{fig:A1758}
\end{figure*}

Abell\,1758 is composed of two main clusters, Abell\,1758N and Abell\,1758S, see Figure~\ref{fig:A1758}. These two clusters appear to be in a pre-merger phase \citep{1998MNRAS.301..328R,2004ApJ...613..831D}. The individual clusters themselves are disturbed and already undergoing their own merger events. The presence of a radio halo in A1758N was first reported by \cite{2001ApJ...548..639K} and further studied by \cite{2009A&A...507.1257G} and \cite{2013A&A...551A..24V}.

\cite{2018MNRAS.478..885B} studied this cluster with LOFAR HBA observations, a subset of the data used here. They measured a size of about 2~Mpc for the A1758N radio halo and discovered a 1.6~Mpc radio halo and a $\sim 0.5$~Mpc relic (labelled R) in A1758S. In addition, a hint of faint emission was detected between A1758N and A1758S. This ``radio bridge'' was subsequently confirmed by \cite{2020MNRAS.499L..11B}. Two bright patches, labelled S1 and S2, were found in A1758N. They are not directly associated with an optical counterpart and possibly trace compressed AGN fossil plasma. 

For the radio halo in A1758N we find a flux of $123\pm25$~mJy based on the elliptical model fit. This value is lower than the $307\pm63$~mJy measured earlier by \cite{2020MNRAS.499L..11B}. The reason for this difference is that we excluded the region around sources A and~B \citep[see][]{2018MNRAS.478..885B} and S1 and S2. This particular case highlights the difficulties in classifying and separating the emission from different components with uncertain identifications. For the southern radio halo we determine $S_{144}= 63\pm14$~mJy from the elliptical model, consistent with the value reported by \cite{2018MNRAS.478..885B}. For a more comprehensive description of the LOFAR findings we refer the reader to \cite{2018MNRAS.478..885B,2020MNRAS.499L..11B}.

\subsection{PSZ2\,G111.75+70.37, Abell\,1697}
\begin{figure*}
\centering
   \includegraphics[width=1.0\columnwidth]{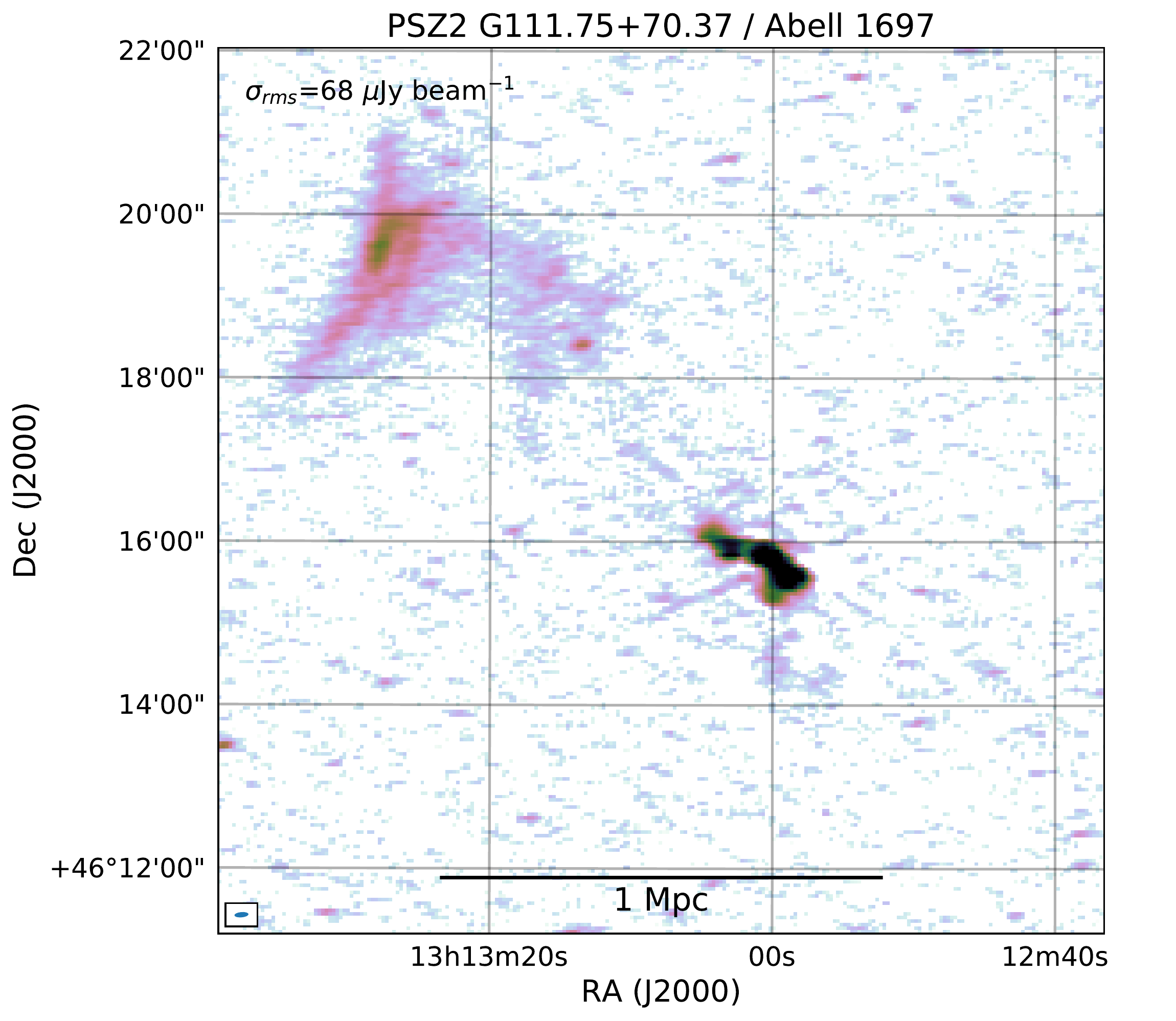}
   \includegraphics[width=1.0\columnwidth]{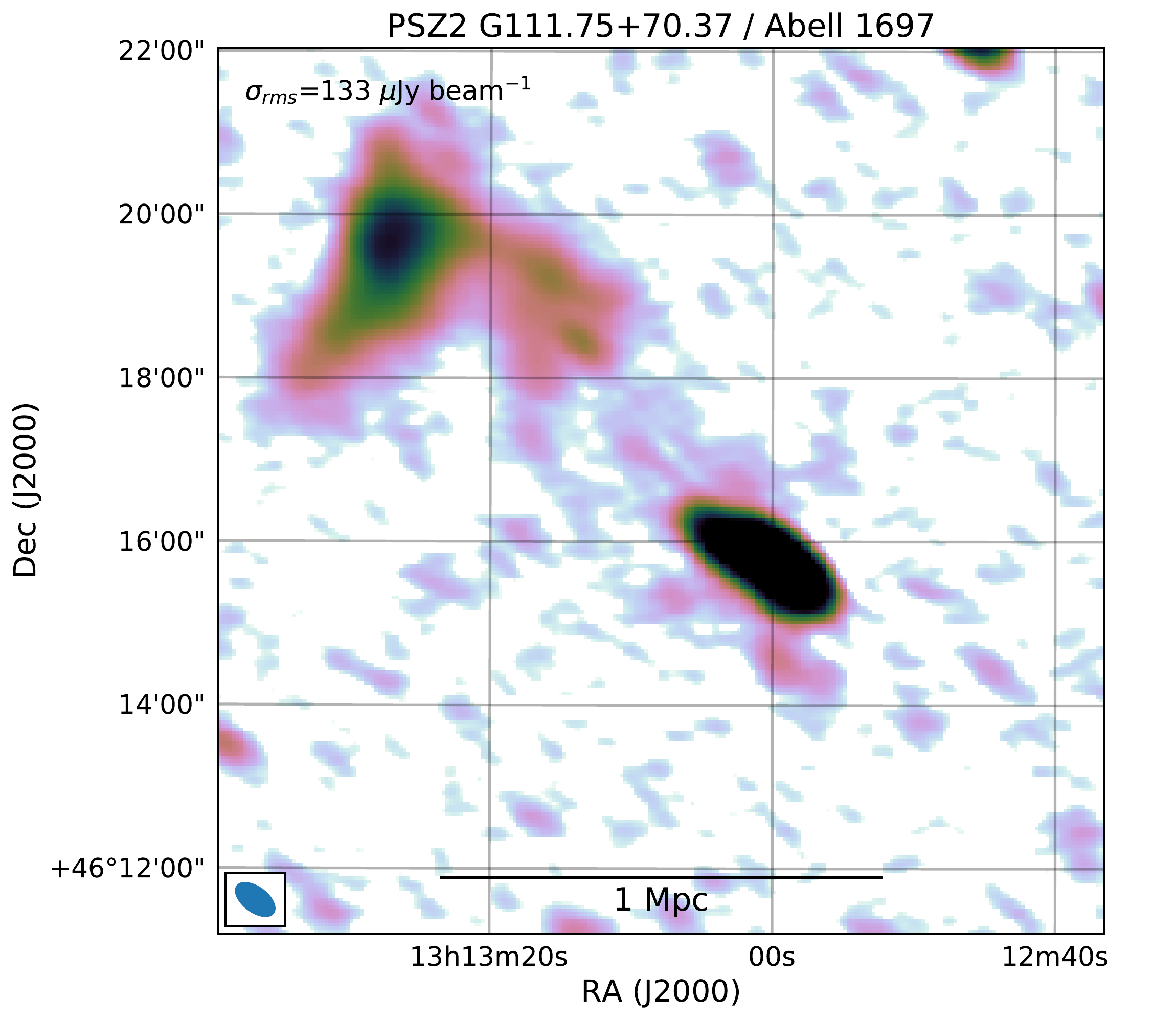}
   \includegraphics[width=1.0\columnwidth]{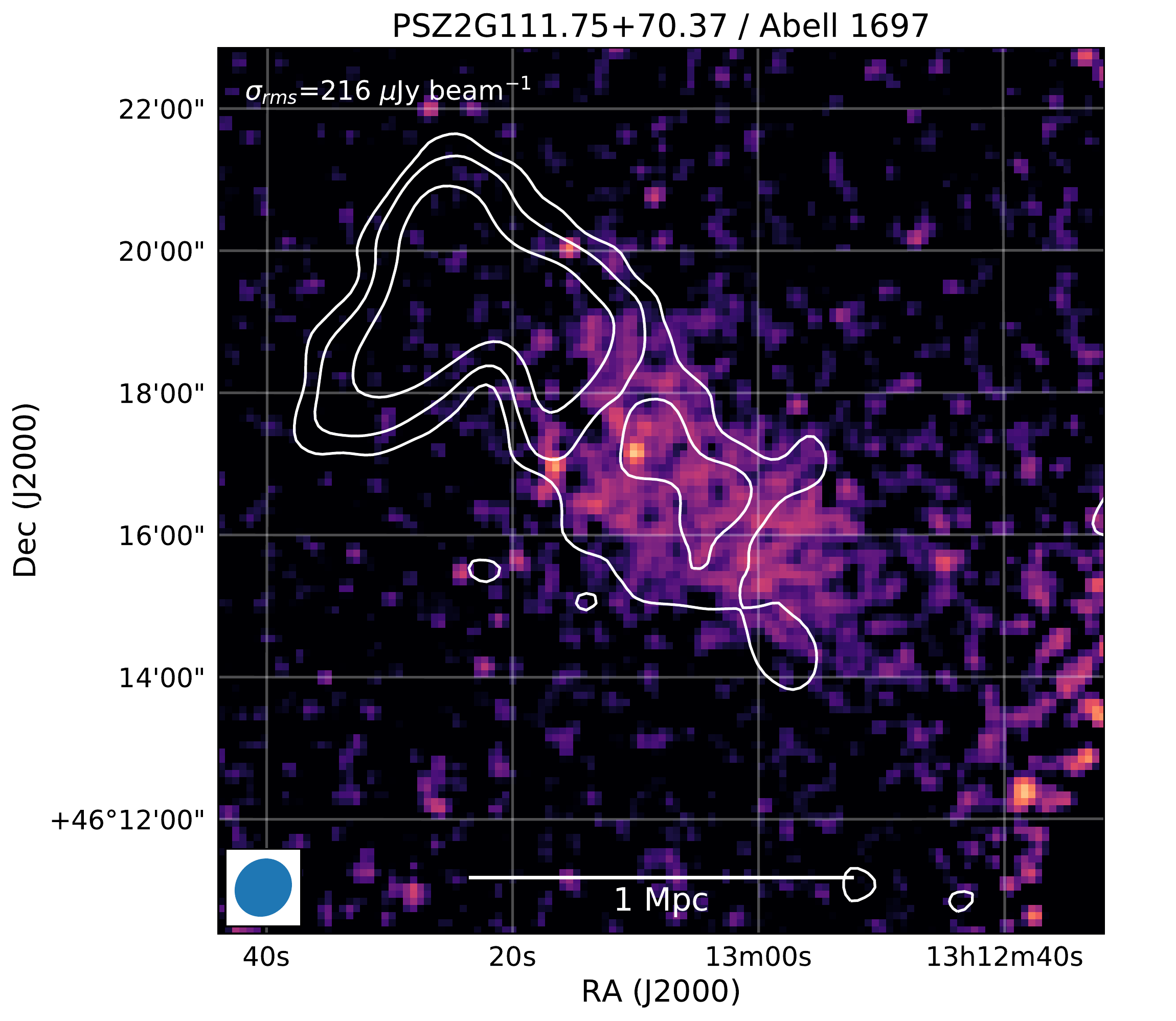}
   \includegraphics[width=1.0\columnwidth]{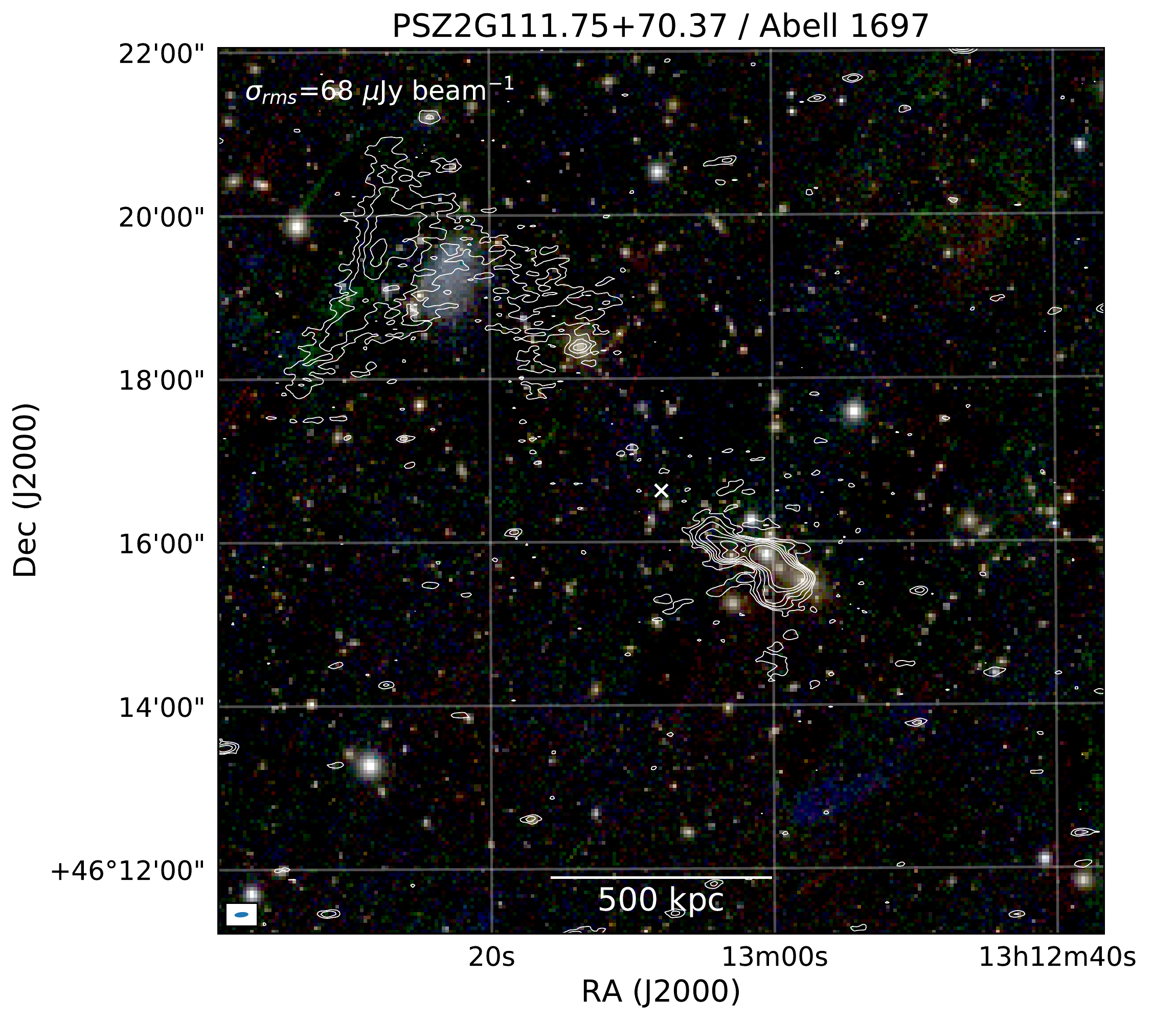}
   \caption{PSZ2\,G111.75+70.37 / Abell\,1697. Top Left: Robust $-0.5$ radio image. Top right: 10\arcsec~tapered radio image. Bottom left: XMM-Newton X-ray image with 30\arcsec~tapered radio contours (compact sources were subtracted). Bottom right: Optical image with Robust $-0.5$ image radio contours. For more details see the caption of Figure~\ref{fig:A2018}.}
   \label{fig:PSZ2G11175}
\end{figure*}

The XMM-Newton X-ray image shows an elongated cluster without a central concentration. The LOFAR images reveal two prominent radio sources in this cluster, see Figure~\ref{fig:PSZ2G11175}. One is located near the southwestern BCG of the cluster. The second source is more extended and located at the NE periphery of the cluster. This emission is also discussed by \cite{2020A&A...633A..59P} based on LoTSS DR1 images \citep{2019A&A...622A...1S}. The source consists of a NW-SE elongated structure with an LLS of about 700~kpc, 
as well as emission trailing SW towards a second BCG. Compact radio emission from this BCG is detected, but no obvious morphological connection with the extended source is visible. The extended source somewhat resembles the Toothbrush relic \citep{2012A&A...546A.124V}. 
We note that the extended radio source partly overlaps with the nearby irregular dwarf galaxy UGC\,8308 \citep[DDO\,167; e.g.,][]{1998A&AS..128..325T}. However, we consider an association of the radio source with the cluster more likely. We therefore classify the source as a radio relic. This classification is also consistent with the elongation of the ICM, suggesting a NE-SW merger event. In the low resolution radio images extended emission is visible in the region between the NE and the SW BCG. This emission is classified as a radio halo since it follows the X-ray emission from the ICM. The flux density of the halo is $27.7\pm6.3$~mJy. We note that this emission is also discussed by \cite{2020A&A...633A..59P} (``trailing relic emission'') and it is concluded that this emission has an ultra-steep spectrum. We note however that the flux density estimate provided by  \cite{2020A&A...633A..59P} is incorrect as the emission is not fully deconvolved in the LoTSS DR1 images and thus measurements directly from those images are complicated. This leads to a large overestimation of the flux density as is discussed by \cite{2019A&A...622A...1S}. This problem is addressed to some extent in the upcoming DR2.

\subsection{PSZ2\,G114.31+64.89, Abell 1703}
\begin{figure*}
\centering
   \includegraphics[width=0.285\paperwidth]{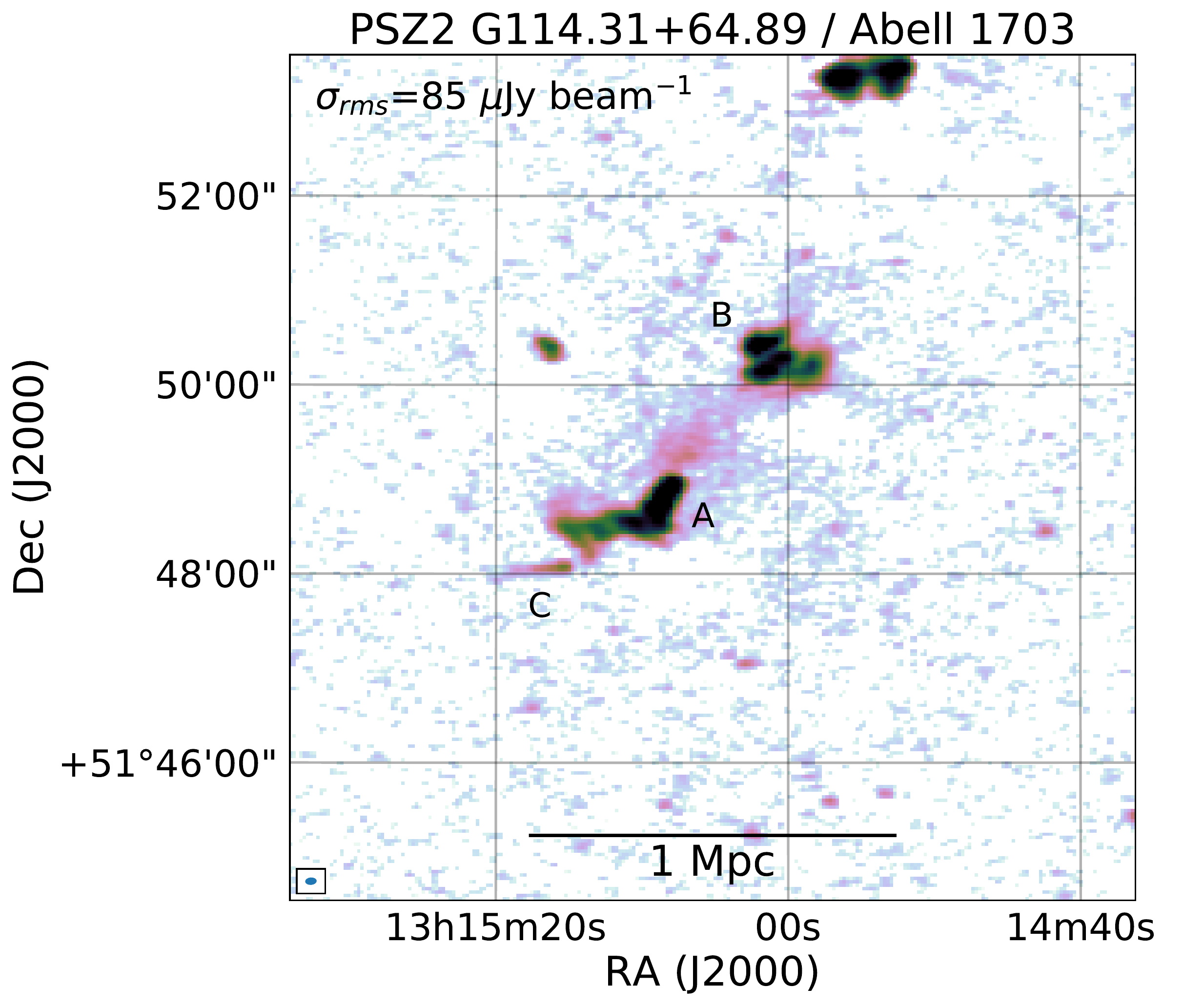}
   \includegraphics[width=0.285\paperwidth]{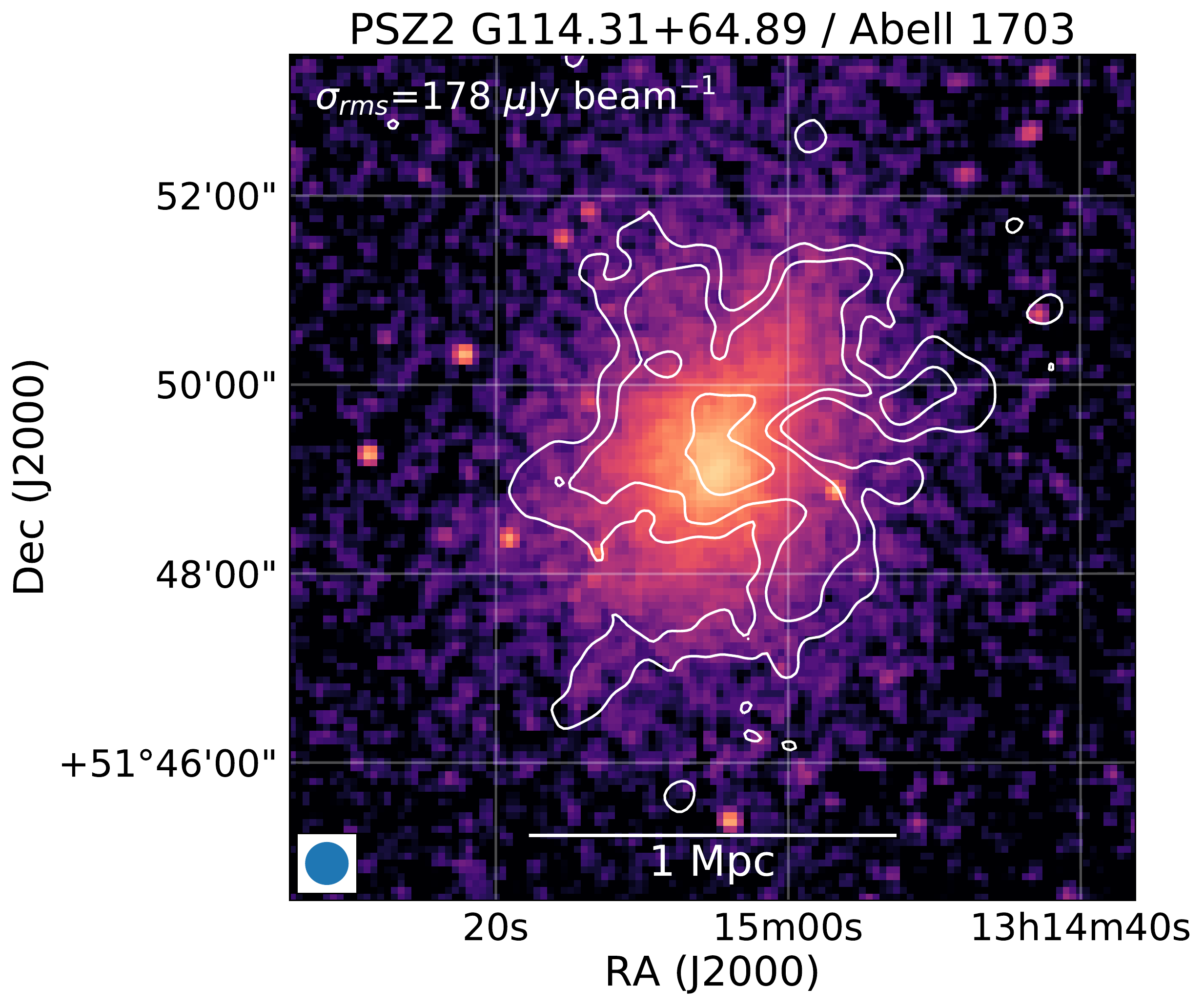}
   \includegraphics[width=0.285\paperwidth]{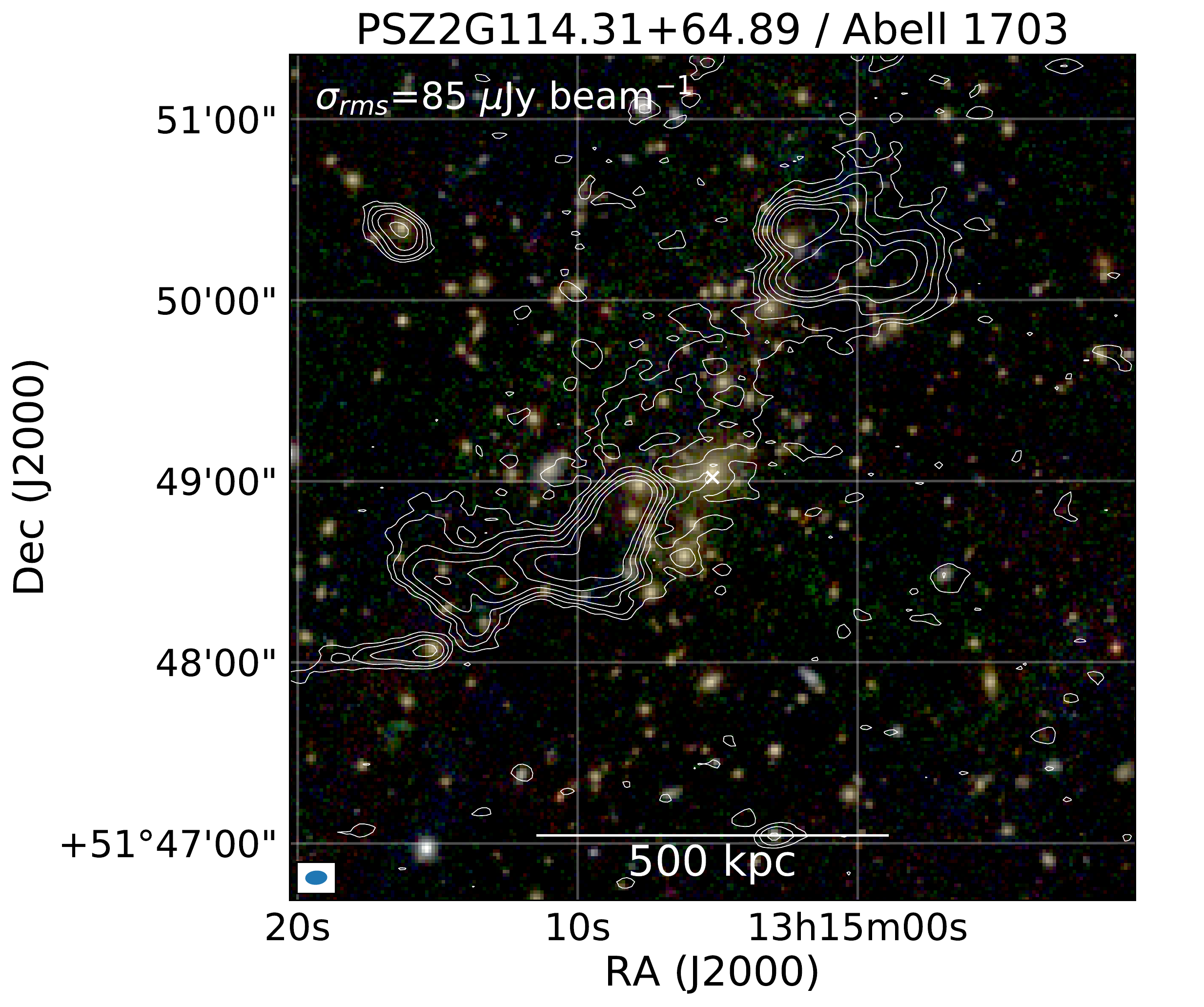}
   \caption{PSZ2\,G114.31+64.89 / Abell\,1703. Left: Robust $-0.5$ radio image. Middle: Chandra X-ray image with 15\arcsec~tapered radio contours (compact sources were subtracted). Right: Optical image with Robust $-0.5$ image radio contours. For more details see the caption of Figure~\ref{fig:A2018}.}
   \label{fig:A1703}
\end{figure*}

The massive cluster Abell 1703~has been extensively studied at optical wavelengths due to its gravitational lensing properties. It is generally classified as a relaxed cluster \citep[e.g.,][]{2011ApJ...729..127U,2009A&A...498...37R}. The Chandra image shows an overall regular appearance with the cluster being somewhat elongated in the NNW-SSE direction, see Figure~\ref{fig:A1703}. \cite{2017ApJ...843...76A} determined the concentration parameters, cuspiness, and central density of the cluster using Chandra X-ray data. These properties, however, point to a non-cool core classification. Furthermore, a recent dynamical analysis from \cite{2020MNRAS.492.2405B} indicates that this is a merging cluster consisting of two or three subclumps. 

The LOFAR image reveals the presence of central diffuse emission, filling the region around between tailed radio source A and a complex compact region of emission which we labelled B. Source B is likely related to AGN activity. A faint tailed source, labelled C, is located in the SE part of the cluster. The total extent of the diffuse emission is difficult to determine because of A and B. Given the central location and extent of at least 0.5~Mpc, we classify the diffuse emission as a radio halo with a integrated flux density of $91.7\pm18.6$~mJy based on an elliptical model fit.

\subsection{PSZ2\,G114.99+70.36, Abell\,1682}
\begin{figure*}
\centering
   \includegraphics[width=0.285\paperwidth]{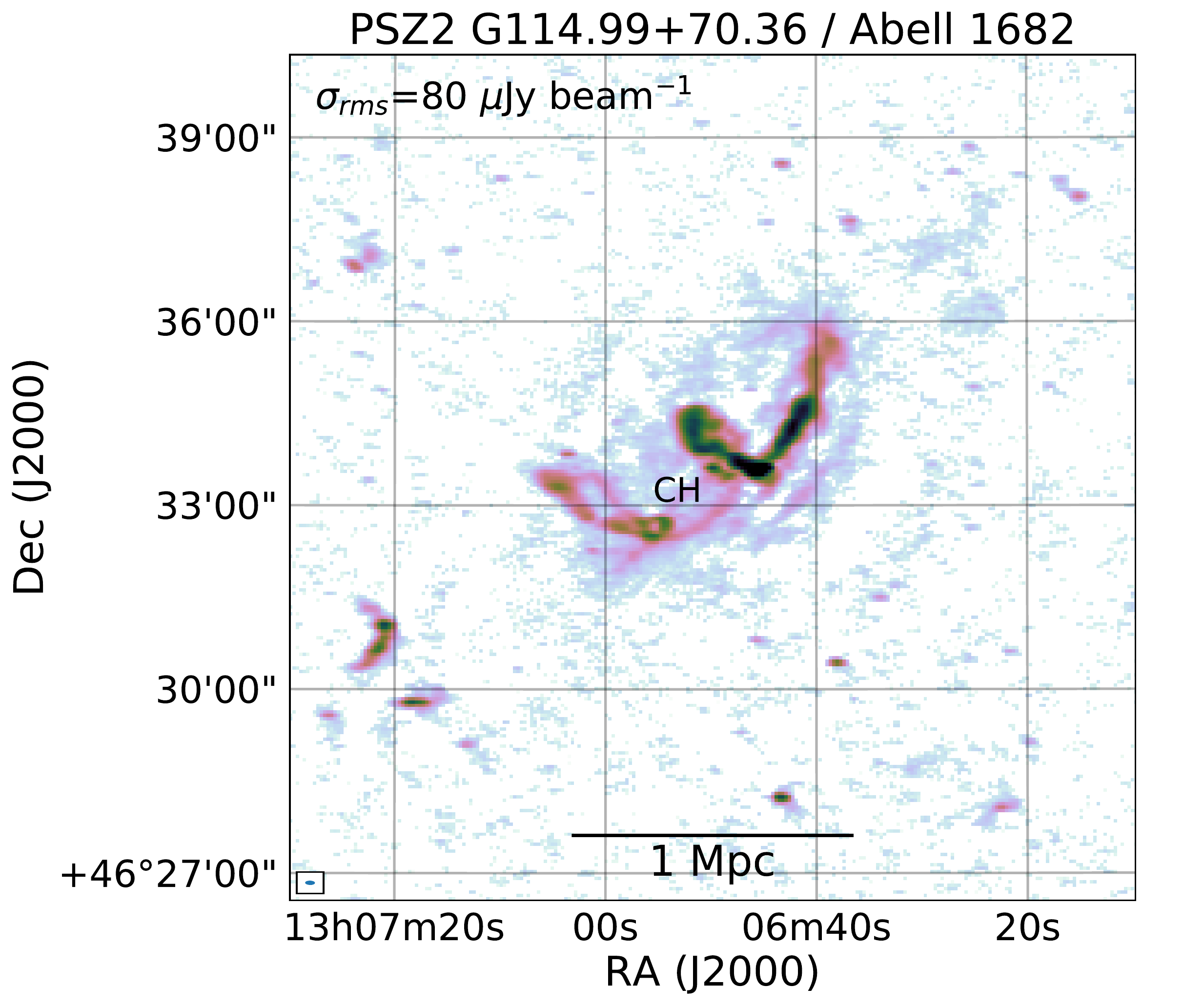}
   \includegraphics[width=0.285\paperwidth]{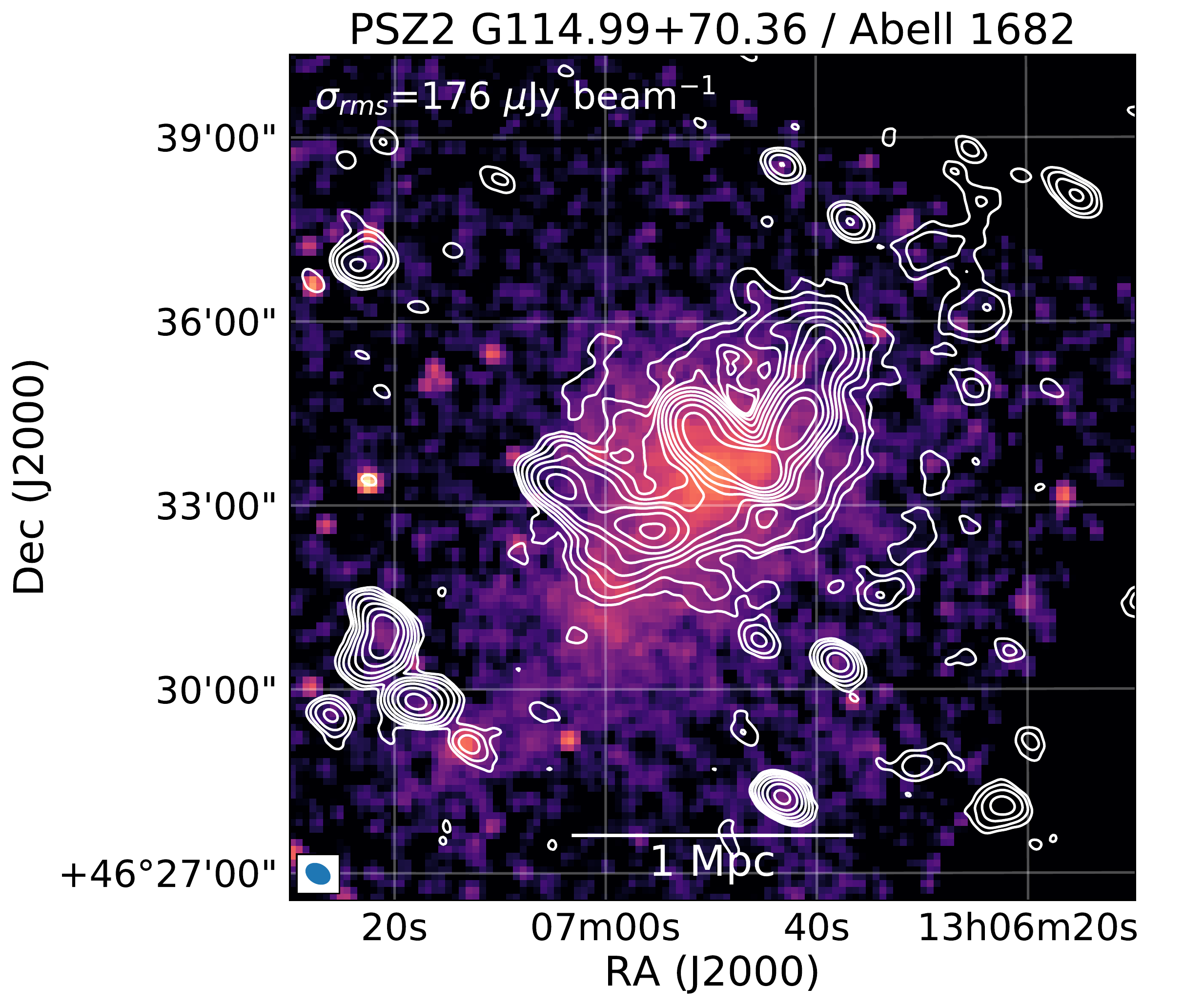}
   \includegraphics[width=0.285\paperwidth]{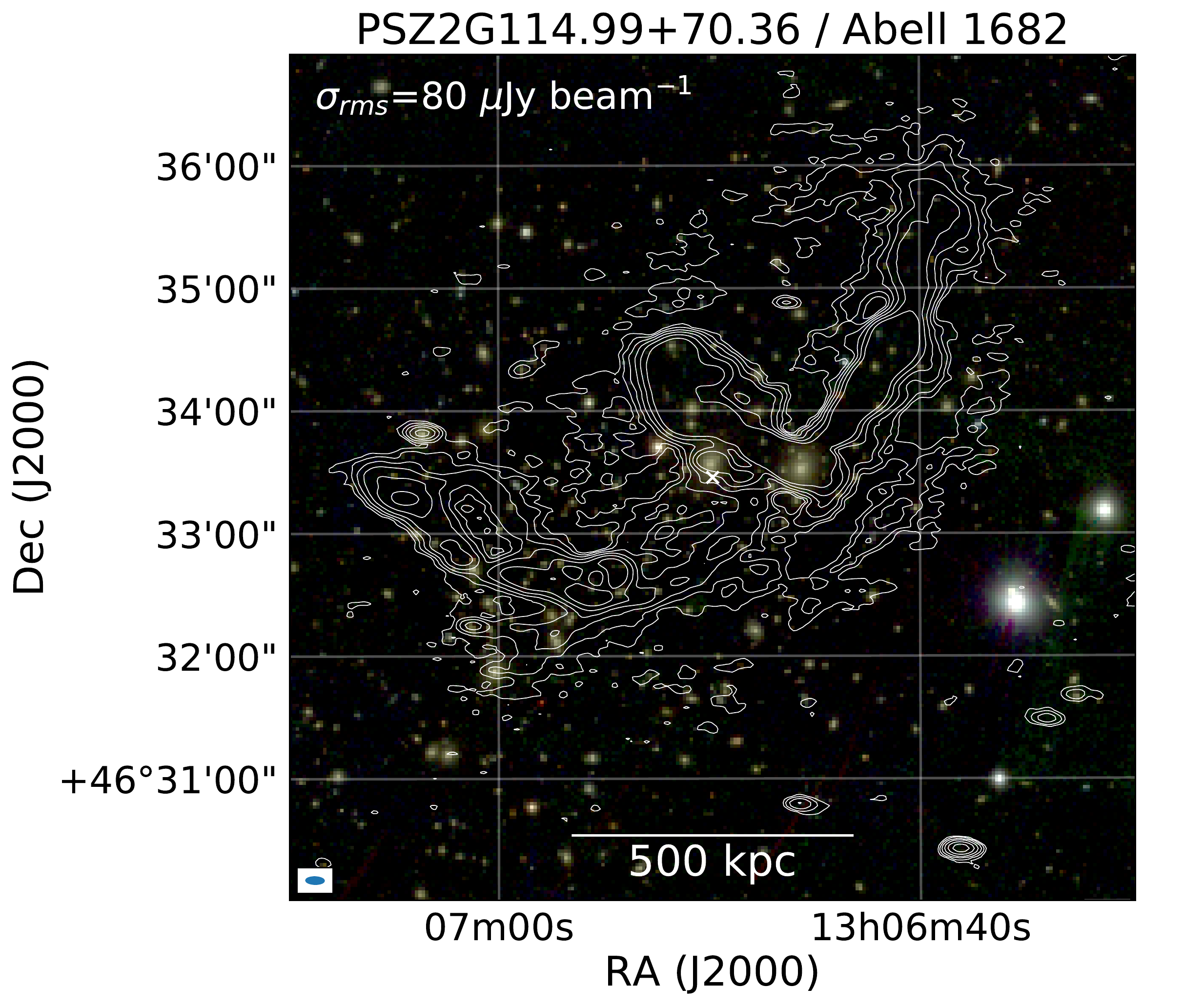}
   \caption{PSZ2\,G114.99+70.36 / Abell\,1682. Left: Robust $-0.5$ radio image. Middle: Chandra X-ray image with 10\arcsec~tapered radio contours. Right: Optical image with Robust $-0.5$ image radio contours. For more details see the caption of Figure~\ref{fig:A2018}. }
   \label{fig:A1682}
\end{figure*}

The presence of diffuse emission in this cluster was first reported by \cite{2008A&A...484..327V} and subsequently studied in more detail by \cite{2011JApA...32..501V,2013AA...551A.141M,2013A&A...551A..24V}. 
LOFAR observations show complex diffuse extended emission extending over a region of more than 1~Mpc, see Figure~\ref{fig:A1682}. The Chandra X-ray image displays a disturbed ICM indicating that the cluster is undergoing a merger event. The LOFAR results have been presented in \cite{2019A&A...627A.176C}.
Some of the structures in the cluster are related to distorted and tailed radio galaxies. A few regions of the diffuse emission show a steep spectral index of $\alpha \sim -2$. These regions could trace AGN fossil plasma that has been re-accelerated (or revived) by shocks and turbulence related to the ongoing merger event. A candidate radio halo is located near the peak of the X-ray emission, labelled (CH). The complexity of the various radio structures in the vicinity prevent us from measuring its properties.

\subsection{PSZ2\,G118.34+68.79} 
\begin{figure*}
\centering
   \includegraphics[width=0.285\paperwidth]{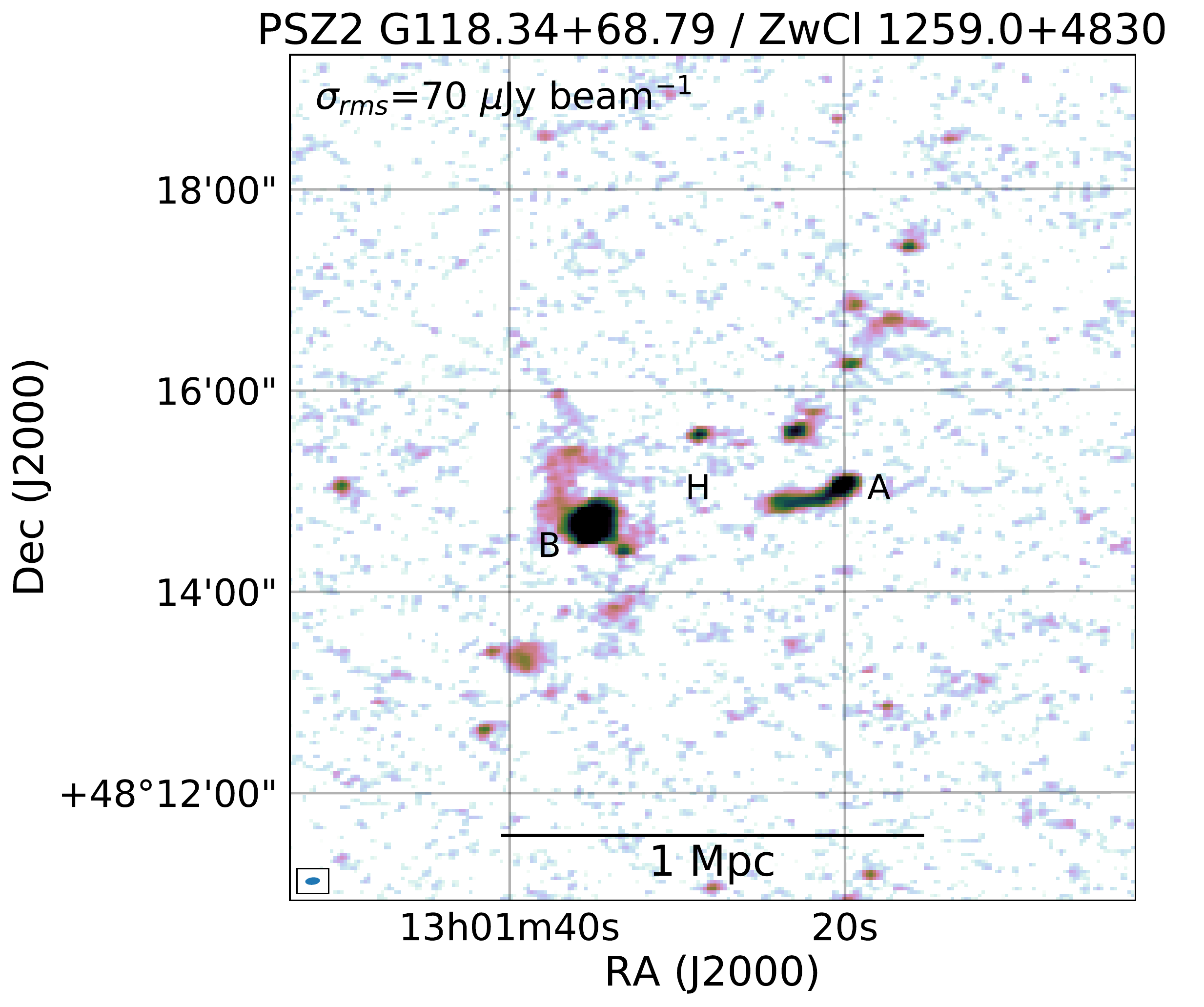} 
   \includegraphics[width=0.285\paperwidth]{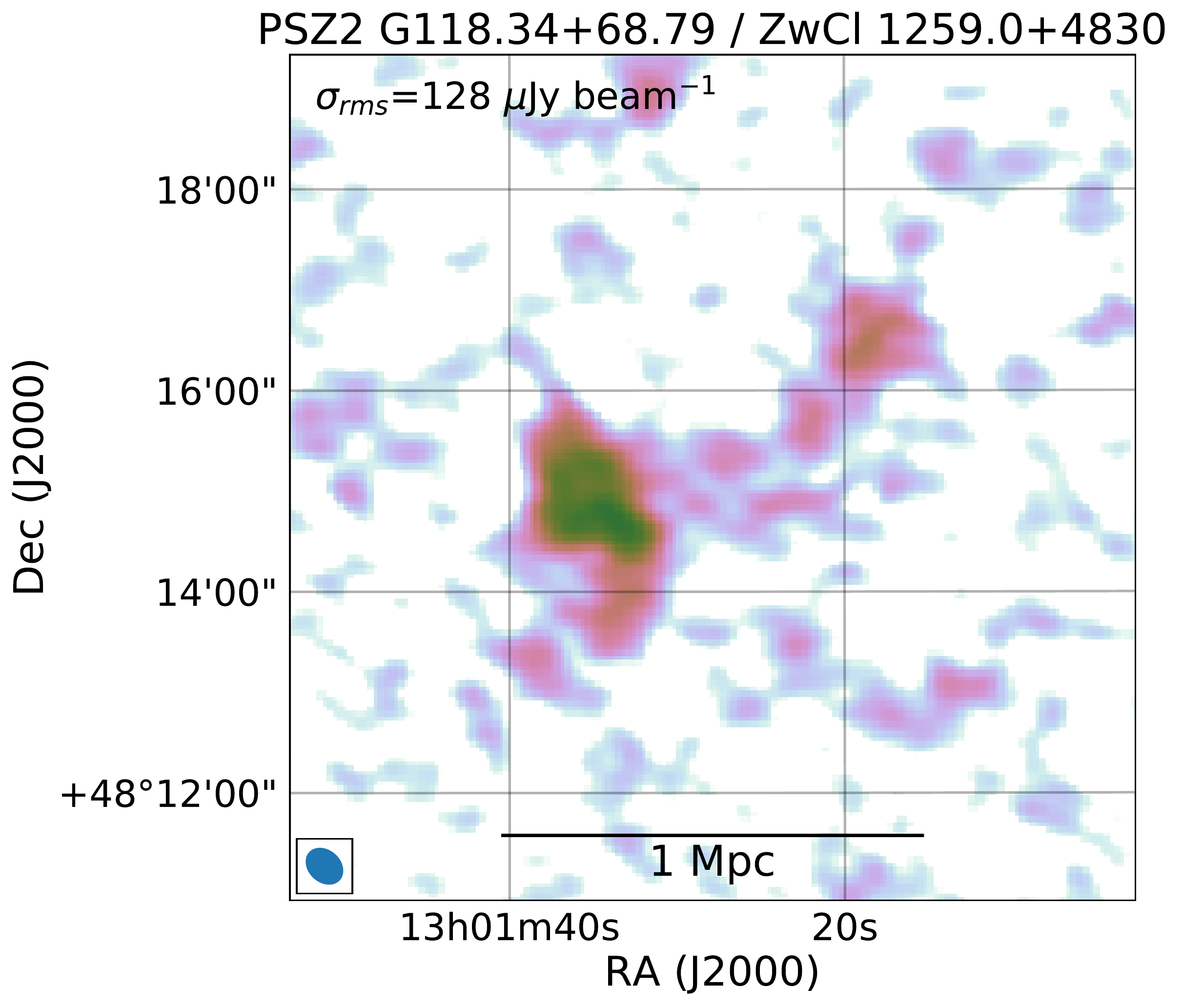}
   \includegraphics[width=0.285\paperwidth]{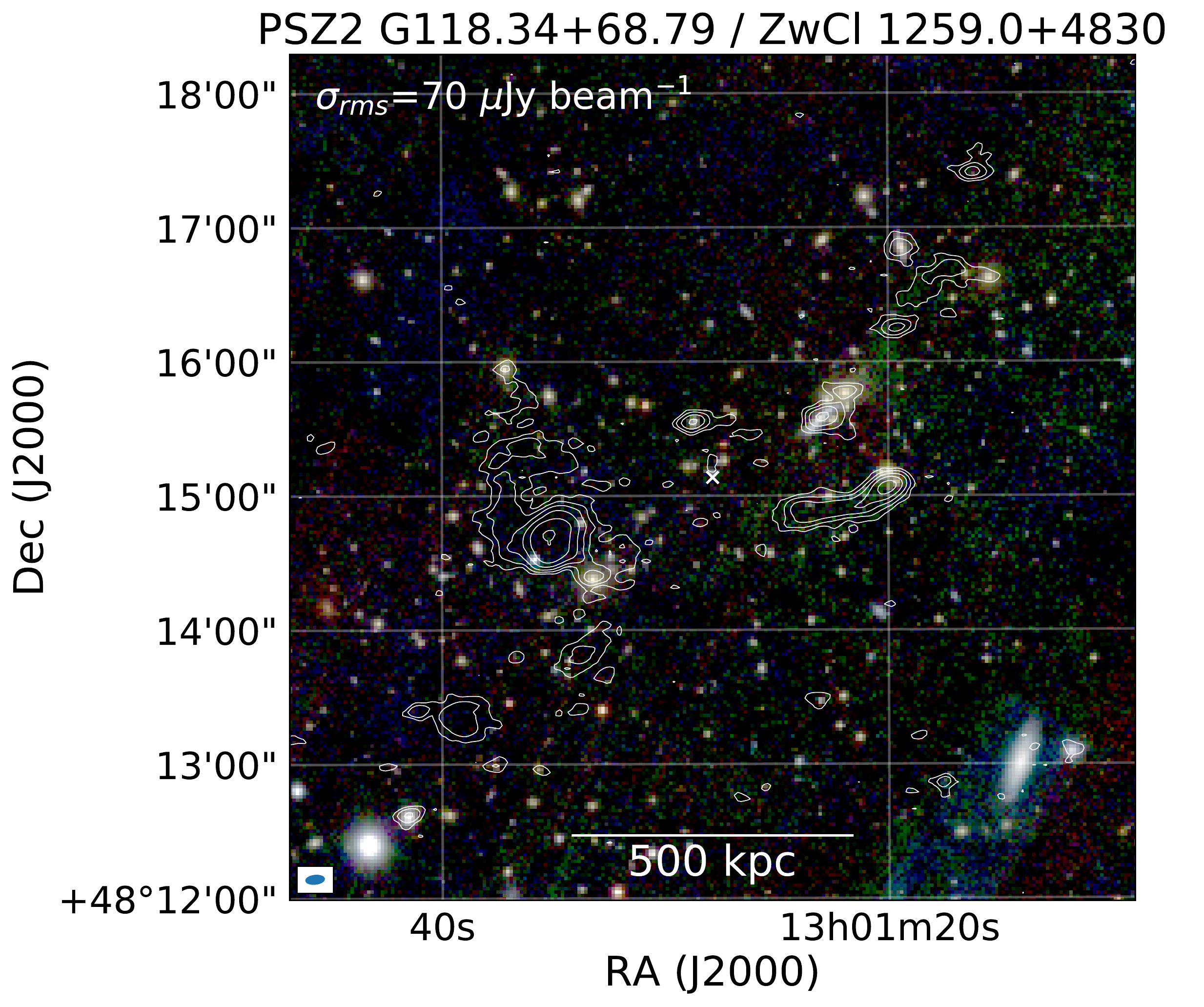}
   \caption{PSZ2\,G118.34+68.79. Left: Robust $-0.5$ radio image. Middle: 15\arcsec~tapered radio image with compact radio sources subtracted. Right: Optical image with Robust $-0.5$ image radio contours. For more details see the caption of Figure~\ref{fig:A2018}.}
   \label{fig:PSZ2G11834}
\end{figure*}

Several radio galaxies and diffuse emission are detected in this cluster, see Figure~\ref{fig:PSZ2G11834}. An optical image shows two BCGs that are located along a SE-NW axis. The cluster hosts a tailed radio galaxy,  labeled A.
To the east, a bright patch of emission (B) is found just above the SE BCG. Additional fainter emission is located around B. Given the high surface brightness of B and several nearby radio AGN, B is likely AGN plasma, possibly revived by the passage of a shock. 

We also find low-level diffuse emission extending on scales of about 0.4~Mpc in the central regions of the cluster, labelled H. However, it is hard to determine its full spatial extent as it partly blends with other extended radio sources in this region. Since the diffuse emission approximately follows the overall galaxy distribution and has a low surface brightness we classify it as a candidate radio halo. Fitting a circular model we estimate $S_{144}=28\pm13$~mJy, where we masked the region near B and west of A.

\subsection{PSZ2\,G133.60+69.04, Abell\,1550}
The existence of diffuse emission in Abell\,1550 ($z=0.2540$) cluster was first reported by \cite{2012A&A...545A..74G}.
Extended radio emission is clearly detected in this cluster by LOFAR. It extends over a significantly larger region than found by \citeauthor{2012A&A...545A..74G}, corresponding to an LLS of 1.8~Mpc. This is explained by the shallower depth of the VLA image used \citeauthor{2012A&A...545A..74G} and/or a steep radio spectrum. 

The Chandra image shows a roughly roundish ICM distribution. Additional faint X-ray emission is observed NE of the main structure. The main part of the diffuse radio emission, labeled H, traces the X-ray emission from the ICM. We therefore classify it as a giant radio halo with a size of about 0.9~Mpc. Based on the elliptical model fit, we obtain a radio halo integrated flux density of $129\pm26$~mJy. 
The radio images also show a bright tailed radio galaxy (A) and several structures in the western part of the cluster (D, E, C) that do not have clear optical counterparts. A diffuse patch of emission, labeled B, is placed near a group of galaxies located at the same redshift as the main cluster. A lower resolution radio image shows that B is connected to the main radio halo. Source B seems to be associated with the NE X-ray extension. Sources E, D, and C look to be relics, possibly related to revived AGN fossil plasma. We note that both D and C are connected to the emission from the main radio halo. Based on the extension of the ICM and placement of the diffuse radio sources, the cluster seems to have undergone a merger event in the NE-SW direction. The location of B in the NE, and C-D-E in the SW could indicate that these structures are related to mergers shocks. If this interpretation is correct the cluster hosts a double radio relic. The combination of a double radio relic and a radio is relatively rare \citep{2017MNRAS.470.3465B}.

\begin{figure*}
\centering
\includegraphics[width=1.0\columnwidth]{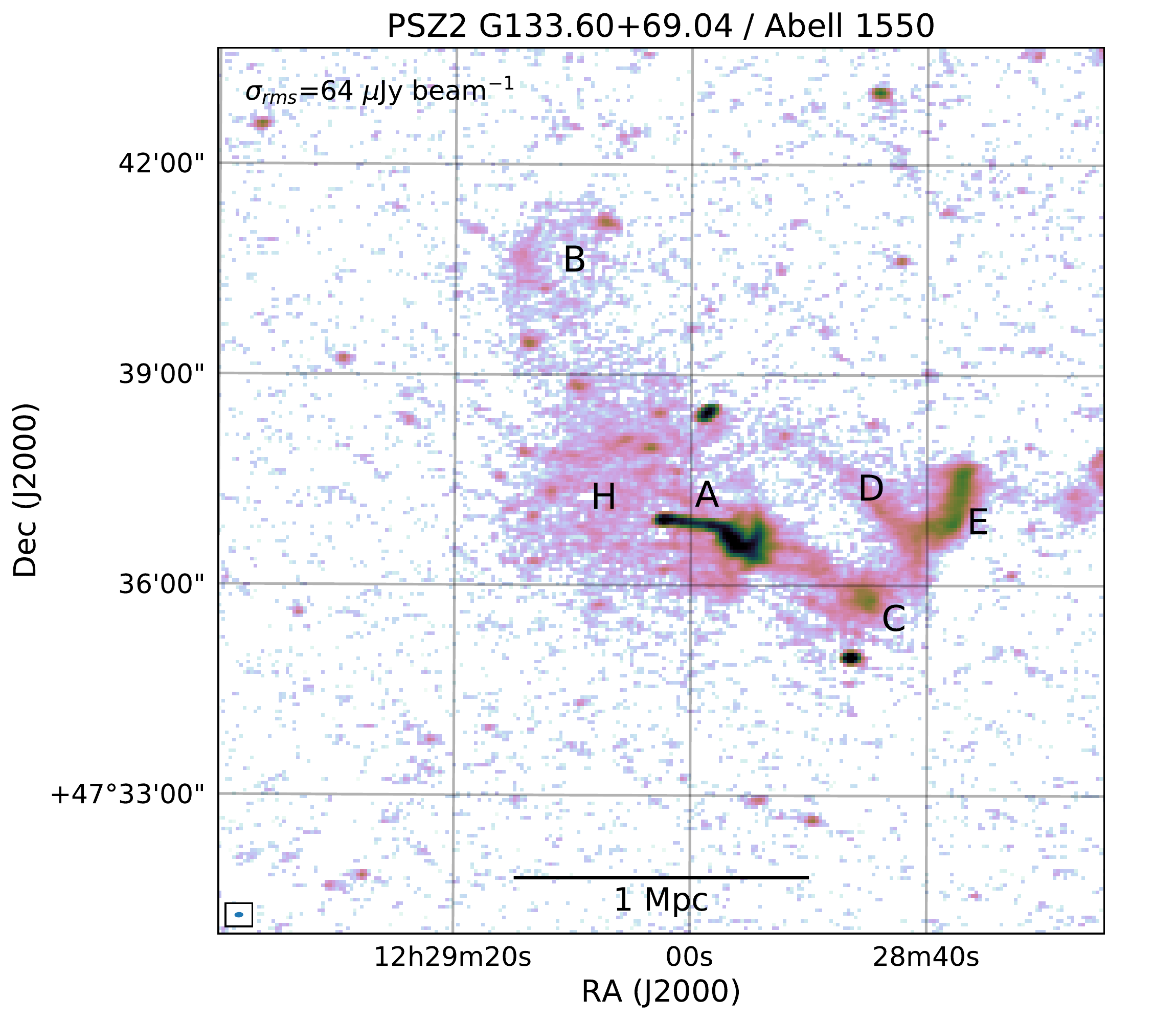}
\includegraphics[width=1.0\columnwidth]{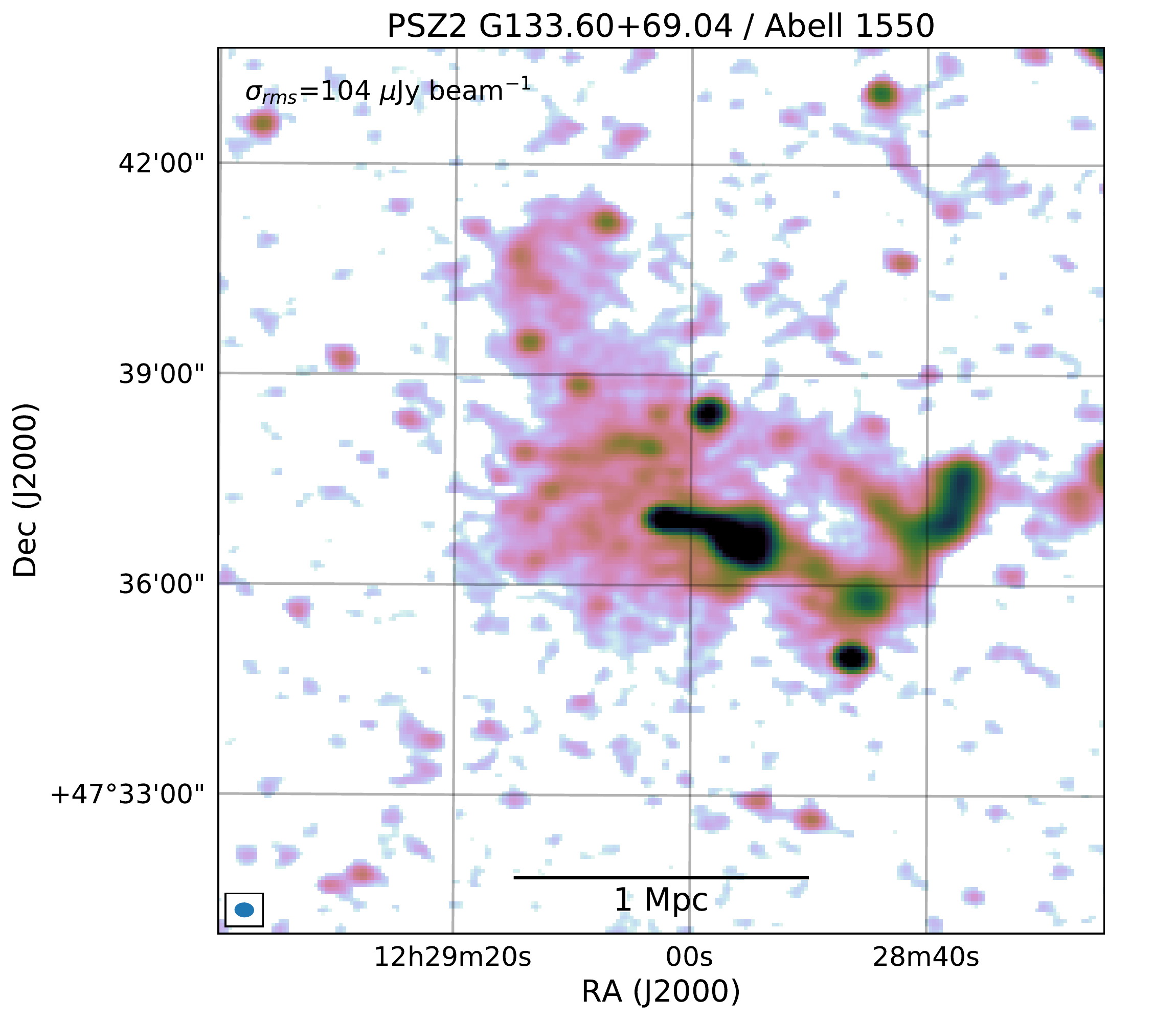}
\includegraphics[width=1.0\columnwidth]{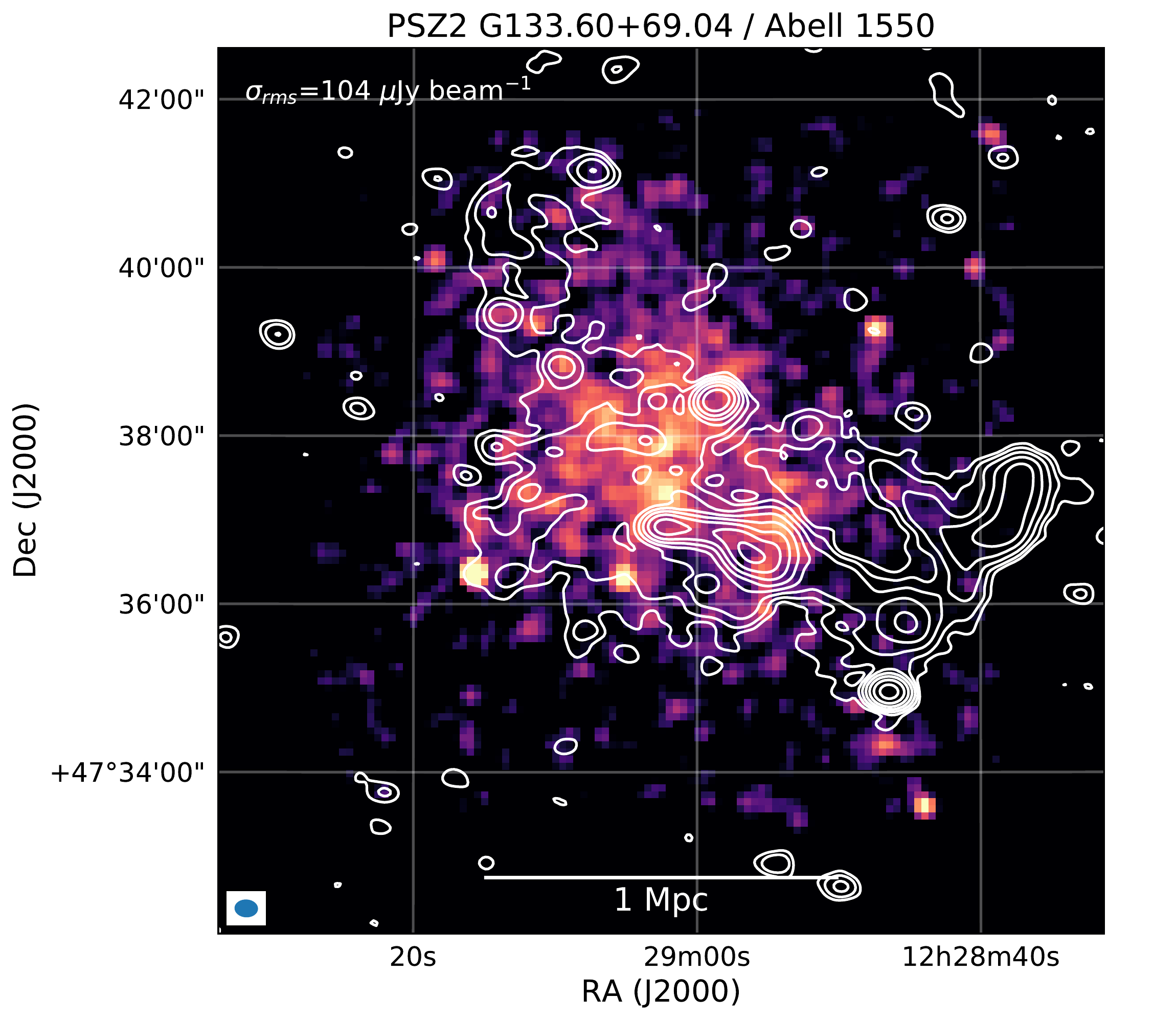}
\includegraphics[width=1.0\columnwidth]{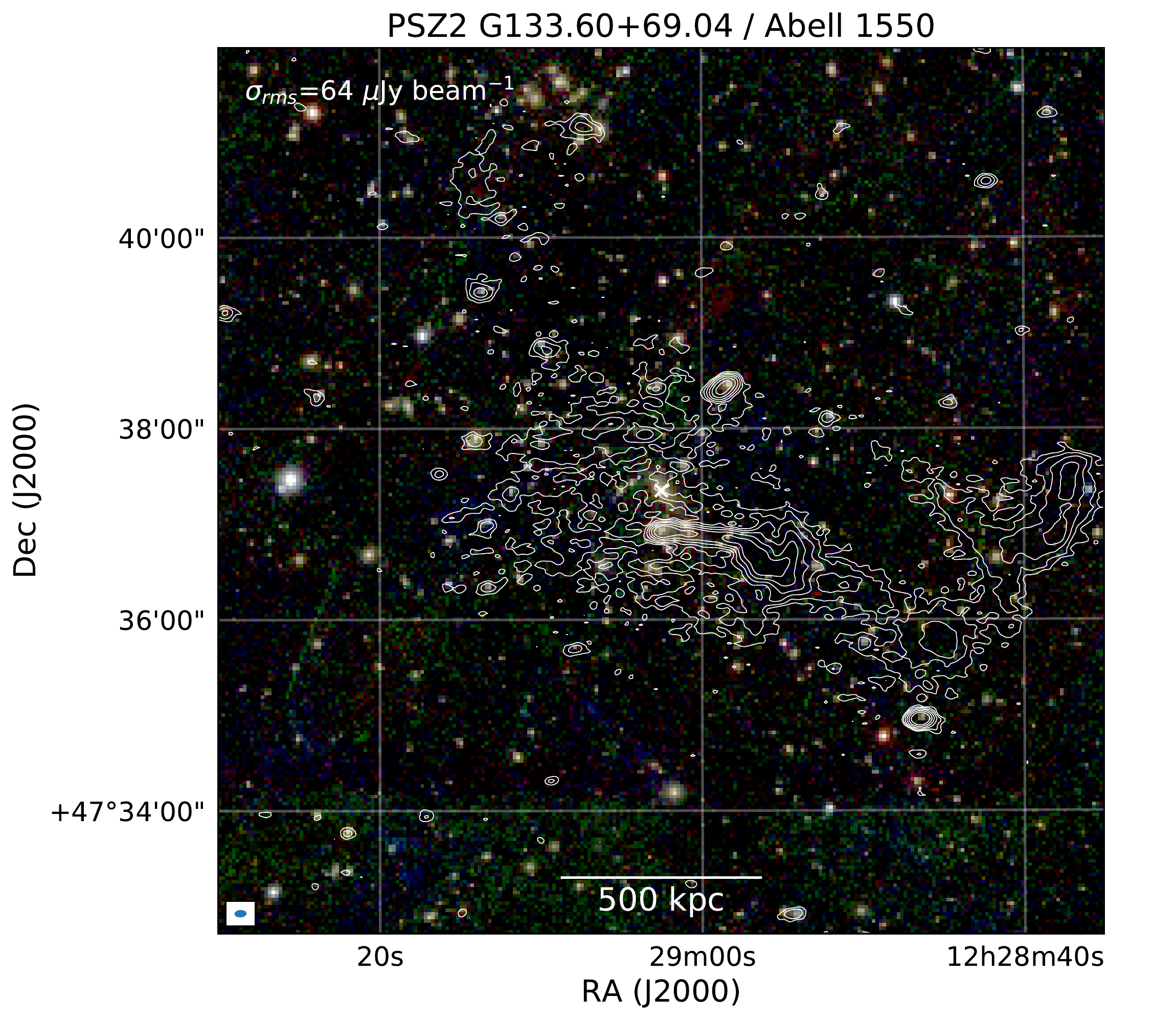}   
\caption{PSZ2\,G133.60+69.04 / Abell\,1550. Top left: Robust $-0.5$ radio image. Top right: 10\arcsec~tapered radio image. Bottom left: Chandra X-ray image with 10\arcsec~tapered radio contours. Bottom right: Optical image with Robust $-0.5$ image radio contours. For more details see the caption of Figure~\ref{fig:A2018}.}
     \label{fig:A1550}
\end{figure*}

\subsection{PSZ2\,G135.17+65.43}
\begin{figure*}
\centering
   \includegraphics[width=0.285\paperwidth]{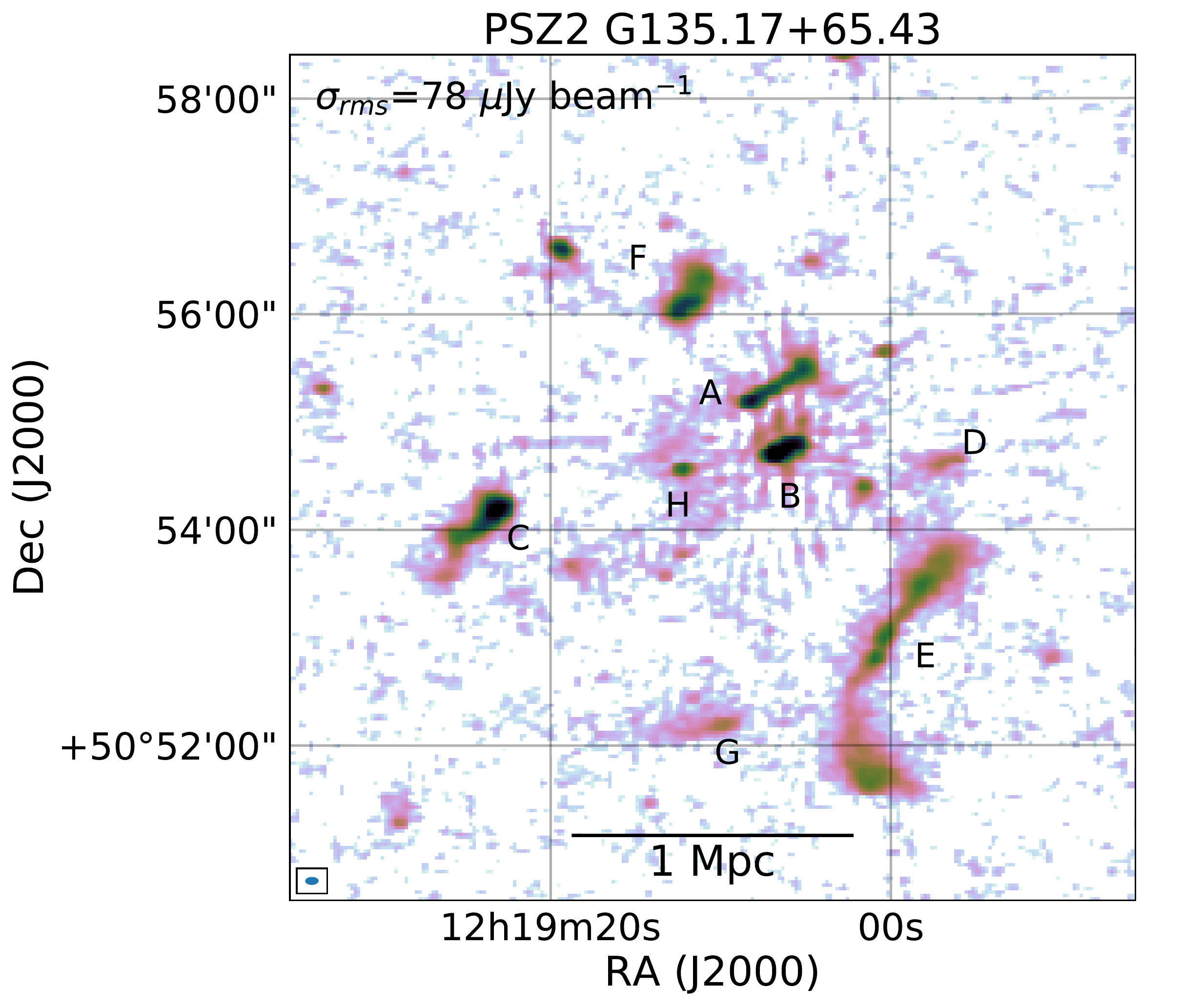} 
   \includegraphics[width=0.285\paperwidth]{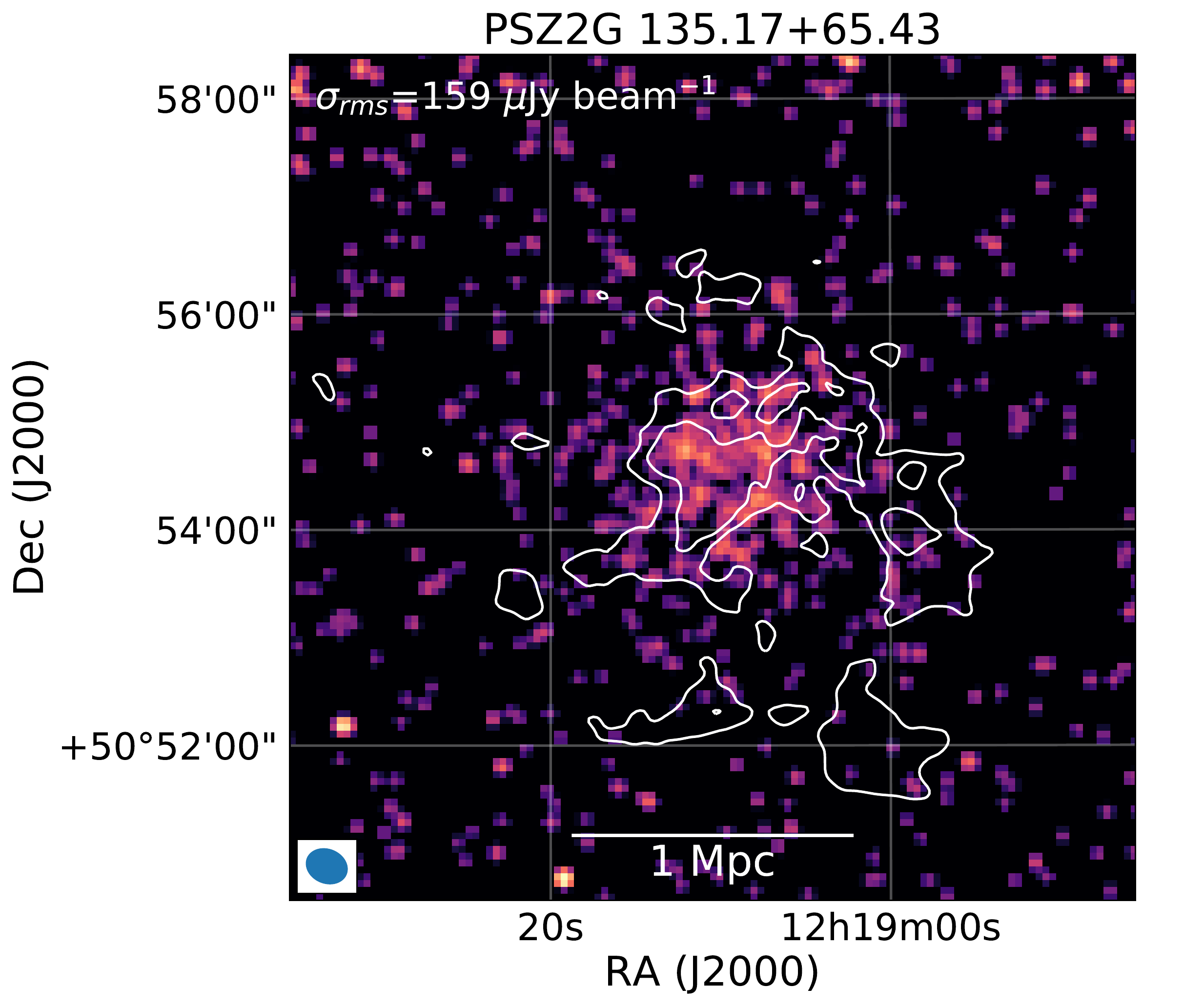}
   \includegraphics[width=0.285\paperwidth]{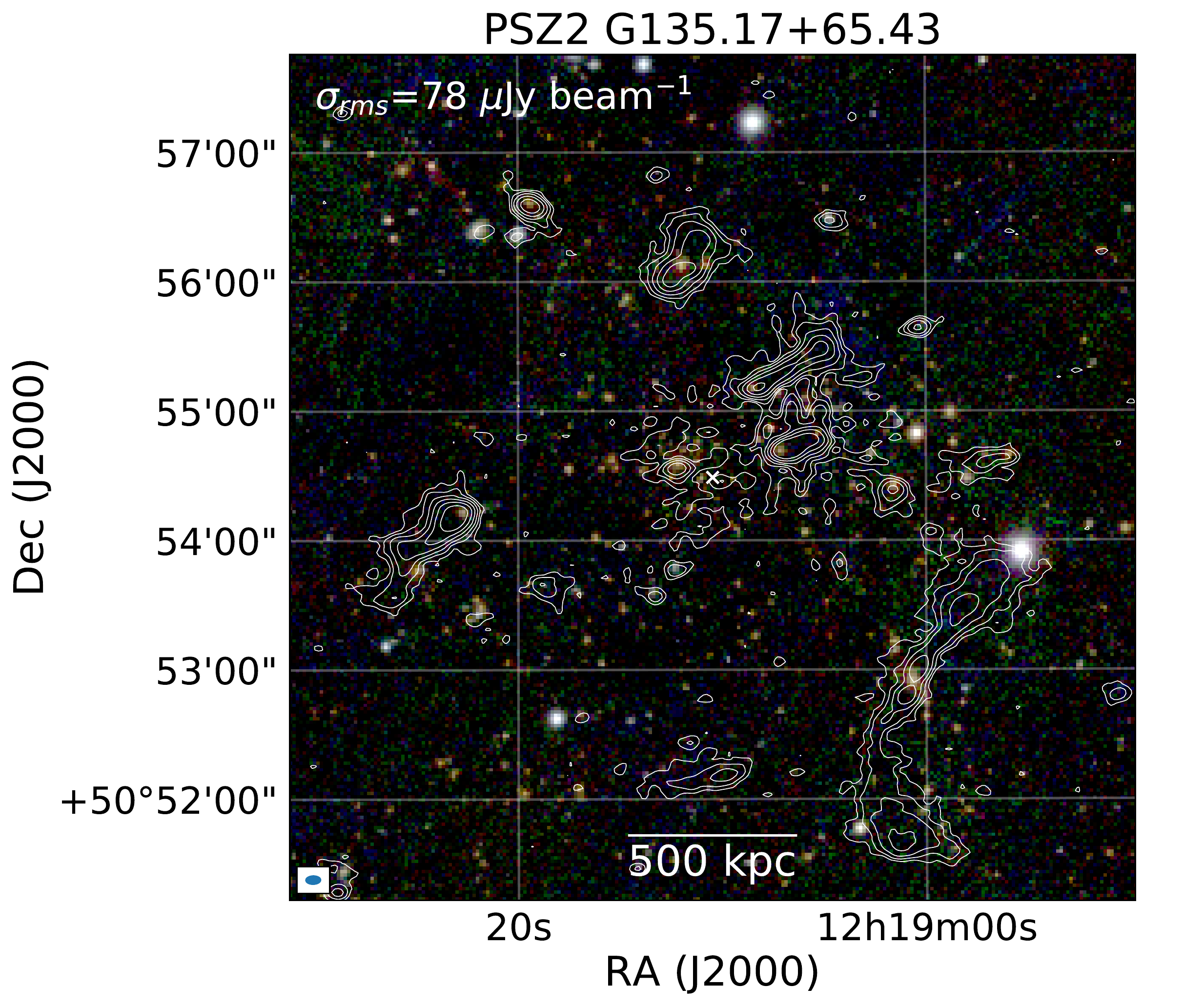}
   \caption{PSZ2\,G135.17+65.43. Left: Robust $-0.5$ radio image. Middle: Chandra X-ray image with 10\arcsec~tapered radio contours (compact sources were subtracted). Right: Optical image with Robust $-0.5$ image radio contours. For more details see the caption of Figure~\ref{fig:A2018}.}
   \label{fig:PSZ2G13517}
\end{figure*}

The LOFAR image uncovers a large number of radio galaxies in this $z=0.5436$ cluster which are labeled A to~F, see Figure~\ref{fig:PSZ2G13517}. One of these, source~E, has a physical extent of about 1~Mpc and we classify it as a giant double-double. Sources A to~D display head-tail morphologies. Source~G is a peripheral source with an extent of about 400~kpc. We do not identify an  optical counterpart for this source and tentatively classify it as a relic or an AGN fossil plasma source. The Chandra image of PSZ2\,G135.17+65.43 indicates a non-relaxed cluster without a clear central peak.
We also detect central diffuse emission in this cluster with a total extent of about 500~kpc, labeled~H. Due to the presence of the tailed radio galaxies A and~B, the full extent of H is hard to determine but the emission approximately follows the thermal ICM. Given its central location and extent, we classify H as a candidate radio halo with a flux density of $S_{144}=29.0\pm6.6$~mJy based on the circular model fit.  


\subsection{PSZ2\,G143.26+65.24, Abell\,1430}
\begin{figure*}
\centering
   \includegraphics[width=1.0\columnwidth]{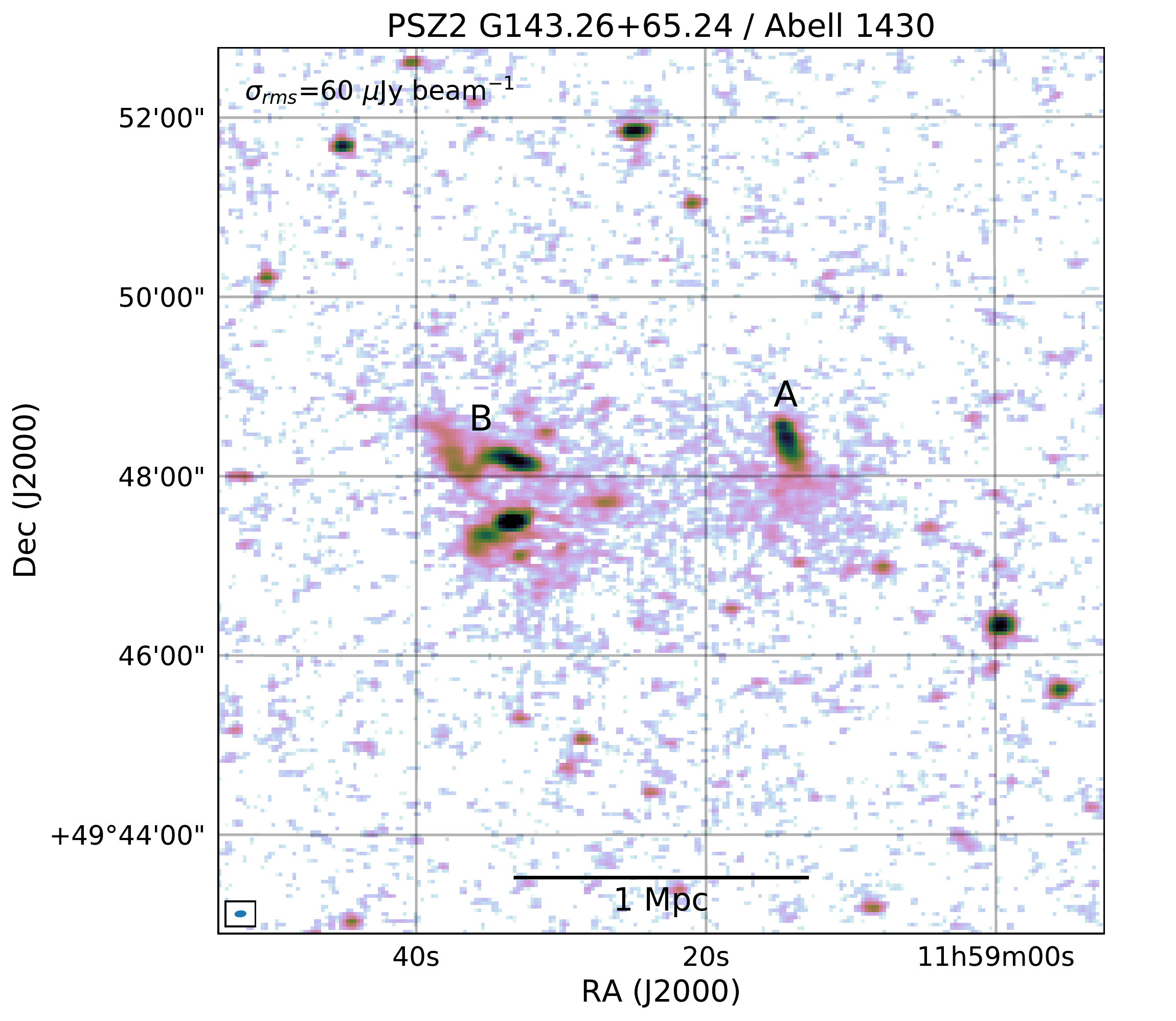} 
   \includegraphics[width=1.0\columnwidth]{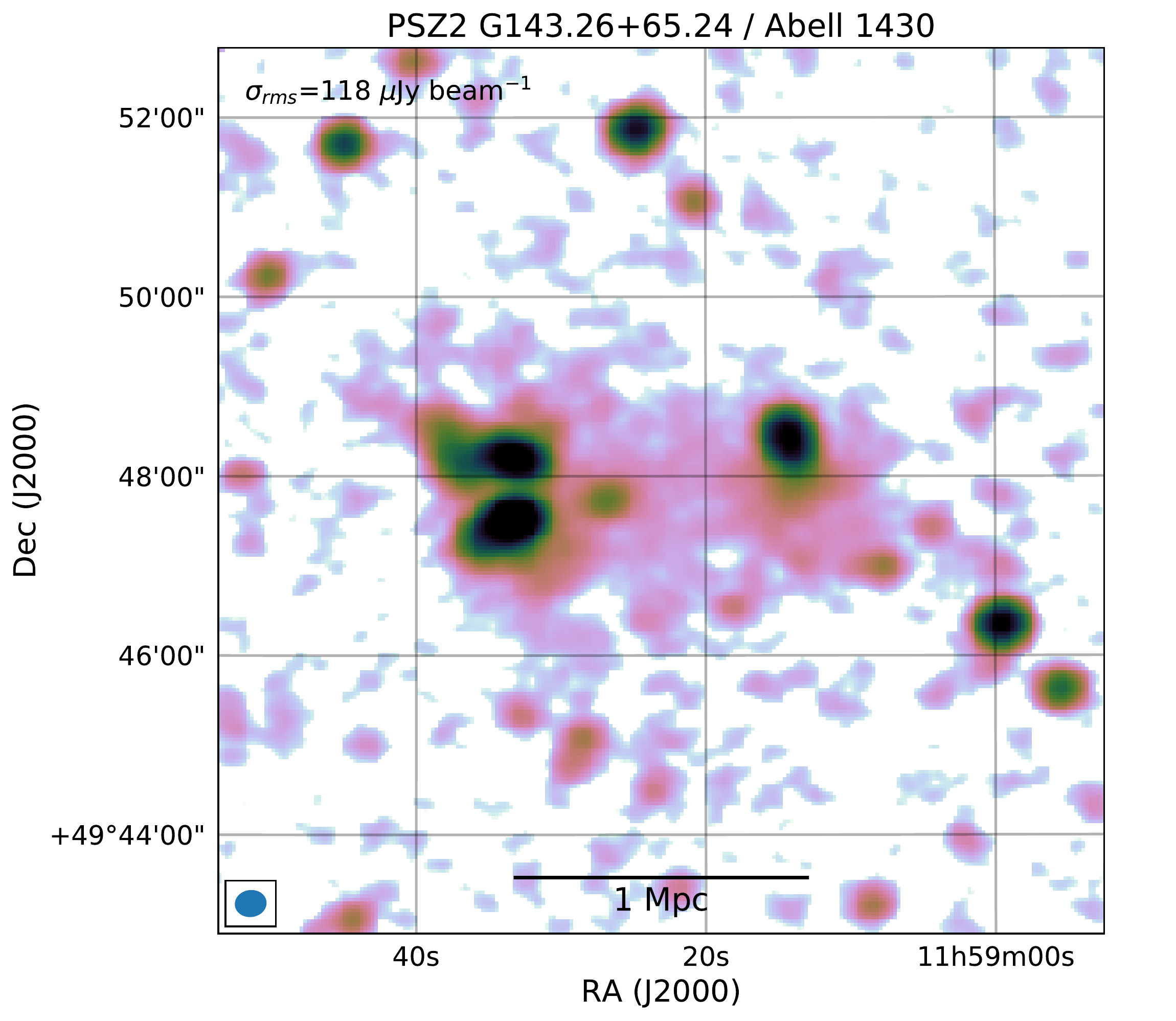}
   \includegraphics[width=1.0\columnwidth]{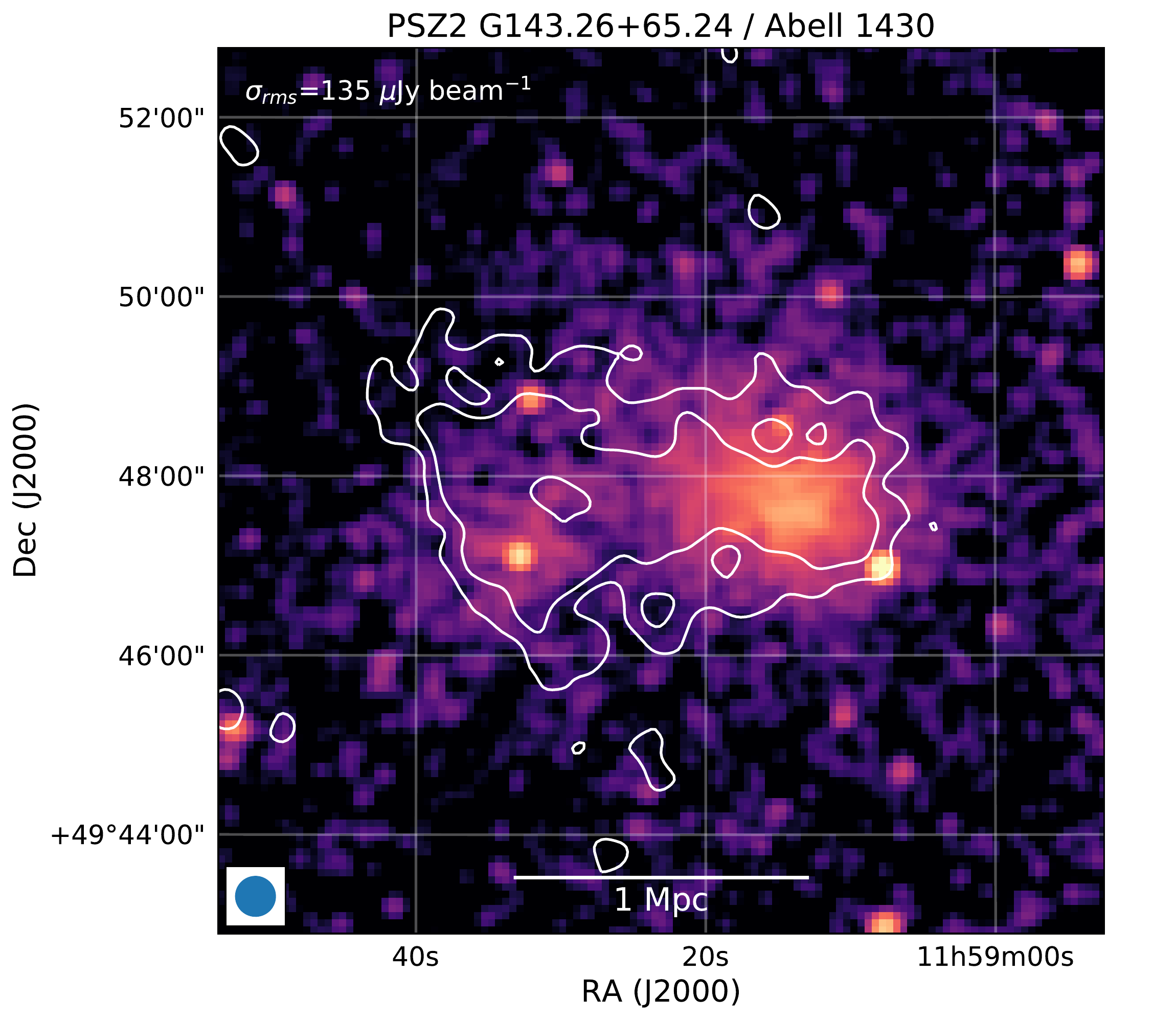}
   \includegraphics[width=1.0\columnwidth]{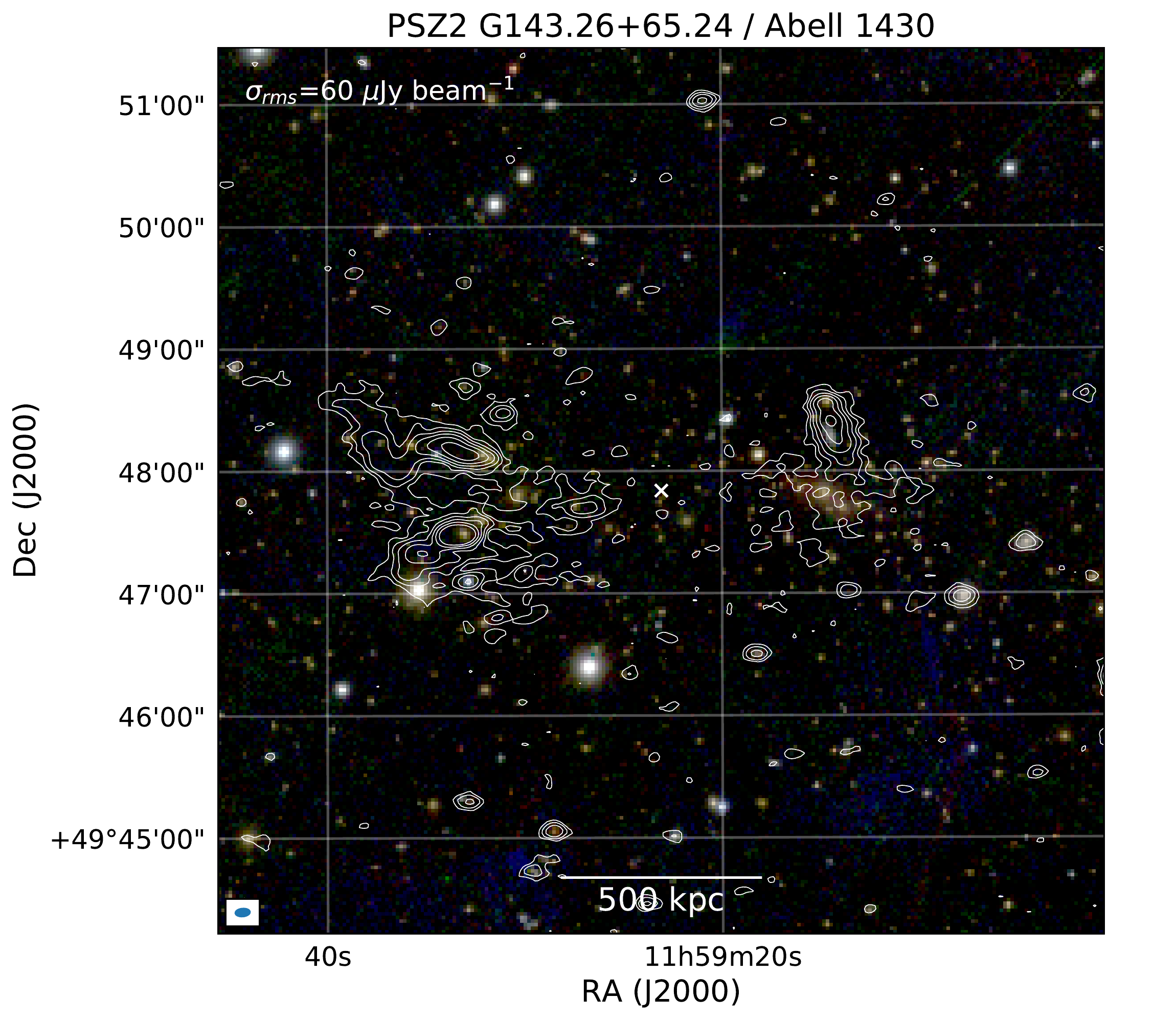}
   \caption{PSZ2\,G143.26+65.24 / Abell\,1430. Top left: Robust $-0.5$ radio image. Top right: 10\arcsec~tapered radio image. Bottom left: Chandra X-ray image with 15\arcsec~tapered radio contours (compact sources were subtracted). Bottom right: Optical image with Robust $-0.5$ image radio contours. For more details see the caption of Figure~\ref{fig:A2018}.}
   \label{fig:A1430}
\end{figure*}

The LOFAR images reveal the presence of diffuse emission in this cluster which is located at $z=0.3634$, see Figure~\ref{fig:A1430}. The LOFAR observations are discussed in more detail in \cite{hoeft}, so we only give a brief overview of the main findings. 
The diffuse emission extends over a region of about 5\arcmin~by 2.5\arcmin, with the emission being elongated in the EW direction. The angular extent corresponds to physical size of 1.5~Mpc by 0.8~Mpc. Our LOFAR image also reveals two tailed radio galaxies, labelled A and B (Figure~\ref{fig:A1430}). An optical image displays two subclusters. The western subcluster corresponds to the main structure seen in the Chandra X-ray image, while the eastern cluster is much fainter in the Chandra image. The radio emission spans the full region between the main western and the smaller eastern subcluster.  Given the clear correspondence between the radio and X-ray emission of the western subcluster and the large extent of the diffuse emission, we classify the western part of the diffuse radio emission as a giant radio halo. We measure $S_{144}=29.8\pm6.6$~mJy for the radio halo from our fitting. We note that this value might be affected by the diffuse emission around the eastern subcluster since it partly overlaps with the radio halo from the western subcluster. The nature of the diffuse emission around the eastern subcluster is not fully clear but it could be a radio bridge \citep[for a discussion on this see][]{hoeft}. The disturbed character of the cluster, both in optical and X-ray images, indicates that the cluster is undergoing a merger event, consistent with the presence of the giant radio halo. A possible source of seed electrons for the radio halo could be the tail of source A, since it blends into the halo emission.


\subsection{PSZ2\,G145.65+59.30, Abell\,1294}
\begin{figure*}
\centering
   \includegraphics[width=0.285\paperwidth]{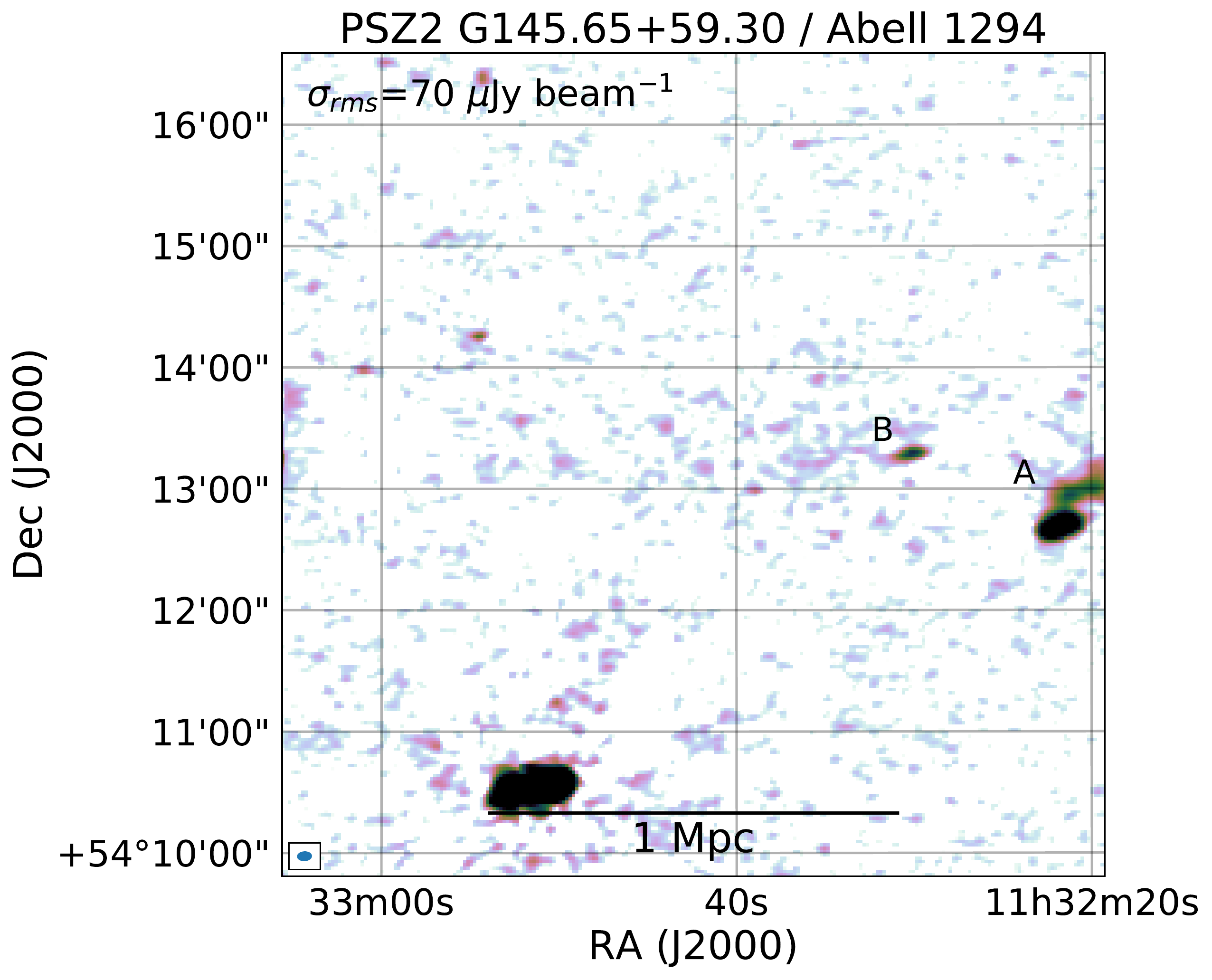}
   \includegraphics[width=0.285\paperwidth]{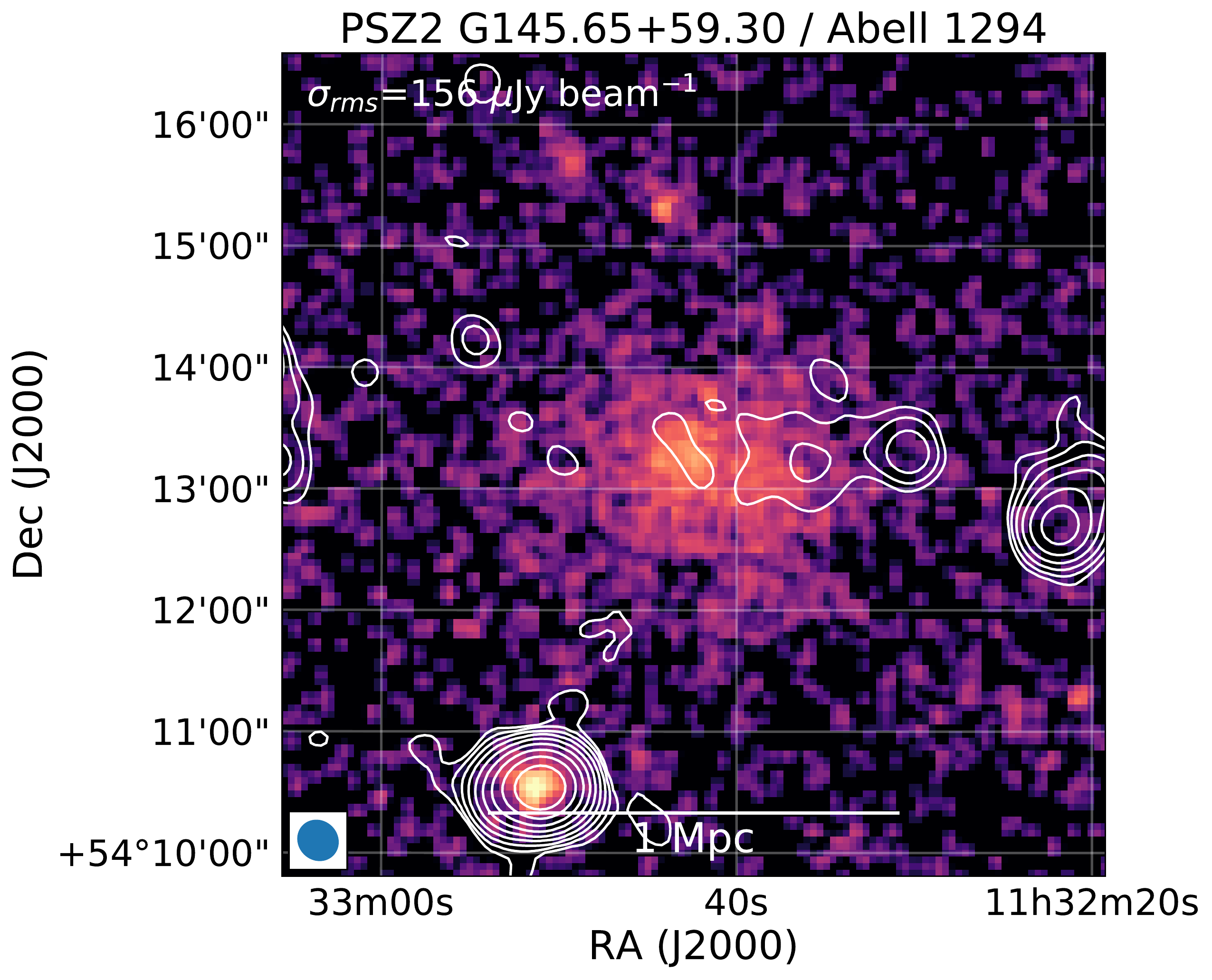}
   \includegraphics[width=0.285\paperwidth]{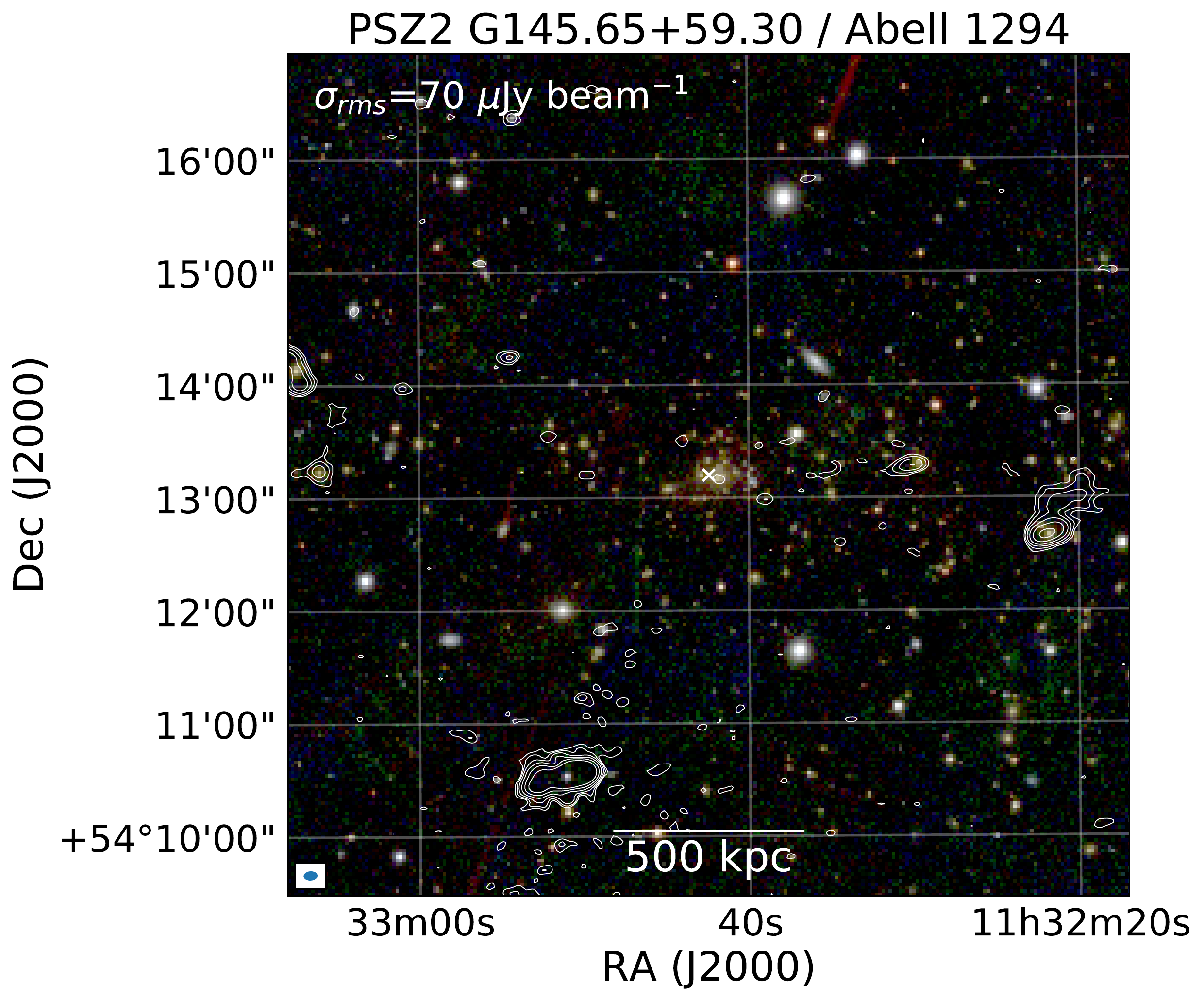}
   \caption{PSZ2\,G145.65+59.30 / Abell\,1294. Left: Robust $-0.5$ radio image. Middle: XMM-Newton X-ray image with 10\arcsec~tapered radio contours. Right: Optical image with Robust $-0.5$ image radio contours. For more details see the caption of Figure~\ref{fig:A2018}.}
   \label{fig:A1294}
\end{figure*}

An XMM-Newton image shows a cluster that is elongated in the EW direction, see Figure~\ref{fig:A1294}. The galaxy distribution shows a similar elongation and extending all the way to a tailed radio galaxy A. Our LOFAR image reveals  faint diffuse emission, with an LSS of about 0.4~Mpc, located at the western part of the cluster. This emission is very faint and close to the detection limit of our observations.  A candidate tailed radio galaxy (B) is located just east of the diffuse radio source. The diffuse emission is difficult to classify with the current data in hand. A possibility is that it is related to revived or re-accelerated fossil plasma from source B. On the other hand, it could also be a radio halo, somewhat similar to the  ``off-axis'' radio halo found in Abell\,1132 \citep{2018MNRAS.473.3536W}. The integrated flux density of the diffuse source is $S_{144}=6.7\pm1.7$~mJy, based on the circular model fit.



\subsection{PSZ2\,G150.56+58.32, MACS\,J1115.2+5320}
\begin{figure*}
\centering
   \includegraphics[width=0.285\paperwidth]{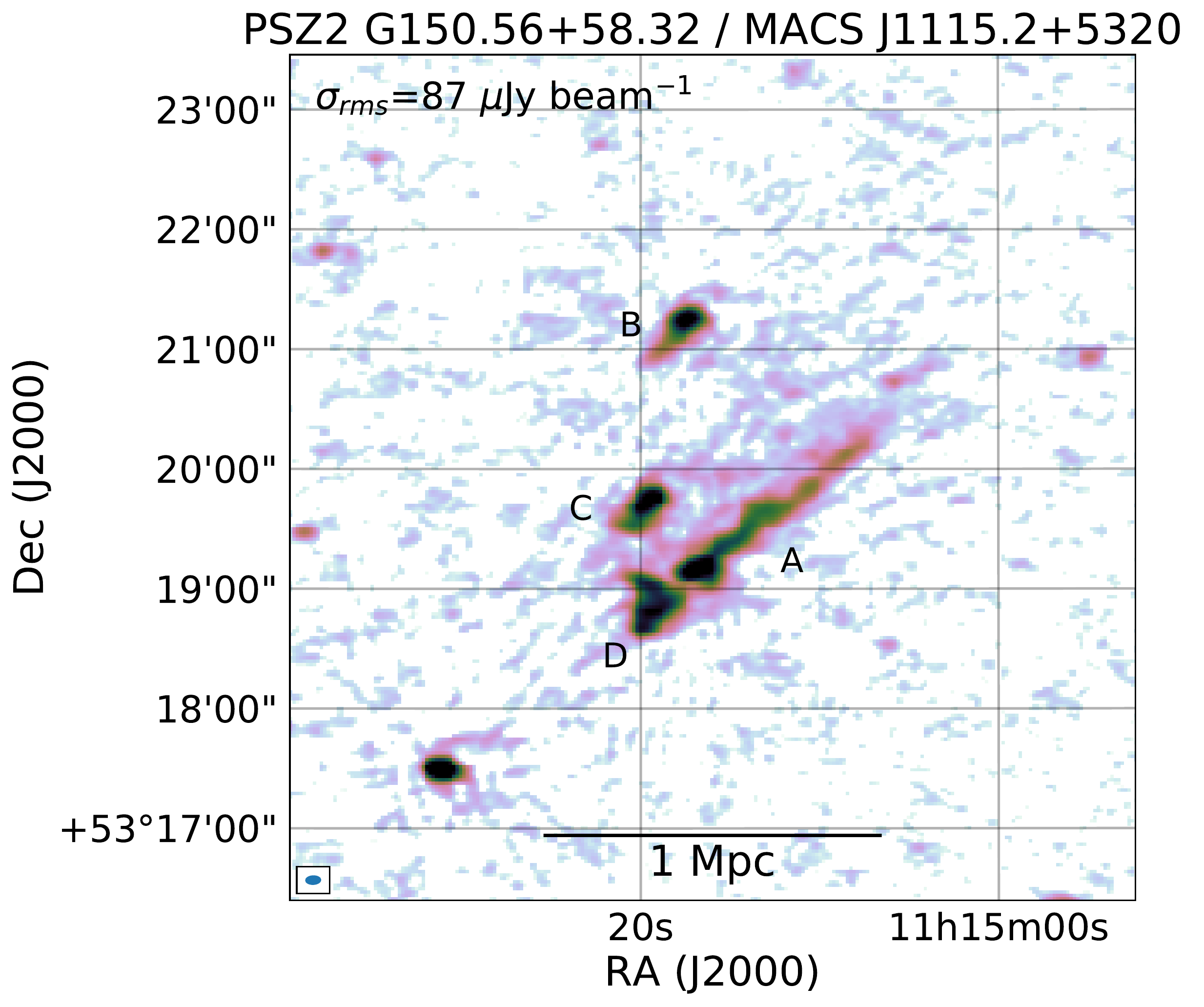}
   \includegraphics[width=0.285\paperwidth]{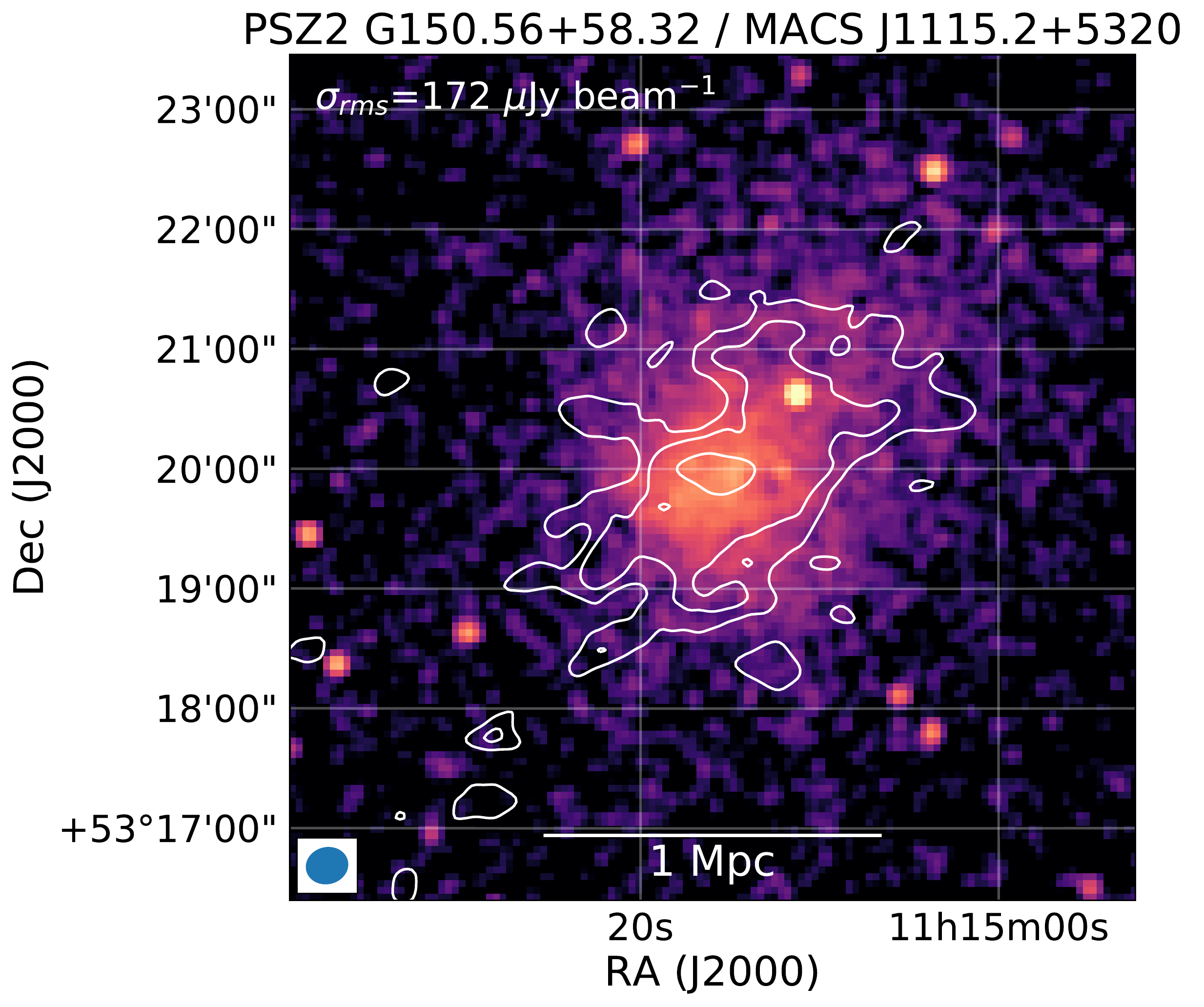}
   \includegraphics[width=0.285\paperwidth]{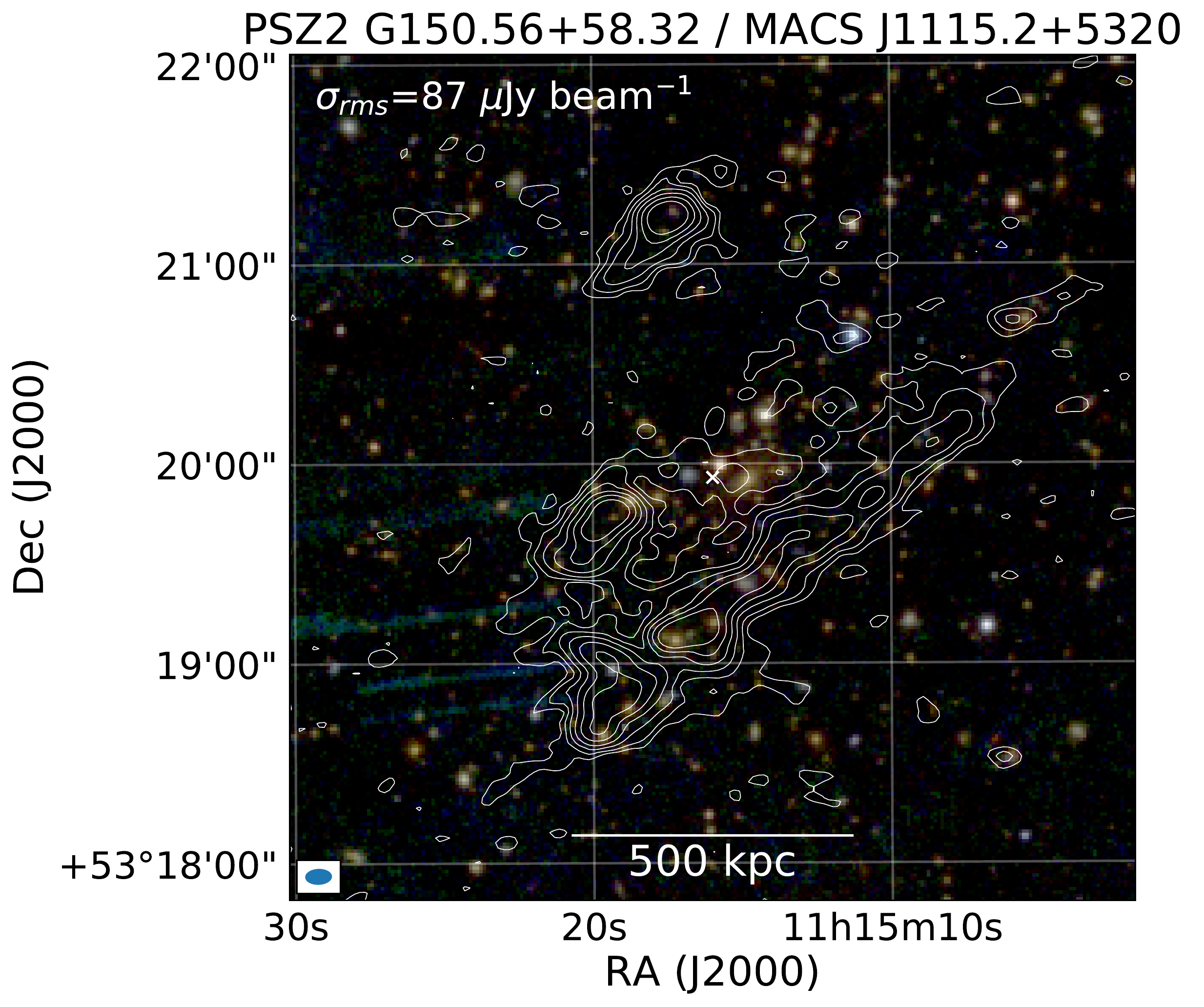}
   \caption{PSZ2\,G150.56+58.32 / MACS\,J1115.2+5320. Left: Robust $-0.5$ radio image. Middle: Chandra X-ray image with 10\arcsec~tapered radio contours (compact sources were subtracted). Right: Optical image with Robust $-0.5$ image radio contours. For more details see the caption of Figure~\ref{fig:A2018}.}
   \label{fig:PSZ2G150}
\end{figure*}

MACS\,J1115.2+5320 is a massive merging cluster located at $z=0.466$. A Chandra X-ray image reveals an ongoing merger event along a NW-SE axis, see Figure~\ref{fig:PSZ2G150}.   \cite{2012MNRAS.420.2120M} classified this system as a possible head-on binary cluster merger. The global temperature of the cluster was measured to be $8.6 \pm 1.1$~keV by \cite{2015MNRAS.450.2261M}.

The LOFAR image shows a long 770~kpc tailed radio galaxy (A), with the tail pointing NW. Two additional tailed radio galaxies (B, C) have their tails pointing to the SE. Source~D is located just SE of the head-tail source~A and has a rather complex morphology in our high-resolution image. Near source~D, a possible merger induced cold front is visible in the Chandra image. Low surface brightness radio emission is detected in the region north of source A. This emission extends over 1~Mpc and given its size we classify it as a giant radio halo. Fitting an elliptical model, we determine $S_{144}=71.2\pm14.5$~mJy for the radio halo.

\subsection{PSZ2\,G156.26+59.64}
\begin{figure*}
\centering
   \includegraphics[width=0.285\paperwidth]{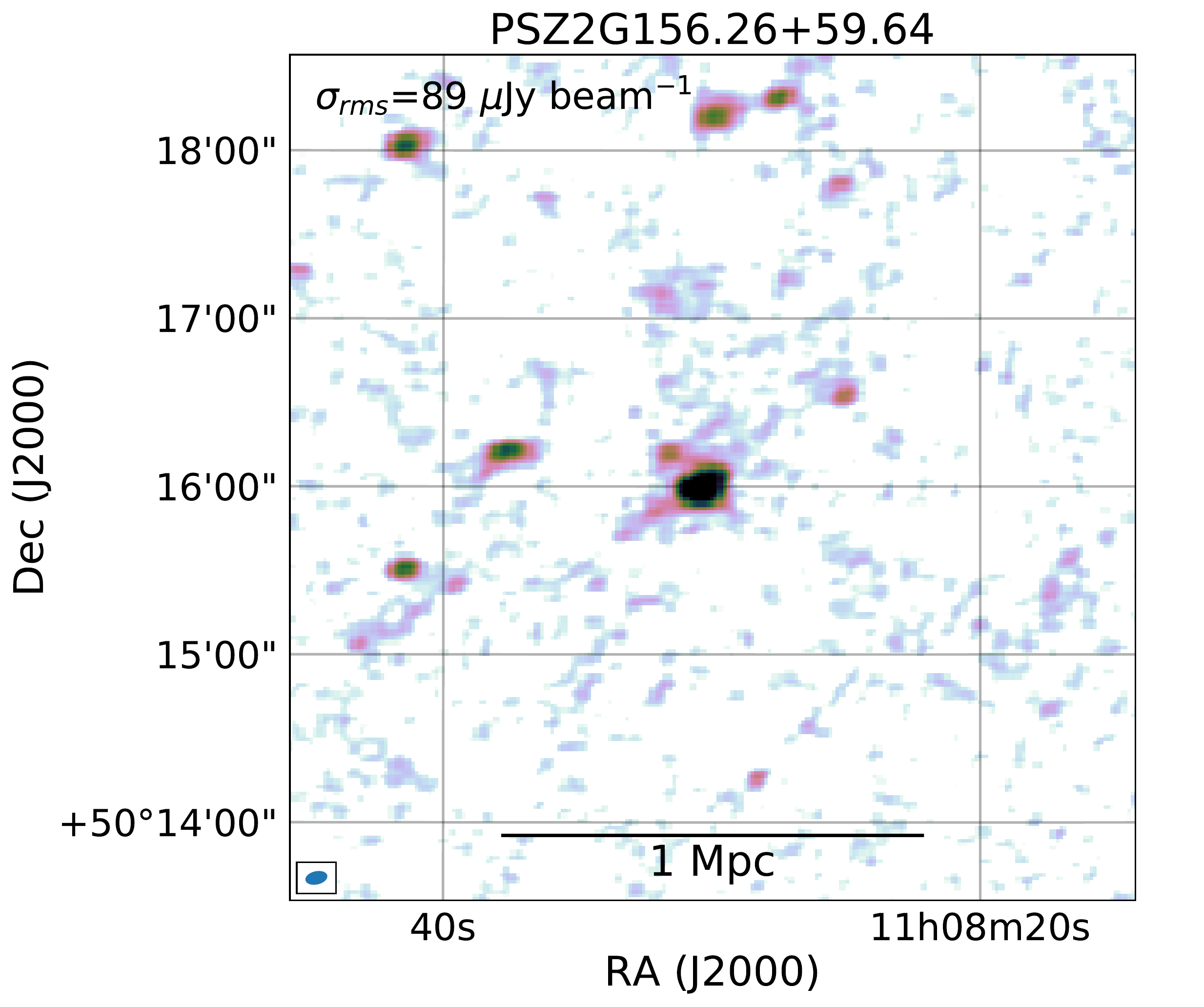} 
    \includegraphics[width=0.285\paperwidth]{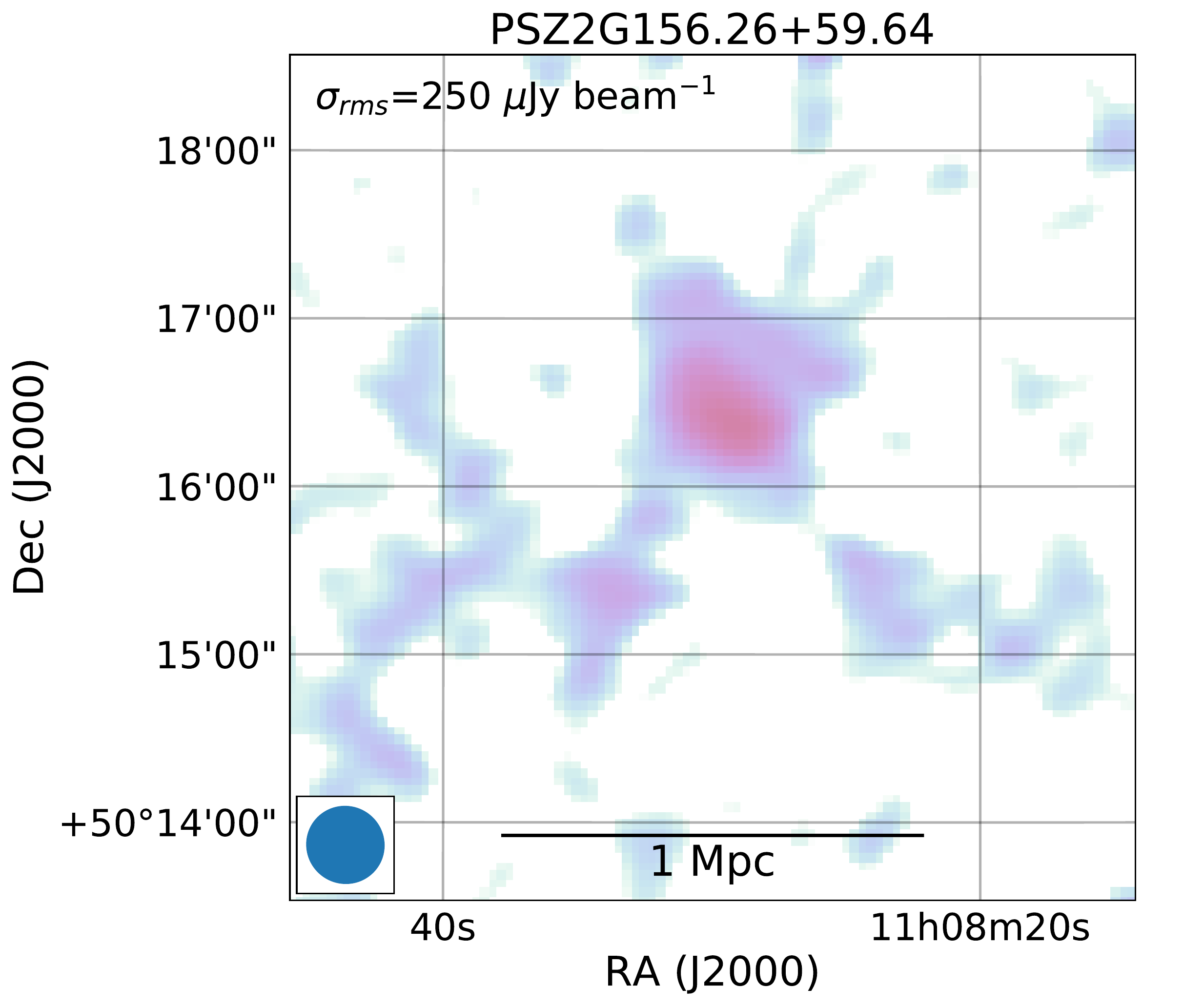}
   \includegraphics[width=0.285\paperwidth]{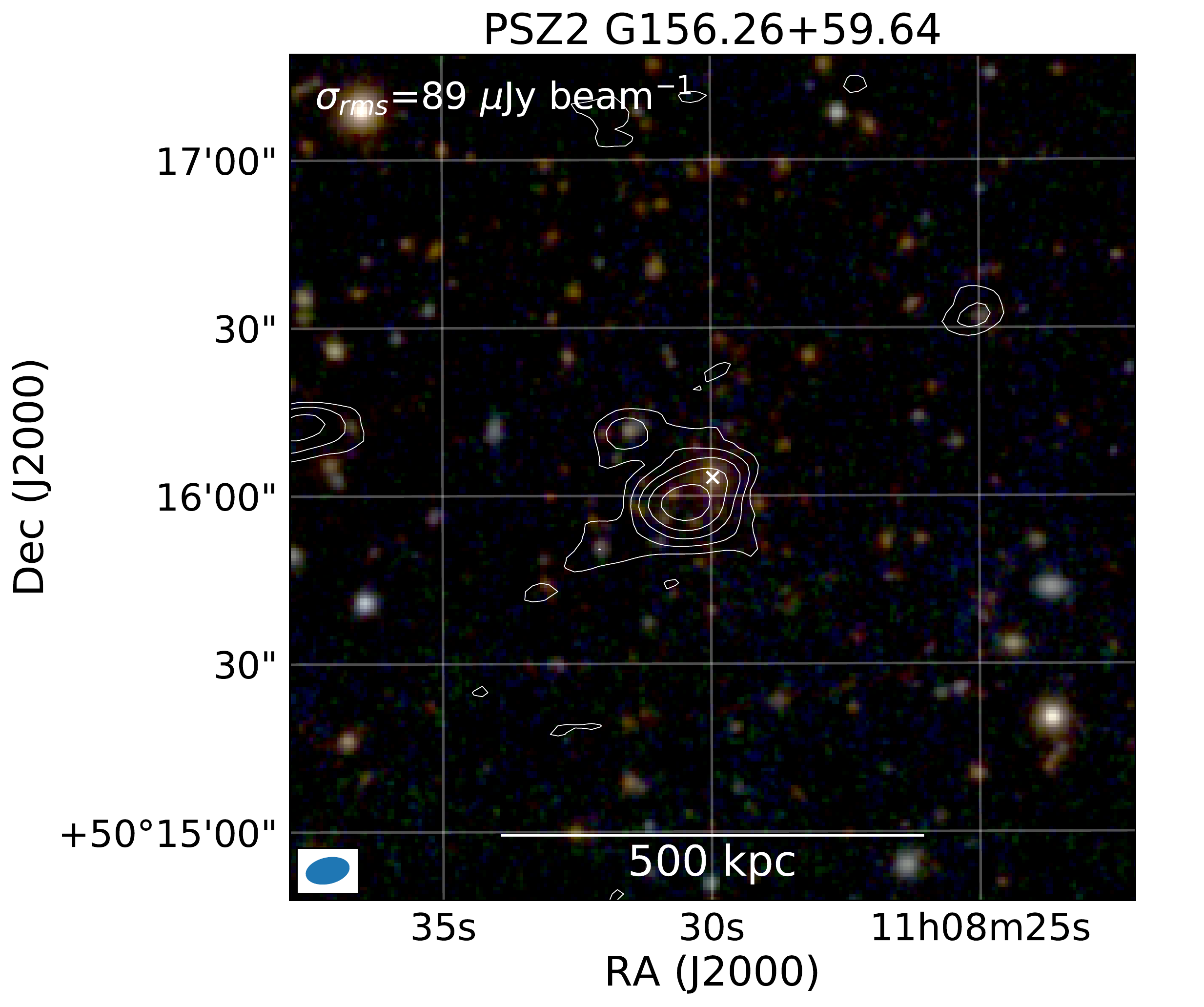}
   \caption{PSZ2\,G156.26+59.64. Left: Robust $-0.5$ radio image. Middle: 15\arcsec~tapered radio image with compact sources subtracted. Right: Optical image with Robust $-0.5$ image radio contours. For more details see the caption of Figure~\ref{fig:A2018}.}
   \label{fig:PSZ2G15626}
\end{figure*}

A bright compact radio source is detected in this  $z=0.6175$ cluster which seems to be associated with a BCG, see Figure~\ref{fig:PSZ2G15626}. In our low-resolution images, with compact sources removed, faint diffuse  emission is detected in the region above the compact radio source. Only a hint of this emission is detected in our high-resolution image. This diffuse emission has a total extent of about 0.5~Mpc. Given its approximate central location, with respect to the cluster member galaxies as judged from the Pan-STARRS images, and large physical size, we classify this source as a candidate radio halo. For the source we determine a flux density of $7.9\pm 3.7$~mJy.

\subsection{RXC\,J1053.7+5452} 
Peripheral diffuse radio emission in this cluster was first reported by \cite{2009ApJ...697.1341R} and subsequently studied by \cite{2011A&A...533A..35V}. Chandra and Suzaku observations were presented by \cite{2017PASJ...69...88I} showing the cluster is undergoing a merger event. The peripheral radio source is classified as a relic.

\begin{figure*}
\centering
   \includegraphics[width=1.0\columnwidth]{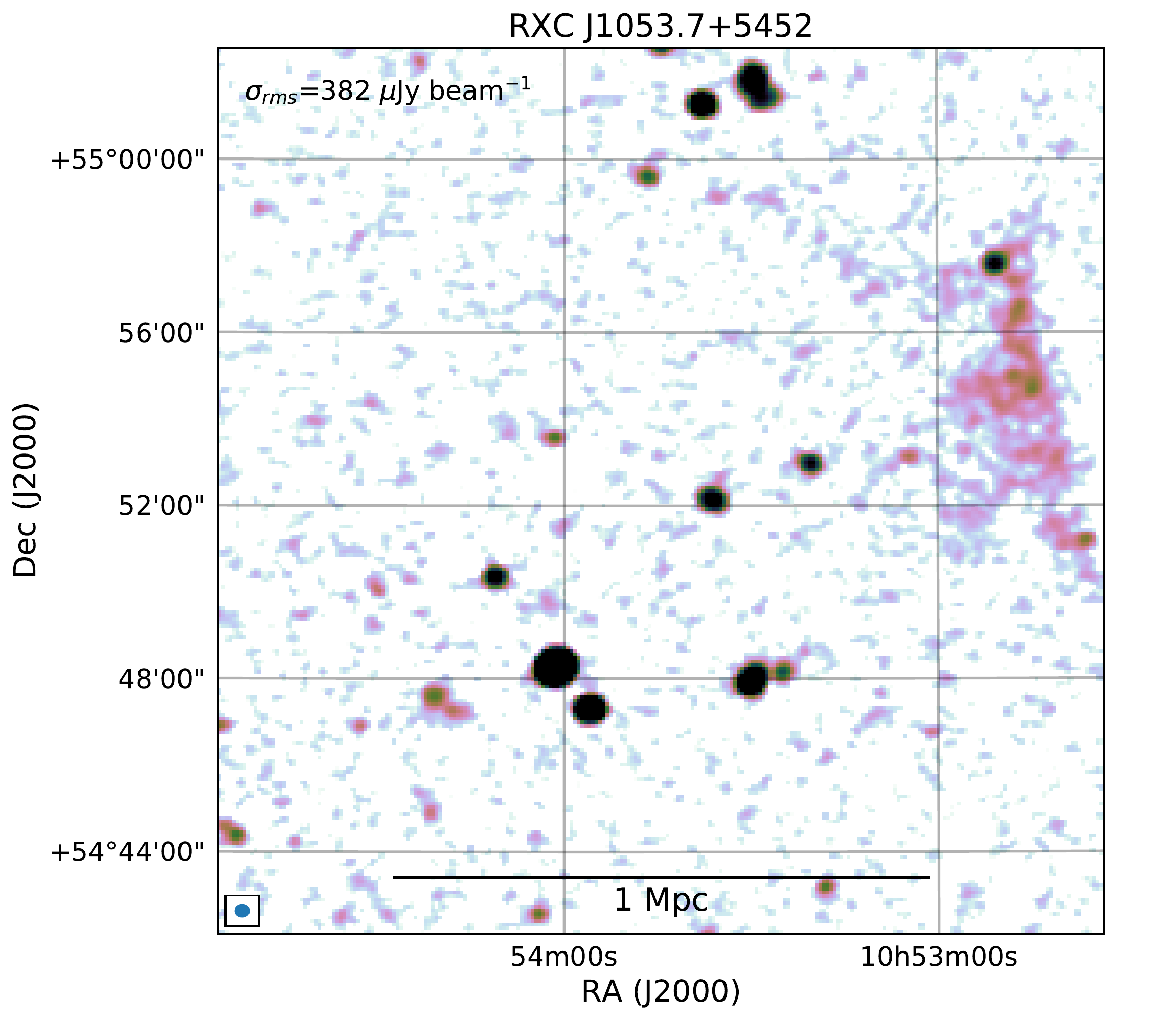}
   \includegraphics[width=1.0\columnwidth]{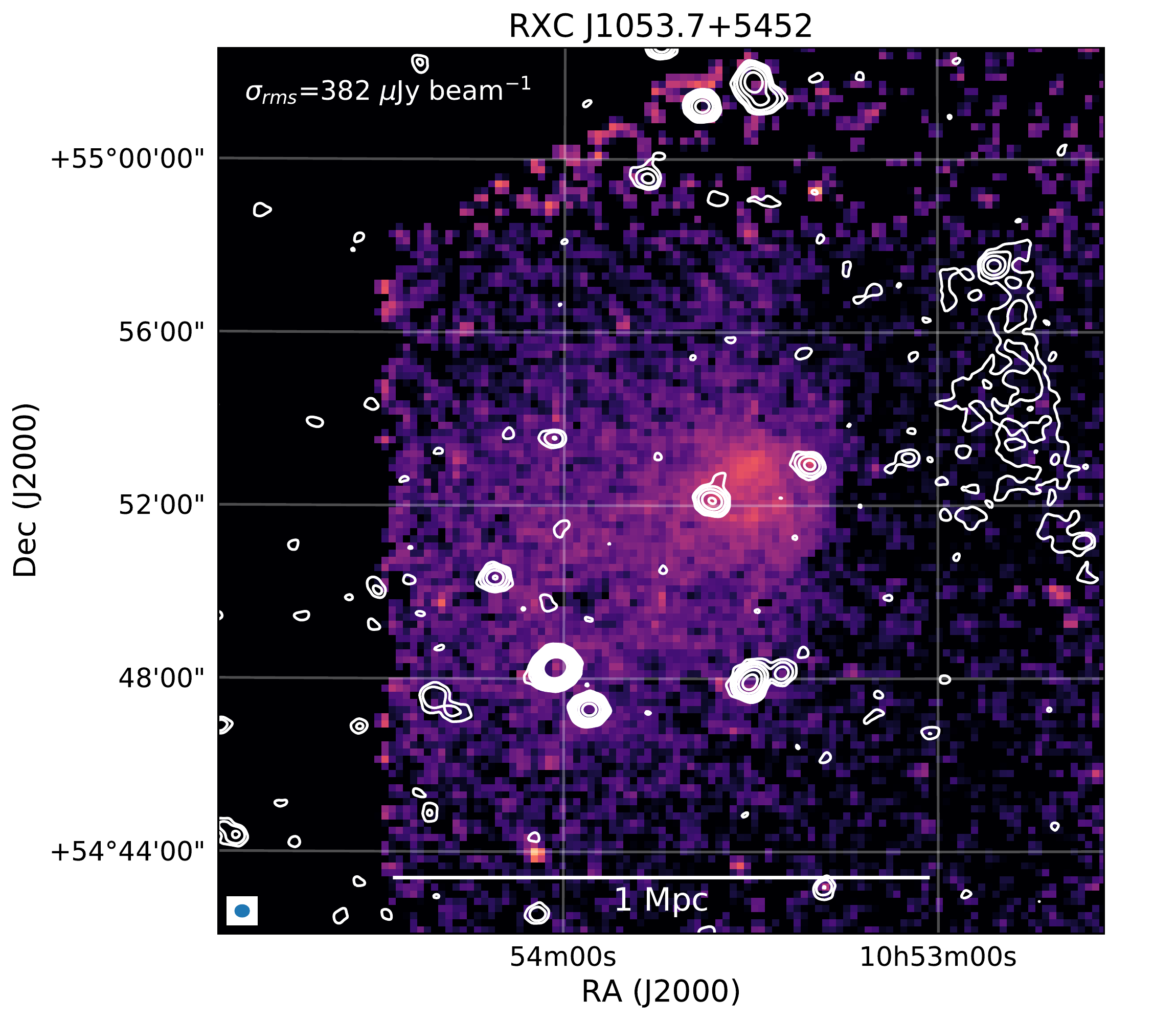}
   \caption{RXC\,J1053.7+5452. Left: Robust $-0.5$ radio image. Right: Chandra X-ray image with 10\arcsec~tapered radio contours. For more details see the caption of Figure~\ref{fig:A2018}.}
   \label{fig:RXC10537}
\end{figure*}

The main LOFAR pointing on this source (P164+55) had to be discarded as it was affected by bad ionospheric conditions. Hence the noise levels in our images are higher than for other clusters.  Despite the higher noise, the relic is clearly detected in our LOFAR images (Figure~\ref{fig:RXC10537}) and the source has a similar appearance as in the WSRT observations presented in \cite{2011A&A...533A..35V}. 
In the LOFAR image, the relic has an LLS of about 0.75~Mpc. A hint of an extension is visible from the northern tip of the relic (near a compact source) towards the west and north. Combining the LOFAR flux density measurement with the one obtained from the WSRT, we obtain $\alpha=-1.17\pm0.11$, which is typical for radio relics. 

\subsection{Abell\,1156}
\begin{figure*}
\centering
   \includegraphics[width=0.285\paperwidth]{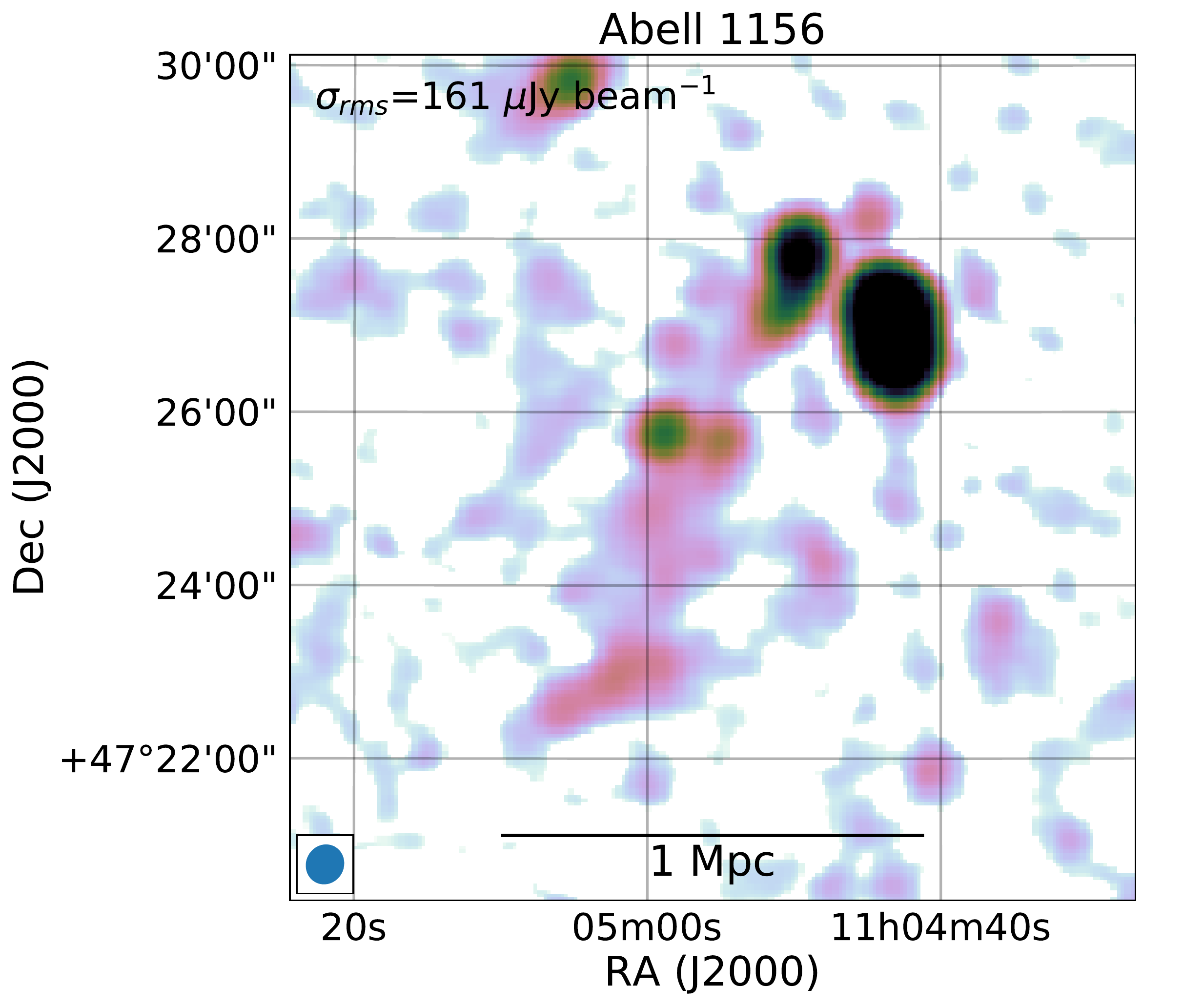}
   \includegraphics[width=0.285\paperwidth]{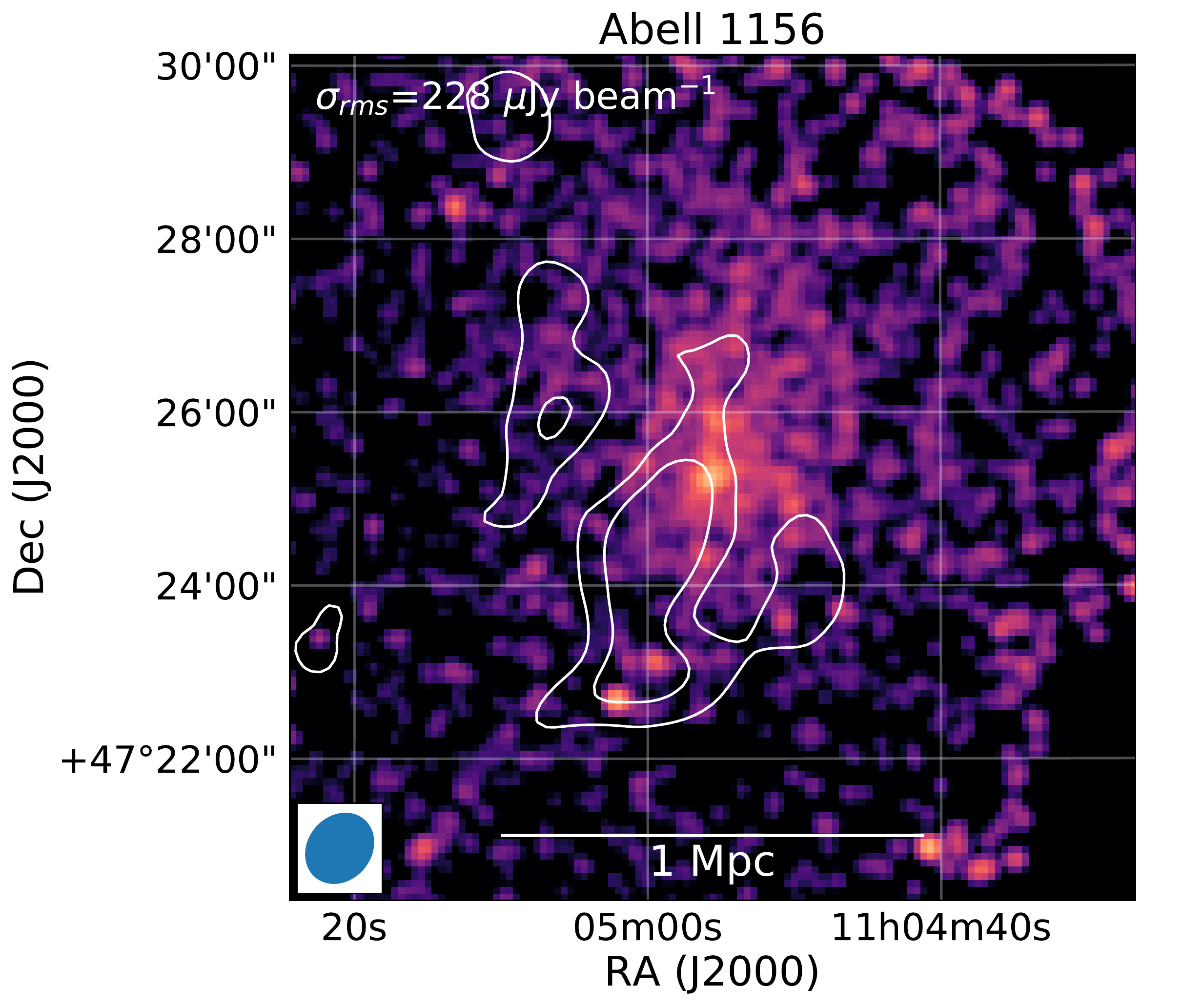}
   \includegraphics[width=0.285\paperwidth]{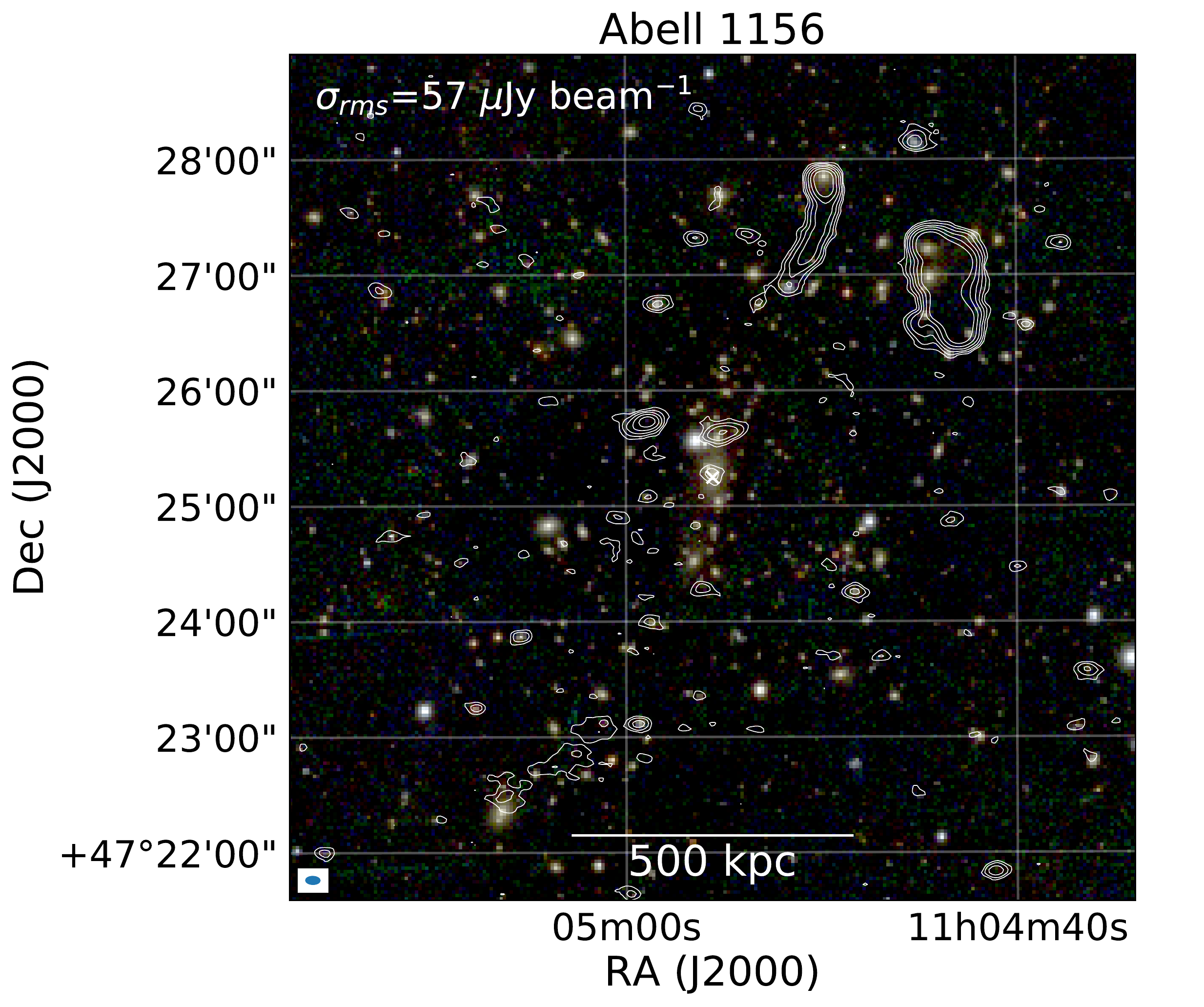}
   \caption{Abell\,1156. Left: 15\arcsec~tapered radio image. Middle: Chandra X-ray image with 30\arcsec~tapered radio contours (compact sources were subtracted). Right: Optical image with Robust $-0.5$ image radio contours. For more details see the caption of Figure~\ref{fig:A2018}.}
   \label{fig:A1156}
\end{figure*}

A Chandra X-ray image displays a cluster that is elongated in the NS direction (Figure~\ref{fig:A1156}). Diffuse  radio emission, also elongated in the NS direction, is detected in this cluster with an LLS of 0.7~Mpc. This radio emission does not peak at the cluster center but south of it. However, some faint diffuse emission is also visible north of the cluster center. We list this source as a candidate radio halo, originating from a possible NS merger event. Additional extended radio emission, with a mostly EW elongation is found in the southern periphery of the cluster, its origin is not fully clear but it might be related to AGN activity as a connection to a cluster member galaxy is suggested. The cluster also hosts a prominent head-tail radio source in the north, with the tail extending south. along the direction of the proposed merger axis. 
By fitting the elliptical model we determine $S_{144}=15.7\pm6.8$~mJy for the candidate radio halo.


\subsection{Abell\,1314}

\begin{figure*}
\centering
   \includegraphics[width=0.285\paperwidth]{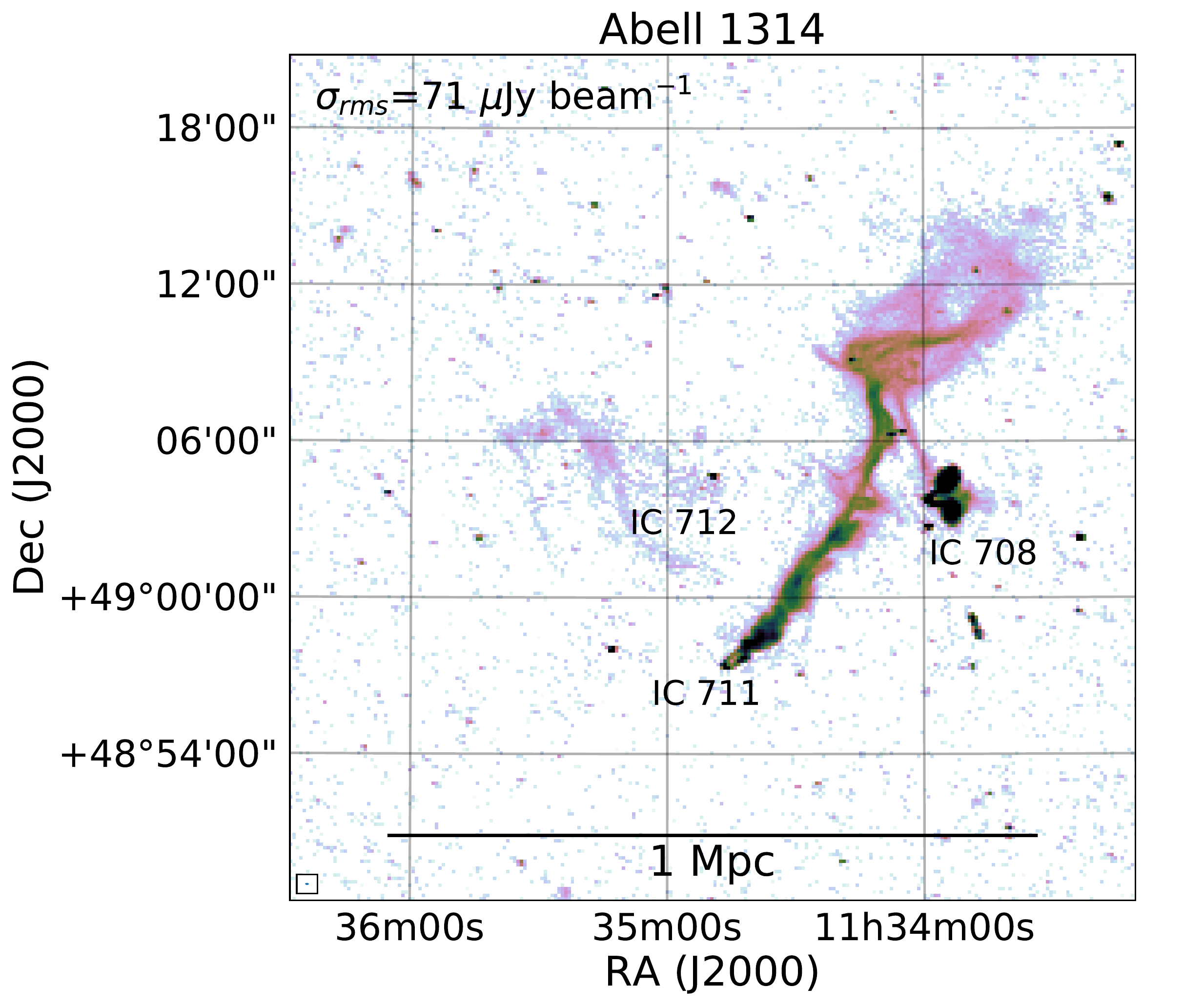}
   \includegraphics[width=0.285\paperwidth]{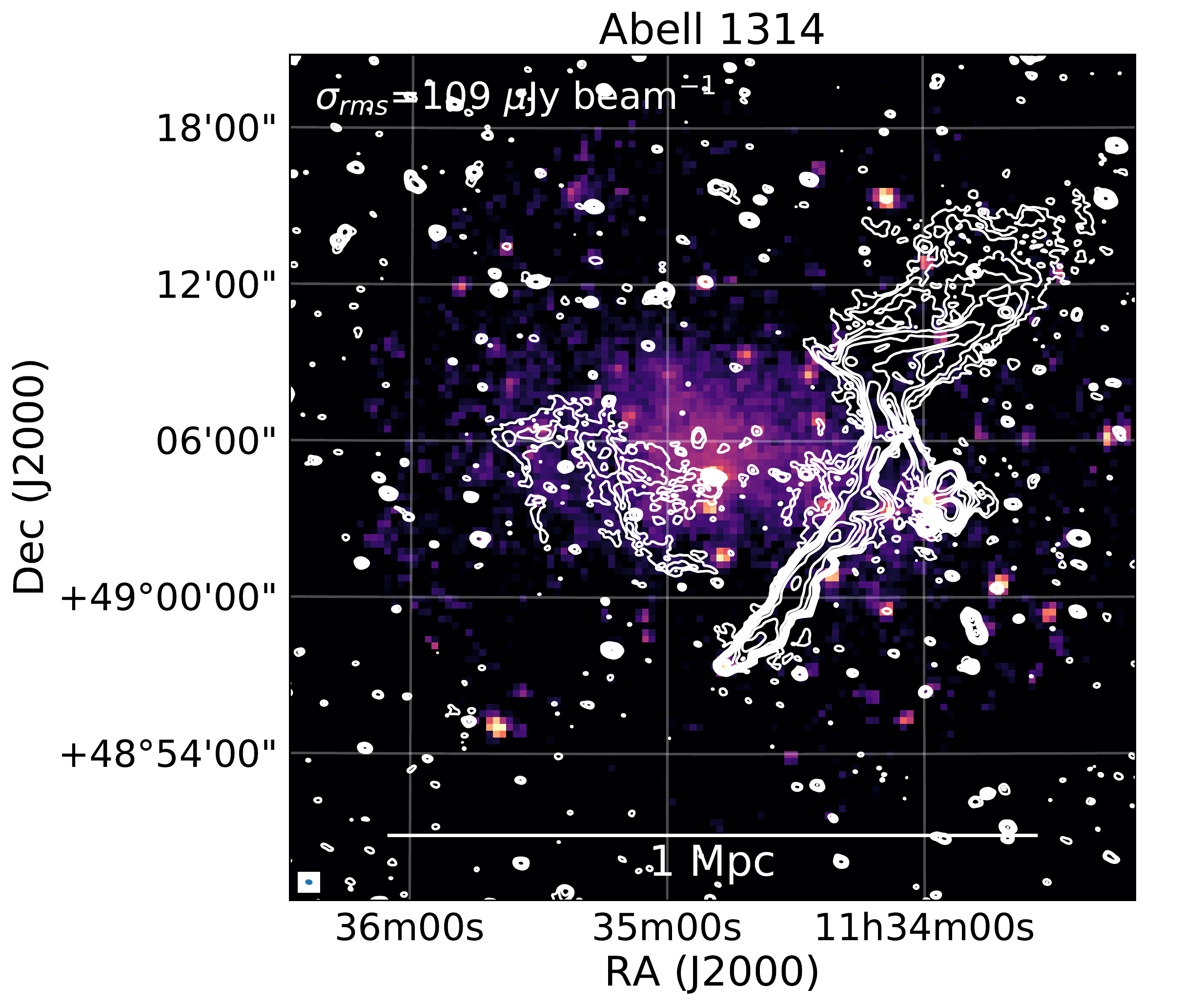}
   \includegraphics[width=0.285\paperwidth]{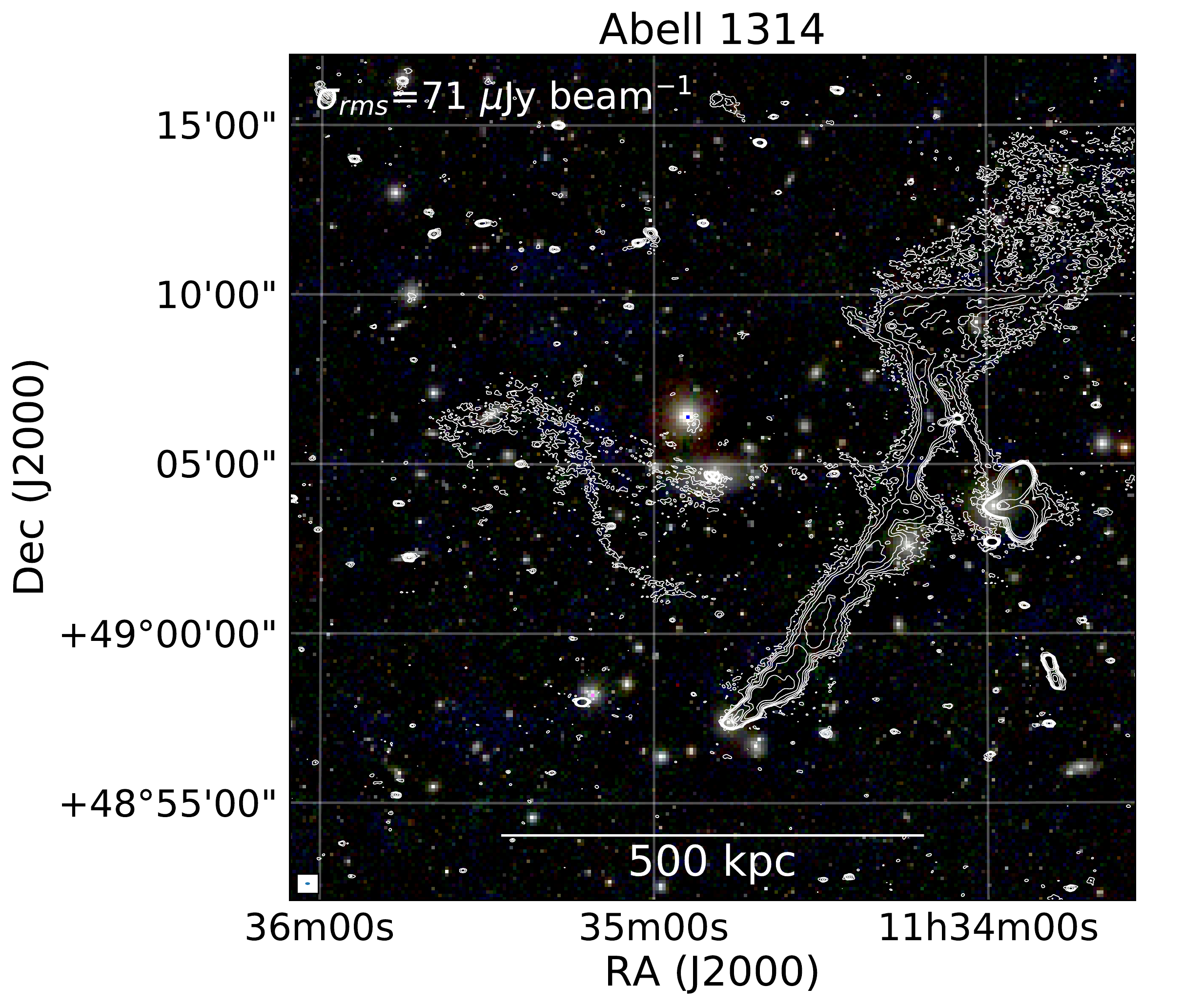}
   \caption{Abell\,1314. Left: Robust $-0.5$ radio image. Middle: XMM-Newton X-ray image with 10\arcsec~tapered radio contours. Right: Optical image with Robust $-0.5$ image radio contours. For more details see the caption of Figure~\ref{fig:A2018}.}
   \label{fig:A1314}
\end{figure*}

This nearby low-mass cluster hosts a bright 800~kpc long tailed radio galaxy associated with the galaxy \object{IC\,711}. The main tails shows a range of complicated linear features, see Figure~\ref{fig:A1314}. LOFAR observations of this source are described \cite{ 2019AA...622A..25W}. GMRT observations have been presented by  \cite{2016arXiv161007783S,2017AJ....154..169S}. Another smaller bright tailed radio galaxy is associated with  \object{IC\,708}. In addition, a filamentary source is detected near the cluster center. This is a candidate radio phoenix related to the central BCG  \object{IC\,712}. Our new LOFAR images of Abell\,1314 reveal some additional details not visible in the previous LOFAR images presented in \cite{ 2019AA...622A..25W} due to the improved calibration. One of these is a thin elongated structure that connects \object{IC\,708} to \object{IC\,711}. Its origin is unclear. In addition, our images show more clearly the filamentary nature of the phoenix source, with an LLS of 0.44~Mpc.

\subsection{NSC\,J143825+463744}
\begin{figure*}
\centering
   \includegraphics[width=1.0\columnwidth]{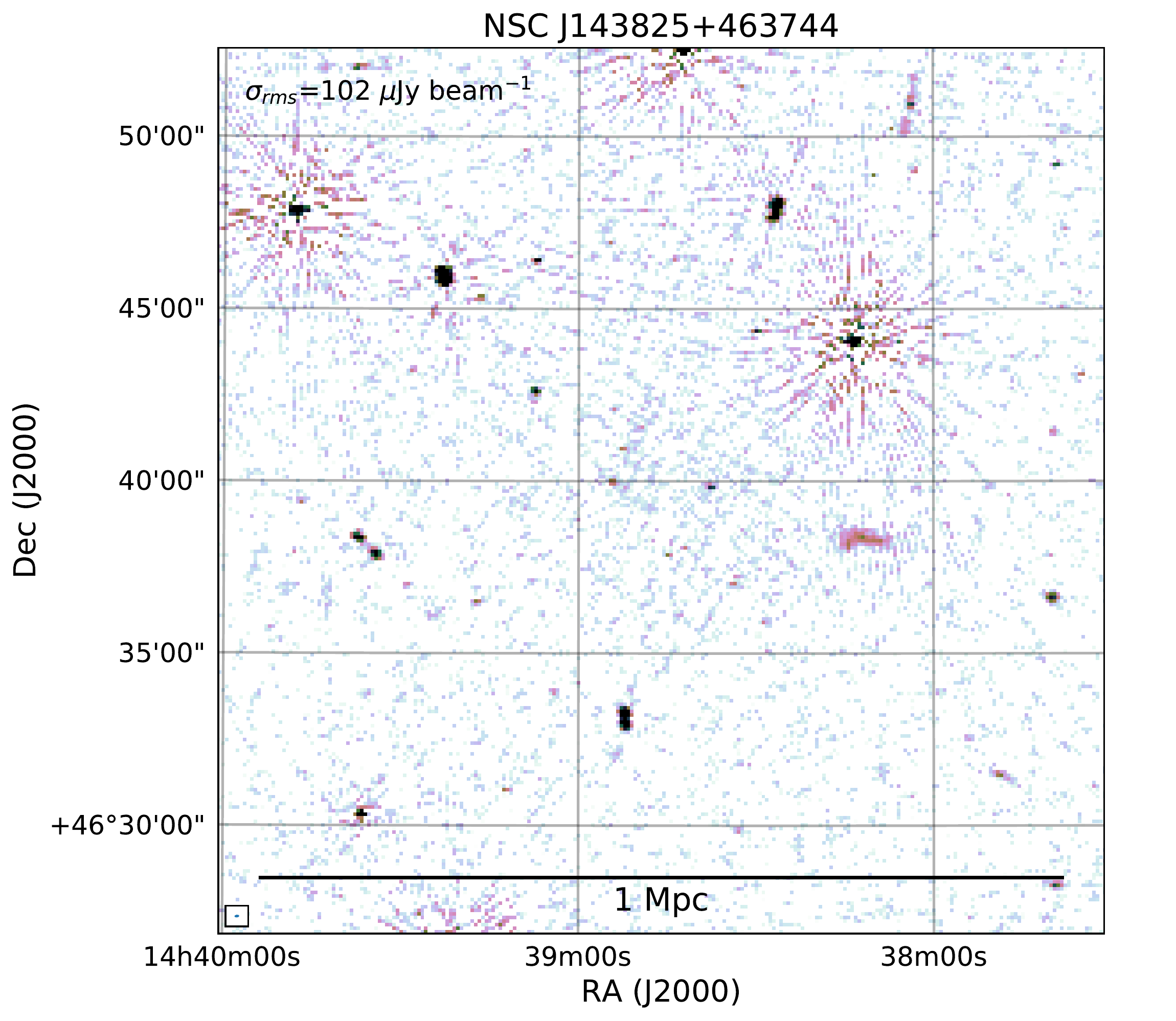}
   \includegraphics[width=1.0\columnwidth]{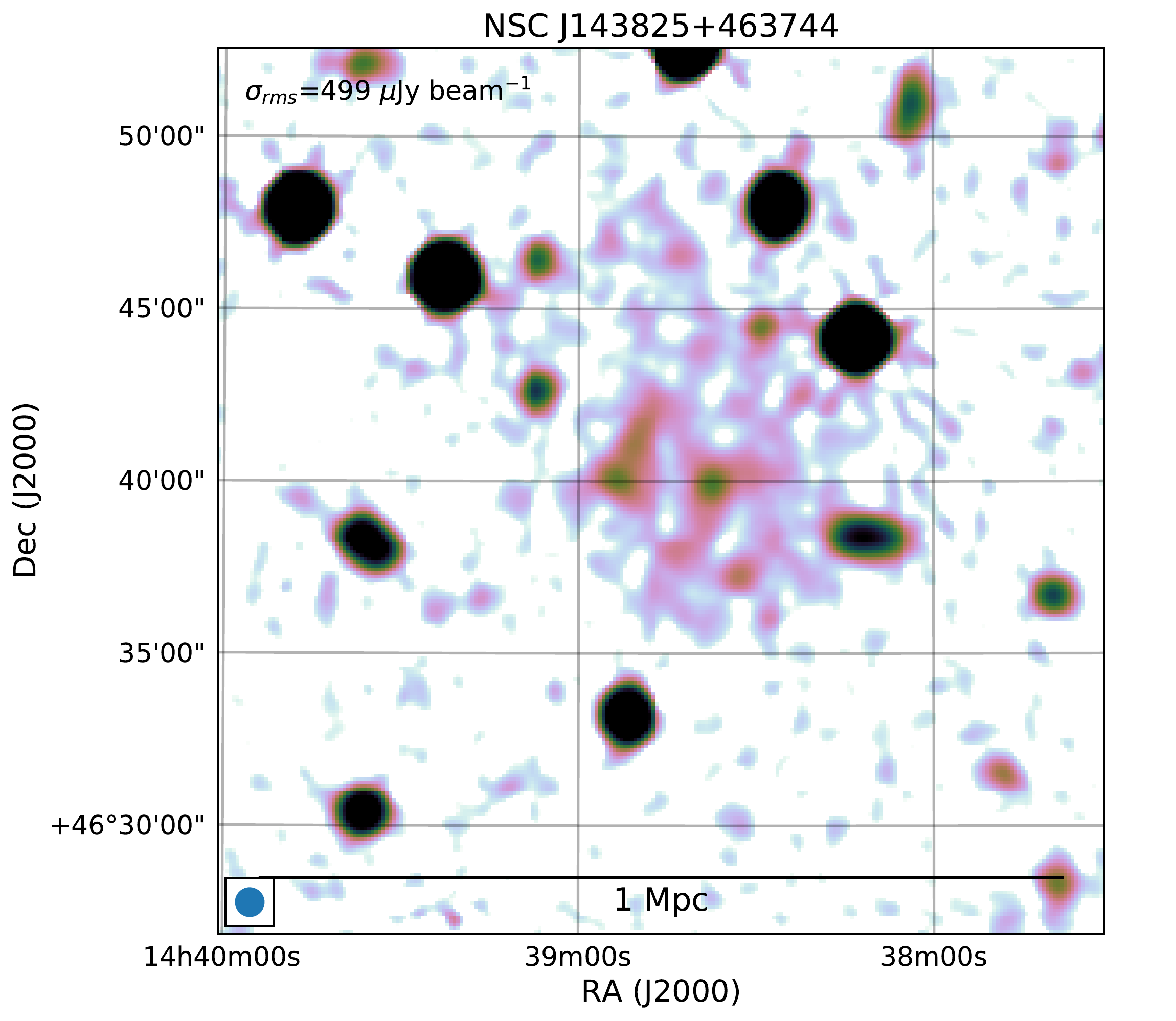}
   \caption{NSC\,J143825+463744.  Left: Robust $-0.5$ radio image. Right: 30\arcsec~tapered radio image. For more details see the caption of Figure~\ref{fig:A2018}.}
   \label{fig:NSCJ1438}
\end{figure*}

The LOFAR observations for NSC\,J143825+463744, a nearby $z=0.03586$ system, are affected by bad ionospheric conditions. Despite of the poor image quality, central extended emission is detected with a size of about$\sim0.4$~Mpc, see Figure~\ref{fig:NSCJ1438}. NSC\,J143825+463744 was  classified as a galaxy group (\object{MLCG\,1495}) by  \cite{2003AJ....125.2064G}. It is composed of two dominant galaxies \object{NGC\,5722} and \object{NGC\,5717}. No X-ray emission is detected from this system by ROSAT, confirming it has a low mass. Given the lack of an ICM detection by ROSAT, low mass of the system, and poor calibration, we list it as unclassified.

Note that a small EW extended source, right of the image center, is associated with \object{NGC\,5714}. NGC\,5714 is foreground galaxy and unrelated to \object{MLCG\,1495}.

\subsection{WHL\,J122418.6+490549}
\begin{figure*}
\centering
   \includegraphics[width=0.285\paperwidth]{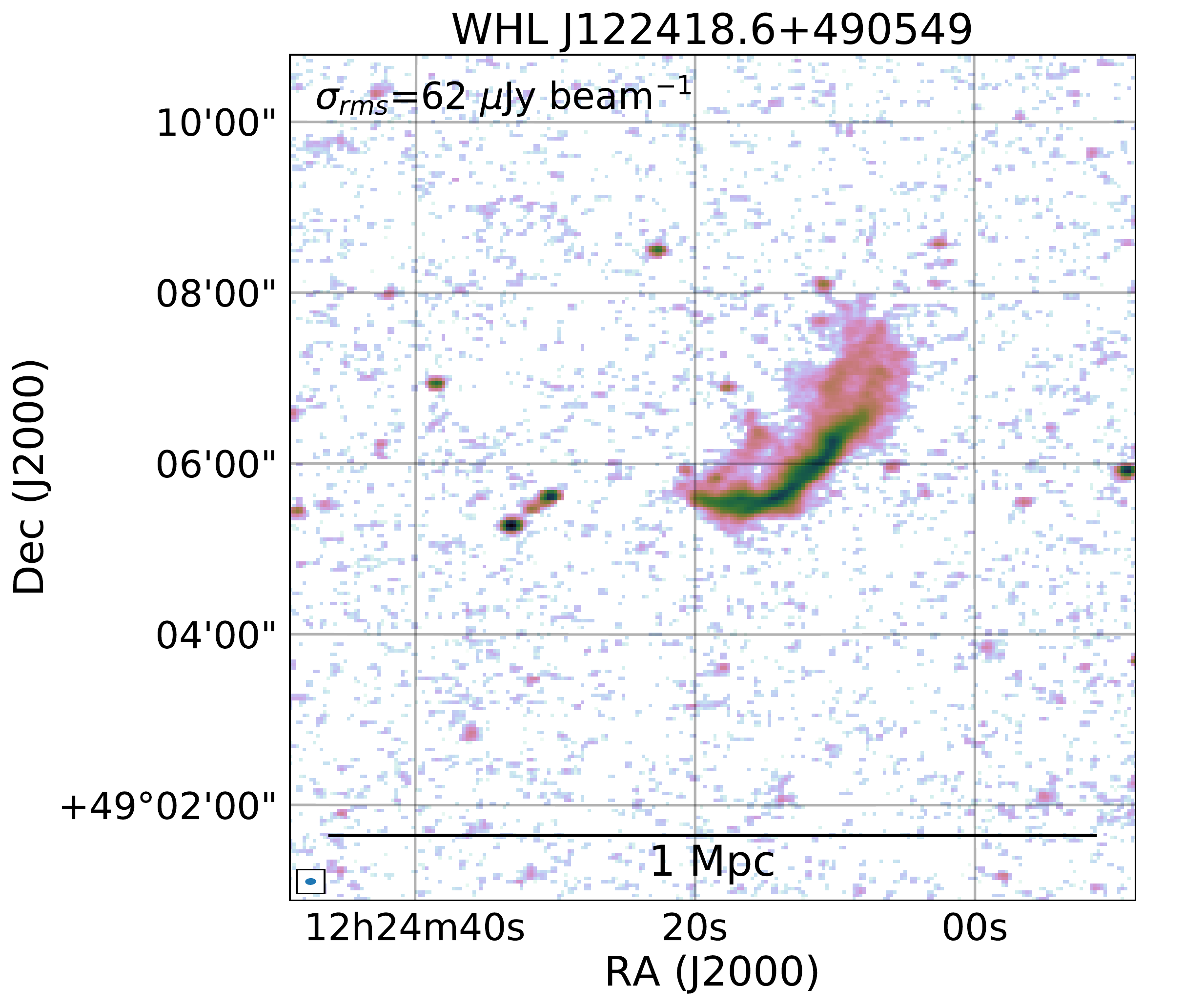}
   \includegraphics[width=0.285\paperwidth]{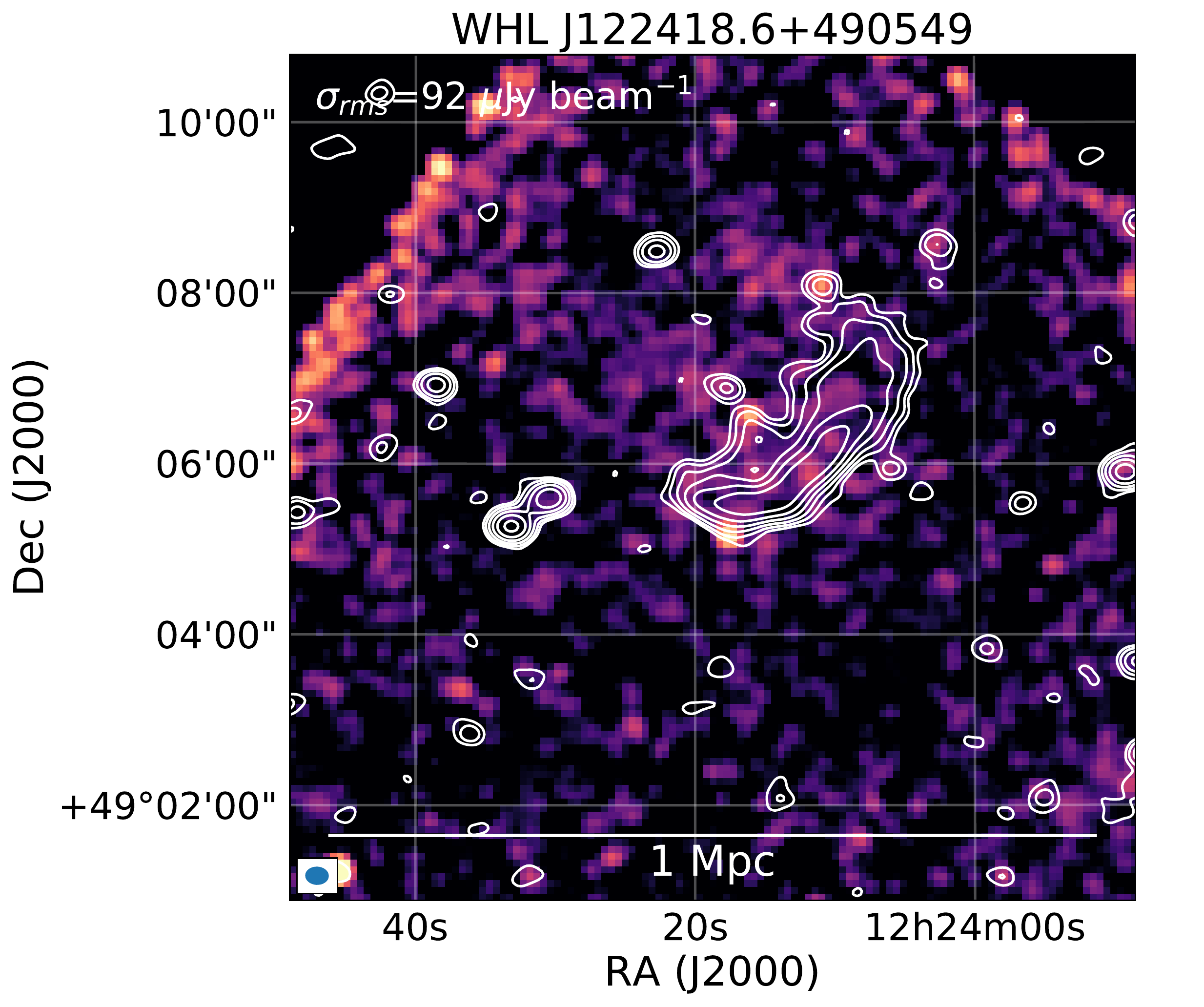}
   \includegraphics[width=0.285\paperwidth]{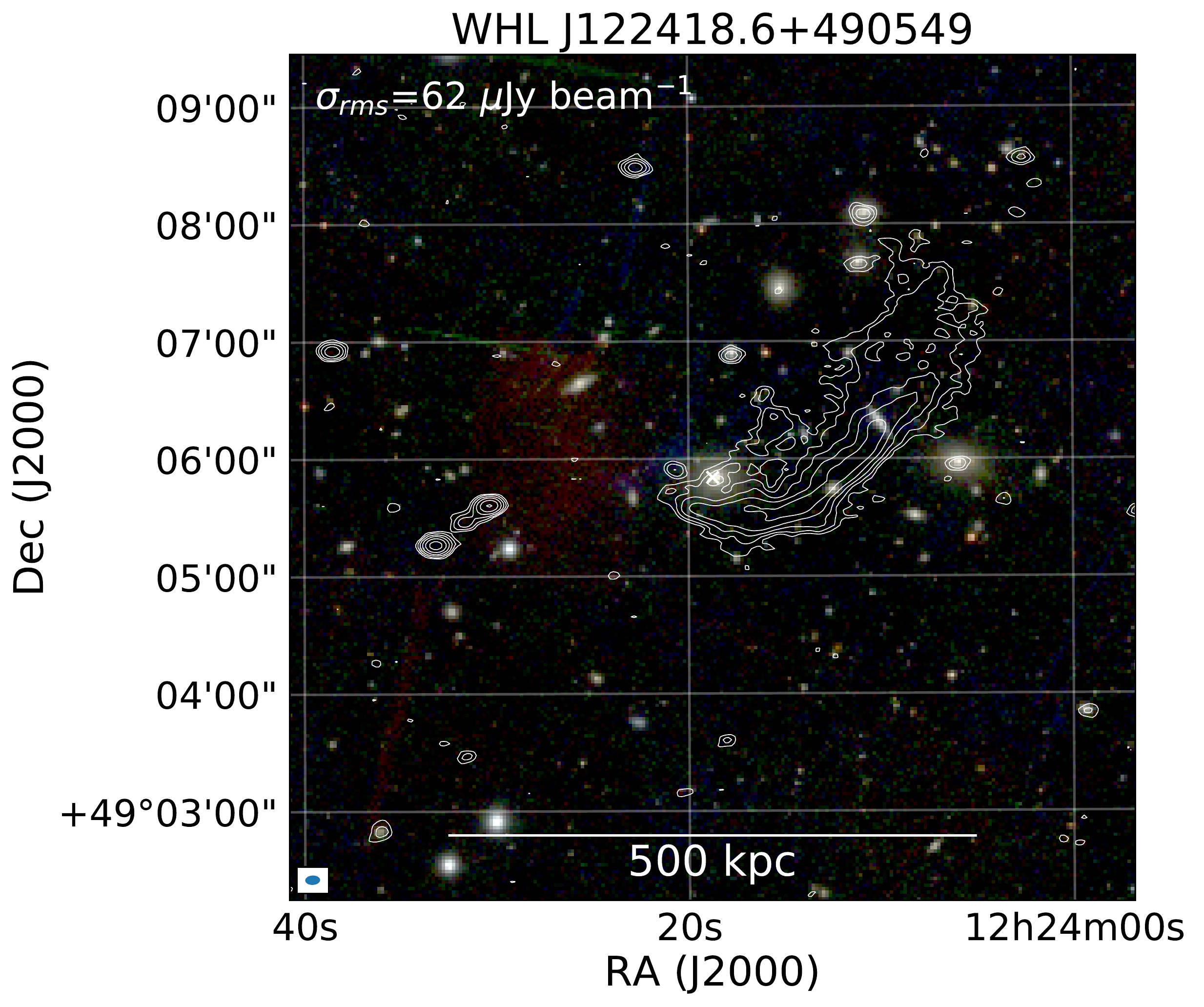}
   \caption{WHL\,J122418.6+490549. Left: Robust $-0.5$ radio image. Middle: Chandra X-ray image with 10\arcsec~tapered radio contours. Right: Optical image with Robust $-0.5$ image radio contours. For more details see the caption of Figure~\ref{fig:A2018}.}
   \label{fig:WHL1224}
\end{figure*}

WHL\,J122418.6+490549 hosts a bright elongated radio source with an LLS of 370~kpc (Figure~\ref{fig:WHL1224}). 
This radio emission could have originated from  the BCG (\object{LEDA\,2333420}), which is located at the southeastern tip of the elongated source.  The cluster is barely detected in a Chandra observation, indicating a low-mass system. The source seems somewhat similar to the revived remnant radio lobe found in low-mass cluster Abell\,1931  \citep{2018MNRAS.477.3461B}.

\subsection{WHL\,J124143.1+490510}
\begin{figure*}
\centering
   \includegraphics[width=1.0\columnwidth]{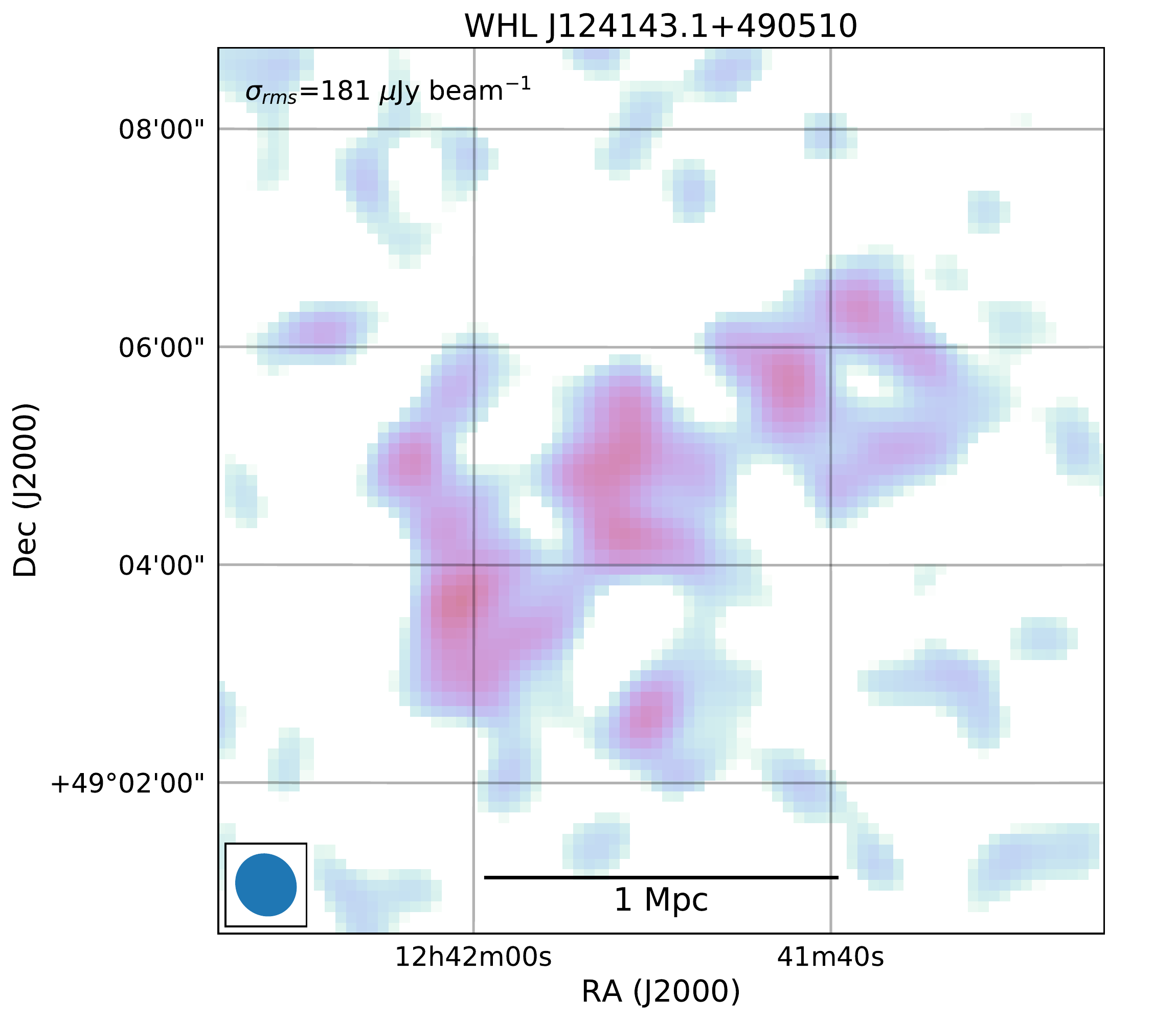}
   \includegraphics[width=1.0\columnwidth]{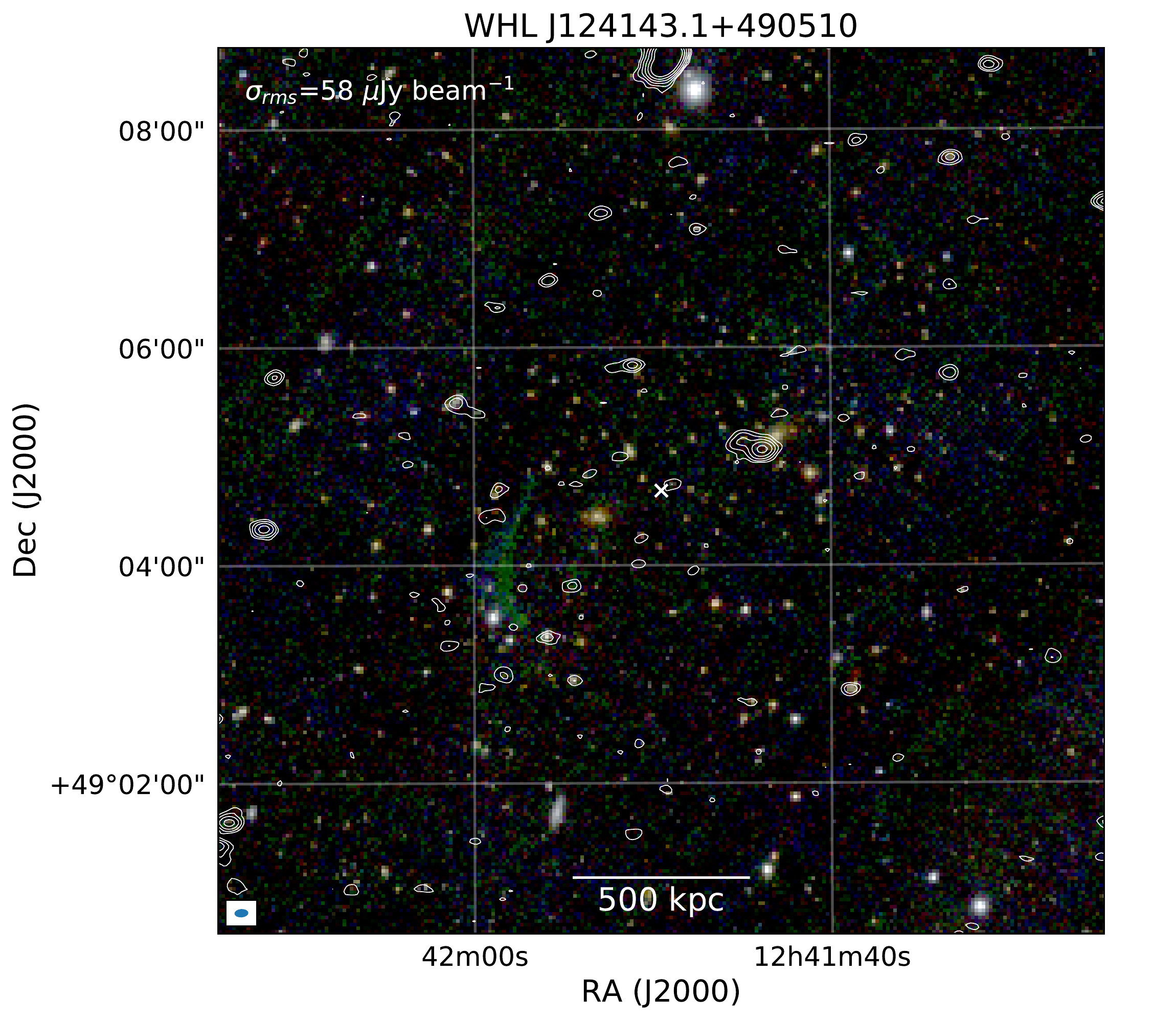}
   \caption{WHL\,J124143.1+490510.  Left: 30\arcsec~tapered radio image with compact sources subtracted.  Right: Optical image with Robust $-0.5$ image radio contours. For more details see the caption of Figure~\ref{fig:A2018}.}
   \label{fig:WHL1241}
\end{figure*}

Faint patchy extended emission is detected in our low resolution images of this cluster, see Figure~\ref{fig:WHL1241}. This emission extends over a region of 1.2~Mpc. Given that the emission is approximately centered on the galaxy distribution we classify it as a candidate radio halo.

\subsection{WHL\,J132226.8+464630}
\begin{figure*}
\centering
   \includegraphics[width=1.0\columnwidth]{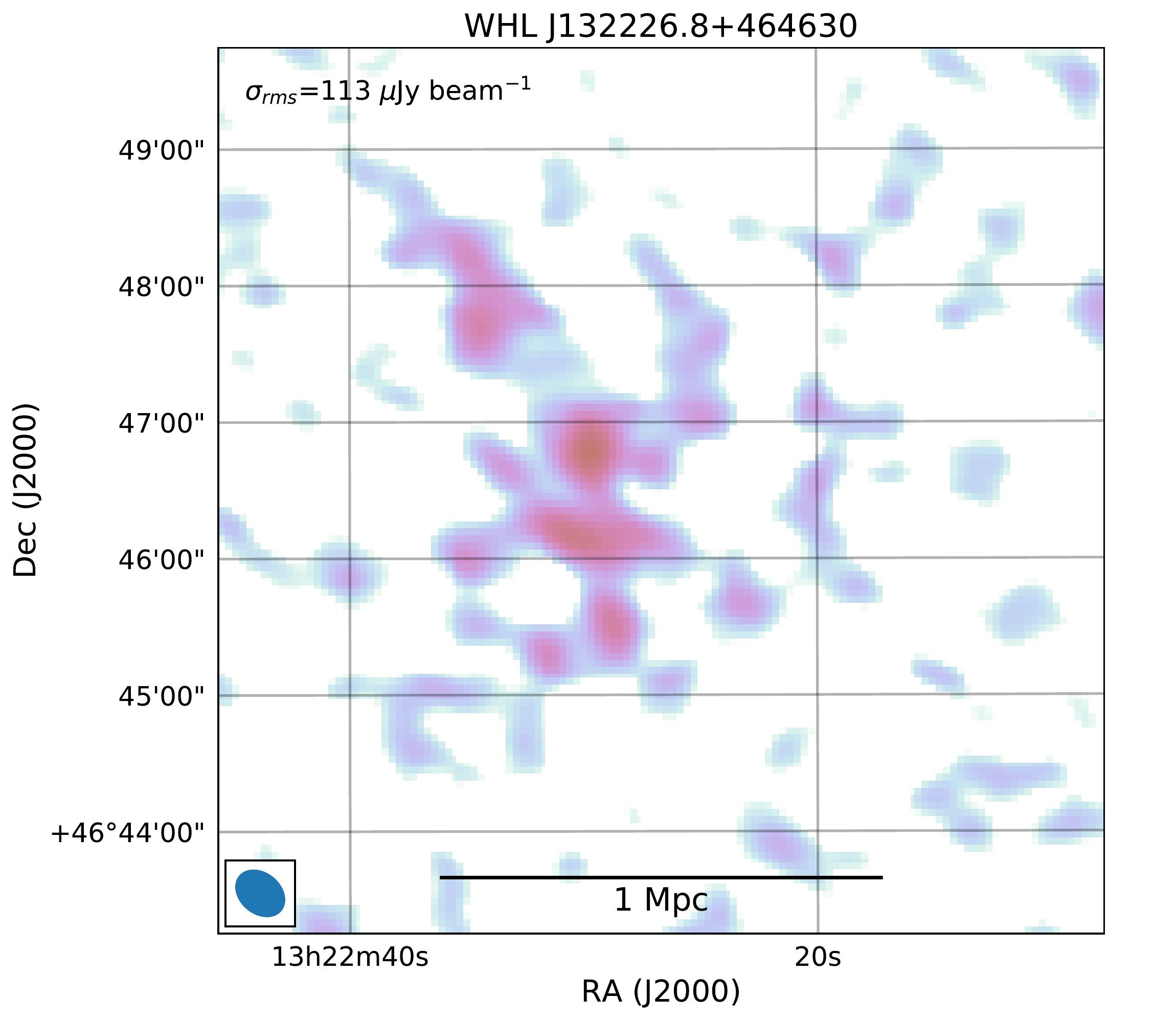}
   \includegraphics[width=1.0\columnwidth]{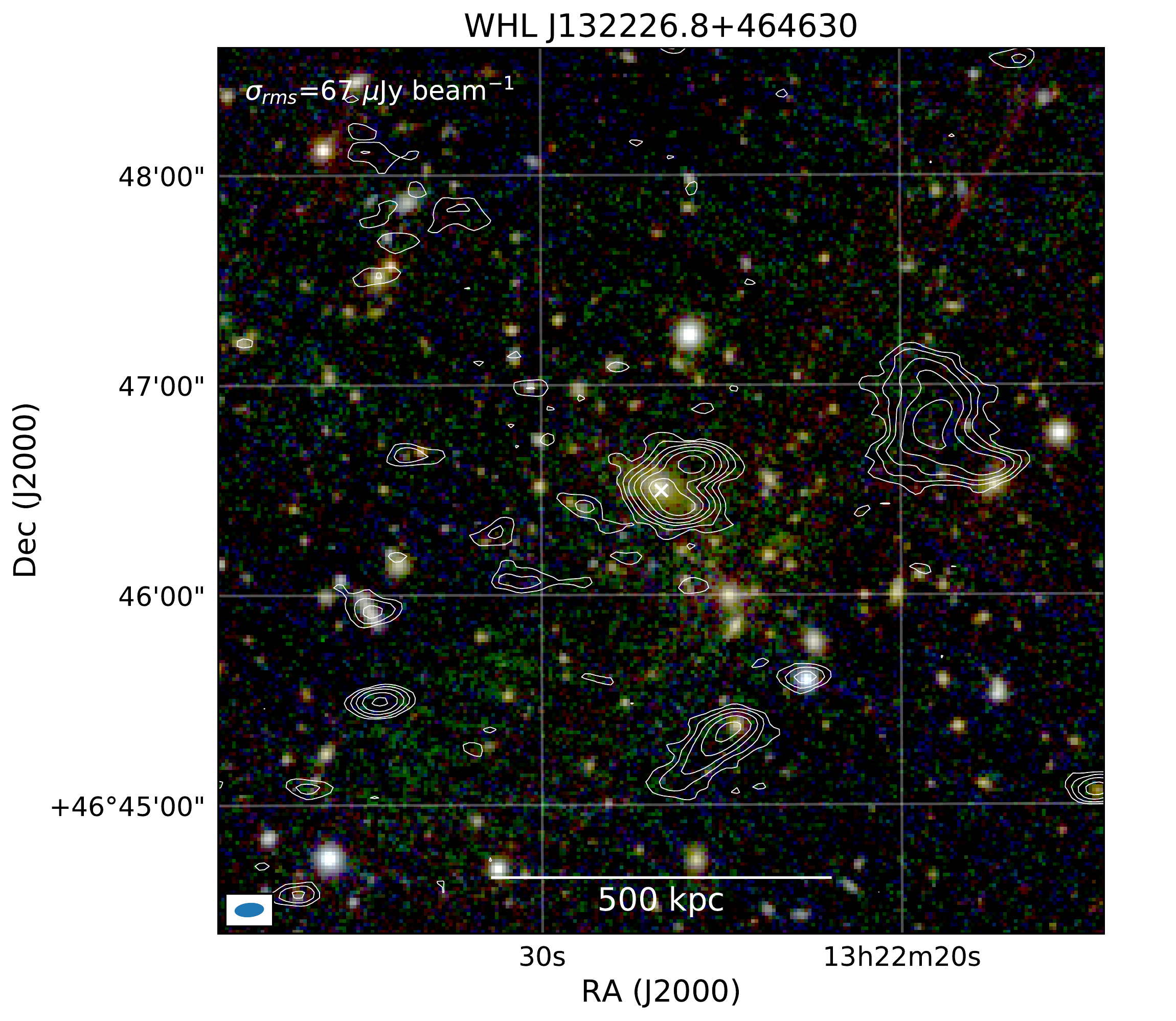}
   \caption{WHL\,J132226.8+464630. Left: 15\arcsec~tapered radio image  with compact sources subtracted. Right: Optical image with Robust $-0.5$ image radio contours. For more details see the caption of Figure~\ref{fig:A2018}.}
   \label{fig:WHL1322}
\end{figure*}

Patchy diffuse radio emission with an extent of about 0.5~Mpc is detected west of the BCG (Figure~\ref{fig:WHL1322}). Given its approximate central location we classify it as a candidate radio halo.

\subsection{WHL\,J133936.0+484859}
\begin{figure*}
\centering
   \includegraphics[width=1.0\columnwidth]{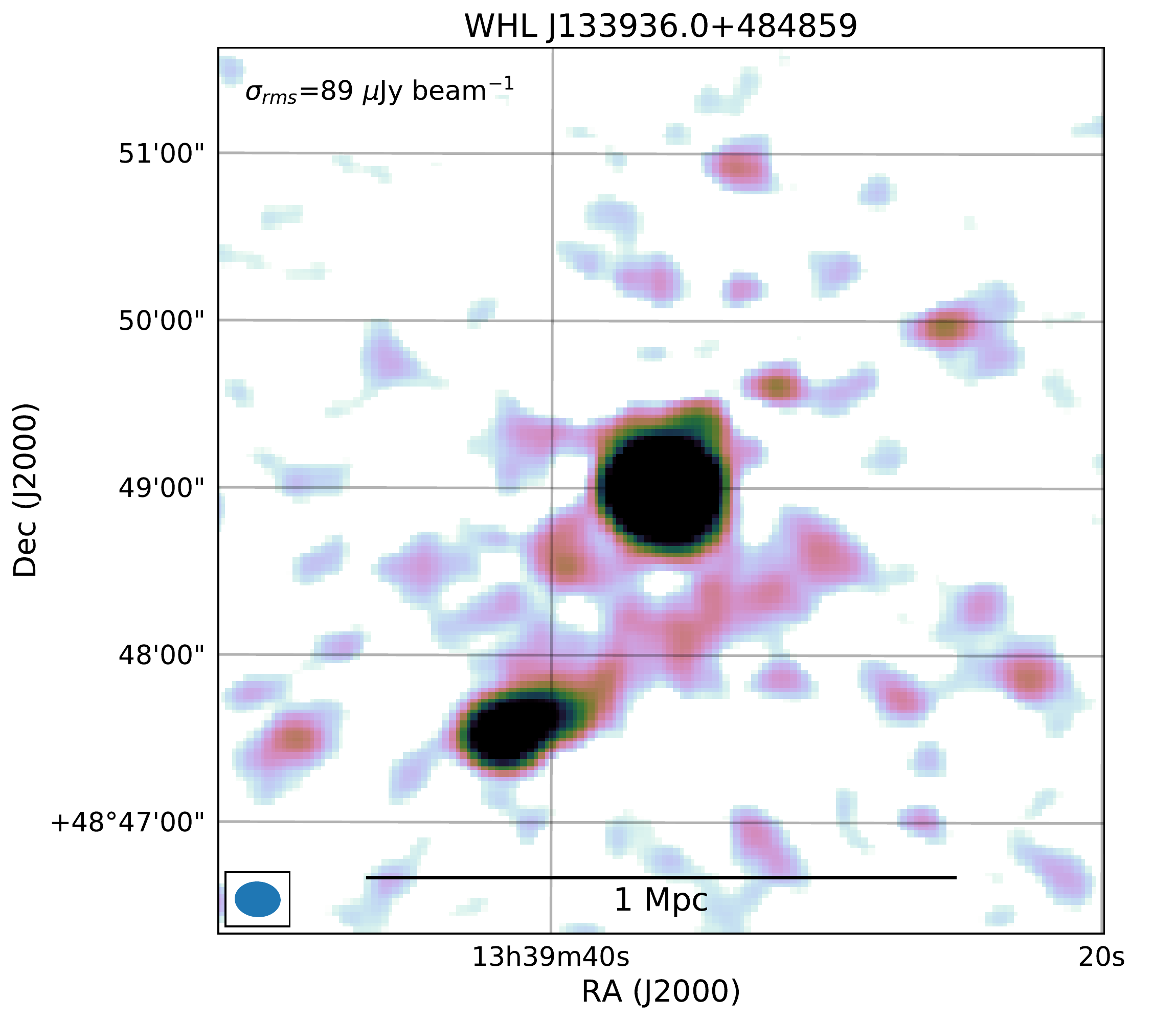}
   \includegraphics[width=1.0\columnwidth]{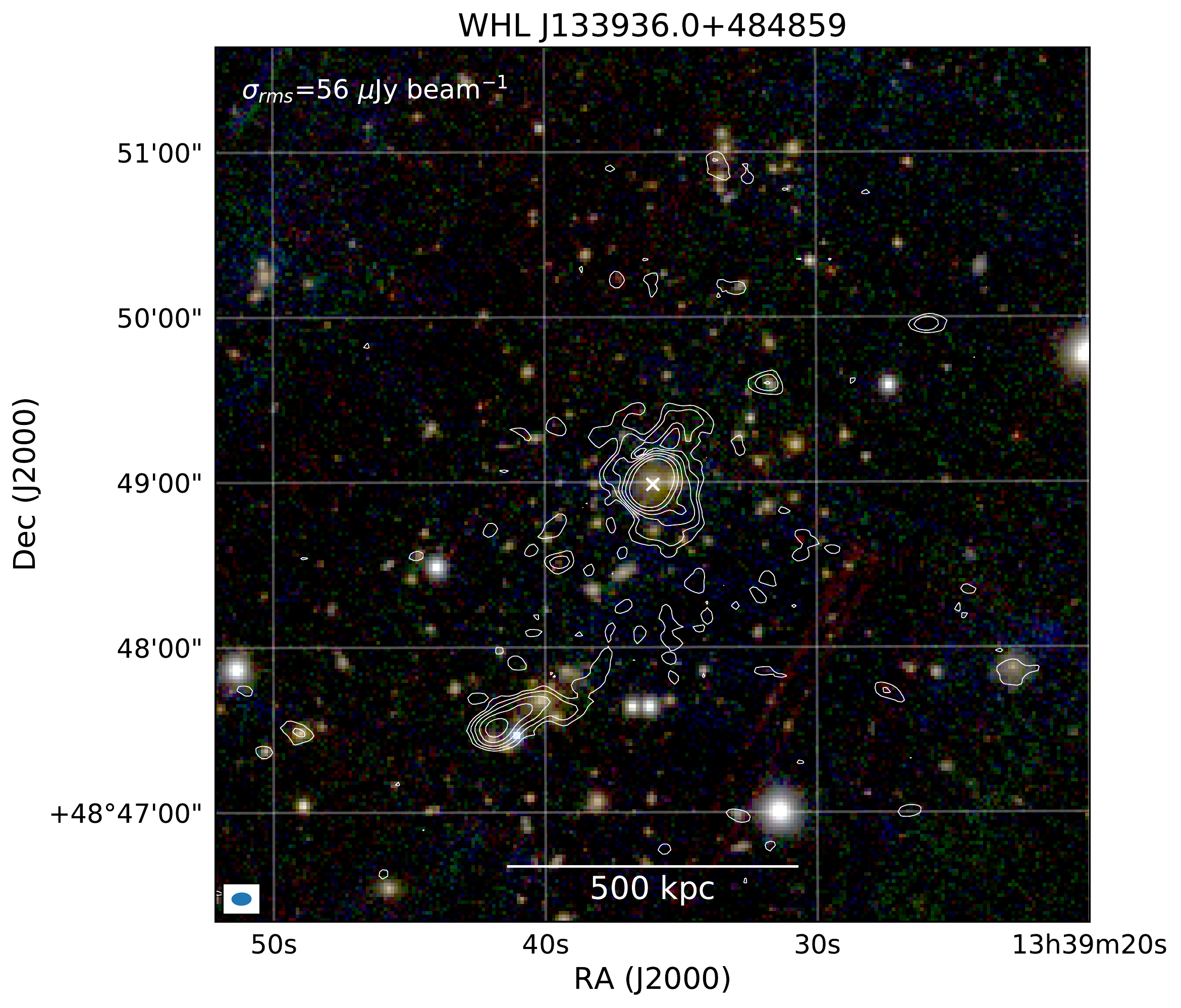}
   \caption{WHL\,J133936.0+484859. Left: 10\arcsec~tapered radio image. Middle: Right: Optical image with Robust $-0.5$ image radio contours. For more details see the caption of Figure~\ref{fig:A2018}.}
   \label{fig:WHL1339}
\end{figure*}

Extended radio emission is detected in this cluster south of the BCG, see Figure~\ref{fig:WHL1339}. This emission seems to be the extension of a tailed radio galaxy.  Additional extended ($\sim300$~kpc) emission surrounds the BCG. This candidate radio (mini-)halo will need to be confirmed with deeper observations.

\subsection{WHL\,J125836.8+440111}
\begin{figure*}
\centering
   \includegraphics[width=1.0\columnwidth]{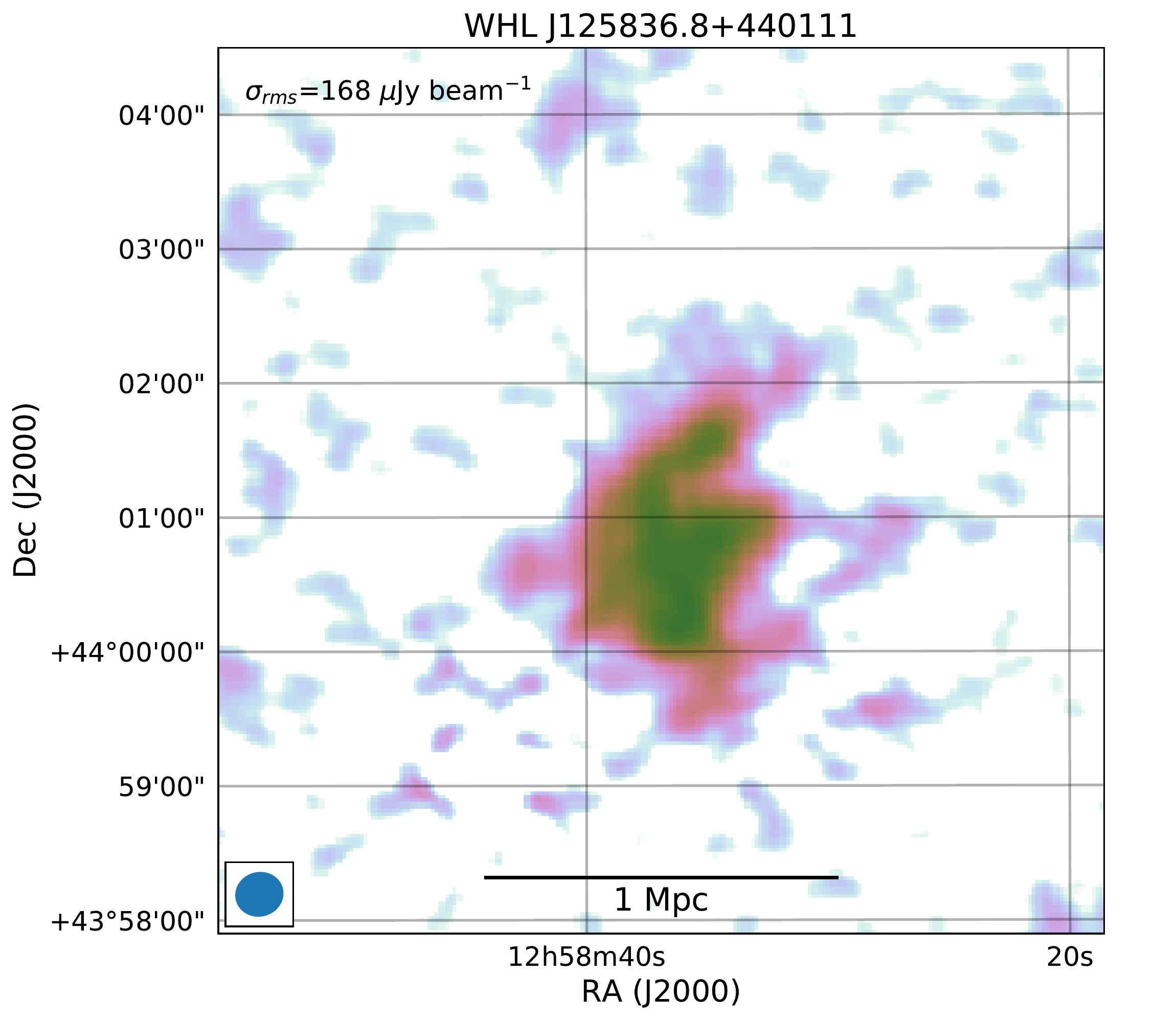}
   \includegraphics[width=1.0\columnwidth]{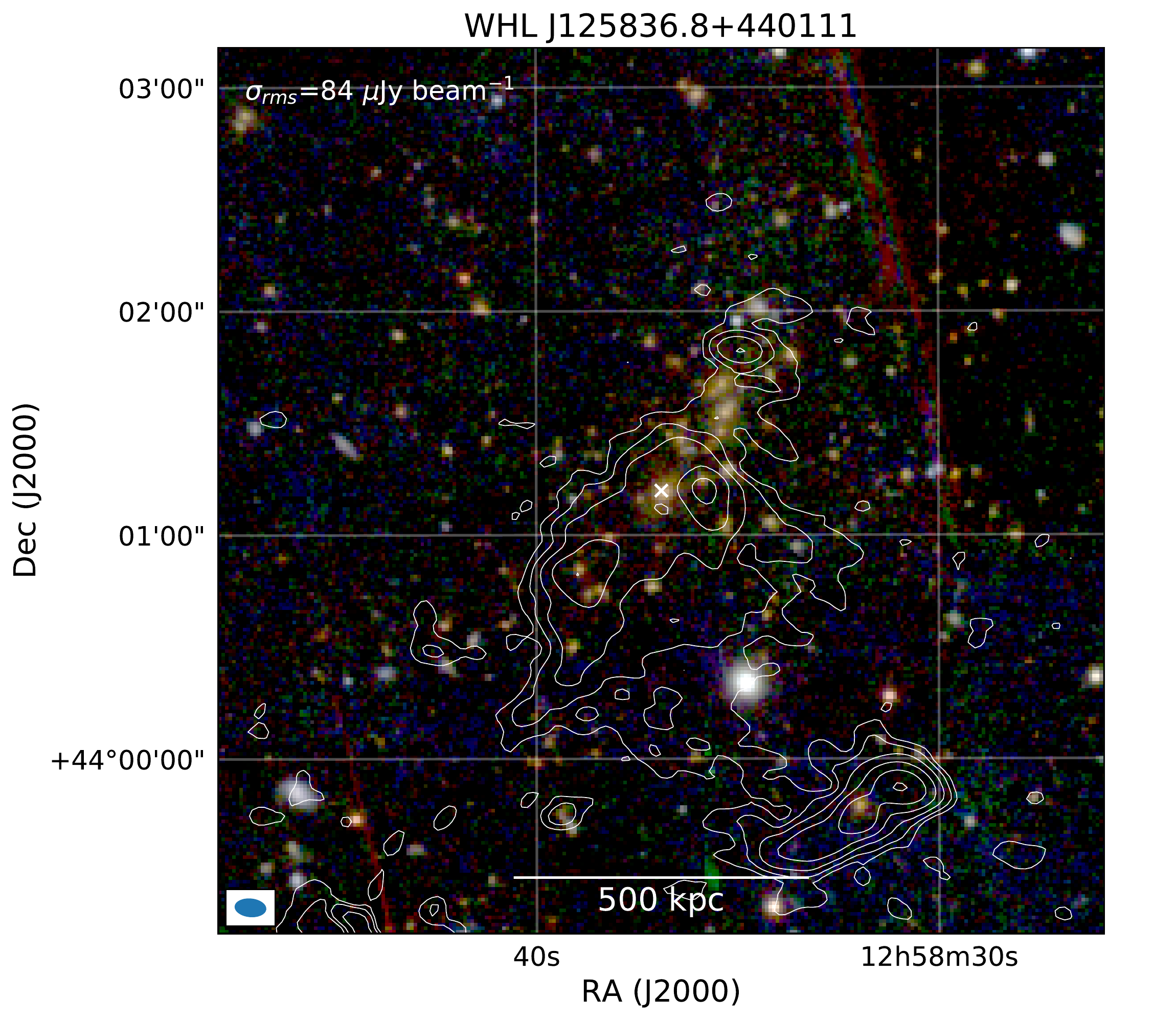}
   \caption{WHL\,J125836.8+440111. 
    Left: 10\arcsec~tapered radio image  with compact sources subtracted.
   Right: Optical image with Robust $-0.5$ image radio contours. For more details see the caption of Figure~\ref{fig:A2018}.}
   \label{fig:WHL125836}
\end{figure*}

In WHL\,J125836.8+440111 we find extended radio emission with an LLS of 800~kpc, see Figure~\ref{fig:WHL125836}. The northern part of the emission seems to originate from two radio galaxies. However, the emission extends further south, all the way to a distorted double lobed source. Given the large area that is covered by the extended emission we classify part of this emission as a  radio halo.


\section{Discussion}  
\label{sec:discussions}

\subsection{Classification of diffuse radio emission}   
Our LOFAR observations emphasize the challenge  of classifying diffuse cluster radio sources in sensitive low-frequency images. In particular, the distinction between diffuse sources and (old) AGN radio plasma is not always clear. The reason for this is that old AGN plasma also has a steep radio spectrum so this plasma brightens significantly in low frequency images. For example, this manifests itself by the longer tails of tailed radio galaxies. Several of the clusters in our sample show extended AGN emission with an LLS of more than 0.5~Mpc (e.g., PSZ2\,G114.99+70.36, PSZ2\,G135.17+65.43, Abell\,1314, and GMBCG\,J211.77332+55.09968, see Appendix~\ref{sec:GMBCGJ211}). AGN with an LLS of a few hundred kpc are even more common. When the emission from tailed and distorted AGN starts to overlap with relics and halos the classification of diffuse emission becomes more difficult.

A second reason for that classification is challenging is related to the origin of the diffuse radio emission itself. Particle re-acceleration models require a source of (mildly) relativistic electrons. The LOFAR images reveal that AGNs provide a rich source of fossil electrons and therefore the observational distinction between plasma directly accelerated by the AGN or in-situ re-accelerated in the ICM becomes more difficult. Spectral measurements and a comparison with the thermal ICM properties will therefore be important to correctly classify the extended radio emission. 
   
\subsection{Projections for the completed LoTSS survey.}

Based on the number of detected radio halos and relics in the HETDEX area, we can make a prediction for the number of radio halos and relics that will be detected in the completed LoTSS survey. We assume here that the LoTSS survey will have uniform sensitivity over the entire northern sky. We ignore any complications due to differences in the mass and redshift distributions between the PSZ2 clusters in the HETDEX DR1 area in the completed LoTSS survey. The uncertainties are computed based on Poisson statistics. 

We first focus on the radio halos in the Planck PSZ2 clusters. We detect a total  of 8 radio halos and 8 candidate halos in this sample of 26 clusters (where we count PSZ2\,G107.10+65.32, consisting of Abell 1758N and Abell 1758S, only once). The number of PSZ2 clusters without any diffuse emission is 7. We should note that for three of these clusters the noise levels were higher because of bad ionospheric conditions. Four PSZ2 clusters host radio relics. The fraction of confirmed radio halos in PSZ2 clusters is $31\pm 11\%$. Considering that there are 595 PSZ2 clusters above a declination of 0 degrees with confirmed redshifts, we estimate that there will be  $183 \pm 65$ radio halos detected in the LoTSS survey in PSZ2 clusters. Also including candidate radio halos, the detection fraction is $62\pm15\%$. The number of PSZ2 halos in the LoTSS survey then becomes $366\pm 92$. The fraction of clusters with some form of diffuse emission is $73\pm17\%$. This suggests that $435\pm99$ PSZ2 clusters with diffuse emission will be detected in LoTSS.

We detect one radio halo in the non-PSZ2 clusters and four candidate radio halos. Considering these numbers, and that the region we surveyed covers 424 deg$^{2}$, we expect to find at least $10^{2}$ radio halos in non-PSZ2 clusters in the full LoTSS. Here we assumed that about half of the candidate radio halos are real. The total number of radio halos (PSZ2 plus non-PSZ2) that will be found in the full LoTSS survey will thus be around 400--500, again assuming that half of the candidates are real.

The number of radio halos that our extrapolation predicts for the LoTSS survey agrees with that predicted from re-acceleration models  \citep{2010AA...509A..68C,2012AA...548A.100C}. These models predict that a significant fraction of halos have steep spectra, especially in clusters with smaller masses and higher redshift, and that the occurrence of halos in clusters declines with cluster mass.  The large statistics that is expected from LoTSS will allow to test the dependence of the occurrence of radio halos with cluster mass and redshift. 

Interestingly, we did not detect clear examples of radio mini-halos. Radio mini-halos are more difficult to classify due to their smaller sizes and the sample studied here is not known to host any strong cool-core clusters. Therefore, the lack of radio mini-halos might be (partly) related to the properties of our sample. However, there is no consensus that SZ-selected samples are biased against cool-core clusters \cite{2011A&A...526A..79E,2017MNRAS.468.1917R,2017ApJ...843...76A}.
The lack of mini-halos in our sample could also be a reflection of the properties of mini-halos at low frequencies. Several low-frequency studies have found the presence of diffuse emission at larger radii, beyond the extent of the mini-halos measured at GHz frequencies. Thus, low-frequency studies would be less likely to report mini-halos if size is used as a criterion.
Whether this extended emission beyond the ``classical'' mini-halo extent can be considered as a part of the mini-halo, or is an unrelated component more similar to that of a giant radio halo, is  unclear
\citep{2014MNRAS.444L..44B,2014IJMPD..2330007B,2016MNRAS.459.2940K,2017AA...603A.125V,2018MNRAS.478.2234S,2019AA...622A..24S,2019MNRAS.486L..80K}. 

A further complication for the detection is mini-halos is the presence of a radio bright BCG. The calibration needs to achieve a dynamic range sufficient for a detection. This can be a challenge at low frequencies, see for example Figure~\ref{fig:A1940}.

The fraction of radio relic hosting clusters in PSZ2 is $15\pm8\%$. We thus expect $92\pm46$ clusters hosting radio relics in LoTSS. One radio relic is detected in a non-PSZ2 cluster. The expected number of radio relics in the LoTSS survey falls about an order of magnitude below the prediction of \cite{2012MNRAS.420.2006N}. Predictions are not available yet for more recent models \cite[e.g.,][]{2017MNRAS.470..240N,2020MNRAS.493.2306B}.

\subsection{$P_{\rm{150~MHz}}$ scaling relations}
\label{sec:scaling}
It is well established that the radio power of giant halos scales with the X-ray luminosity of  clusters \citep[e.g.,][]{2000ApJ...544..686L,2006MNRAS.369.1577C,2013ApJ...777..141C,2015A&A...579A..92K}. A similar scaling exists for the integrated Compton Y parameter which traces the ICM integrated pressure along the line of sight \citep{2012MNRAS.421L.112B,2013ApJ...777..141C}. Both the X-ray and SZ measurements are proxies of clusters mass. Therefore the  suggestion is that the observed correlations originate from an underlying relation between cluster mass and radio power. The explanation for this relation is that a fraction of the gravitational energy released during a merger event, which scales with cluster host mass, is channeled into the re-acceleration of cosmic rays via turbulence \citep[e.g.,][]{2005MNRAS.357.1313C,2004JKAS...37..589C}. 
  
Traditionally, the radio power is computed at a rest-frame frequency of 1.4~GHz ($P_{\rm{1.4~GHz}}$) and all correlation studies so far have used this quantity. With the new LOFAR radio halo detections it becomes feasible to study this relation at a rest-frame frequency of 150~MHz, with $P_{\rm{150~MHz}}$ given by 

\begin{equation}
P_{\rm{150~MHz}} = \frac{ 4\pi D^{2}_{L} S_{\rm{150~MHz}} }  {\left(1+z\right)^{\alpha+1}  }   \mbox{ ,}
\end{equation}
with $D_{L}$ the luminosity distance of the cluster. Because our integrated flux density measurements are obtained at 144~MHz, $P_{\rm{150~MHz}}$ is only marginally affected by the  adopted (unknown) radio spectral index.

In this work, we determine the M$_{500}$--$P_{\rm{150~MHz}}$ scaling relation for radio halos using clusters from the PSZ2 catalog which provides an SZ-based mass estimate. SZ-based mass selected samples should be less affected by the cluster's dynamical state compared to X-ray selected sample which are biased towards relaxed cool-core clusters \citep[e.g.,][]{2011A&A...526A..79E,2017MNRAS.468.1917R,2017ApJ...843...76A}. We complement our new LOFAR measurements with halo detections in the 120--180~MHz range from the literature when reliable measurements are available, see Table~\ref{tab:litP150}. 
These literature values come mostly from previous LOFAR and GMRT studies. A few of them are taken from Murchison Widefield Array (MWA) studies. We do not include all MWA detections, only those with high-quality measurements, as indicated from good spectral fits with low ($<2$) reduced $\chi^{2}$ values \citep{2017MNRAS.467..936G}, and from halos not significantly affected by the uncertainties from compact source subtraction.

\begin{table*}
\begin{center}
\caption{Literature sample for $P_{\rm{150~MHz}}$ scaling relations.}
\begin{tabular}{lllllll}
\hline\hline
cluster & alternative name & redshift & flux density & frequency  & reference \\
        &                  &          & mJy         & MHz        \\
\hline
PSZ2\,G108.17$-$11.56 & PSZ1\,G108.18$-$11.53 & 0.336 & $124 \pm 11$ & 147  & \cite{2015MNRAS.453.3483D} \\
PSZ2\,G057.80+88.00 & Coma Cluster  & 0.0231 & $7200\pm800$ &  150 & \cite{1985MNRAS.215..437C} \\
PSZ2\,G149.22+54.18 & Abell\,1132  & 0.1351 & $178 \pm 27$ & 144 & \cite{2018MNRAS.473.3536W} \\
PSZ2\,G058.29+18.55 & RXC\,J1825.3+3026  & 0.065 &  $163 \pm 47$ &   144 & \cite{2019BLyra} \\ 
PSZ2\,G151.19+48.27 & Abell\,959 & 0.2894 & $94 \pm 14$ &  143 & \cite{2019MNRAS.487.4775B} \\
PSZ2\,G071.39+59.54 & RXC\,J1501.3+4220 & 0.2917 & $20.2 \pm 2.0$ &  144 &  \cite{2019AA...622A..25W} \\
PSZ2\,G139.62+24.18 & PSZ1\,G139.61+24.20 &  0.2671 & $30 \pm 4$ &  144 &\cite{2018MNRAS.478.2234S} \\
PSZ2\,G049.22+30.87	& RX\,J1720.1+2638 & 0.164 & $165 \pm 25$&   144 & \cite{2019AA...622A..24S} \\
PSZ2\,G138.32-39.82 & RXC\,J0142.0+2131 & 0.280 &  $32\pm6$ & 144 & \cite{2019AA...622A..24S} \\
PSZ2\,G226.18+76.79 & Abell\,1413 & 0.143 &  $40 \pm 7$ &144 & \cite{2019AA...622A..24S}\\
PSZ2\,G055.59+31.85 & Abell\,2261 &  0.224 &  $165\pm 25^{a}$ & 144 &\cite{2019AA...622A..24S} \\
PSZ2\,G180.25+21.03 & MACS\,J0717.5+3745 & 0.546 &$370\pm 60$& 147 &  \cite{2018MNRAS.478.2927B} \\
PSZ2\,G195.75$-$24.32 & Abell\,520 & 0.201 &$229.7 \pm 34.8$ & 145& \cite{2019AA...622A..20H} \\
PSZ2\,G100.14+41.67 & Abell\,2146 & 0.232 &  $19.8 \pm 5.0^{b}$&   144& \cite{2019AA...622A..21H} \\
PSZ2\,G186.37+37.26 & Abell\,697 & 0.282 &  $135\pm27$ & 153 & \cite{2013AA...551A.141M} \\
PSZ2\,G208.80$-$30.67 & Abell\,521 & 0.247 & $328\pm 66$ & 153 &\cite{2013AA...551A.141M} \\
PSZ2\,G008.94$-$81.22 & Abell\,2744 & 0.308 & $415\pm42$ &150& \cite{2017MNRAS.467..936G} \\
PSZ2\,G175.69$-$85.98 & Abell\,141 &  0.230 & $110\pm 11$ & 168 &\cite{2017arXiv170703517D} \\
\hline
\hline
\end{tabular}
\end{center}
$^a$ No flux density uncertainty provided, adopted an uncertainty of 15\%.\\
$^b$ A range of flux densities was provided for the radio halo by \cite{2019AA...622A..21H}. Adopted the average value of that range, with the uncertainty reflecting the provided range. 
\label{tab:litP150}
\end{table*}

\begin{figure*}
\centering
   \includegraphics[width=1.0\columnwidth]{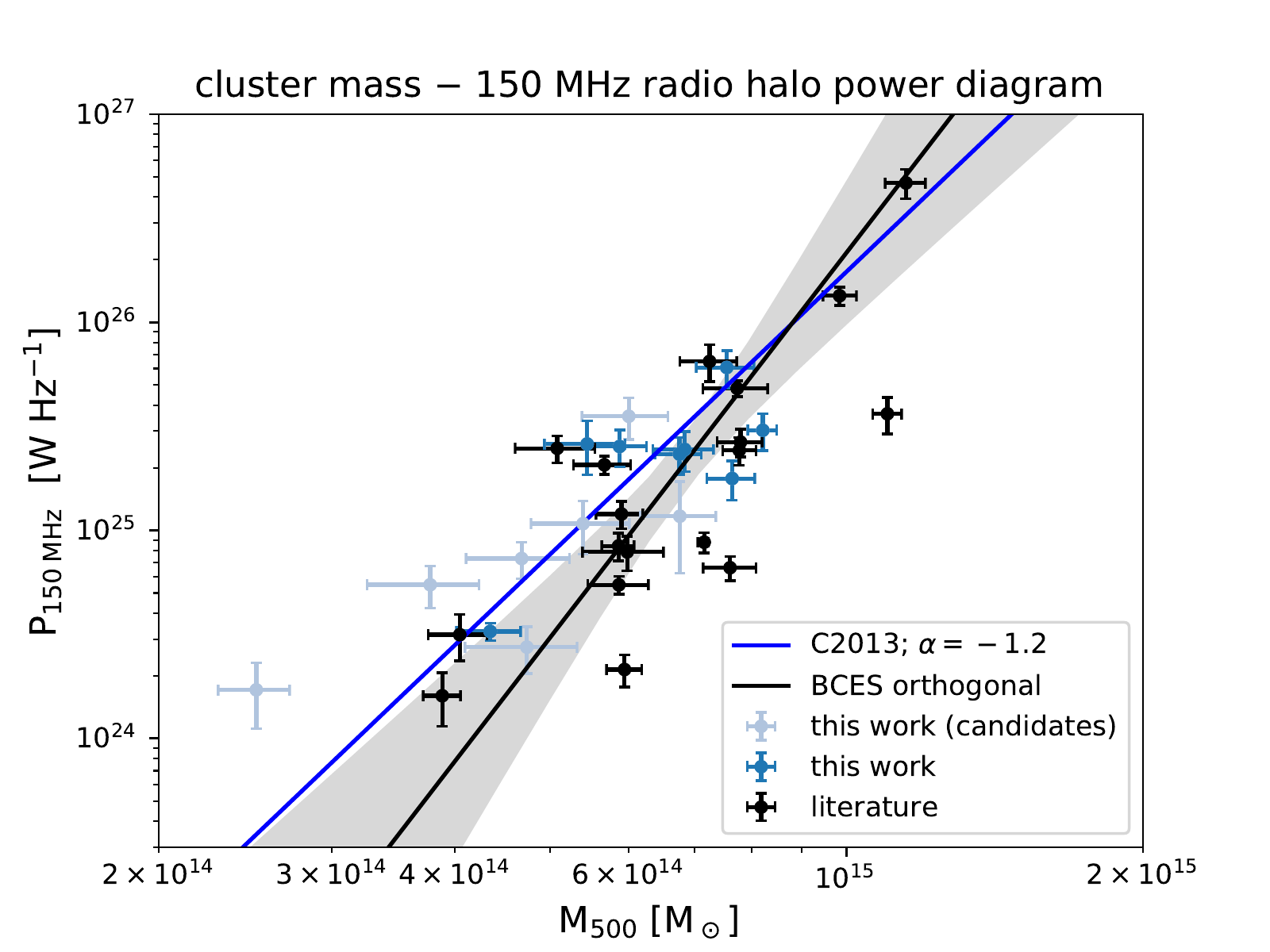}
   \includegraphics[width=1.0\columnwidth]{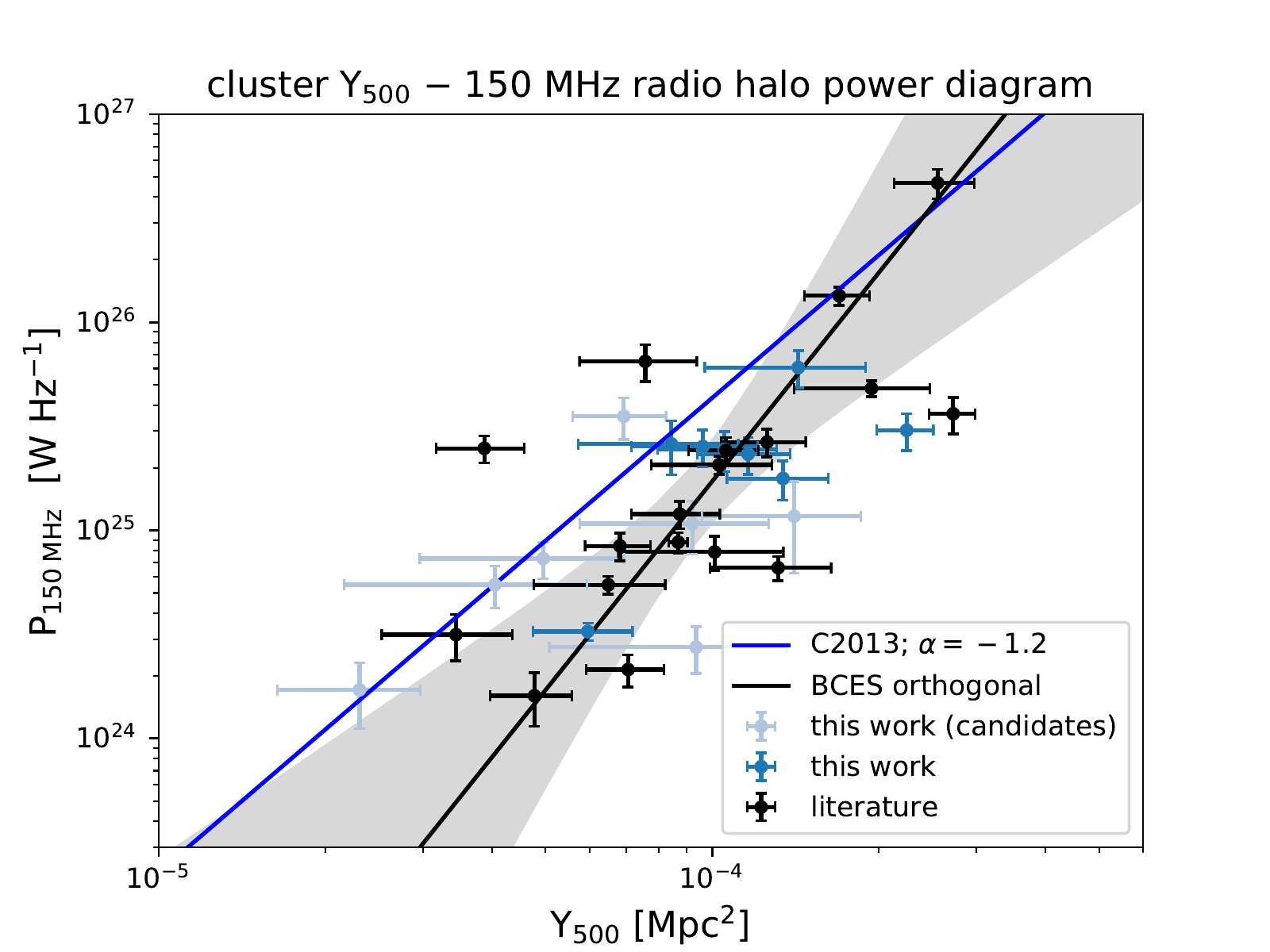}
   \caption{Left: Distribution of clusters in the mass ($M_{500}$) 150~MHz radio power ($P_{\rm{150~MHz}}$) plane. The black solid line displays the BCES orthogonal fit (candidate halos were excluded). The shaded region shows the $3\sigma$ (99.7\% confidence) region of the fit.
   The blue line is the BCES orthogonal fit from \cite{2013ApJ...777..141C} scaled with a spectral index of $-1.2$. Right: The distribution of clusters in the $Y_{500}$ 150~MHz radio power plane. The plotted symbols and lines represent the same datasets as in the left panel.}
   \label{fig:MP150}
\end{figure*}

Following \cite{2013ApJ...777..141C} we use the following relation
\begin{equation}
\log_{10}{\left(\frac{P_{150~\rm{MHz}}}{10^{24.5}\rm{ W } \rm{ Hz}^{-1}}\right)  } = B \log_{10}{ \left(\frac{M_{500}}{10^{14.9} \rm{ M}_\odot}\right)  } + A  
\end{equation}
between radio power and cluster mass. The best fitting parameters are found using the BCES orthogonal regression algorithm \citep{1996ApJ...470..706A,2012Sci...338.1445N}. Past work has adopted by default BCES bisector fits (which give consistently flatter slopes), although, as is mentioned by \cite{2010arXiv1008.4686H}, bisector fits are not recommended. For comparison with previous work we also report the results from the BCES bisector and $Y|X$ (with $X$ the independent variable) fits. Our sample  approximately covers the mass range  $3 - 10 \times 10^{14}$~M$_{\odot}$. The results are shown in Figure~\ref{fig:MP150} (left panel) and Table~\ref{tab:scaling}. As is the case for $P_{\rm{1.4~GHz}}$, there is also a clear correlation between between M$_{500}$ and $P_{\rm{150~MHz}}$ and we find a slope of $B = 6.13 \pm 1.11$ (BCES orthogonal). With the candidate halos included, we find somewhat flatter slopes (see Table~\ref{tab:scaling}), but the differences are not significant considering the uncertainties.

We determine the Y$_{500}$--$P_{\rm{150~MHz}}$ radio halo scaling relation using the PSZ2 Compton~Y parameter. For that we convert the Y$_{5R_{500}}$ values from \cite{2016A&A...594A..27P} to Y$_{500}$ using Y$_{500}$ = 0.56 Y$_{5R_{500}}$ \citep{2010A&A...517A..92A}. We also convert the units from arcmin$^{2}$ to Mpc$^{2}$ to facilitate the comparison with \cite{2013ApJ...777..141C}. We apply the same fitting methods as for the M$_{500}$--$P_{\rm{150~MHz}}$ correlation, using a relation of the form
\begin{equation}
\log_{10}{\left(\frac{P_{150~\rm{MHz}}}{10^{24.5}\rm{ W } \rm{ Hz}^{-1}}\right)  } = B \log_{10}{ \left(\frac{Y_{500}}{10^{-4} \rm{ Mpc}^2}\right)  } + A \mbox{ .}   
\end{equation}
The results are plotted in Figure~\ref{fig:MP150} (right panel) and are reported in Table~\ref{tab:scaling}. The slope of the best fit BCES orthogonal relation is $B = 3.32 \pm 0.65$. The slopes are slightly flatter when including the candidate radio halos, but the differences are not significant considering the uncertainties (see Table~\ref{tab:scaling}).


The slopes of the 150~MHz scaling relations we find are steeper than the ones obtained at 1.4~GHz by \cite{2013ApJ...777..141C}. For Y$_{500}$--$P_{\rm{150~MHz}}$ they reported $B = 2.28 \pm 0.35$ from  at 1.4~GHz, compared to  $B = 3.32 \pm 0.65$ in this work. For M$_{500}$--$P_{\rm{150~MHz}}$ they reported a slope of  $B = 4.51 \pm 0.78$, compared to our value of $B = 6.13 \pm 1.11$ in this work. However, considering the uncertainties, the slopes at 150~MHz are still consistent with the ones reported at 1.4~GHz.

Statistical models employing turbulent re-acceleration \citep{2013ApJ...777..141C} predict $P_{\rm{1.4\,GHz}} \propto M_{500}^{4}$, or steeper depending on the ICM magnetic field strength with respect to the equivalent magnetic strength of the cosmic microwave background. 
It should be noted that these models make a number of simplifying assumptions. The most important simplification is that there is no spatial dependence of the magnetic field and acceleration efficiency, see for example \cite{2010AA...509A..68C} about these ``homogeneous models''. \cite{2010AA...517A..10C} predicts that the slope of the $L_{X}$--$P_{\rm{1.4~GHz}}$ scaling relation should steepen by about~$0.4$ due to the  ultra-steep spectrum halos visible at low frequencies associated with intermediate mass galaxy clusters. Using the fact that $L_{\rm{X}}\propto M_{500}^2$, we thus expect a steepening of about~0.8 for the M$_{500}$--$P_{\rm{150~MHz}}$ scaling relation. In addition, the scatter around the scaling relations should increase. Interestingly, the slopes we find at 150~MHz are indeed steeper than at 1.4~GHz, in line with this prediction. That said, the uncertainties one the determined slopes are still too large for any firm claim. For this reason, extending the sample size and mass range will be crucially important to confirm this result.

One of the limitations of our presented analysis is that due to the small sample size we did not apply any cut in mass or redshift. For a more detailed comparison with the \cite{2013ApJ...777..141C}, a similar redshift and mass cut should be applied. We also note that our sample contains three clusters (Abell\,1413, PSZ1\,G139.61+24.20 and RX\,J1720.1+2638) with diffuse sources that were previously classified as mini-halos. Since the diffuse emission in these clusters turns to be more extended\footnote{This also includes  Abell~1413 based on new LoTSS data} at low frequencies than the typical scale of mini-halos, we have included them in our sample. Further investigations are also required of how the low-frequency scaling relations and occurrence rates depend on the cluster dynamical state. Another limitation of our derived scaling relations is the inclusion of literature values that were obtained in a variety of ways, adopting different methods to measure the integrated flux density. That said, the literature values do not seem to be strongly biased either high or low when compared to the measurements obtained in this work for clusters with similar masses, and hence this should not have a large effect on the derived slopes. This limitation will be removed in future work when larger LOFAR samples will become available.

A comparison between the 150~MHz and 1.4~GHz relations allows use to investigate the average spectral index of radio halos in this frequency range. We find that when using $\alpha=-1.2$ a good  match is obtained between the two M$_{500}$--$P_{\rm{150~MHz}}$ scaling relations. For  Y$_{500}$--$P_{\rm{150~MHz}}$ a slightly flatter spectral index seems to be preferred (the blue line in Figure~\ref{fig:MP150} (right panel) is located mostly above our best fit). \cite{2009A&A...507.1257G} obtained a medium spectral index of $-1.3$ for radio halos between 325~MHz and 1.4~GHz, in reasonable agreement our results. However, for a better assessment of the radio halo spectral indices, detailed cluster-to-cluster comparisons are required. In particular, the used scaling relation from \cite{2013ApJ...777..141C} did not include USSRHs. This could have a direct effect on the derived spectral indices when comparing to a low-frequency sample which potentially contains a significant number of USSRH. 

\begin{table}
\begin{center}
\caption{BCES fitted scaling relations for radio halos.}
\begin{tabular}{lllll}
\hline\hline
method & B & $\sigma_{B}$ & A  &$\sigma_{A}$\\
\\
\hline
\hline
M$_{500}$--$P_{\rm{150\,MHz}}$ \\
\hline

orthogonal  & 6.13 & 1.11 & 1.22 & 0.12\\
bisector  & 4.67 & 0.62 & 1.11 & 0.09\\
$Y|X$  & 3.84 & 0.69 & 1.05 & 0.09\\

\hline
M$_{500}$--$P_{\rm{150\,MHz}}$ with candidates \\
\hline
orthogonal  & 5.00 & 0.87 & 1.21 & 0.10\\
bisector  & 4.05 & 0.55 & 1.11 & 0.08\\
$Y|X$  & 3.40 & 0.53 & 1.04 & 0.09\\
\hline
\hline
Y$_{500}$--$P_{\rm{150\,MHz}}$ \\
\hline
orthogonal  & 3.32 & 0.65 & 0.74 & 0.12\\
bisector  & 2.60 & 0.53 & 0.74 & 0.09\\
$Y|X$   & 2.10 & 0.67 & 0.74 & 0.08\\

\hline
Y$_{500}$--$P_{\rm{150\,MHz}}$ with candidates \\
\hline
orthogonal  & 3.05 & 0.68 & 0.78 & 0.10\\
bisector  & 2.43 & 0.42 & 0.76 & 0.08\\
$Y|X$   & 1.98 & 0.48 & 0.74 & 0.07\\
\hline
\hline
\end{tabular}
\end{center}
\label{tab:scaling}
\end{table}

\section{Conclusions}
\label{sec:conclusions}
We have presented a method to extract and re-calibrate targets of interest from the LoTSS survey based on the {\tt DDF-pipeline}. This method allows joint imaging of data from multiple pointings and an improvement in calibration accuracy. It also enables fast imaging of targets without covering the full FoV of a LOFAR observation.

We applied the above scheme to 26 Planck PSZ2 and 15 other clusters located in the HETDEX spring region. In total we detect 10 radio halos. Five clusters host radio relics. We also report 12 candidate radio halos. The occurrence fraction of radio halos in PSZ2 clusters is $31\pm11\%$, and $62\pm15\%$ if we include the candidate radio halos. The fraction of clusters with some form of diffuse radio emission in PSZ2 clusters is $73\pm17\%$.  The relatively large number of candidate radio halos results from the difficulties of unambiguously classifying the emission. Based on the above numbers, we expect to find at least $183 \pm 65$ radio halos from the   
analysis of PSZ2 clusters in the LoTSS survey. Considering all clusters and candidate radio halos, we expect that 400--500 halos will be found in the completed LoTSS survey.

We determined for the first time the radio halo scaling relations between cluster mass, $Y_{\rm{SZ}}$, and the 150~MHz radio power ($P_{\rm{150~MHz}}$). The slopes for these scaling relations are slightly steeper than those determined at 1.4~GHz, in line with predictions, but considering the uncertainties this is not a statistically significant result. In a future work, using a larger sample, we will present a more detailed statistical analysis of the properties of the diffuse radio sources in the LoTSS survey.

\begin{acknowledgements}

This paper is based on data obtained with the International LOFAR Telescope (ILT) under project codes LC2\_038 and LC3\_008. LOFAR \citep{2013A&A...556A...2V} is the LOw Frequency ARray designed and constructed by ASTRON. It has observing, data processing, and data storage facilities in several countries, which are owned by various parties (each with their own funding sources) and are collectively operated by the ILT foundation under a joint scientific policy. The ILT resources have benefited from the following recent major funding sources: CNRS-INSU, Observatoire de Paris and Universit\'e d'Orl\'eans, France; BMBF, MIWF-NRW, MPG, Germany; Science Foundation Ireland (SFI), Department of Business, Enterprise and Innovation (DBEI), Ireland; NWO, The Netherlands; The Science and Technology Facilities Council, UK; Ministry of Science and Higher Education, Poland; The Istituto Nazionale di Astrofisica (INAF), Italy. This research made use of the Dutch national e-infrastructure with support of the SURF Cooperative (e-infra 180169) and the LOFAR e-infra group. The Jülich LOFAR Long Term Archive and the GermanLOFAR network are both coordinated and operated by the Jülich Supercomputing Centre (JSC), and computing resources on the supercomputer JUWELS at JSC were provided by the Gauss Centre for Supercomputinge.V. (grant CHTB00) through the John von Neumann Institute for Computing (NIC). This research made use of the University of Hertfordshire high-performance computing facility (\url{http://uhhpc. herts.ac.uk}) and the LOFAR-UK computing facility located at the University of Hertfordshire and supported by STFC [ST/P000096/1], and of the Italian LOFAR IT computing infrastructure supported and operated by INAF, and by the Physics Department of Turin University (under an agreement with Consorzio Interuniversitario per la Fisica Spaziale) at the C3S Supercomputing Centre, Italy. This research made use of Astropy\footnote{http://www.astropy.org}, a community-developed core Python package for Astronomy \citep{2013A&A...558A..33A,2018AJ....156..123A}.

RJvW and ABott acknowledge support from the VIDI research programme with project number 639.042.729, which is financed by the Netherlands Organisation for Scientific Research (NWO). GB, RC, and FG acknowledge  support  from  INAF  mainstream  project ``Galaxy Clusters Science with LOFAR'' 1.05.01.86.05. ABon and EB ackowledge support from the ERC-Stg DRANOEL n. 714245 and from the Italian MIUR grant FARE ``SMS''.
      
The Pan-STARRS1 Surveys (PS1) and the PS1 public science archive have been made possible through contributions by the Institute for Astronomy, the University of Hawaii, the Pan-STARRS Project Office, the Max-Planck Society and its participating institutes, the Max Planck Institute for Astronomy, Heidelberg and the Max Planck Institute for Extraterrestrial Physics, Garching, The Johns Hopkins University, Durham University, the University of Edinburgh, the Queen's University Belfast, the Harvard-Smithsonian Center for Astrophysics, the Las Cumbres Observatory Global Telescope Network Incorporated, the National Central University of Taiwan, the Space Telescope Science Institute, the National Aeronautics and Space Administration under Grant No. NNX08AR22G issued through the Planetary Science Division of the NASA Science Mission Directorate, the National Science Foundation Grant No. AST-1238877, the University of Maryland, Eotvos Lorand University (ELTE), the Los Alamos National Laboratory, and the Gordon and Betty Moore Foundation.    

This research has made use of data obtained from the Chandra Data Archive and the Chandra Source Catalog, and software provided by the Chandra X-ray Center (CXC) in the application packages CIAO, ChIPS, and Sherpa. Based on observations obtained with XMM-Newton, an ESA science mission with instruments and contributions directly funded by ESA Member States and NASA.

\end{acknowledgements}

\bibliographystyle{aa}
\bibliography{aa}

\begin{appendix}
 \section{Radio images of clusters without diffuse emission}
 \label{sec:nodiffuse}

\subsection{PSZ2\,G087.39+50.92}
\begin{figure}
\centering
  \includegraphics[width=1.0\columnwidth]{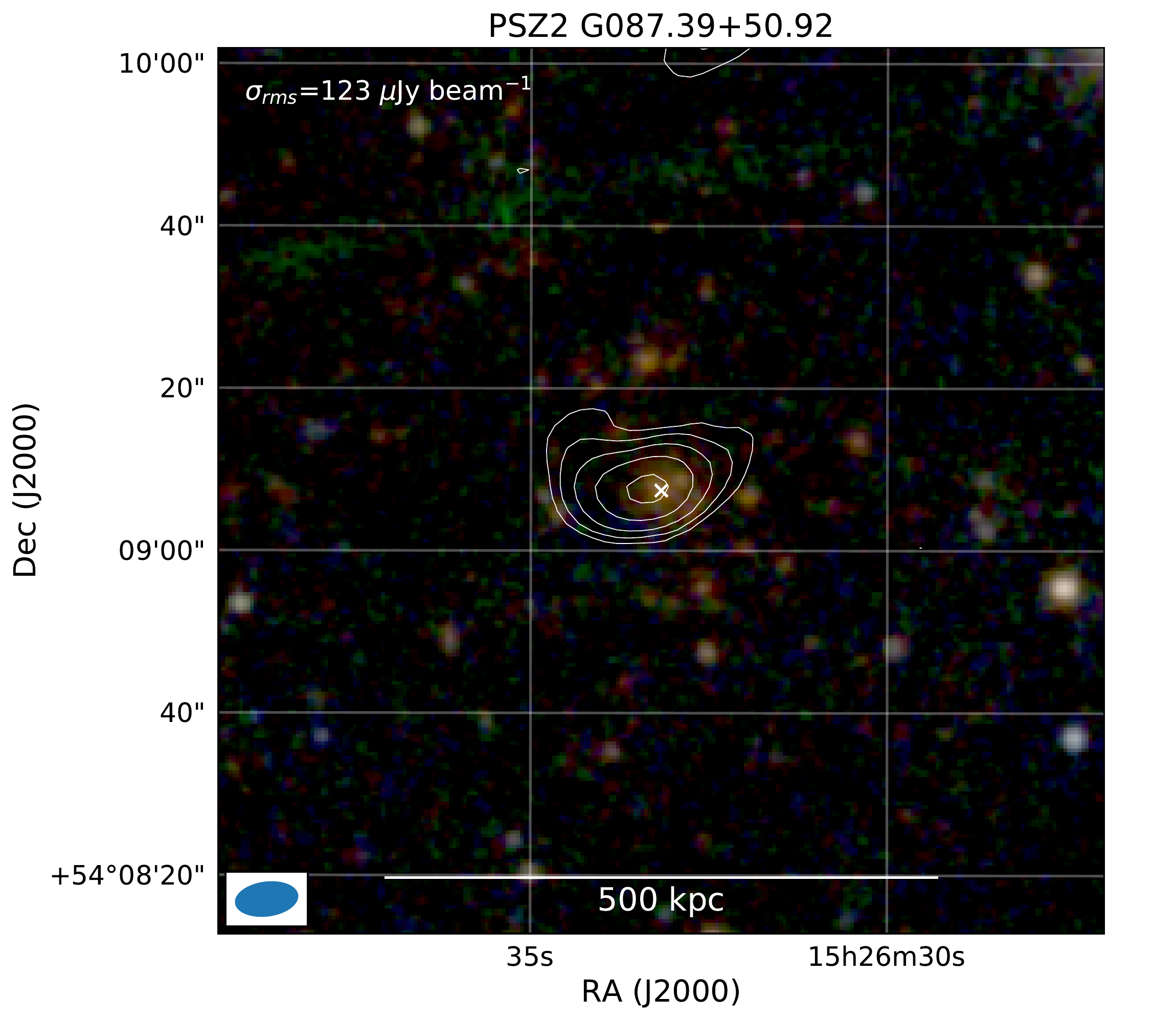}
   \caption{PSZ2\,G087.39+50.92. Optical image with Robust $-0.5$ image radio contours. For more details see the caption of Figure~\ref{fig:A2018}.}
   \label{fig:PSZ2G08739}
\end{figure}

The LOFAR image show a compact AGN associated with central BCG of this distant cluster (Figure~\ref{fig:PSZ2G08739}). No diffuse radio emission is detected.
We note that one of the LoTSS observations had to be discarded due to bad ionospheric conditions. 

\subsection{PSZ2\,G088.98+55.07}
\begin{figure}
\centering
   \includegraphics[width=1.0\columnwidth]{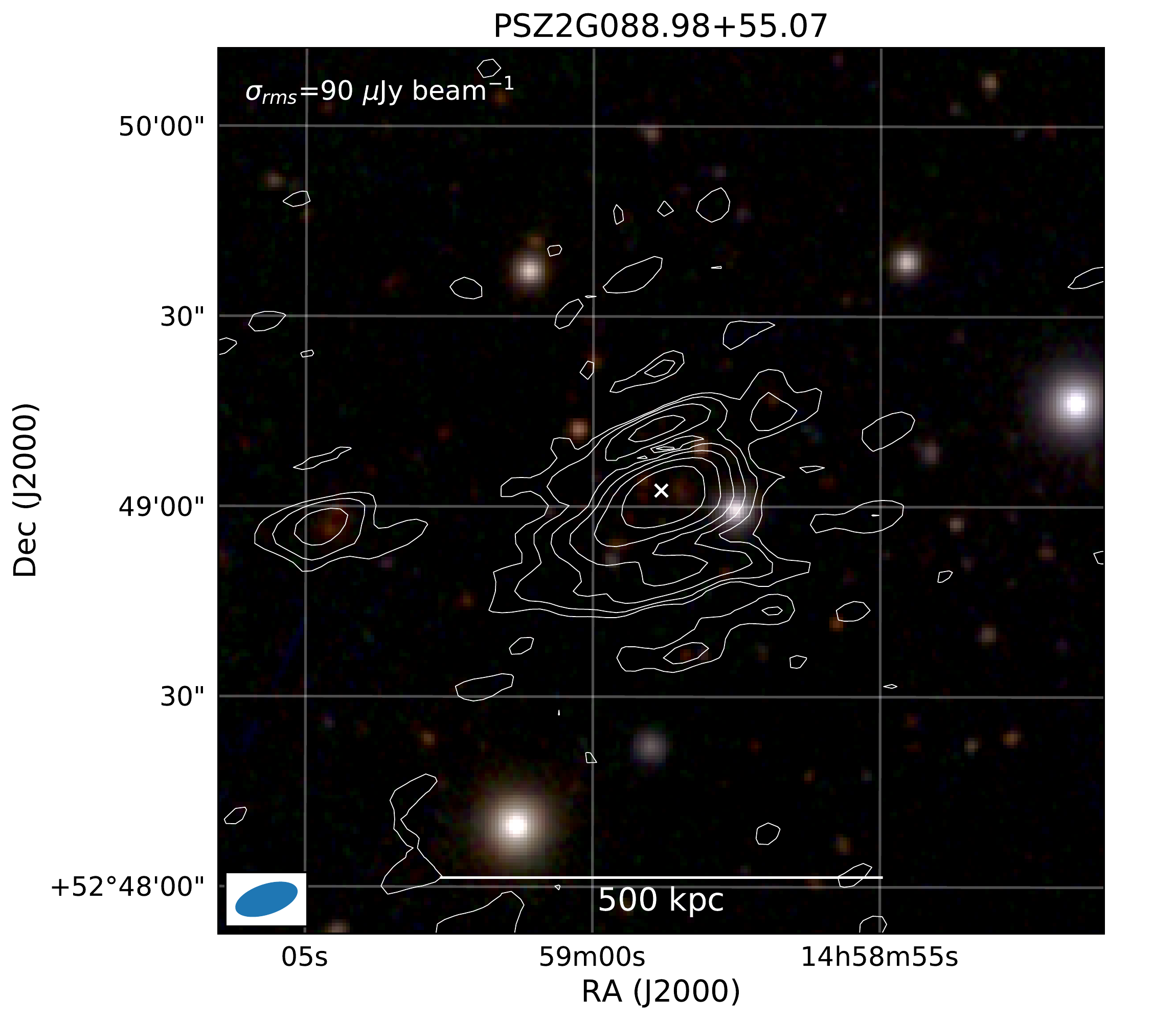}
   \caption{PSZ2\,G088.98+55.07. Optical image with Robust $-0.5$ image radio contours. For more details see the caption of Figure~\ref{fig:A2018}.}
   \label{fig:PSZ2G08898}
\end{figure}

A compact radio source is detected at the center of this distant $z=0.7023$ cluster. An optical image with radio contours overlaid is shown in Figure~\ref{fig:PSZ2G08898}. We do not find evidence for diffuse emission in this cluster. However, we note that the LOFAR observations were affected by bad ionospheric conditions.

\subsection{PSZ2\,G096.14+56.24, Abell\,1940}
\begin{figure}
\centering
  \includegraphics[width=1.0\columnwidth]{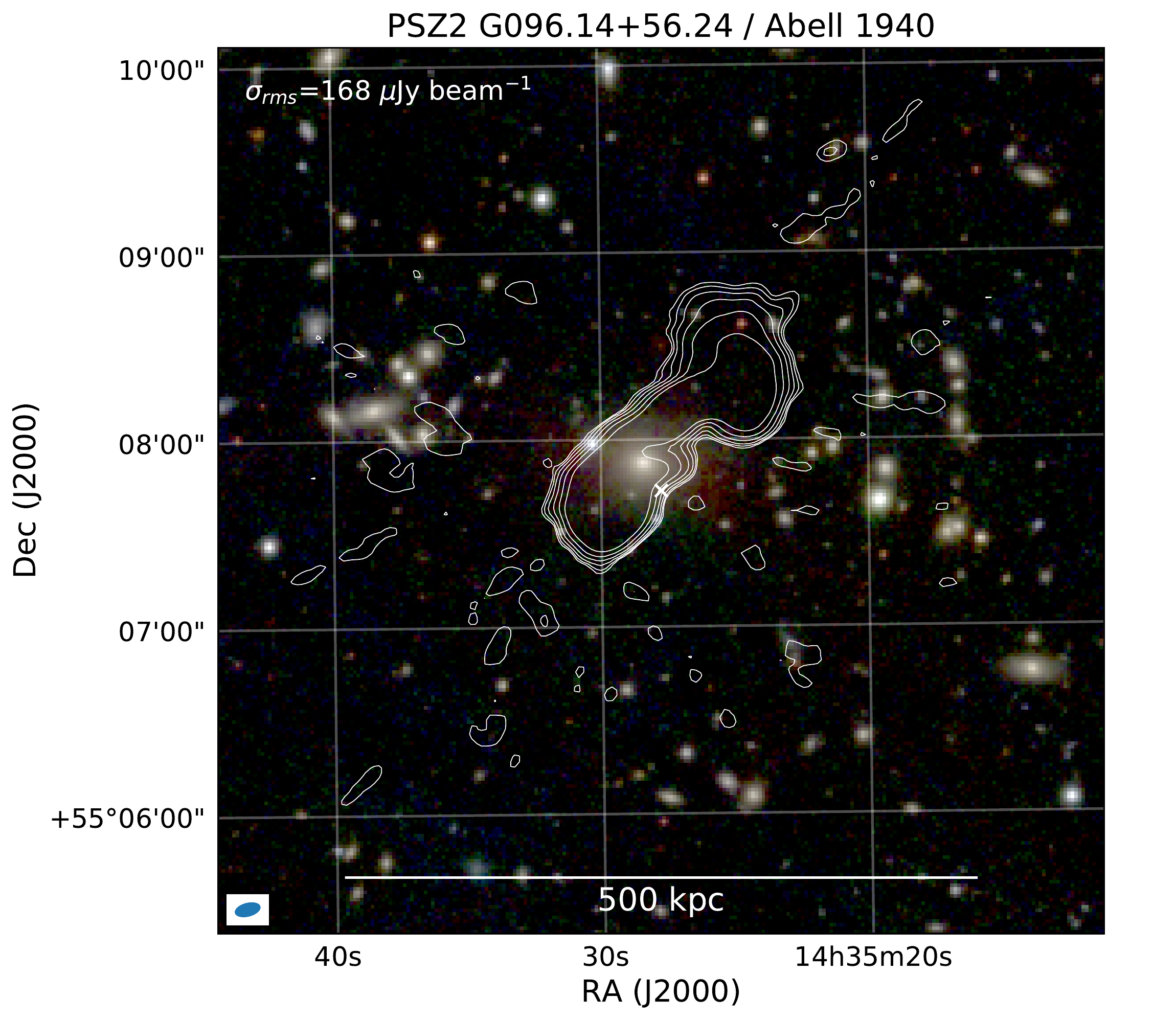}
   \caption{PSZ2\,G096.14+56.24 / Abell\,1940. Optical image with Robust $-0.5$ image radio contours. For more details see the caption of Figure~\ref{fig:A2018}.}
   \label{fig:A1940}
\end{figure}

No diffuse radio emission is detected  Abell\,1940. The image dynamic range is limited by the bright double lobed radio source associated with the BCG. An optical image with radio contours overlaid is shown in Figure~\ref{fig:A1940}.

\subsection{PSZ2\,G098.44+56.59, Abell\,1920}

\begin{figure}
\centering
   \includegraphics[width=1.0\columnwidth]{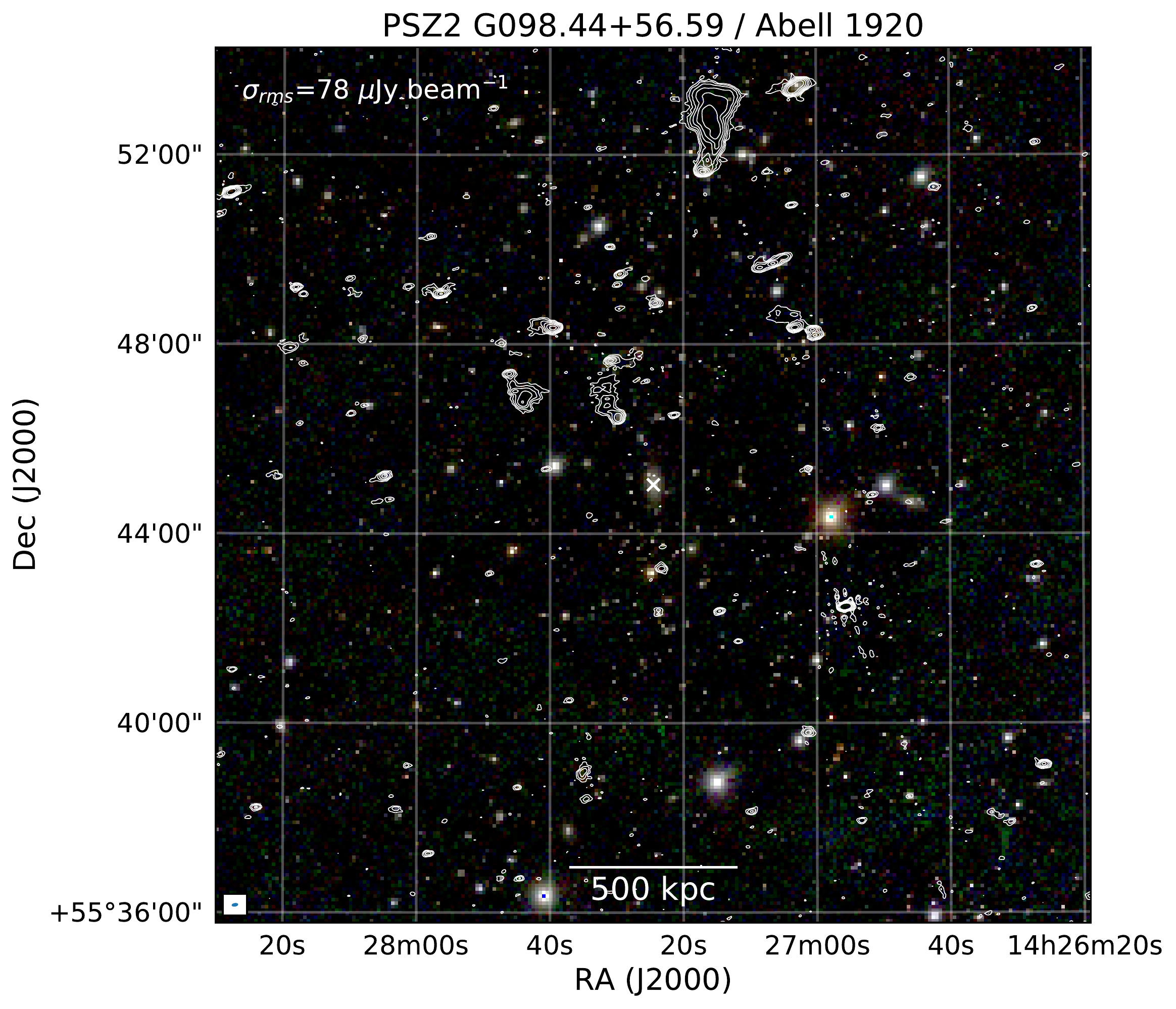}
   \caption{PSZ2\,G098.44+56.59,  Abell\,1920. Optical image with Robust $-0.5$ image radio contours. For more details see the caption of Figure~\ref{fig:A2018}.}
   \label{fig:PSZ2G09844}
\end{figure}

Several tailed radio galaxies are visible in the cluster region, but no diffuse emission is detected (see Figure~\ref{fig:PSZ2G09844}).

\subsection{PSZ2\,G123.66+67.25, Abell\,1622}
\begin{figure*}
\centering
   \includegraphics[width=0.285\paperwidth]{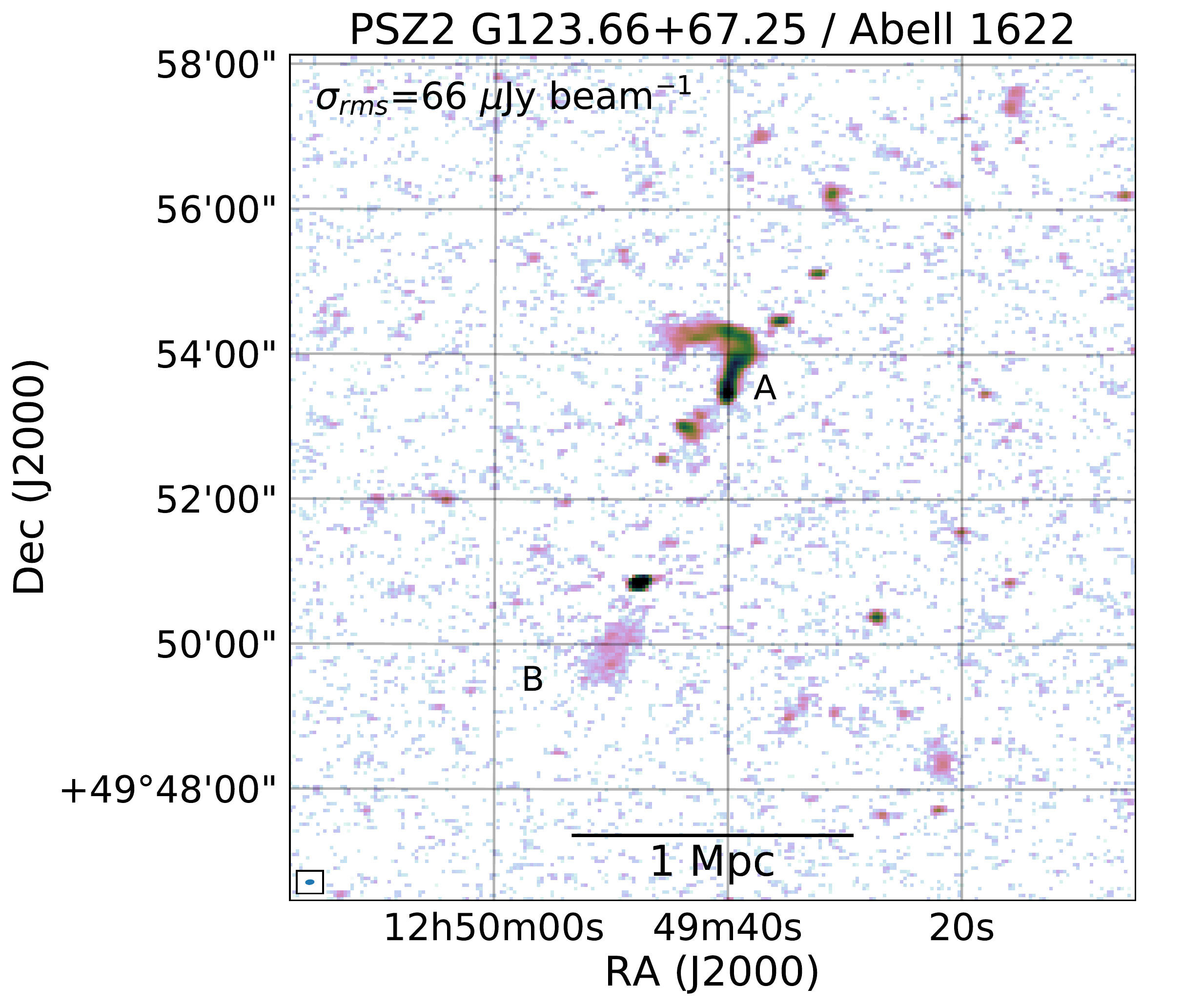}
   \includegraphics[width=0.285\paperwidth]{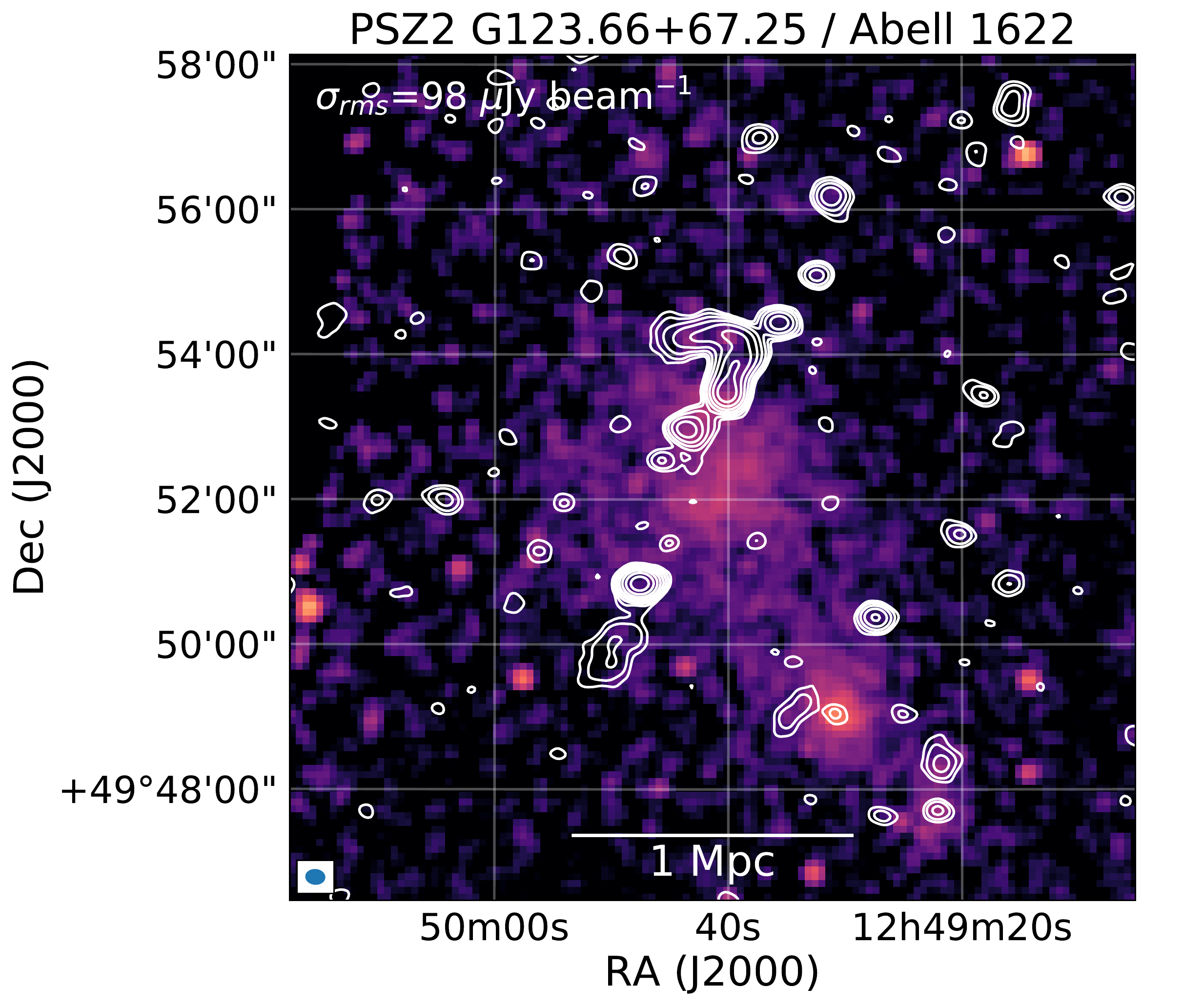}
   \includegraphics[width=0.285\paperwidth]{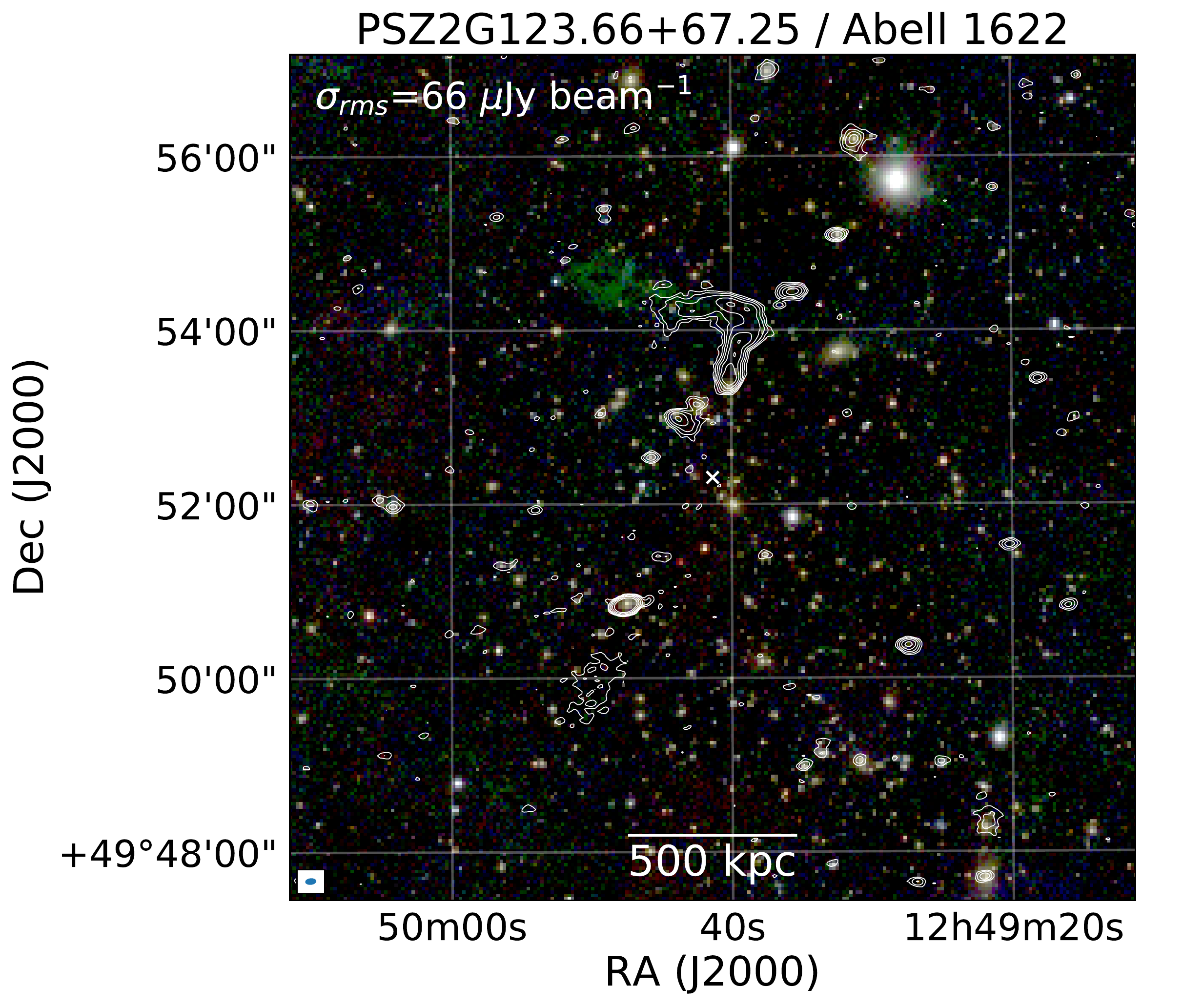}
   \caption{PSZ2\,G123.66+67.25 / Abell\, 1622. Left: Robust $-0.5$ radio image. Middle: Chandra X-ray image with 10\arcsec~tapered radio contours. Right: Optical image with Robust $-0.5$ image radio contours. For more details see the caption of Figure~\ref{fig:A2018}.}
   \label{fig:A1622}
\end{figure*}

 The Chandra X-ray image shows that the cluster consists of two subclusters, see Figure~\ref{fig:A1622}. The main subcluster has a large X-ray extent ($\sim1$~Mpc) and low surface brightness. A smaller more concentrated subcluster is located south of it. Given the overall roundish morphology of both components, the cluster is likely in a pre-merging state. A additional small substructure is seen to the SW of the southern subcluster. PSZ2\,G123.66+67.25  has a temperature of $4.76 \pm 0.87$~keV and is classified as a non-cool core by \cite{2015MNRAS.450.2261M}.
 
 No Mpc-scale diffuse emission is detected in this $z=0.2838$ cluster. A bright tailed radio galaxy (A) is found in the northern part of the main subcluster, with the tail bending east about 200~kpc north of its optical counterpart. A patch of radio plasma (B) with an LLS of 250~kpc is found to the south of the northern subcluster. A compact radio source is located just north of this diffuse source, but there is no clear connection visible in our high-resolution images. We speculate that this the diffuse source is AGN fossil plasma that originated from the compact radio source.

\subsection{PSZ2\,G136.92+59.46,  Abell\,1436}
\begin{figure*}
\centering
   \includegraphics[width=1.0\columnwidth]{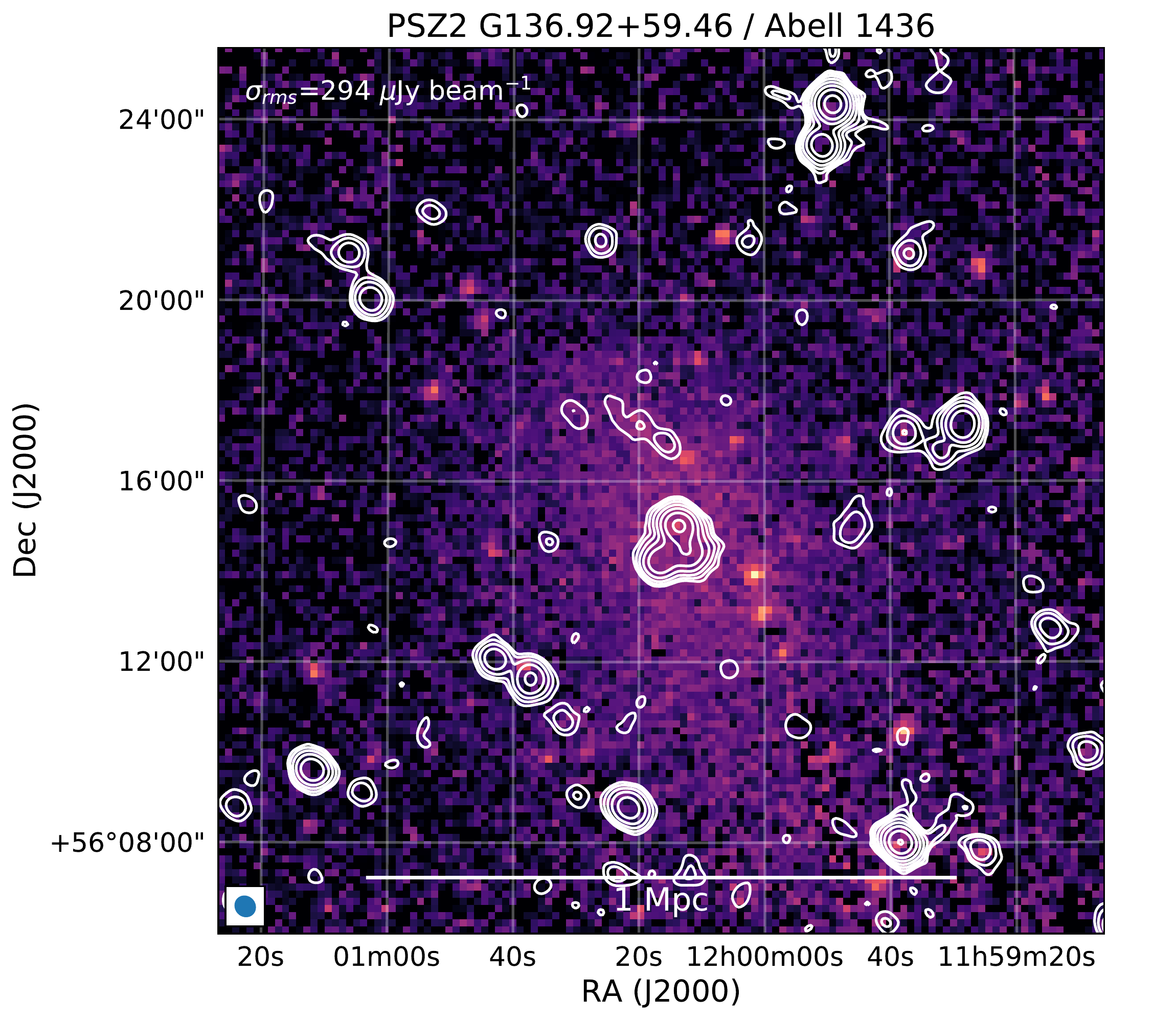}
   \includegraphics[width=1.0\columnwidth]{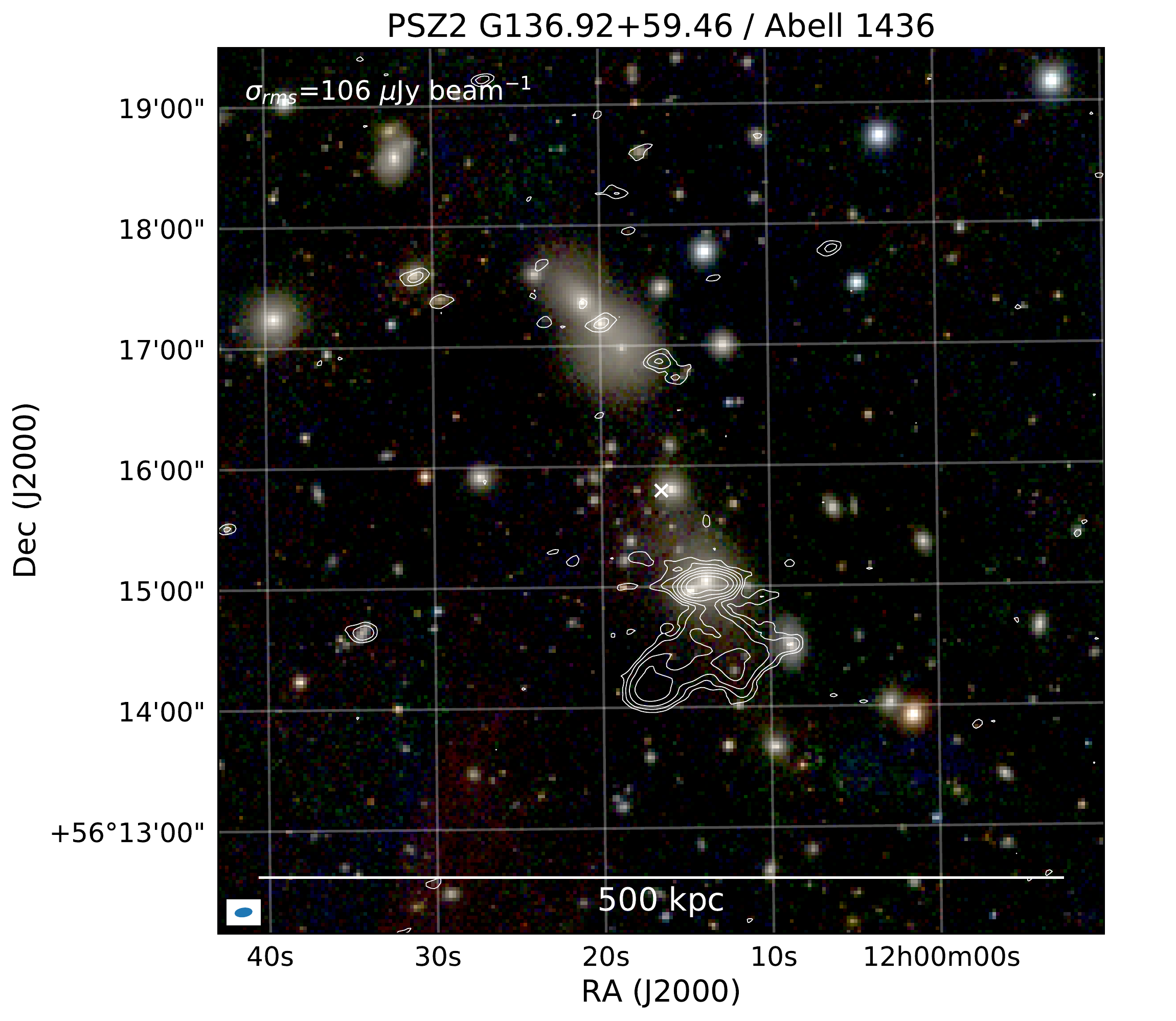}
   \caption{PSZ2\,G136.92+59.46 / Abell\,1436. Left: XMM-Newton X-ray image with 10\arcsec~tapered radio contours. Right: Optical image with Robust $-0.5$ image radio contours. For more details see the caption of Figure~\ref{fig:A2018}.}
   \label{fig:A1436}
\end{figure*}

The LOFAR image displays compact radio emission associated with the BCG. The XMM-Newton image reveals an elongated low surface brightness cluster, see Figure~\ref{fig:A1436}.
Optical Pan-STARRS images show the BCG to have a double nucleus. Additional radio emission is detected south of the BCG. Given that this emission connects to the BCG it likely originated from the BCG. No diffuse radio emission is detected in this cluster. 

\subsection{PSZ2\,G144.33+62.85, Abell\,1387}
No diffuse radio emission is detected in this cluster. Two BCGs seem to be present in this cluster based on optical Pan-STARRS images. A distorted tailed radio galaxy is associated with one of the two BCGs.

\begin{figure}
\centering
   \includegraphics[width=1.0\columnwidth]{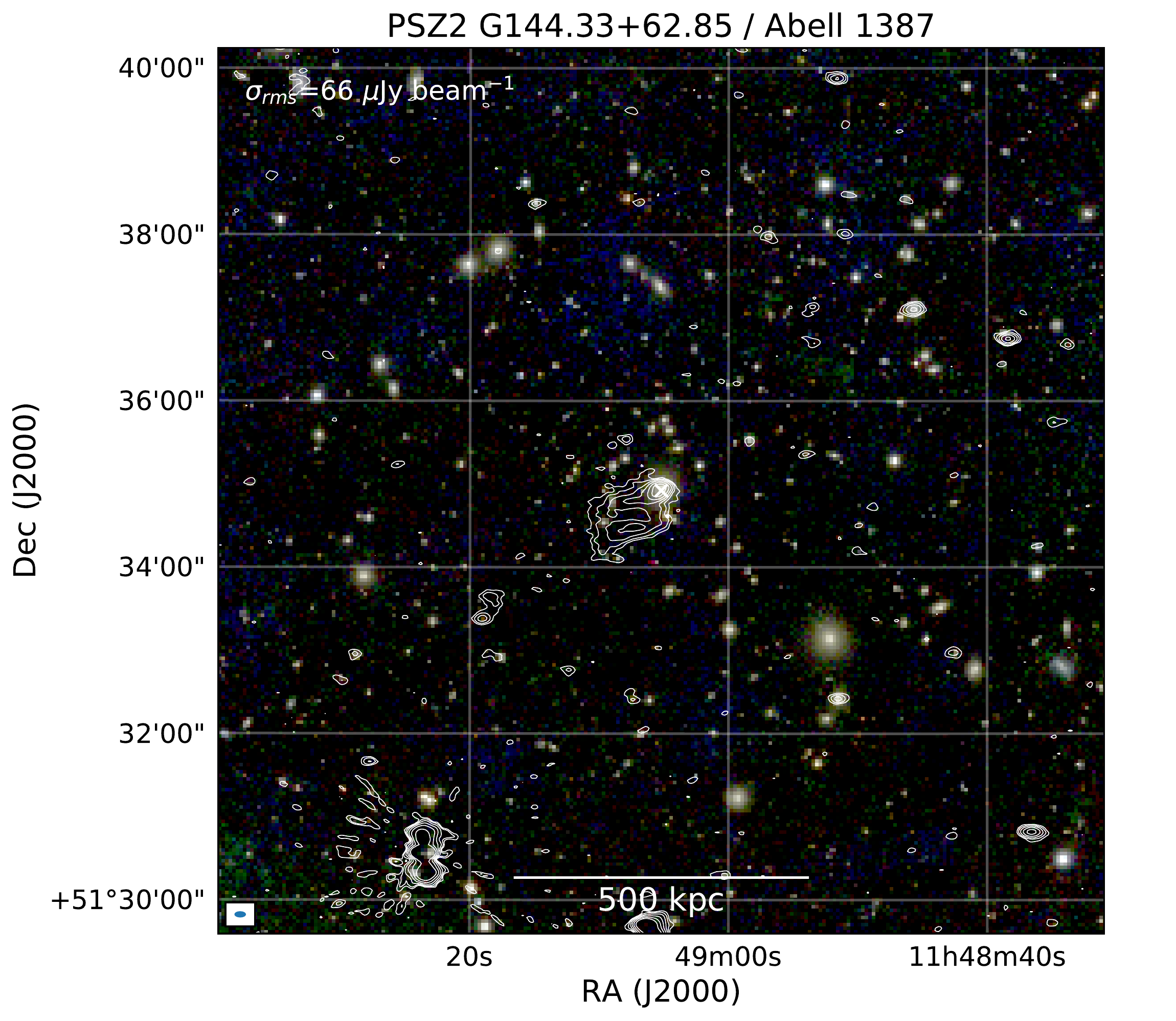}
   \caption{PSZ2\,G144.33+62.85, Abell\,1387. Optical image with Robust $-0.5$ image radio contours. For more details see the caption of Figure~\ref{fig:A2018}.}
   \label{fig:PSZ2G14433}
\end{figure}

\subsection{PSZ2\,G151.62+54.78}
\begin{figure}
\centering
   \includegraphics[width=1.0\columnwidth]{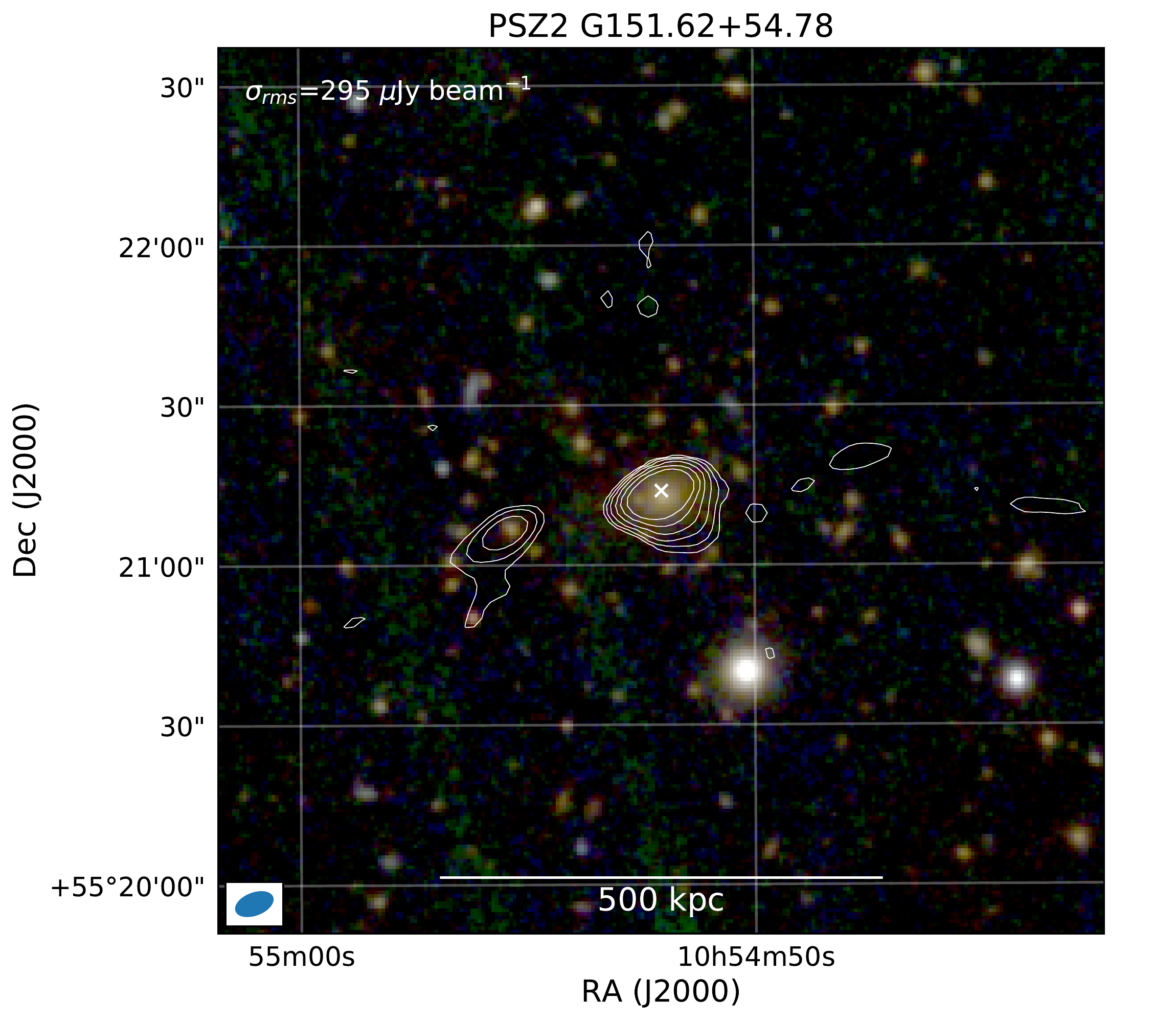}
   \caption{PSZ2\,G151.62+54.78. Optical image with Robust $-0.5$ image radio contours. For more details see the caption of Figure~\ref{fig:A2018}.}
   \label{fig:PSZ2G15162}
\end{figure}

No diffuse emission is detected in this cluster, see Figure~\ref{fig:PSZ2G15162}. However, the observations for this cluster were affected by bad ionospheric conditions, compromising the image quality.

\subsection{Abell\,1615}
\begin{figure}
\centering
   \includegraphics[width=1.0\columnwidth]{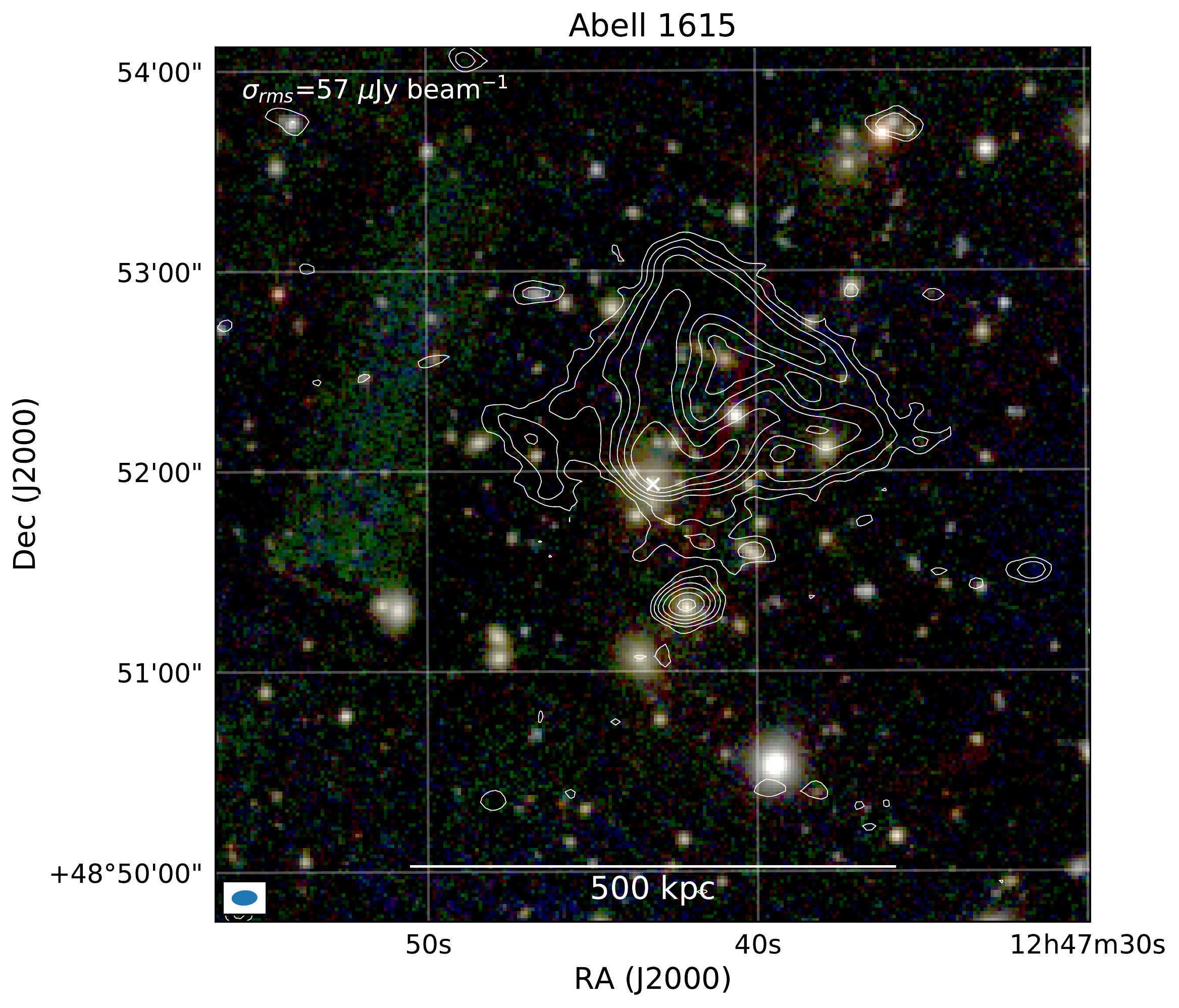}
   \caption{Abell\,1615. Optical image with Robust $-0.5$ image radio contours. For more details see the caption of Figure~\ref{fig:A2018}.}
   \label{fig:A1615}
\end{figure}

This cluster host a complex distorted tailed radio galaxy related to the BCG. This source has a largest extended of 450~kpc. An optical image with radio contours overlaid is shown in Figure~\ref{fig:A1615}.

\subsection{GMBCG\,J211.77332+55.09968}
\label{sec:GMBCGJ211}
\begin{figure}
\centering
   \includegraphics[width=1.0\columnwidth]{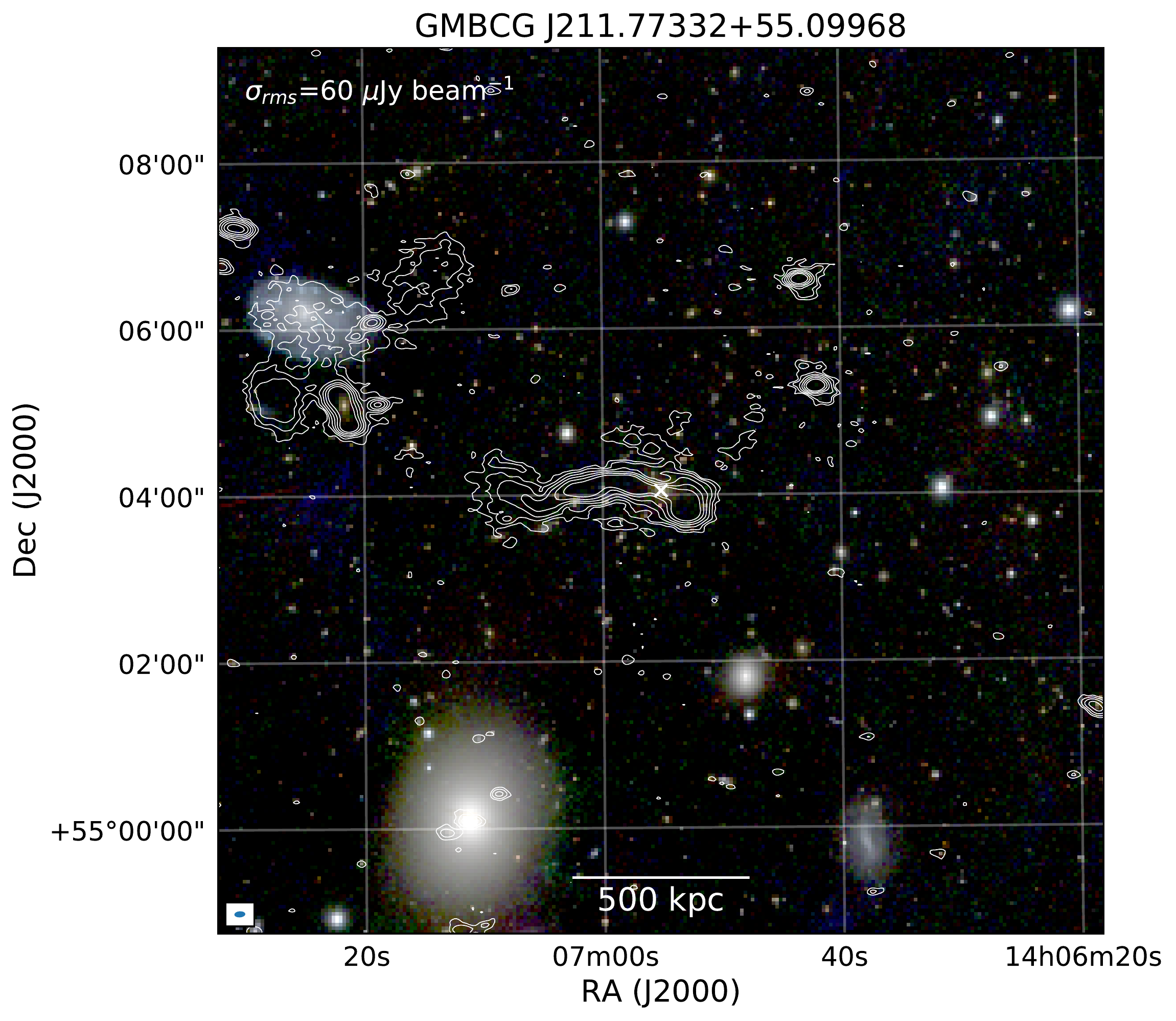}
   \caption{GMBCG\,J211.77332+55.09968.  Optical image with Robust $-0.5$ image radio contours. For more details see the caption of Figure~\ref{fig:A2018}.}
   \label{fig:GMBCGJ211}
\end{figure}

This cluster hosts large 600~kpc asymmetric radio galaxy associated with the BCG. Extended radio emission is also found NW of the BCG, see Figure~\ref{fig:GMBCGJ211}.
This emission is composed of three distinct radio sources, a compact double lobed source, an extended double lobed source, and a foreground spiral galaxy.

\subsection{MaxBCG\,J173.04772+47.81041}
\begin{figure}
\centering
   \includegraphics[width=1.0\columnwidth]{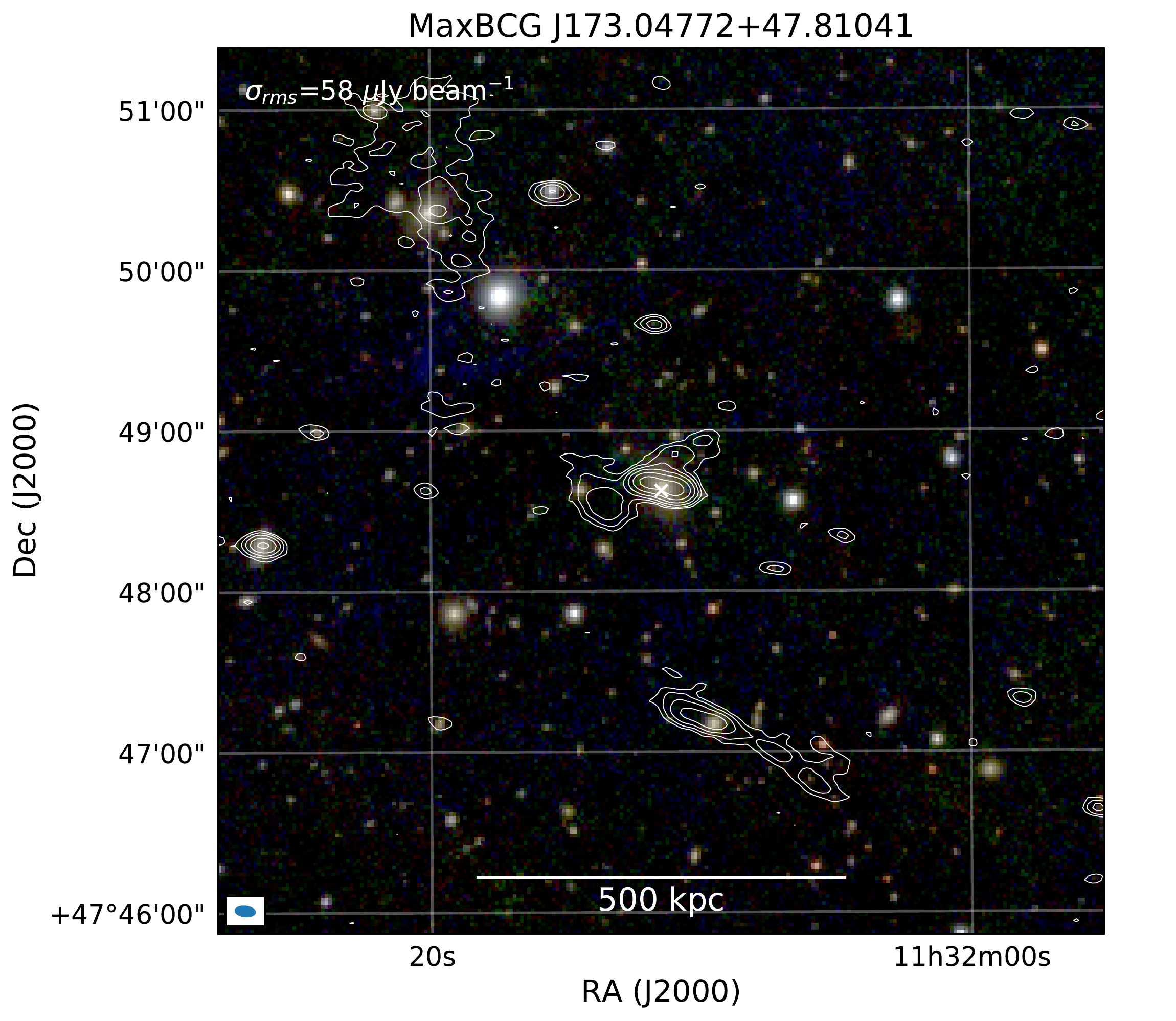}
   \caption{MaxBCG\,J173.04772+47.81041. Optical image with Robust $-0.5$ image radio contours. For more details see the caption of Figure~\ref{fig:A2018}.}
   \label{fig:MaxBCGJ173}
\end{figure}

Extended radio emission, likely related to AGN activity from an elliptical galaxy, is detected in the northern part of the cluster.  An optical image with radio contours overlaid is shown in Figure~\ref{fig:MaxBCGJ173}.

\subsection{WHL\,J132615.8+485229}
\begin{figure*}
\centering
   \includegraphics[width=1.0\columnwidth]{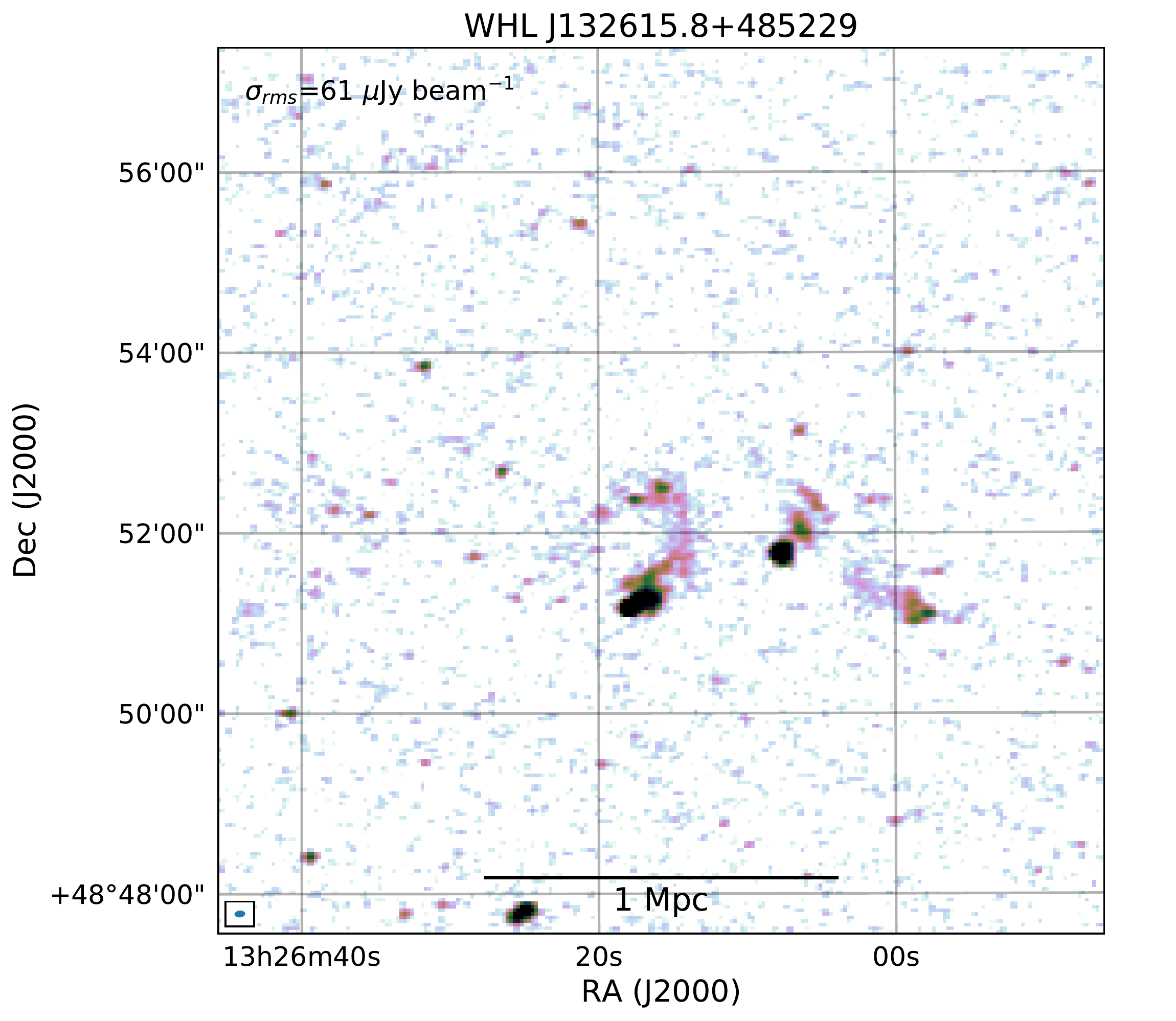}
   \includegraphics[width=1.0\columnwidth]{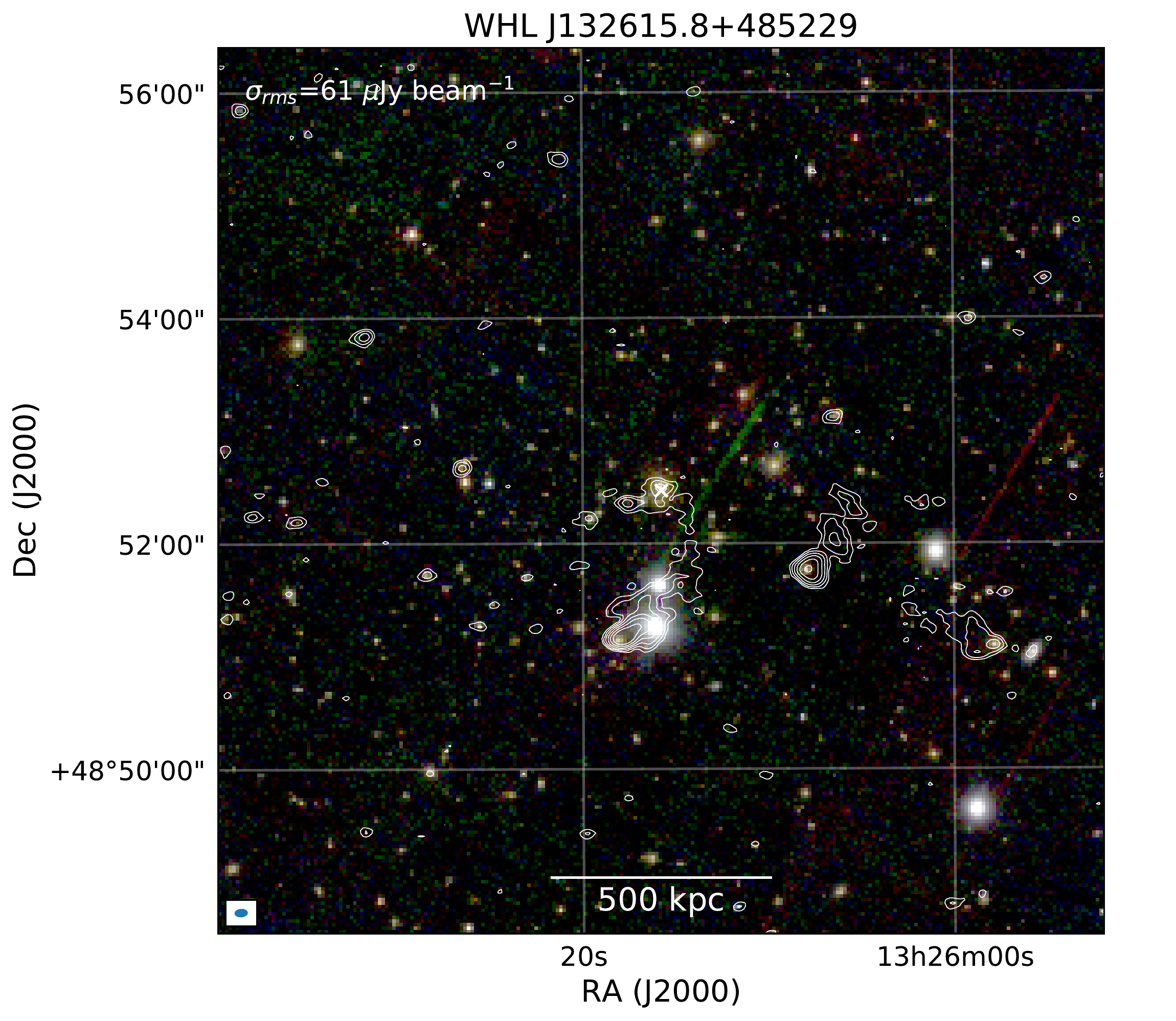}
   \caption{WHL\,J132615.8+485229. Left: Robust $-0.5$ image radio image. Right: Optical image with Robust $-0.5$ image radio contours. For more details see the caption of Figure~\ref{fig:A2018}.}
   \label{fig:WHL1326}
\end{figure*}

Elongated radio emission is detected that seems to originate from a tailed radio galaxy  south of the cluster center (see Figure~\ref{fig:WHL1326}). This emission extends all the way to the BCG.

\subsection{WHL\,J134746.8+475214}
\begin{figure}
\centering
   \includegraphics[width=1.0\columnwidth]{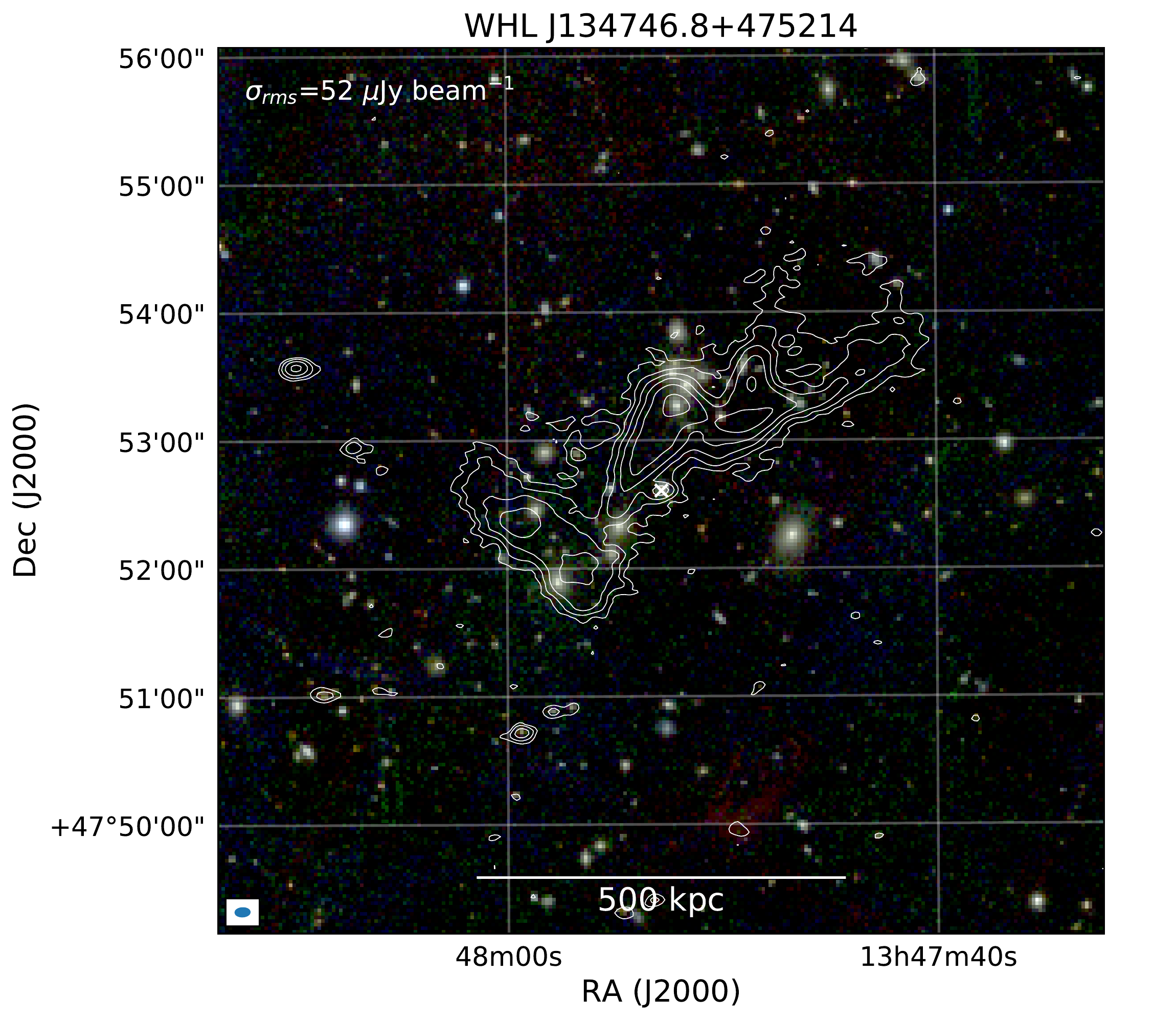}
   \caption{WHL\,J134746.8+475214. Optical image with Robust $-0.5$ image radio contours. For more details see the caption of Figure~\ref{fig:A2018}.}
   \label{fig:WHL1347}
\end{figure}

This cluster host complex extended radio emission on scales of 500~kpc which seems to be related to AGN activity. An optical image with radio contours overlaid is shown in Figure~\ref{fig:WHL1347}.

 \end{appendix}

\end{document}